\documentclass[conference,10pt]{IEEEtran}
\usepackage{mathtools}
\usepackage{amsmath}
\usepackage{algorithm}
\usepackage{environ}
\usepackage[noend]{algpseudocode}
\usepackage{booktabs}
\usepackage{cite}
\usepackage{color}
\usepackage{graphicx}
\usepackage{epstopdf}
\usepackage{epsfig}
\usepackage{ifpdf}
\usepackage{graphicx}
\usepackage{tabularx}
\usepackage{rotating}
\usepackage{hhline}
\usepackage{times}
\usepackage{comment}
\usepackage{amssymb}
\usepackage{amsthm}
\usepackage{pifont}
\usepackage{url}
\usepackage{multirow}
\usepackage{footnote}
\usepackage{hhline}
\usepackage[bookmarks=false]{hyperref}
\usepackage[dvipsnames,table]{xcolor}
\usepackage{xspace} 
\usepackage{setspace}
\usepackage{bm}
\usepackage[export]{adjustbox}
\usepackage{dblfloatfix} 

\ifCLASSOPTIONcompsoc
  \usepackage[caption=false,font=normalsize,labelfont=sf,textfont=sf]{subfig}
\else
  \usepackage[caption=false,font=footnotesize]{subfig}
\fi

\hypersetup{
colorlinks=true, 
linktoc=page,     
linkcolor=black,  
urlcolor=black,
citecolor=black,
}

\usepackage{xcolor}

\newcommand{\ie}{i.e., }
\newcommand{\eg}{e.g., }

\theoremstyle{definition}
\newtheorem{definition}{Definition}

\usepackage{listings}

\definecolor{darkgray}{rgb}{.4,.4,.4}

\def\sys{\textsc{VeriVis}\xspace}

\lstdefinelanguage{CheckerLang}{
  keywords={typeof, new, true, false, catch, function, return, null, catch, switch, var, if, in, while, do, else, case, break, float, int, assert, all},
  keywordstyle=\color{blue}\bfseries,
  ndkeywords={class, export, boolean, throw, implements, import, this, extends, predict, maketform, imtransform},
  ndkeywordstyle=\color{darkgray}\bfseries,
  identifierstyle=\color{black},
  sensitive=false,
  comment=[l]{//},
  morecomment=[s]{/*}{*/},
  commentstyle=\color{purple}\ttfamily,
  stringstyle=\color{red}\ttfamily,
  morestring=[b]",
}

\begin{document}\sloppy
\title{Towards Practical Verification of Machine Learning: \\ The Case of Computer Vision Systems}
\author{Kexin Pei$^*$, Linjie Zhu$^*$, Yinzhi Cao$^\dag$, Junfeng Yang$^*$, Carl Vondrick$^*$, Suman Jana$^*$\\$^*$Columbia University, $^\dag$Johns Hopkins University}
\maketitle
\thispagestyle{plain}
\pagestyle{plain}

\begin{abstract}
Due to the increasing usage of machine learning (ML) techniques in security- and safety-critical domains, such as autonomous systems and medical diagnosis, ensuring correct behavior of ML systems, especially for different corner cases, is of growing importance. In this paper, we propose a generic framework for evaluating security and robustness of ML systems using different real-world safety properties. We further design, implement, and evaluate \sys, a scalable methodology that can verify a diverse set of safety properties for state-of-the-art computer vision systems with only blackbox access. By leveraging different input space reduction techniques, \sys is able to find thousands of safety violations in fifteen state-of-the-art computer vision systems including ten Deep Neural Networks (DNNs) such as Inception-v3 and Nvidia's Dave self-driving system with thousands of neurons as well as five commercial third-party vision APIs including Google vision and Clarifai for twelve different safety properties. Furthermore, \sys can successfully verify these safety properties, on average, for around 31.7\% of the test images. \sys finds up to $64.8\times$ more violations than existing gradient-based methods that, unlike \sys, cannot ensure non-existence of any violations. Finally, we show that retraining using the safety violations detected by \sys can reduce the average number of violations up to 60.2\%.

\end{abstract}

\section{Introduction}
\label{sec:intro}

Recent advances in Machine Learning (ML) techniques like Deep Learning (DL) have resulted in an impressive performance boost for a broad spectrum of complex, real-world tasks including object recognition, image segmentation, and speech recognition. ML systems are increasingly getting deployed in security- and safety-critical domains such as self-driving cars~\cite{bojarski2016end}, automated passenger screening~\cite{dhschallenge}, and medical diagnosis~\cite{gulshan2016development}. Several such systems have already either achieved or surpassed human-level performance on curated test sets.

However, security- and safety-critical systems, besides correctly handling the common cases, must also demonstrate correct behavior for rare corner cases. However, despite their significant progress, machine learning systems often make dangerous and even potentially fatal mistakes. For example, a Tesla autonomous car was recently involved in a fatal crash that resulted from the system's failure to detect a white truck against a bright sky with white clouds~\cite{tesla-accident}. Such incidents demonstrate the need for rigorous testing and verification of ML systems under different settings (\eg different lighting conditions for self-driving cars) to ensure the security and safety of ML systems.

Most existing testing methods for ML systems involve measuring the accuracy and loss using manually-labeled randomly-chosen test samples~\cite{witten2016data}. Unfortunately, similar to traditional software, such random testing approaches are not effective in finding erroneous corner-case behaviors~\cite{pei2017deepxplore,tian2017deeptest}. Moreover, unlike traditional software, the ML decision logic is learned from data and is often opaque even to their designers, which makes the corner-case behaviors more unpredictable than traditional software. Therefore, verifying the security, safety, and reliability of ML systems for different corner cases is critical for wide deployment of ML systems. 

Further, existing approaches for checking the security and robustness of ML systems cannot provide strong guarantees about the absence of different types of erroneous behaviors in real-world-sized ML systems~\cite{pei2017deepxplore, tian2017deeptest, pulina2010abstraction, huang2017safety, katz2017reluplex, szegedy2013intriguing}. Moreover, popular techniques for measuring and improving robustness of ML systems like adversarial input generation operate under very strong attacker models (e.g., the attacker can arbitrarily change all  pixels of an input image) which is unrealistic in many settings~\cite{szegedy2013intriguing, goodfellow2014explaining, nguyen2015deep, liu2016delving, papernot2016limitations, wilber2016can, sharif2016accessorize, papernot2017practical, carlini2017towards}.

In this paper, we propose a generic framework for evaluating the security and robustness of ML systems with different real-world safety properties that are designed to defend against attackers with different capabilities (\eg changing brightness/contrast, rotation, and blurring for vision systems). Note that even though detailed specifications describing an ML system’s internal states are hard to write, safety properties involving input-output behaviors are intuitive and easy to specify. For example, while it is extremely hard (if not impossible) to recreate the logic of a human driver, it is easy to envision safety properties like a self-driving car’s steering angle should not change significantly for the same road under different lighting conditions.

We further design, implement, and evaluate \sys, to the best of our knowledge, the first methodology for verifying a wide range of realistic safety properties (e.g., invariance of a self-driving car’s steering angle under different brightness conditions or rotation invariance for image classifiers) on state-of-the-art computer vision systems. \sys uses a blackbox approach that uses different search space reduction techniques to efficiently search for safety property violations. Our approach is conceptually similar to explicit-state model checking of traditional software. The main benefit of our approach over whitebox approaches is that the verification effort primarily depends on the type of safety property, input size, and input domain but not on the complexity of the ML system itself. 

In this paper, we demonstrate that the search spaces for different real-world safety properties can be reduced significantly for efficient enumeration. The key insight behind our approach is that ML, unlike traditional software, commonly operates on discretized inputs such as images, video, speech (\eg image pixels can only have integer values between 0 and 255). Moreover, most state-of-the-art ML systems operate on relatively small and fixed size inputs for better training and inference performance (\eg $299\times299\times3$ images for Inception-V3~\cite{szegedy2016rethinking}, $640\times480$ images for Google Cloud Vision API~\cite{google-size}, or audio speech input using only one out of three original frames for speech recognition~\cite{sak2015fast}). 

These two properties allow us to significantly reduce the search space for different realistic safety properties. We present a generic decomposition and search space reduction framework for different real-world image transformations used in safety properties (e.g., rotation, brightness change, translation, etc.). Using our framework, we are further able to show that the size of the search spaces for a wide range of real-world safety properties is polynomial in the size of the input demonstrating the scalability of our approach with increasing input size. In this paper, we primarily focus on designing and evaluating \sys to find violations of safety properties in different computer vision systems but the underlying principles are generic and can be applied to other ML domains as well.

We evaluate \sys with twelve different safety properties involving real-world transformations like rotation, contrast, and brightness for fifteen state-of-the-art computer vision systems: (1) six image classification DNNs trained on the imagenet dataset (VGG-16\cite{simonyan2014very}, VGG-19~\cite{simonyan2014very}, MobileNet~\cite{howard2017mobilenets}, Xception~\cite{chollet2016xception}, InceptionV3~\cite{szegedy2016rethinking} and ResNet50~\cite{he2015deep}), which achieved the state-of-the-art performances in ILSVRC~\cite{ILSVRC15} competitions; (2) five commercial third-party vision APIs provided by Google~\cite{google-vision-api}, Clarifai~\cite{clarifai}, IBM~\cite{ibm}, Microsoft~\cite{microsoft}, and Amazon~\cite{amazon}; (3) four popular self-driving car models including Rambo, one of the top performers in Udacity self-driving car challenge, and other three are based on Nvidia Dave self-driving systems. For all of these systems, \sys found thousands of violations for different input images\textemdash$60\times$ more than existing gradient-based methods. For a small number of images (on average 31.7\% of all tested images), \sys was to able to verify that the tested ML systems satisfy the tested safety properties. Our results further demonstrate that by specifically retraining on the violated images generated by \sys, the robustness of the corresponding Ml systems can be significantly improved by reducing up to 60.2\% of the violations.

\begin{figure*}[!t]
\centering
\captionsetup[subfloat]{labelformat=empty, captionskip=-.01cm, justification=centering}

\subfloat[Turn right]{
\includegraphics[width=0.11\textwidth, height=0.11\textwidth]{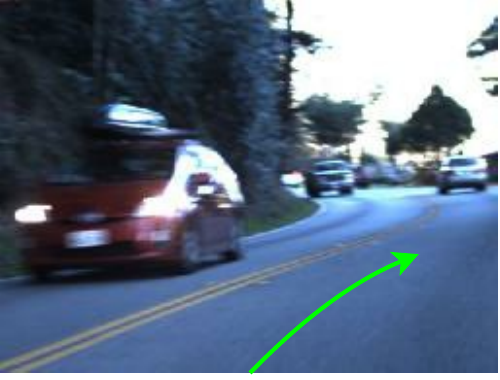}
\label{subfig:blur_orig}}
\subfloat[Amusement ride]{
\includegraphics[width=0.11\textwidth, height=0.11\textwidth]{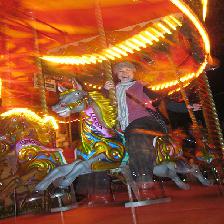}
\label{subfig:rotate_orig}}
\subfloat[Medicine chest]{
\includegraphics[width=0.11\textwidth, height=0.11\textwidth]{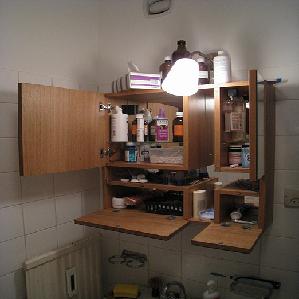}
\label{subfig:occl_orig}}
\subfloat[Moving van]{
\includegraphics[width=0.11\textwidth, height=0.11\textwidth]{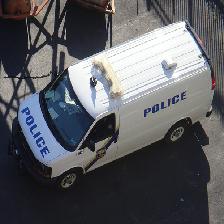}
\label{subfig:shft_orig}}
\subfloat[Soup bowl]{
\includegraphics[width=0.11\textwidth, height=0.11\textwidth]{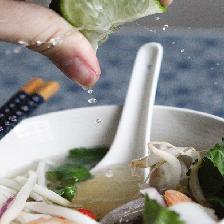}
\label{subfig:illu_orig}}
\subfloat[Bassinet]{
\includegraphics[width=0.11\textwidth, height=0.11\textwidth]{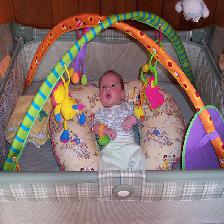}
\label{subfig:contrast_orig}}
\subfloat[Desktop computer]{
\includegraphics[width=0.11\textwidth, height=0.11\textwidth]{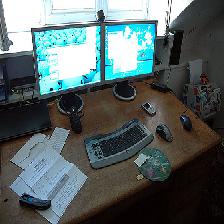}
\label{subfig:shear_orig}}
\subfloat[Tandem bicycle]{
\includegraphics[width=0.11\textwidth, height=0.11\textwidth]{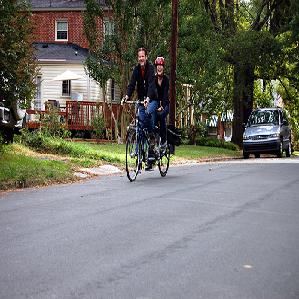}
\label{subfig:scale_orig}}

\vspace{-0.41cm}
\subfloat[Go forward \textbf{Rambo~\cite{rambo}}][Go forward \\ \textbf{Smoothing}\\ \textbf{Rambo~\cite{rambo}}]{
\includegraphics[width=0.11\textwidth, height=0.11\textwidth]{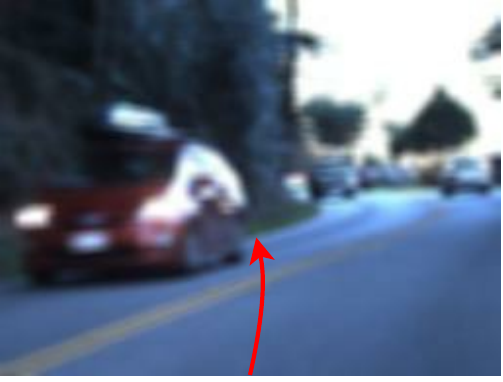}
\label{subfig:blur}}
\subfloat[Organism \textbf{Google API~\cite{google-vision-api}}][Organism \\ \textbf{Rotation} \\ \textbf{Google API~\cite{google-vision-api}}]{
\includegraphics[width=0.11\textwidth, height=0.11\textwidth]{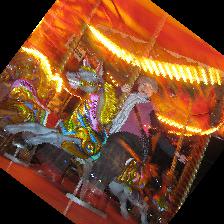}
\label{subfig:rotate}}
\subfloat[Microwave \textbf{Xception~\cite{chollet2016xception}}][Microwave \\ \textbf{Occlusion} \\ \textbf{Xception~\cite{chollet2016xception}}]{
\includegraphics[width=0.11\textwidth, height=0.11\textwidth]{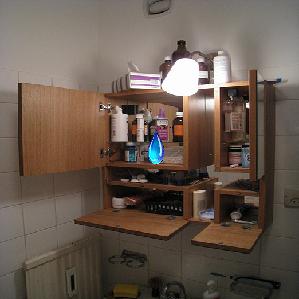}
\label{subfig:occl}}
\subfloat[Space shuttle \textbf{VGG-16~\cite{simonyan2014very}}][Space shuttle \\ \textbf{Translation} \\ \textbf{VGG-16~\cite{simonyan2014very}}]{
\includegraphics[width=0.11\textwidth, height=0.11\textwidth]{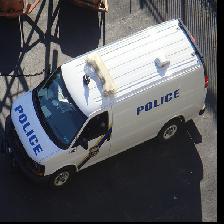}
\label{subfig:shft}}
\subfloat[Ice cream \textbf{VGG-19~\cite{simonyan2014very}}][Ice cream \\ \textbf{Brightness} \\ \textbf{VGG-19~\cite{simonyan2014very}}]{
\includegraphics[width=0.11\textwidth, height=0.11\textwidth]{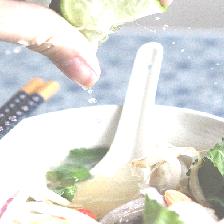}
\label{subfig:illu}}
\subfloat[Jigsaw puzzle \textbf{ResNet-50~\cite{he2015deep}}][Jigsaw puzzle \\ \textbf{Contrast} \\ \textbf{ResNet-50~\cite{he2015deep}}]{
\includegraphics[width=0.11\textwidth, height=0.11\textwidth]{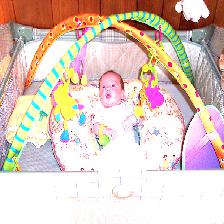}
\label{subfig:contrast}}
\subfloat[Mouse \textbf{MobileNet~\cite{howard2017mobilenets}}][Mouse \\ \textbf{Shear} \\ \textbf{MobileNet~\cite{howard2017mobilenets}}]{
\includegraphics[width=0.11\textwidth, height=0.11\textwidth]{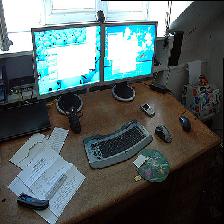}
\label{subfig:shear}}
\subfloat[Horse cart \textbf{InceptionV3~\cite{szegedy2016rethinking}}][Horse cart \\ \textbf{Scale} \\ \textbf{InceptionV3~\cite{szegedy2016rethinking}}]{
\includegraphics[width=0.11\textwidth, height=0.11\textwidth]{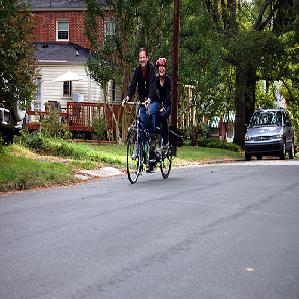}
\label{subfig:scale}}

\caption{Upper row shows the original inputs. Lower row shows the violations generated by different transformations for different computer vision systems as identified by \sys.}
\label{fig:wrong_demonstration}
\end{figure*}

Our main contributions are:

\begin{itemize}
\item We define a general framework for specifying realistic safety properties of ML systems modeling different types of attacker capabilities.
 \item We present a novel decomposition and state space reduction methodology for efficient verification different types of safety properties for computer vision systems.
\item We implement our techniques as part of \sys and use it to verify twelve safety properties on fifteen state-of-the-art vision systems including Imagenet classifiers, self-driving cars, and commercial vision APIs. \sys found thousands of violations in all tested systems (see Figure~\ref{fig:wrong_demonstration} for some samples). 
\item The number of safety violations of the computer vision systems can be reduced by up to 60.2\% by retraining them with the violations detected by \sys.
\end{itemize}








The rest of the paper is organized as follows. We define different types of ML safety properties in Section~\ref{sec:framework}. We provide an overview of our methodology  in Section~\ref{sec:overview}. The details of how \sys verifies different safety properties and complexity analysis of the corresponding search spaces are described in Section~\ref{sec:reduction}. Section~\ref{sec:impl} outlines our experiment setup and Section~\ref{sec:eval} presents a detailed evaluation of \sys. We discuss the related work in Section~\ref{sec:related} and conclude the paper in Section~\ref{sec:conclusion}. 

\section{A Framework for Verifying ML Robustness and Security}
\label{sec:framework}
In this section, we present a general framework for verifying the robustness and security of ML systems against attackers with different capabilities. First, we describe the differences between verifying traditional software and ML systems. Next, we compare and contrast the safety properties of ML systems with those of traditional software. Finally, we present formal definitions and a taxonomy of different ML safety properties. 

\vspace{.1cm}\noindent{\bf Verifying traditional software vs. ML.} A principled way of verifying a program against different kinds of attacks is to check the behavior of the program against a complete set of functional specifications and ensure that the program always behaves according to these specifications. For most traditional software, writing complete specifications, while effort intensive, is conceptually feasible as the developers write the code themselves and therefore have detailed knowledge about the internal states and the desired behavior of the software under different settings. For example, a file format parser is supposed to accept all inputs according to the rules of the corresponding grammar or a protocol implementation should follow the protocol state machine. 

By contrast, the logic of an ML system is automatically learned from data and is opaque even to the designers of the ML system. Therefore, writing complete specifications of ML systems is not feasible as the rules and internal states involved are too complex for a human to enumerate. For example, creating a complete specification for the correct behavior of a self-driving car under different driving conditions essentially involves recreating the logic of a human driver, which is computationally infeasible and not practical.

\vspace{.1cm}\noindent{\bf Safety properties.} An alternative way of ensuring software security and robustness is to verify a software against a set of safety properties describing unwanted behaviors that the software should never exhibit, \ie any violation of the safety properties will indicate a security/robustness issue. 

This approach has several advantages over checking against complete functional specifications. First, safety properties require less manual effort to create than complete specifications. Second, unlike functional specifications, even verification of an incomplete set of safety properties can potentially provide meaningful security and robustness guarantees under different threat models. 

\vspace{.1cm}\noindent{\bf ML safety properties.} Safety property of traditional software is usually expressed in terms of unsafe program states (e.g., the length of a string cannot be negative). This is not feasible for ML systems as the states are opaque even to their designers. However, unlike traditional software, ML systems usually take fixed size discrete input and produces output within a finite range. For example, the state-of-the-art Xception~\cite{chollet2016xception} and inception-v3~\cite{szegedy2016rethinking} object detection systems take $299\times 299$ images as input. ML system designers can specify safety properties involving certain groups of inputs and how the corresponding outputs will be related. Therefore, the safety property for ML systems will operate in the input-output space of the system. Note that for traditional software input-output-based safety specifications are often not very practical due to the large and potentially unbounded input space (e.g., strings of arbitrary length).

Input-output based safety properties are applicable to a wide range of ML systems.  For example, consider a safety property specifying that a self-driving car's steering angle should remain similar for the same road under different lighting conditions. In this setting, even though it is hard for an ML developer to predict the safe steering angles for the self-driving car under different scenarios, it is easy to specify such safety properties. Similarly, a safety property for a speech recognition system can ensure that the recognized phrases/sentences will not change under different background noises. For a machine translation system, the translated output should not change significantly if the input words are substituted with other synonyms. Malware detection systems should not change their classifications from malware to benign due to different types of code obfuscation/transformation techniques that do not affect malicious functionality~\cite{xu2016automatically}.

Moreover, different ML safety properties can be designed to model attackers with different capabilities. For example, a safety property can check that an attacker that can occlude a small part of an input image cannot cause a self-driving car to change its steering angle significantly. Similarly, another safety property might ensure that an attacker who can change the lighting of an input image cannot cause a crash. 

A key difference between verification of ML and traditional safety properties is that the problem domains of ML systems, unlike that of traditional software, tend to be ambiguous and therefore can easily tolerate minor deviations in the results. For example, a car can be safely driven on the road with many slightly different but similar steering angles. Similarly, some errors in image classification are less serious than others. Consider the case where an image of an elephant gets incorrectly classified as an African elephant instead an Asian elephant. In many settings, such errors may not be critical. Therefore, unlike traditional software, safety properties for ML system must have more flexibility built into them. We provide a taxonomy and formal definitions of such ML safety properties that can specify ambiguity below. 

\vspace{.1cm}\noindent{\bf A taxonomy of ML safety properties.} An ML model can be thought of as a function $f$ mapping input $\bm{x}\in\mathbb{X}$ to output $\bm{y}\in\mathbb{Y}$, \ie $f: \mathbb{X}\rightarrow \mathbb{Y}$. 
Depending on the type of task (\ie classification or regression), the ML model produces either continuous or discrete output $\mathbb{Y}$. 
For classification tasks like object recognition, $\mathbb{Y}$ is a set of discrete labels. 
By contrast, $\mathbb{Y}$ is a continuous range for regression tasks like driving an autonomous vehicle that outputs steering angles. 

The safety properties of a ML system based on its input-output behaviors can be defined as follows. Consider a transformation function $\mathcal{T}(\cdot;c)$ parameterized by $c \in \mathbb{C}$ ($\mathbb{C}$ is transformation parameter space) that transforms an input $\bm{x}$ to $\bm{x'}$ and sends it to the underlying ML model $f$. 

Below, we define two types of general safety properties for classification and regression tasks respectively. 

{\it \underline{Locally $k$-safe classification}.} Given a classification model $f$, an input $\bm{x}$, a parameterized transformation function $\mathcal{T}(\cdot;c)$ where $c \in \mathbb{C}$, we define that $f$ is $k$-safe with respect to $\mathcal{T}(\cdot;c)$, $\mathbb{C}$, and $\bm{x}$ if and only if $\forall c\in\mathbb{C}$, $f(\mathcal{T}(\bm{x};c),1)\subseteq f(\bm{x},k)$ where $f(\bm{x},k)$ denotes the top-k prediction by $f$ for $\bm{x}$. 

{\it \underline{Globally $k$-safe classification}.} A classification model $f$ is considered globally k-safe if $\forall \bm{x} \in \mathbb{X}$, it is locally k-safe.

{\it \underline{Locally $t$-safe regression}.} Given a regression DNN $f$, an input $\bm{x}$, a parameterized transformation function $\mathcal{T}(\cdot;c)$ where $c \in \mathbb{C}$, we define that $f$ is locally $t$-safe with respect to $\mathcal{T}(\cdot;c)$, $\mathbb{C}$, and $\bm{x}$ if and only if $\forall c\in\mathbb{C}$, $|f(\mathcal{T}(\bm{x};c))-f(\bm{x})|\leq t$.

{\it \underline{Globally $t$-safe regression}.} A regression model $f$ is considered globally k-safe if $\forall \bm{x}\in\mathbb{X}$, it is locally t-safe.

Note that robustness of ML models to adversarial inputs~\cite{goodfellow2014explaining} can also be easily expressed as safety properties where the transformation function $\mathcal{T}(\bm{x};\bm{c})=\bm{x}+\bm{c}$. In this case, the transformation parameter $\bm{c}$ is an attacker-crafted perturbation (a tensor variable that have the same size with the input $\bm{x}$). The transformation parameter space includes all $\bm{c}$ such that different types of norms of $\bm{c}$ is bounded by a user-defined budget $\sigma$, \eg $\lVert c \rVert_0<\sigma$~\cite{papernot2016limitations}, $\lVert c \rVert_1<\sigma$~\cite{xu2016automatically}, $\lVert c \rVert_2<\sigma$~\cite{szegedy2013intriguing}, or $\lVert c \rVert_{\infty}<\sigma$~\cite{goodfellow2014explaining,xu2016automatically}. The search space of the adversarial perturbations are significantly larger than real-world transformations like rotation, changing brightness, etc. In this paper, we primarily focus on verifying local safety-critical properties involving real-world transformations~\cite{pei2017deepxplore, sharif2016accessorize}.

\vspace{.1cm}\noindent{\bf Verification of ML safety properties.} 
Verifying that an ML system satisfies a particular safety property $\phi$ is essentially a search problem over the parameter space $\mathbb{C}$ of a transformation $\mathcal{T}(\cdot;c)$ (\ie searching all possible values of $c\in \mathbb{C}$) for violations of the safety property. The size of the search space for a given ML safety property varies widely based on the property and input size/domain. For example, a local safety property involving the image inversion operation for image classification systems will have a search space of only one transformed image for any given image. By contrast, a transformation function that simulates different lighting conditions along with the different shadows for an object in a given image will have a significantly larger search space~\cite{belhumeur1998set, zhang2013toward}. In many cases, such large search spaces may only be checked probabilistically for safety property violations.

The verification techniques for checking ML safety properties can use either static, symbolic, or dynamic approach. However, existing static/symbolic techniques for traditional software does not scale well for ML systems due to the highly non-linear nature of the ML systems. In this paper, we primarily focus on dynamic and blackbox verification techniques for ML systems. However, our approach can be easily augmented with whitebox approaches, if needed, to further reduce the search space.

\section{Methodology}
\label{sec:overview}

\begin{figure}[!t]
\centering
\includegraphics[width=0.9\columnwidth]{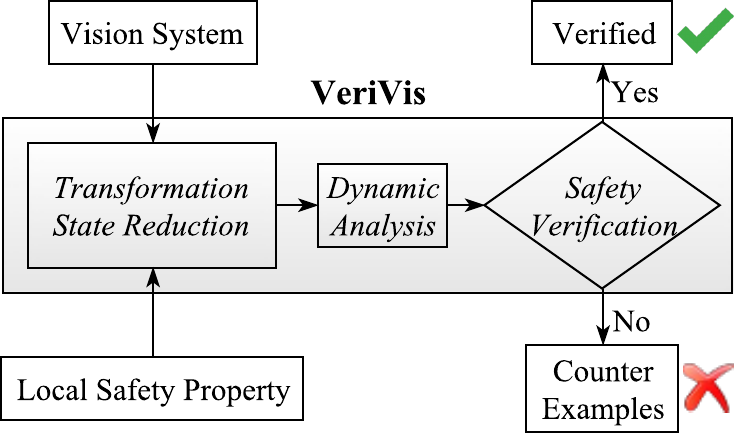}
\caption{Overall verification architecture of \sys.}
\label{fig:procedure}
\end{figure}

\subsection{\sys overview}
In this paper, we design, implement, and evaluate \sys, an instance of the verification framework described in Section~\ref{sec:framework} specifically tailored for checking different local safety properties of computer vision systems such as self-driving cars and image classifiers. Figure~\ref{fig:procedure} shows the high-level structure of \sys. \sys takes a computer vision system (\eg self-driving car or image classifier) along with a local safety property as inputs and either verifies that the computer vision system satisfies the safety property or produces a counter-example violating the safety property. Even though we primarily focus on verifying computer vision systems in this paper, the underlying principles are applicable to other types of ML systems operating on discrete and bounded input domains as well (e.g., malware detection). 

\sys uses blackbox dynamic analysis for verifying safety properties of computer vision systems without knowing its internal details. For example, \sys can even verify cloud vision APIs provided by companies like Google or Microsoft without any information about their internals (see Section~\ref{sec:eval} for details). Specifically, \sys applies different input space reduction techniques to skip unrealizable inputs and efficiently verify a computer vision system by only checking for the inputs that can be feasibly generated by the image transformation specified in the safety property. Such input space reduction is possible because even though the image transformation parameters are continuous (e.g., floating point values), their output domains (e.g., the pixel values) are discrete. Therefore, multiple different values of a transformation parameter may lead to the same image and therefore can be safely skipped without affecting the verification guarantees.  For example, while the degree of image rotation can be a floating point value, the pixel values of the output image are limited and discrete (e.g., integers ranging between $0$ and $255$). 

In general, we show that the constraints of a wide range of realistic image transformations (\eg rotation, changing lighting condition) imply that the space of unique images that can be generated by these transformations is polynomial in the input image size for a fixed parameter range. Therefore, \sys can verify a broad spectrum of safety properties while scaling up to the largest state-of-the-art computer vision systems. \sys's approach is conceptually similar to explicit-state model checking~\cite{holzmann1997model} for traditional software where feasible and unique states of a traditional program are explicitly enumerated for safety property violations.
 
\subsection{Reducing input spaces of image transformations}
\label{subsec:reduce_output}

Each safety property supported by \sys has a corresponding parameterized image transformation associated with the property. Similar to the transformations described in Section~\ref{sec:framework}, a parameterized image transformation takes an input image $\bm{I}$, a parameter value $c$, and produces an output image $O=\mathcal{T}(\bm{I};c)$. Essentially, an image transformation computes the value of each output pixel based on the values of the input pixels and the parameter. In theory, for an arbitrary image transformation, each output pixel may depend on all of the input pixels. However, in practice, an output pixel's value usually only depends on a small number of neighboring pixels in the input image for most image transformations designed to generate realistic images (\eg rotation, changing brightness/contrast, erosion). This property allows us to drastically reduce the search space for safety property verification. Note that this is an inherent property of most image transformations designed to produce realistic images as most physically realizable changes (\ie changes in camera orientation, lighting conditions, etc.) tend to satisfy this property.

\vspace{.1cm}\noindent{\bf Decomposition framework.} Before describing the details of the input space reduction process, we first define a generic decomposition framework for reasoning about the space of all distinct output images for a given parameterized image transformation and a parameter space. Our framework relies on the fact that most image processing operations can be decomposed into a multi-step pipeline of stencil operations, where each point in the two-dimensional space is updated based on weighted contributions from a subset of its neighbors~\cite{Ragan-Kelley:2013:HLC:2491956.2462176}. Our framework decomposes parameterized image transformations into a sequence of parameterized stencil operations for efficient search space reduction. Even though decomposition of image processing code into stencil computations have been used to optimize the performance of the code by Ragan-Kelley et al.~\cite{Ragan-Kelley:2013:HLC:2491956.2462176}, to the best of our knowledge, we are the first ones to reduce the output space of a parameterized image transform using such techniques.

\begin{figure}[!t]
\centering
\includegraphics[width=0.9\columnwidth]{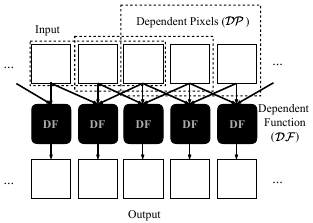}
    \caption{Decomposition of a simple one-dimensional transformation (e.g., one-dimensional blurring) into Dependent Pixels ($\mathcal{DP}$) and Dependence Function ($\mathcal{DF}$). White squares represent the pixels.}
\label{fig:decompose}
\end{figure}

Specifically, we describe two pixel-specific parameterized functions that can express the relations between the input and output pixels of different realistic image transformations: \textit{Dependent Pixels} ($O_{DP}=\mathcal{DP}(\bm{I}, \langle i, j \rangle;c)$) and \textit{Dependence Function} ($O_{DF}=\mathcal{DF}(\bm{I}(O_{DP});c)$) where $\bm{I}(O_{DP})$ denotes the pixel values of coordinates output by $\mathcal{DP}$ in image $\bm{I}$. For a given input image $\bm{I}$ and a coordinate $\langle i,j \rangle$, dependent pixels ($\mathcal{DP}$) return a list of pixel coordinates in the output image whose values are dependent on the input pixel's value for a given parameter value $c$. Dependent function ($\mathcal{DF}$) takes the pixel values of input image $\bm{I}$ on coordinates $O_{DP}$ as input and computes the value of the corresponding output pixels based on the input pixel values for a given parameter value $c$. Essentially, all possible output images for an image transformation $\mathcal{T}(\bm{I};c)$ of an image $\bm{I}$ of size $W\times H$ and $c\in\mathbb{C_{\phi}}$ can be enumerated by evaluating Equation~\ref{eqnoutspace} ($\mathbb{C}_{\phi}$ represents the user-specified parameter space of a transformation for safety property $\phi$). 

\begin{equation}
\label{eqnoutspace}
\bigcup\limits_{\forall c\in\mathbb{C}_{\phi}}\bigcup\limits_{i=0}^{W-1}\bigcup\limits_{j=0}^{H-1} \mathcal{DF}(\bm{I}(\mathcal{DP}(\bm{I}, \langle i,j \rangle;c));c))
\end{equation}

Figure~\ref{fig:decompose} shows a sample decomposition of a simple one-dimensional transformation into $\mathcal{DP}$ and $\mathcal{DF}$. In this case, an output pixel always depends only on three neighboring input pixels. However, for an arbitrary image transformation, $\mathcal{DP}$, for any given input pixel, might potentially produce all output pixels, i.e., any output pixel might be dependent on all input pixels. Similarly, $\mathcal{DF}$ can also produce all possible values (\eg $0$ to $255$) for each output pixel. Therefore, in the worst case, the number of all unique output images that can be generated by an arbitrary transformation can be $256^{W\times H}$ for a gray-scale input image with size $W\times H$. Even for simple networks designed to work on small images like the MNIST hand-written digits each with $28\times28$ pixels, the resulting number of output images ($256^{28\times 28}$) will be too large to enumerate exhaustively. 

However, for most realistic image transformations, $\mathcal{DP}$ produces only a few dependent output pixels for each input pixel and $\mathcal{DF}$ outputs only a subset of all possible output pixel values (e.g., $0-255$). For a wide range of real-world transforms, we demonstrate that the number of unique output images for a given input image and parameter range is polynomial in the size of the image as shown in Section~\ref{sec:reduction} and Table~\ref{tab:bounds}. \sys enumerates these output images efficiently and therefore is able to verify a wide range of safety properties in a scalable manner.  

To understand how $\mathcal{DP}$ and $\mathcal{DF}$ look for realistic image transformations, consider a simple parameterized brightness transformation that brightens an image by adding a constant parameter value to all pixel values. For such a transformation, the output pixel coordinate of $\mathcal{DP}$ will be same as the input pixel coordinate as an output pixel's value only depends on the input pixel's value with the same coordinate. Similarly, $\mathcal{DF}$ will simply add the parameter value with the current value of the input pixel. 

\vspace{.1cm}\noindent{\bf Reducing the parameter space.}
As most the parameters of most image transformations are continuous floating point values, enumerating all possible parameter values in a given parameter space is not feasible.  For example, a safety property might specify that an image can be rotated to any arbitrary angle between $-10^\circ$ and $10^\circ$. Enumerating all possible rotation angles within the specified parameter range is too slow to be practical. 

The key insight behind our input space reduction technique is that the output of $\mathcal{DP}$ and $\mathcal{DF}$ both are discrete, bounded integers (\eg $\mathcal{DP}$ must return valid pixel coordinates and $\mathcal{DF}$ must return valid pixel values) even though the parameter values are continuous floating point numbers. The discreteness of the output allows us to only enumerate a small finite set $\mathbb{C}_{critical}\in\mathbb{C}$ of {\emph critical parameter values} that can cover all possible unique output images.  We formally define the critical parameter values below.

\begin{definition}
\label{def:critical_states} 
Critical parameter values in a parameter space $\mathbb{C}_{\phi}$ and transformation $\mathcal{T}$ in safety property $\phi$ is a monotonic increasing finite sequence $(c_i\in\mathbb{C}_{\phi})_{i=1}^n$, \ie $\forall i\in\{1,2,...,n-1\}$, $c_{i+1}>c_i$, where the following holds. 
$\forall c_i, c_{i+1}$ where $i\in\{0,...,n-1\}$, $\forall c\in\mathbb{C}_{\phi}$ where $c_i<c<c_{i+1}$, $\mathcal{T}(\bm{I};c) = \mathcal{T}(\bm{I};c_i)$ or $\mathcal{T}(\bm{I};c) = \mathcal{T}(\bm{I};c_{i+1})$.
\end{definition}

Definition~\ref{def:critical_states} ensures that for any continuous $c$ between two critical parameter values, the transformation $\mathcal{T}$ specified in a safety property $\phi$ will not generate a new output other than those already generated by the immediately preceding and following critical parameter values. For example, consider the image translation operation where shift step is the parameter. Translation shifts an image to the desired direction by the desired amount. If the shift step is set to any floating point number, the translated coordinates will be rounded (up) to the nearest integer values. Therefore, all $\langle i,j \rangle$ pairs where $0\le i \le W-1$ and $0 \le j \le H-1$ will be critical parameter values for the translation operation. 

\sys explicitly enumerates critical parameter values $\mathbb{C}_{critical}=(c_i)_{i=1}^n$ ignoring the continuous values that lie between the critical parameter values. 
Algorithm~\ref{alg:findcs} shows the detailed procedure of finding the critical parameter values for $\mathcal{DP}$ and $\mathcal{DF}$ for a given transformation $\mathcal{T}$ and parameter space $\mathbb{C}_\phi$. $\mathcal{DP}^{-1}$ and $\mathcal{DF}^{-1}$ indicate inverses of $\mathcal{DP}$ and $\mathcal{DF}$ respectively, \ie they compute $c$ given input and output pixels. However, as discussed in Section~\ref{subsec:reduce_output}, more than one values of $c$ can often  map to same $O_{DP}$ or $O_{DF}$. Therefore, we assume that $\mathcal{DP}^{-1}$ and $\mathcal{DF}^{-1}$ randomly samples one value from the set of all candidates for $c$.

Note that the critical parameter values for more complicated image transformations like rotation, unlike translation, may not be equidistant from each other. We describe how $\mathcal{DP}$, $\mathcal{DF}$, $\mathcal{DP}^{-1}$ and $\mathcal{DF}^{-1}$ are computed for different transformations in detail in Section~\ref{sec:reduction}. 

\begin{algorithm}
\footnotesize
\renewcommand{\arraystretch}{.9}
\setlength{\tabcolsep}{2pt}
	\caption{\footnotesize Algorithm for computing critical parameter values in parameter space $\mathbb{C}_{\phi}$ for transformation $\mathcal{T}$ in property $\phi$.}
\label{alg:findcs}
\begin{tabular}{lp{2.9in}}
\textbf{Input}:
    & $\mathbb{C}_{\phi}$
\end{tabular}

\begin{spacing}{1.1}
\begin{algorithmic}[1]

\State $\mathbb{C}_{critical}\leftarrow \{\}$
	\For{all coordinates $(i,j)$ in $\bm{I}$}
	\For{all feasible $O_{DP}$ $\in$ $\mathcal{DP}(\bm{I},\langle i,j \rangle;\mathbb{C}_{\phi})$}
    	\State $c = \mathcal{DP}^{-1}(\bm{I},\langle i,j\rangle;O_{DP})$
		\If{$c\in \mathbb{C}_{\phi}$}
    		\State $\mathbb{C}_{critical}=\mathbb{C}_{critical}\cup c$
        \EndIf
        \For{all feasible $O_{DF}$ $\in$ $\mathcal{DF}(\bm{I}(O_{DP});c)$}
            \State $c = \mathcal{DF}^{-1}(\bm{I}(O_{DP});O_{DF})$
            \If{$c\in\mathbb{C}_{\phi}$}
                \State $\mathbb{C}_{critical}=\mathbb{C}_{critical}\cup c$
            \EndIf
        \EndFor
    \EndFor
\EndFor

\State return sorted($\mathbb{C}_{critical}$)

\end{algorithmic}
\end{spacing}
\end{algorithm}

\section{Decomposition \& Analysis of Real-world Image Transformations}
\label{sec:reduction}

In this section, we describe how \sys supports verification of a wide range of safety properties with different real-world image transformations. Specifically, we describe twelve different image transformations corresponding to twelve safety properties ($\phi_1,...,\phi_{12}$) summarized in Table~\ref{tab:properties}. These image transformations and their compositions can simulate a wide range of real-world distortions, noises, and deformations that most security-critical vision systems must handle correctly. The transformation parameters for each transformation are shown in the third column of Table~\ref{tab:properties}. 

These transformations can be broadly categorized into three groups: convolutions, point transformations, and geometric transformations. Convolution-based transformations like blurring (e.g., $\phi_1$ to $\phi_4$) applies a convolution kernel on the input image and produce the output images such that each pixel value is determined by its local neighbors and the corresponding kernel weights.  By contrast, for point transformations (e.g., $\phi_5$ and $\phi_6$), each pixel's new value is only decided by its original value in the input. Finally, geometric transformations (e.g., $\phi_7$ to $\phi_{12}$) shuffles the pixel values based on different geometric constraints.

As described in Section~\ref{sec:overview}, \sys reduces the input space of these twelve transformations by decomposing them into $\mathcal{DP}$ and $\mathcal{DF}$ as defined in Equation~\ref{eqnoutspace} and finding the corresponding critical parameter values using Algorithm~\ref{alg:findcs}. We describe the decomposition process for each transformation in detail below. We also perform the verification complexity analysis for each transformation and demonstrate that safety properties related to all of these transformations can be verified in polynomial time with respect to the input image size as shown in Table~\ref{tab:complexity}.



\begin{table}[!t]
\setlength{\tabcolsep}{2.5pt}
\footnotesize
\centering
\renewcommand{\arraystretch}{1}
\caption{A list of safety properties and corresponding transformations that can simulate a wide range of common real-world image distortions and deformations.
}
\label{tab:properties}
\begin{tabular}{c|c|c|c}\hline
\textbf{Property} & \textbf{Transformation} & \textbf{Parameters} & \textbf{Example} \\ \hhline{=|=|=|=}
$\bm{\phi_1}$ & Average smoothing & Kernel size & \includegraphics[width=0.3\columnwidth,valign=c]{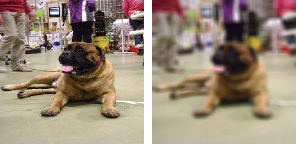} \\
$\bm{\phi_2}$ & Median smoothing & Kernel size & \includegraphics[width=0.3\columnwidth,valign=c]{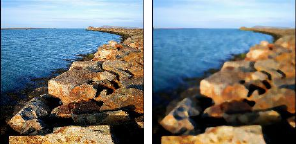} \\
$\bm{\phi_3}$ & Erosion & Kernel size & \includegraphics[width=0.3\columnwidth,valign=c]{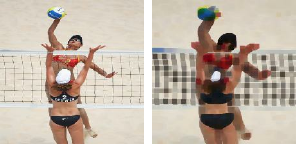} \\
$\bm{\phi_4}$ & Dilation & Kernel size & \includegraphics[width=0.3\columnwidth,valign=c]{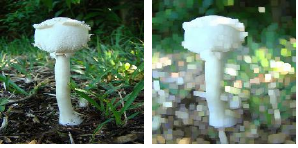} \\
$\bm{\phi_5}$ & Contrast & Gain & \includegraphics[width=0.3\columnwidth,valign=c]{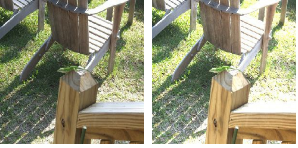} \\
$\bm{\phi_6}$ & Brightness & Bias & \includegraphics[width=0.3\columnwidth,valign=c]{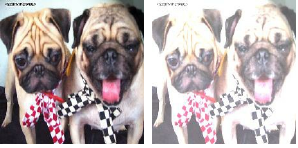} \\
$\bm{\phi_7}$ & Occlusion & Coordinate & \includegraphics[width=0.3\columnwidth,valign=c]{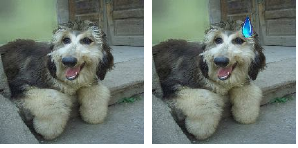} \\
$\bm{\phi_8}$ & Rotation & Rotation angle & \includegraphics[width=0.15\columnwidth,valign=c]{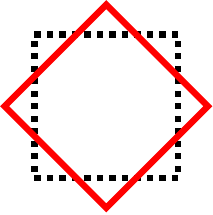} \\
$\bm{\phi_9}$ & Shear & Proportion & \includegraphics[width=0.15\columnwidth,valign=c]{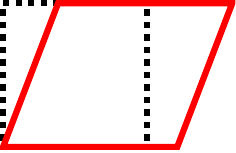} \\
$\bm{\phi_{10}}$ & Scale & Scalar & \includegraphics[width=0.15\columnwidth,valign=c]{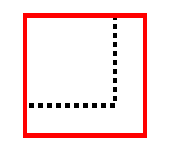} \\
$\bm{\phi_{11}}$ & Translation & Shift step & \includegraphics[width=0.15\columnwidth,valign=c]{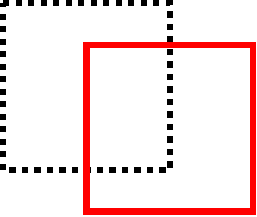} \\ 
$\bm{\phi_{12}}$ & Reflection & Direction & \includegraphics[width=0.2\columnwidth,valign=c]{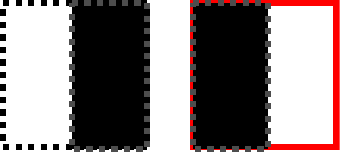} \\ \hline
\end{tabular}
\end{table}

\begin{table}[!t]
\setlength{\tabcolsep}{3pt}
\footnotesize
\centering
\renewcommand{\arraystretch}{1.1}
\caption{Verification complexity of different transformations with respect to the input size $n$.}
\label{tab:complexity}
\begin{tabular}{|c|c||c|c|}
\hline
Transformations & Complexity & Transformations & Complexity \\ \hline 
Avg. smoothing ($\phi_1$) & $O(n)$ & Med. smoothing ($\phi_2$) & $O(n)$ \\ \hline
Erosion ($\phi_3$) & $O(n)$ & Dilation ($\phi_4$) & $O(n)$ \\ \hline
Contrast ($\phi_5$) & $O(1)$ & Brightness ($\phi_6$) & $O(1)$ \\ \hline
Occlusion ($\phi_7$) & $O(n)$ & Rotation ($\phi_8$) & $O(n^2)$ \\ \hline
Shear ($\phi_9$) & $O(n^3)$ & Scale ($\phi_{10}$) & $O(n^2)$ \\ \hline
Translation ($\phi_{11}$) & $O(n)$ & Reflection ($\phi_{12}$) & $O(1)$ \\ \hline
\end{tabular}
\end{table}

\subsection{Convolutions}

\vspace{.1cm}\noindent\textbf{Decomposition.}
For all convolution-based transformations with a kernel of size $c$, $\mathcal{DP}(\bm{I},\langle i,j \rangle;c) = \{\langle k,l \rangle  |  i-c/2<k<i+c/2$ $and$ $j-c/2<l<j+c/2\}$ defining a square area surrounding the pixel at $\langle i,j \rangle$. By contrast, $\mathcal{DF}$ will depend on the actual operation of the transformation. For  $\bm{\phi_1}$, $\bm{\phi_2}$, $\bm{\phi_3}$, and $\bm{\phi_4}$,  $\mathcal{DF}$ computes the average, median, minimum, and maximum of the pixel values of the coordinates returned by $\mathcal{DP}$ ($O_{\mathcal{DP}}$), respectively.

\vspace{.1cm}\noindent\textbf{Critical parameter values.}
The possible sizes of a convolution kernel (a square) for an image with width $W$and height $H$ can vary from $2\times 2$ to $S\times S$ where $S=min(W,H)$ because kernel cannot be larger than the input image. As kernel sizes have to be integers, the output of $\mathcal{DP}$ on each pixel $\langle i,j \rangle$ can be $2\times 2$, $3\times 3$,...,$S\times S$ pixels (i.e., $S-1$ different values) surrounding the input pixel. Moreover, $\mathcal{DF}$ for $\phi_1$ to $\phi_4$ does not dependent on the kernel size. Therefore, these convolution-based transformations ($\phi_1$ to $\phi_4$ ) have $S-1$ critical parameter values, i.e., $\mathbb{C}_{critical}=\{c\in \mathbb{N}:2\leq c\leq min(W,H)\}$.

\vspace{.1cm}\noindent\textbf{Verification complexity.}
As the number of different convolution kernel sizes can only be integer values and is bounded by the image size, the number of unique output images is $O(n)$ where the input image size is $n$.

\subsection{Point transformations}

\vspace{.1cm}\noindent\textbf{Decomposition.}
Point transformations (e.g., $\phi_5$ and $\phi_6$) are simple pixel-space operations. Therefore, $\mathcal{DP}$ for all these cases outputs the input coordinate itself, i.e., $\mathcal{DP}(\bm{I},\langle i,j \rangle;c) = \{\langle i,j \rangle\}$. Essentially, $\mathcal{DP}$ is an identity function for these transformations. 
By contrast, $\mathcal{DF}$ depends on the functionality of each transformation. For $\bm{\phi_5}$, $\mathcal{DF}(\bm{I}(\langle i,j \rangle);c) = c\cdot \bm{I}(\langle i,j\rangle)$, where $c$ is the gain used to adjust the contrast of the input image~\cite{contrastlighting}. Similarly, for $\bm{\phi_6}$, $\mathcal{DF}(\bm{I}(\langle i,j \rangle);c) = c+\bm{I}(\langle i,j \rangle)$, where $c$ is the bias used to adjust the brightness of the input image~\cite{szeliski2010computer}.

\vspace{.1cm}\noindent\textbf{Critical parameter values.}
As noted above, $\mathcal{DP}$ for both $\phi_5$ and $\phi_6$ is an identity function and is independent of $c$. Therefore, $\mathcal{DP}$ does not affect the number of critical parameters. 

For $\bm{\phi_5}$, $\mathcal{DF}$ change the image contrast by multiplying $c$ to each pixel value. Therefore, $PF$ is a function mapping any pixel value in $[0,255]$ to a new value in $[0,255]$ by multiplying $c$. It is easy to see that at most $|\mathbb{C}_{critical}|=256\times 256$ number of critical parameter values are enough to cover all such unique mappings. Specifically, $\mathbb{C}_{critical} \leq \bigcup\limits_{m=0}^{255}\bigcup\limits_{n=0}^{255}\frac{m}{n}$ where critical parameter values resulting in invalid (\eg division by zero) or duplicate values can be further reduced.

Similarly, $\mathcal{DF}$ for $\phi_6$ is a function mapping any pixel value in $[0,255]$ to $[0,255]$ by adding $c$. Therefore $\mathbb{C}_{critical}$ for $\phi_6$ is $\{-255,-254,...,254,255\}$.

\vspace{.1cm}\noindent\textbf{Verification complexity.}
As shown in the earlier analysis, the number of critical parameter values for $\phi_5$ and $\phi_6$ does not depend on the input image size $n$. Therefore, the total number of critical parameter values $\mathbb{C}_{critical}$ has $O(1)$ complexity with respect to the input size.

\subsection{Geometric transformations}

\vspace{.1cm}\noindent\textbf{Decomposition.}
In this paper, we analyze five types of geometric transformations \textemdash occlusion with a predefined mask image ($\phi_7$), rotation ($\phi_8$), shear ($\phi_9$), scale ($\phi_{10}$), translation ($\phi_{11}$), and reflection ($\phi_{12}$). For all of these transformations, $\mathcal{DP}$ maps one coordinate to another within the image, \ie $\mathcal{DP}(\bm{I},\langle i,j \rangle;c) = \{\langle i',j' \rangle\}$, and $\mathcal{DF}$ is an identity function, \ie $\mathcal{DF}(\bm{I}, \langle i',j' \rangle;c)=\bm{I}[i', j']$. We describe the individual $\mathcal{DP}$ function for each transformation below.

For occlusion ($\phi_7$), $\mathcal{DP}(\bm{I},\langle i,j \rangle;c) = \{\langle i, j \rangle\}$ if $i\notin[c_W,c_W+W_{OcclMask}]$ and $j\notin[c_H,c_H+H_{OcclMask}]$. If a pixel's coordinates are within this range, its value is decided by the occlusion mask and is independent of any pixel value in the input image. Here, $(c_W, c_H)$ denotes the coordinate of the upper-left corner of the image where the occlusion mask is applied and $W_{OcclMask}$ and $H_{OcclMask}$ denote the widths and heights of the occlusion mask respectively.

For rotation ($\phi_8$), $\mathcal{DP}(\bm{I},\langle i,j \rangle;c)=\{\langle i\cdot\cos c-j\cdot\sin c,i\cdot\sin c+j\cdot\cos c \rangle\}$ where $c$ is the rotation degree. Note that we only consider rotation around the center of the image here but $\mathcal{DP}$ for rotation around arbitrary points can also be constructed in the same manner.

Similarly, for shear ($\phi_9$), $\mathcal{DP}(\bm{I},\langle i,j \rangle;c)=\{\langle i+jc_W,ic_H+j \rangle\}$, where $c=(c_W,c_H)$ are the horizontal and vertical shear parameters.

For scale ($\phi_{10})$, $\mathcal{DP}(\bm{I},\langle i,j \rangle;c)=\langle ic_W,jc_H \rangle$, where $c=(c_W,c_H)$ are the horizontal and vertical scale parameters.

For translation ($\phi_{11}$), $\mathcal{DP}(\bm{I},\langle i,j \rangle;c)=\langle i+c_W,j+c_H \rangle$, where $c=(c_W,c_H)$ are the horizontal and vertical shifting parameters.

Finally, for reflection ($\phi_{12}$), $\mathcal{DP}(\bm{I},\langle i,j \rangle;c)=\langle ic_W,jc_H \rangle$, where $c = (c_W, c_H) \in \{(-1,1), (1,-1), (-1,-1)$, represent three types of reflections (horizontal,vertical, and central).

\vspace{.1cm}\noindent\textbf{Critical parameter values.}
As the size and content of the occlusion mask for $\phi_7$ are fixed, the number of different outputs of $\mathcal{DP}$ depend on where the occlusion mask is applied as decided by $c=(c_W,c_H)$. Therefore, $\mathbb{C}_{critical}=\{\langle i,j \rangle:i\in[0,W-W_{OcclMask}],j\in[H-H_{OcclMask}]\}$, which are all possible coordinates where  $OcclMask$ can be applied. As the coordinates have to be integers, the number of critical parameter values is $(W-W_{OcclMask}+1)\times (H-H_{OcclMask}+1)$.

We describe the computation of the critical parameter values for $\bm{\phi_8}$, the most complex one among $\phi_8$ to $\phi_{12}$, in detail below and skip the details for the other transformations ($\bm{\phi_9}$-$\bm{\phi_{12}}$) as they are similar to $\bm{\phi_8}$. 

\begin{figure}[!t]
\centering
\includegraphics[width=0.99\columnwidth]{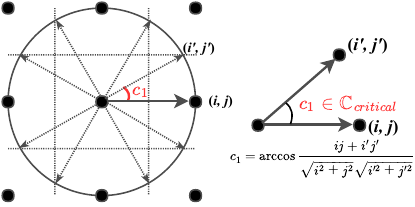}
\caption{Illustration of how rotation shuffles the coordinates around for a $3\times3$ image. Black dots represent the coordinate of each pixel. The circle shows the trajectory of coordinate $\langle i,j \rangle$ during rotation with different angles. Any intersection of the circular trajectory with a dotted line will be rounded to the nearest coordinate. For a new rounded coordinate $\langle i',j' \rangle$, the formula for calculating the corresponding critical rotation angle is shown on the right.}

\label{fig:rotation_proof}
\end{figure}

Figure~\ref{fig:rotation_proof} shows the movement of the coordinates (along the circle) while rotating an image around the center of the image. 
Given any coordinate $\langle i,j \rangle$, the output of $\mathcal{DP}$ will only output new coordinates when the circular trajectory intersects with the dotted lines. For example, $\mathcal{DP}$ will output seven new coordinates in Figure~\ref{fig:rotation_proof}).
All the rotation angles that correspond to the parts of the trajectory between any two adjacent dotted lines with arrows will always be rounded to the same coordinate. 
Therefore, given any $\langle i,j \rangle$, we compute how many dotted lines intersect with the corresponding trajectory. We calculate the rotation angles ($c$) for each intersect coordinate $\langle i',j' \rangle$ using the following equation.

\begin{equation*}\label{eq:degree}
    c = \arccos\frac{i j + i' j'}{\sqrt{i^2+j^2}\sqrt{i'^2+j'^2}}
\end{equation*}

Finally, we compute the union of all possible rotation degrees of each coordinate that can be mapped to a new one by $\mathcal{DP}$ following Algorithm~\ref{alg:findcs}, where the resulting union is the final critical parameter values for rotation ($\mathbb{C}_{critical}$).

\vspace{.1cm}\noindent\textbf{Verification complexity.}
The occlusion mask can only be applied on integer coordinates within the input image. Therefore, as described above, the number of critical parameter values for $\phi_7$ is  $(W-W_{OcclMask}+1)\times(H-H_{OcclMask}+1)$. Therefore, the verification complexity is $O(n)$.

As shown in Figure~\ref{fig:rotation_proof}, for $\phi_8$, each trajectory of a coordinate can only be changed to at most $(w-1)\times(h-1)$ number of new coordinates.
As the input has $w\times h$ number of distinct coordinates, the number of possible changes is $O(w^2\cdot h^2)=O(n^2)$. 

For $\phi_9$, $\mathcal{DP}$ outputs $\langle i+jc_W, ic_H+j \rangle$ for each $\langle i, j \rangle$. Note that $i+jc_W$ can have at most $w\cdot h\cdot w$ number of valid values ($w$ possible values for $i$, $h$ for $j$, and $w$ for $i+jc_W$). The same analysis applies to possible values of $ic_H+j$ as well, i.e., the number of valid values is $w\cdot h\cdot h$. In total, there are $w^3\cdot h^3=O(n^3)$ total possible pairs of $(c_W,c_H)$.

For $\phi_{10}$, $\mathcal{DP}$ outputs  $\langle ic_W, jc_H \rangle$ for input $\langle i, j \rangle$. For a given $i$ and $j$, $ic_W$ can have $w$ distinct values of $ic_W$ and $jc_h$ can also have $h$ distinct values. Since there is $w*h$  numbers of possible $\langle i, j \rangle$ pairs, the verification complexity will be $w^2\cdot h^2=O(n^2)$. 

For $\phi_{11}$, the transformation can only shift the image within the image size and all the critical parameter values must be integers. Therefore, $\phi_{11}$ has $O(n)$ verification complexity.

Finally, $\phi_{12}$ is a special case where the reflection operation on input can only have 3 types and thus has $O(1)$ verification complexity.

\section{Implementation}
\label{sec:impl}
Our implementation of \sys consists of 8,057 lines of Python code. \sys uses OpenCV, the popular image processing and vision library, to implement efficient image transformations. 
All of our experiments are run on a Linux laptop with Ubuntu 16.04 (one Intel i7-6700HQ 2.60 GHz processor with 4 cores, 16 GB memory, and a NVIDIA GTX 1070 GPU). 
To significantly cut down verification time by paralleling the verification process, we also implemented batch prediction~\cite{chollet2015keras}, using both GPU and CPU, to make the target computer vision system predict multiple images in parallel.

We evaluate \sys with $12$ different safety properties with different transformations. We use \sys to verify $15$ total vision systems including $10$ popular pre-trained DNNs performing tasks like object recognition, autonomous driving, etc., and $5$ third-party blackbox image recognition services using API access. Table~\ref{tab:dataset_dnn} shows a summary of all $15$ computer vision systems and the corresponding datasets used for verification in our experiments. These systems can be categorized into three groups based on the  tasks they perform and the type of access they provide (e.g., API access vs. the trained model). 
We describe them in detail below.

\begin{table*}[!htb]
\setlength{\tabcolsep}{8pt}
\footnotesize
\centering
\renewcommand{\arraystretch}{1.1}
\caption{Details of the computer vision systems and the corresponding datasets used for evaluating \sys.}
\label{tab:dataset_dnn}
\begin{tabular}{|c|c|c|c|c|c|c|c|}
\hline
\multicolumn{1}{|c|}{\textbf{Task Description}} & \multicolumn{1}{c|}{\textbf{Test Dataset}} & \multicolumn{1}{c|}{\textbf{ID}} & \multicolumn{1}{c|}{\textbf{\begin{tabular}[c]{@{}c@{}}Underlying\\ Architecture\end{tabular}}} & \multicolumn{1}{c|}{\textbf{\begin{tabular}[c]{@{}c@{}}Input\\ Size$^*$\end{tabular}}} & \multicolumn{1}{c|}{\textbf{\begin{tabular}[c]{@{}c@{}}Top-5\\ Loss\end{tabular}}} & \multicolumn{1}{c|}{\textbf{\begin{tabular}[c]{@{}c@{}}Top-1\\ Loss\end{tabular}}} & \multicolumn{1}{c|}{\textbf{\begin{tabular}[c]{@{}c@{}}Pred. time\\ (ms/img)\end{tabular}}} \\ \hline
\multirow{6}{*}{\begin{tabular}[c]{@{}c@{}}Detect ILSVRC\\ 1000 class labels\end{tabular}} & \multirow{6}{*}{\begin{tabular}[c]{@{}c@{}}ImageNet \cite{deng2009imagenet}\\ provided in \\ ILSVRC \cite{ILSVRC15}\end{tabular}} & IMG\_C1 & VGG-16~\cite{simonyan2014very} & 224$\times$224 & 0.1 & 0.295 & 102.9 \\ \cline{3-8} 
 &  & IMG\_C2 & VGG-19~\cite{simonyan2014very} & 224$\times$224 & 0.09 & 0.273 & 103.9 \\ \cline{3-8} 
 &  & IMG\_C3 & MobileNet~\cite{howard2017mobilenets} & 224$\times$224 & 0.105 & 0.293 & 45.3 \\ \cline{3-8} 
 &  & IMG\_C4 & Xception~\cite{chollet2016xception} & 299$\times$299 & 0.055 & 0.21 & 54 \\ \cline{3-8} 
 &  & IMG\_C5 & Inception-v3~\cite{szegedy2016rethinking} & 299$\times$299 & 0.059 & 0.218 & 87.9 \\ \cline{3-8} 
 &  & IMG\_C6 & ResNet-50~\cite{he2015deep} & 224$\times$224 & 0.071 & 0.242 & 50.8 \\ \hline
\multirow{5}{*}{\begin{tabular}[c]{@{}c@{}}Detect categories\\ of general images\end{tabular}} & \multirow{5}{*}{\begin{tabular}[c]{@{}c@{}}ImageNet \cite{deng2009imagenet}\\ provided in \\ ILSVRC \cite{ILSVRC15}\end{tabular}} & API\_C1 & Google Vision~\cite{google-vision-api} & 224$\times$224 & -$^{**}$ & -$^{**}$ & 904.87 \\ \cline{3-8} 
 &  & API\_C2 & Clarifai Tagging~\cite{clarifai} & 224$\times$224 & -$^{**}$ & -$^{**}$ & 957.98 \\ \cline{3-8}
 \multicolumn{1}{|c|}{} & \multicolumn{1}{c|}{} & \multicolumn{1}{c|}{API\_C3} & \multicolumn{1}{c|}{IBM Vision~\cite{ibm}} & 224$\times$224 & \multicolumn{1}{c|}{-$^{**}$} & \multicolumn{1}{c|}{-$^{**}$} & \multicolumn{1}{c|}{689.14} \\ \cline{3-8}
  \multicolumn{1}{|c|}{} & \multicolumn{1}{c|}{} & \multicolumn{1}{c|}{API\_C4} & \multicolumn{1}{c|}{Microsoft Vision~\cite{microsoft}} & 224$\times$224 & \multicolumn{1}{c|}{-$^{**}$} & \multicolumn{1}{c|}{-$^{**}$} & \multicolumn{1}{c|}{496.68} \\ \cline{3-8}
  \multicolumn{1}{|c|}{} & \multicolumn{1}{c|}{} & \multicolumn{1}{c|}{API\_C5} & \multicolumn{1}{c|}{Amazon Rekognition~\cite{amazon}} & 224$\times$224 & \multicolumn{1}{c|}{-$^{**}$} & \multicolumn{1}{c|}{-$^{**}$} & \multicolumn{1}{c|}{795.91} \\ \hline
\multirow{4}{*}{\begin{tabular}[c]{@{}c@{}}Predict steering angle\\ for each frame\\ captured from car's\\ front scene\end{tabular}} & \multirow{4}{*}{\begin{tabular}[c]{@{}c@{}}Driving images\\ provided by\\ Udacity autonomous\\ car challenge \cite{udacity:challenge}\end{tabular}} & DRV\_C1 & Rambo~\cite{rambo} & 192$\times$256 & 0.058$^+$ & 0.058$^+$ & 33.6 \\ \cline{3-8} 
 &  & DRV\_C2 & Dave-orig~\cite{bojarski2016end,dave-orig} & 100$\times$100 & 0.091$^+$ & 0.091$^+$ & 31.6 \\ \cline{3-8} 
 &  & DRV\_C3 & Dave-norminit~\cite{dave-norminit} & 100$\times$100 & 0.053$^+$ & 0.053$^+$ & 31.1 \\ \cline{3-8} 
 &  & DRV\_C4 & Dave-dropout~\cite{dave-dropout} & 100$\times$100 & 0.084$^+$ & 0.084$^+$ & 31.1 \\ \hline
\multicolumn{8}{l}{\scriptsize \begin{tabular}[c]{@{}l@{}}$^*$ We specify only image width and height for one channel. Color images have 3 channels with same height and width.\\ $^{**}$ The third-party blackbox APIs do not disclose their performance on any public test dataset.\\ $^+$ We use MSE to measure the performance of self-driving car DNNs. Therefore, top-5 and top-1 loss have the same value for those DNNs.\end{tabular}}
\end{tabular}
\end{table*}

\vspace{.1cm}\noindent\textbf{1000-class ImageNet classification.} 
This group of vision systems use DNNs trained using the ImageNet~\cite{deng2009imagenet} dataset. All of these DNNs achieved state-of-the-art image classification performance in ILSVRC~\cite{ILSVRC15} competitions. Specifically, we verify the following six pre-trained DNNs: VGG-16~\cite{simonyan2014very}, VGG-19~\cite{simonyan2014very}, MobileNet~\cite{howard2017mobilenets}, Xception~\cite{chollet2016xception}, Inception-v3~\cite{szegedy2016rethinking} and ResNet50~\cite{he2015deep}. All these DNNs are considered major breakthroughs in DNN architectures as they improved the state-of-the-art performances during each year of ILSVRC~\cite{ILSVRC15} competitions. We also use test images from the test set provided by ILSVRC for verifying the local safety properties.

\vspace{.1cm}\noindent\textbf{Third-party image classification services.}
We also evaluate \sys on five blackbox commercial image classification APIs provided by Google~\cite{google-vision-api}, Clarifai~\cite{clarifai}, IBM~\cite{ibm}, Microsoft~\cite{microsoft}, and Amazon~\cite{amazon}. We use the same test images from ILSVRC as discussed above.

\vspace{.1cm}\noindent\textbf{Self-driving cars.} For verifying computer vision systems performing regression tasks, we use four self-driving car DNNs that control the steering angle based on the input images captured by a front camera. In particular, we adopt Rambo~\cite{rambo}, which is one of the top-ranked models in the Udacity self-driving car challenge~\cite{udacity:challenge}, and three other open-source implementations~\cite{dave-orig, dave-norminit, dave-dropout} based on the Nvidia's Dave-2 self-driving system~\cite{bojarski2016end}. We use the testing set from the Udacity challenge~\cite{udacity:challenge-dataset} for verifying local safety properties.

\section{Evaluation}
\label{sec:eval}

\subsection{Results}
\label{subsec:violations_found}

\begin{table*}[!htb]
\setlength{\tabcolsep}{8pt}
\footnotesize
\centering
\renewcommand{\arraystretch}{1.1}
\caption{Average number of violations found by \sys for each test input in different state-of-the-art ImageNet classifiers and self-driving systems. The results are averages over 10 input images for each safety property.}
\label{tab:violations1}
\begin{tabular}{|c|c|c|c|c|c|c|c|c|c|c|c|c|}
\hline
\textbf{} & $\bm{\phi_1}$ & $\bm{\phi_2}$ & $\bm{\phi_3}$ & $\bm{\phi_4}$ & $\bm{\phi_5}$ & $\bm{\phi_6}$ & $\bm{\phi_7}$ & $\bm{\phi_8}$ & $\bm{\phi_9}$ & $\bm{\phi_{10}}$ & $\bm{\phi_{11}}$ & $\bm{\phi_{12}}$ \\ \hline
\textbf{IMG\_C1} & 5 & 2.7 & 1.3 & 2 & 5359.1 & 39.4 & 4635.4 & 8782.8 & 710.5 & 20863 & 56.2 & 1.3 \\ \hline
\textbf{IMG\_C2} & 2.5 & 2 & 2.5 & 1.4 & 3496.2 & 42.3 & 1657.2 & 25307.7 & 61592.9 & 2455.8 & 49.7 & 1.2 \\ \hline
\textbf{IMG\_C3} & 3.7 & 1.8 & 2 & 2.8 & 5207.6 & 23.2 & 3960.5 & 11459.3 & 10.1 & 1239.7 & 61 & 1.4 \\ \hline
\textbf{IMG\_C4} & 1.7 & 2 & 2.3 & 1.8 & 3218.7 & 21.6 & 5253.5 & 10603.8 & 63613.5 & 13684 & 197 & 1.5 \\ \hline
\textbf{IMG\_C5} & 4 & 1.5 & 1.8 & 1.4 & 6724.5 & 39 & 1805 & 45642.8 & 724 & 368.5 & 40 & 1.6 \\ \hline
\textbf{IMG\_C6} & 6 & 2.1 & 1.7 & 1.7 & 8596 & 43 & 1699 & 9757.8 & 6224 & 47312 & 19 & 2.1 \\ \hhline{|=|=|=|=|=|=|=|=|=|=|=|=|=|}
\textbf{DRV\_C1} & 3.6 & 1.7 & 1.4 & 1.4 & 11935.9 & 89.7 & 65.4 & 7818.5 & 5561.2 & 4286.1 & 23.6 & N/A* \\ \hline
\textbf{DRV\_C2} & 2.4 & 0.7 & 2 & 2.4 & 11294.2 & 35.6 & 2530.5 & 1207 & 51.4 & 303.5 & 108.8 & N/A* \\ \hline
\textbf{DRV\_C3} & 1.6 & 1.8 & 2.3 & 3.6 & 18452.1 & 79 & 539.5 & 1722.2 & 258.1 & 783.3 & 42.6 & N/A* \\ \hline
\textbf{DRV\_C4} & 1.1 & 2.5 & 1.6 & 1.9 & 5768.6 & 56.1 & 1866.8 & 372.8 & 504.6 & 7.6 & 123.3 & N/A* \\ \hline
\multicolumn{13}{l}{\scriptsize *Safety property involving reflected images is not realistic for self-driving cars.} \\
\end{tabular}
\end{table*}

\vspace{.1cm}\noindent\textbf{Summary.}
\sys found thousands of violations of different tested safety properties in all of the tested vision systems. Table~\ref{tab:violations1} and \ref{tab:violations2} summarize the number of violations that \sys found for imagenet classifiers, self-driving cars, and third-party image classification services. Table~\ref{tab:violations_imagenet} and \ref{tab:violations_car} in Appendix~\ref{subsec:appendx}) show some of sample inputs found by \sys that violates safety properties of these vision systems.

Due to high network latency and cost (\$0.001 per query), for third-party image classification services, we only verify with relative properties with small number of critical parameters and report the result in Table~\ref{tab:violations2}. Note that each number is averaged from the results from 10 random seed images. For t- and k-safety properties, as described in Section~\ref{sec:framework}, we set $t=0.1$ and $k=1$ for these experiments. Table~\ref{tab:bounds} shows the parameter spaces ($\mathbb{C}_{\phi}$) that we use for generating these violations.


\begin{table}[!htb]
\setlength{\tabcolsep}{6pt}
\footnotesize
\centering
\renewcommand{\arraystretch}{1.1}
\caption{Average number of violations found by \sys for each test input in the third-party image recognition APIs. The results are averaged over 10 images for each property. Due to high network latency and cost, we only tested for properties with relatively low verification complexity. }
\label{tab:violations2}
\begin{tabular}{|c|c|c|c|c|c|c|c|}
\hline
\textbf{} & $\bm{\phi_1}$ & $\bm{\phi_2}$ & $\bm{\phi_3}$ & $\bm{\phi_4}$ & $\bm{\phi_6}$ & $\bm{\phi_7}$ & $\bm{\phi_{12}}$ \\ \hline
\textbf{API\_C1} & 3.1 & 3.1 & 2.6 & 1.8 & 81.9 & 187.8 & 1.5 \\ \hline
\textbf{API\_C2} & 6 & 2.8 & 1.5 & 1.5 & 25.4 & 105.4 & 1 \\ \hline
\textbf{API\_C3} & 5.6 & 2.1 & 2.8 & 2.3 & 44 & 164.9 & 2.4 \\ \hline
\textbf{API\_C4} & 0.5 & 1.8 & 2 & 2 & 100.5 & 75.6 & 2.2 \\ \hline
\textbf{API\_C5} & 3.5 & 1.6 & 1.7 & 2.2 & 52.5 & 187 & 0.9 \\ \hline
\end{tabular}
\end{table}

\begin{table}[!htb]
\setlength{\tabcolsep}{5pt}
\centering
\footnotesize
\renewcommand{\arraystretch}{1.1}
\caption{The number of critical parameter values for different input sizes for each safety property $\phi$ and the corresponding parameter space tested with \sys. 
}
\label{tab:bounds}
\begin{tabular}{cccccc}
\hline
\multicolumn{1}{|c|}{\multirow{2}{*}{\textbf{Property}}} & \multicolumn{1}{c|}{\multirow{2}{*}{\textbf{$\bm{\mathbb{C}_{\phi}}$}}} & \multicolumn{4}{c|}{\textbf{$\bm{|\mathbb{C}_{critical}|}$ for each input size}} \\ \cline{3-6} 
\multicolumn{1}{|c|}{} & \multicolumn{1}{c|}{} & \multicolumn{1}{c|}{\textit{224}} & \multicolumn{1}{c|}{\textit{299}} & \multicolumn{1}{c|}{\textit{192$\times$256}} & \multicolumn{1}{c|}{\textit{100}} \\ \hline
\multicolumn{1}{|c|}{$\bm{\phi_1}$} & \multicolumn{1}{c|}{{[}2, 10{]}} & \multicolumn{1}{c|}{9} & \multicolumn{1}{c|}{9} & \multicolumn{1}{c|}{9} & \multicolumn{1}{c|}{9} \\ \hline
\multicolumn{1}{|c|}{$\bm{\phi_2}$} & \multicolumn{1}{c|}{{[}2, 10{]}} & \multicolumn{1}{c|}{4$^\star$} & \multicolumn{1}{c|}{4$^\star$} & \multicolumn{1}{c|}{4$^\star$} & \multicolumn{1}{c|}{4$^\star$} \\ \hline
\multicolumn{1}{|c|}{$\bm{\phi_3}$} & \multicolumn{1}{c|}{{[}2, 5{]}} & \multicolumn{1}{c|}{4} & \multicolumn{1}{c|}{4} & \multicolumn{1}{c|}{4} & \multicolumn{1}{c|}{4} \\ \hline
\multicolumn{1}{|c|}{$\bm{\phi_4}$} & \multicolumn{1}{c|}{{[}2, 5{]}} & \multicolumn{1}{c|}{4} & \multicolumn{1}{c|}{4} & \multicolumn{1}{c|}{4} & \multicolumn{1}{c|}{4} \\ \hline
\multicolumn{1}{|c|}{$\bm{\phi_5}$} & \multicolumn{1}{c|}{{[}0.5, 2{]}} & \multicolumn{1}{c|}{32512} & \multicolumn{1}{c|}{32512} & \multicolumn{1}{c|}{32512} & \multicolumn{1}{c|}{32512} \\ \hline
\multicolumn{1}{|c|}{$\bm{\phi_6}$} & \multicolumn{1}{c|}{{[}-100, 100{]}} & \multicolumn{1}{c|}{200} & \multicolumn{1}{c|}{200} & \multicolumn{1}{c|}{200} & \multicolumn{1}{c|}{200} \\ \hline
\multicolumn{1}{|c|}{$\bm{\phi_7}$} & \multicolumn{1}{c|}{-$^\dag$} & \multicolumn{1}{c|}{33856} & \multicolumn{1}{c|}{67081} & \multicolumn{1}{c|}{40592} & \multicolumn{1}{c|}{6400} \\ \hline
\multicolumn{1}{|c|}{$\bm{\phi_8}$} & \multicolumn{1}{c|}{{[}-2, 2{]}} & \multicolumn{1}{c|}{95496} & \multicolumn{1}{c|}{225552} & \multicolumn{1}{c|}{106722} & \multicolumn{1}{c|}{8370} \\ \hline
\multicolumn{1}{|c|}{$\bm{\phi_9}$} & \multicolumn{1}{c|}{{[}-0.01, 0.01{]}} & \multicolumn{1}{c|}{250000} & \multicolumn{1}{c|}{810000} & \multicolumn{1}{c|}{238824} & \multicolumn{1}{c|}{5140} \\ \hline
\multicolumn{1}{|c|}{$\bm{\phi_{10}}$} & \multicolumn{1}{c|}{{[}0.99, 1.01{]}} & \multicolumn{1}{c|}{244036} & \multicolumn{1}{c|}{788544} & \multicolumn{1}{c|}{230580} & \multicolumn{1}{c|}{10000} \\ \hline
\multicolumn{1}{|c|}{$\bm{\phi_{11}}$} & \multicolumn{1}{c|}{{[}-10, 10{]}} & \multicolumn{1}{c|}{400} & \multicolumn{1}{c|}{400} & \multicolumn{1}{c|}{400} & \multicolumn{1}{c|}{400} \\ \hline
\multicolumn{1}{|c|}{$\bm{\phi_{12}}$} & \multicolumn{1}{c|}{-$^\ddag$} & \multicolumn{1}{c|}{3} & \multicolumn{1}{c|}{3} & \multicolumn{1}{c|}{3} & \multicolumn{1}{c|}{3} \\ \hline
\multicolumn{6}{l}{\scriptsize $^\star$ OpenCV only supports odd box sizes and thus smaller $\mathbb{C}_{\phi}$ than $\phi_1$.} \\
\multicolumn{6}{l}{\scriptsize $^\dag$ Position of occlusion mask is bounded by the image size.} \\
\multicolumn{6}{l}{\scriptsize $^\ddag$ Reflection has only three critical parameter values.}
\end{tabular}
\end{table}

\vspace{.1cm}\noindent\textbf{Verified images with no violations.}
We found that the number of verified input images, i.e., images for which a computer vision system do not violate a given safety property, varies widely based on the verification complexity of the safety property. The number of verified images decrease with higher verification complexity. However, even for properties with low verification complexity, we find that the number of verified images is very low (on average 31.7\%).

Table~\ref{tab:verified_imgs} reports the number of verified images for properties $\phi_2$ (median smooth), $\phi_3$ (erosion), $\phi_4$ (dilation), and $\phi_{12}$ (reflection) for IMG\_C3 on all  test images ($100,000$) from ILSVRC ~\cite{ILSVRC15}. As shown in Table~\ref{tab:verified_imgs}, the original top-1 test accuracy of IMG\_C3 is around 70.7\% which is significantly higher than the percentage of verified inputs.

\begin{table}[!tbp]
\setlength{\tabcolsep}{10pt}
\footnotesize
\centering
\renewcommand{\arraystretch}{1.1}
\caption{The number and percentage of verified images with respect to the total number of images in ILSVRC test set for safety properties $\phi_2$, $\phi_3$, $\phi_4$, and $\phi_{12}$ for IMG\_C3 (MobileNet). The last column shows the original accuracy of IMG\_C3 on this test set.}
\label{tab:verified_imgs}
\begin{tabular}{|c|c|c|c|}
\hline
{\bf Property} & \multicolumn{2}{c|}{\bf \# verified inputs} & {\bf Original Accuracy} \\ \hline
$\bm{\phi_2}$ & 28,509 & 28.5\% & 70.7\% \\ \hline
$\bm{\phi_3}$ & 34,644 & 34.6\% & 70.7\% \\ \hline
$\bm{\phi_4}$ & 30,979 & 31\% & 70.7\% \\ \hline
$\bm{\phi_{12}}$ & 32,817 & 32.8\% & 70.7\% \\ \hline
\end{tabular}
\end{table}

\vspace{.1cm}\noindent\textbf{Comparison of \sys with gradient-based methods.}
Adversarial ML inputs, gradient-based approaches for finding violations of based on stochastic gradient descent is one of the most widely-used techniques in prior works to find violations to given ML systems~\cite{goodfellow2014explaining, papernot2016limitations}. These gradient-based approaches do not provide any guarantee about absence of erroneous inputs. However, in order to empirically estimate how many erroneous cases they miss, we compare \sys with gradient-based approaches in terms of number of violations found by both of these techniques. Specifically, we leverage the projected gradient descent approach described by Pei et al.~\cite{pei2017deepxplore} to change the brightness of an image ($\phi_6$) and compare the number of violations found against those found by \sys. We use the same values of the parameter space ($\mathbb{C}_{\phi}$), $k$, and $t$ as described in Section~\ref{subsec:violations_found}. 

\begin{figure}
\centering
\captionsetup[subfloat]{labelformat=empty}
\subfloat[MobileNet (IMG\_C3)]{
\includegraphics[width=0.48\columnwidth]{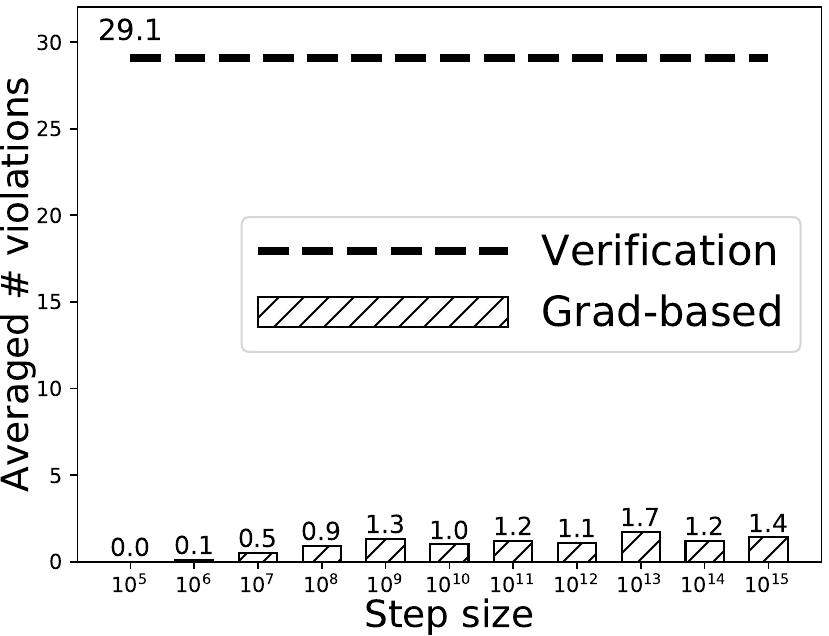}
\label{subfig:mobilenet_grad}}
\subfloat[Dave-orig (DRV\_C2)]{
\includegraphics[width=0.48\columnwidth]{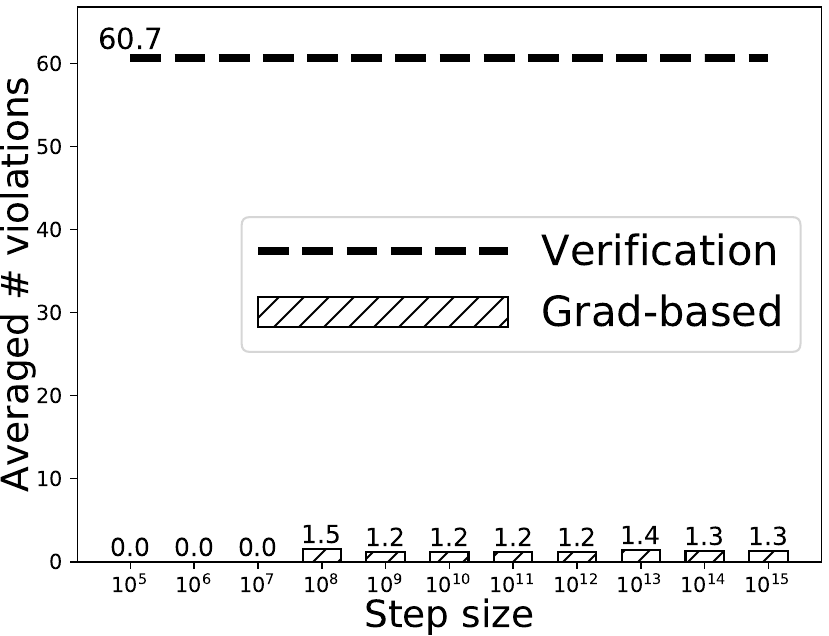}
\label{subfig:dave-orig_grad}}

\caption{Comparison of average numbers of violations found by gradient-based methods and \sys for property $\phi_6$ (changing brightness). \sys finds $64.8\times$ more violations than gradient based approach.}
\label{fig:grad_compare}
\end{figure}

Figure~\ref{fig:grad_compare} shows that \sys finds up to $64.8$ times more violations than the gradient-based approach.  This demonstrates that gradient-based approaches often miss a large number of safety violations in computer vision systems.

\begin{figure*}
\captionsetup[subfloat]{captionskip=-.15cm, labelformat=empty}

{\bf \scriptsize \hspace{0.025in} Altering $k$:
}

\subfloat[]{
\includegraphics[width=0.16\textwidth, height=0.14\textwidth]{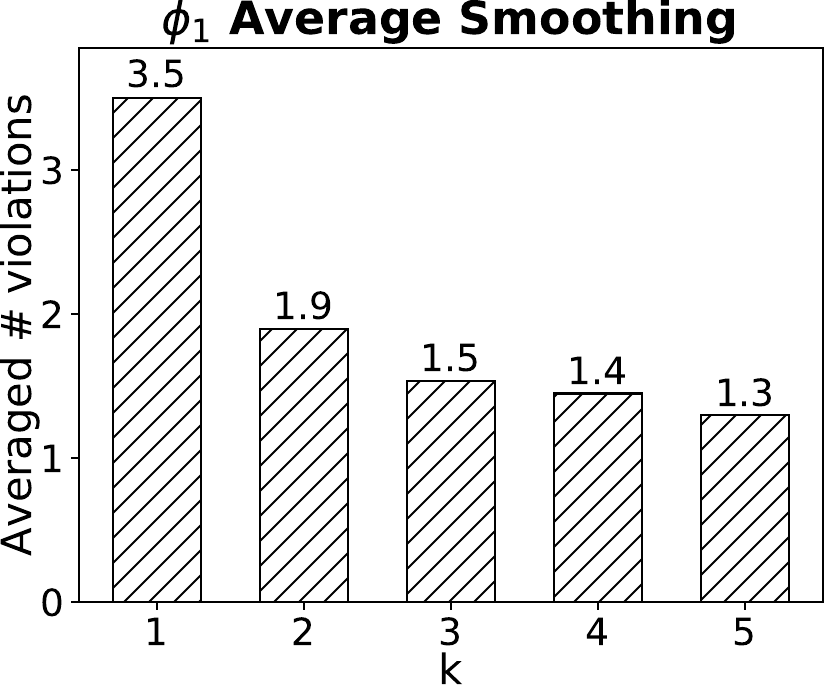}
\label{subfig:k_violation1}}
\hspace{-.7em}
\subfloat[]{
\includegraphics[width=0.16\textwidth, height=0.14\textwidth]{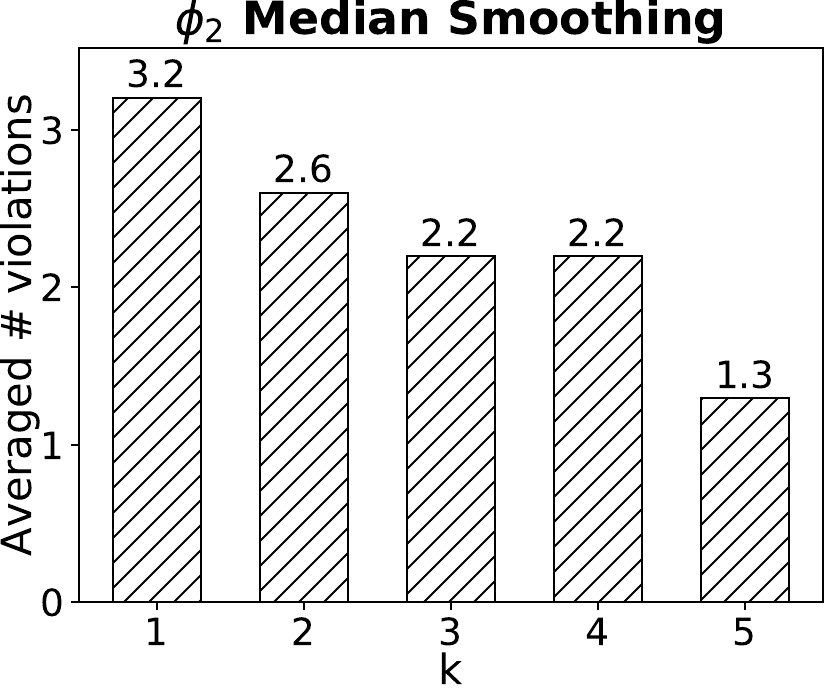}
\label{subfig:k_violation2}}
\hspace{-.7em}
\subfloat[]{
\includegraphics[width=0.16\textwidth, height=0.14\textwidth]{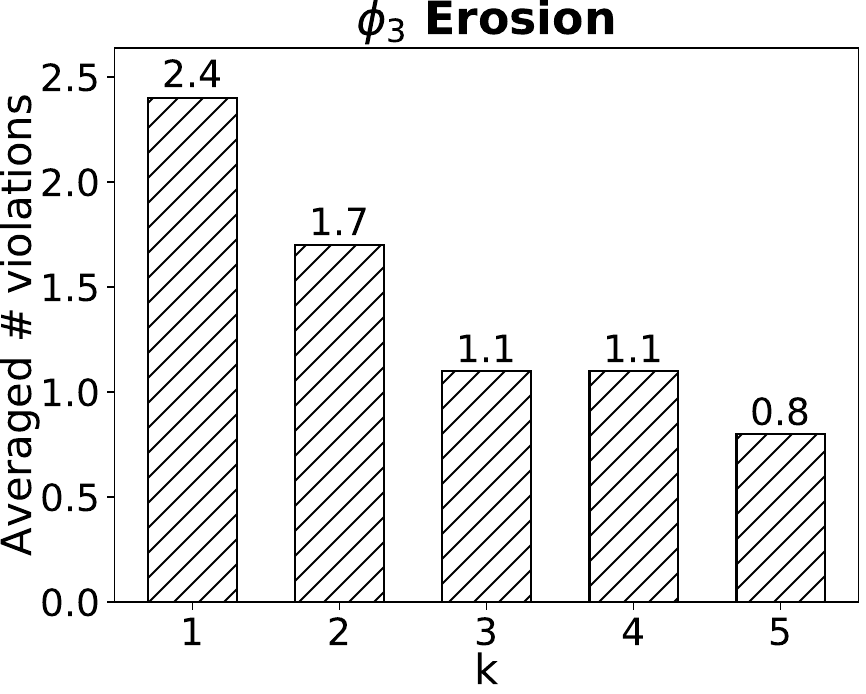}
\label{subfig:k_violation3}}
\hspace{-.7em}
\subfloat[]{
\includegraphics[width=0.16\textwidth, height=0.14\textwidth]{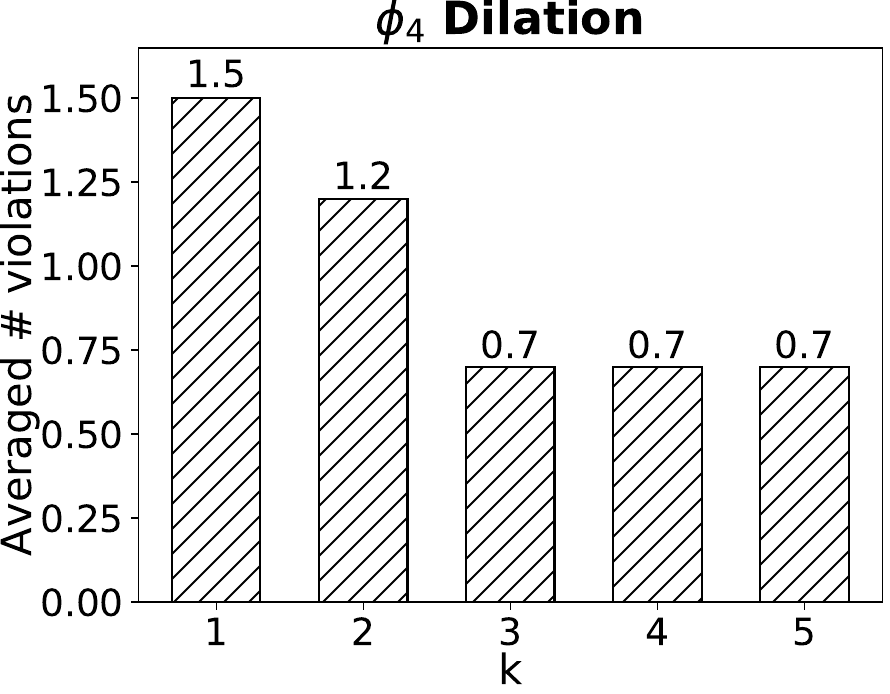}
\label{subfig:k_violation4}}
\hspace{-.7em}
\subfloat[]{
\includegraphics[width=0.16\textwidth, height=0.14\textwidth]{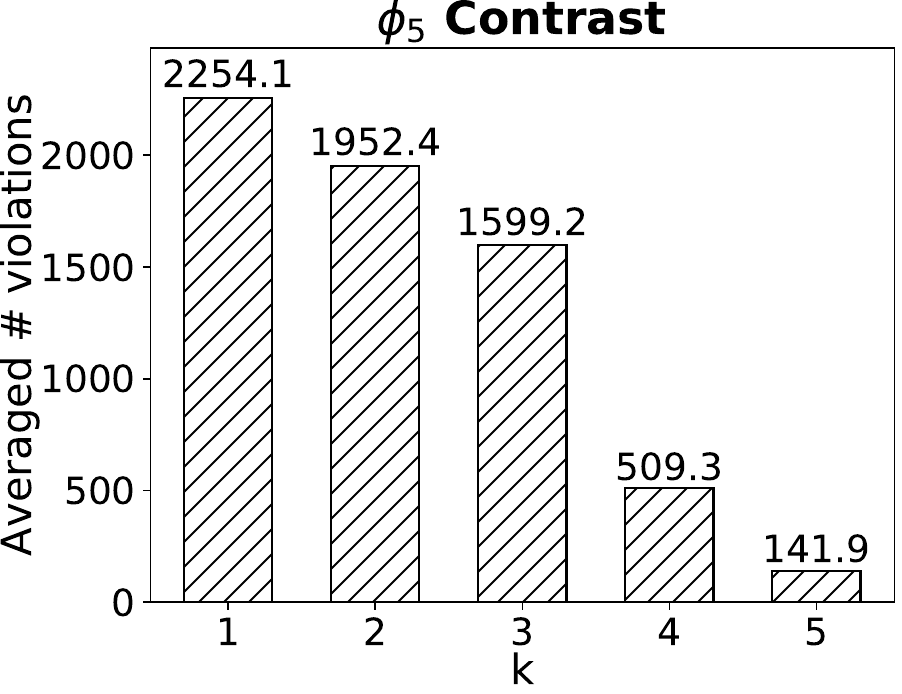}
\label{subfig:k_violation5}}
\hspace{-.7em}
\subfloat[]{
\includegraphics[width=0.16\textwidth, height=0.14\textwidth]{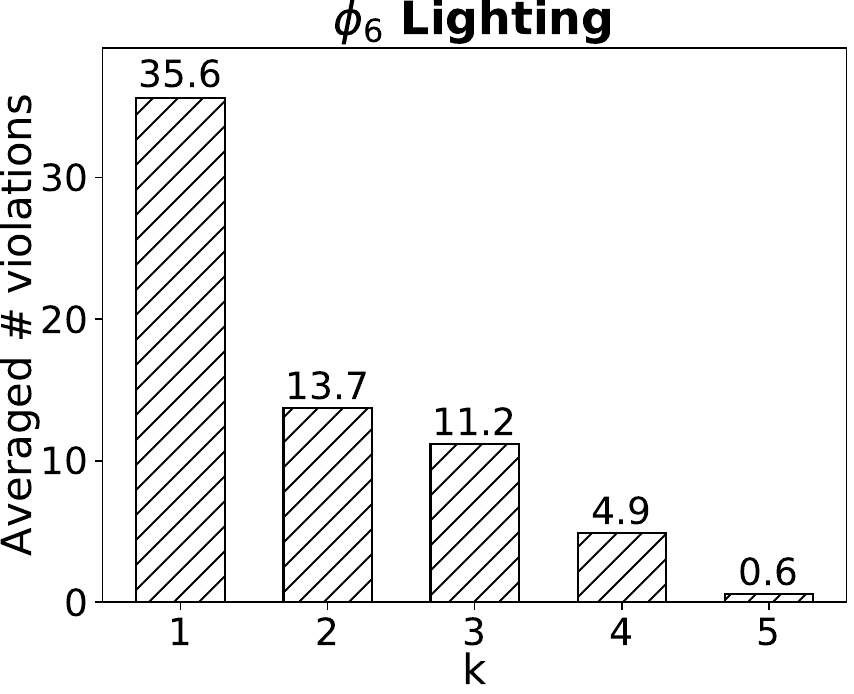}
\label{subfig:k_violation6}}

\subfloat[]{
\includegraphics[width=0.16\textwidth, height=0.14\textwidth]{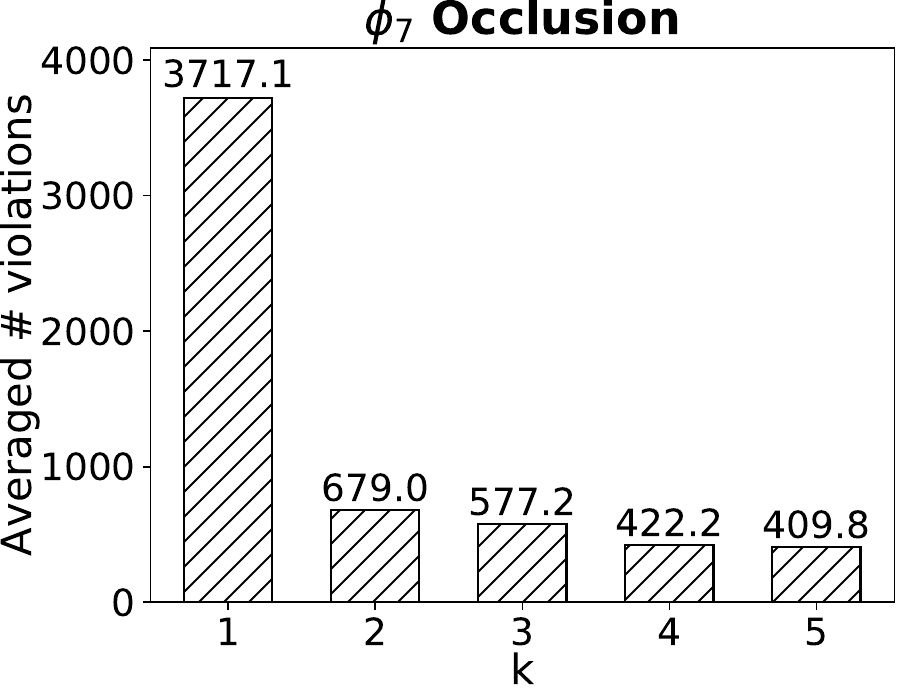}
\label{subfig:k_violation7}}
\hspace{-.7em}
\subfloat[]{
\includegraphics[width=0.16\textwidth, height=0.14\textwidth]{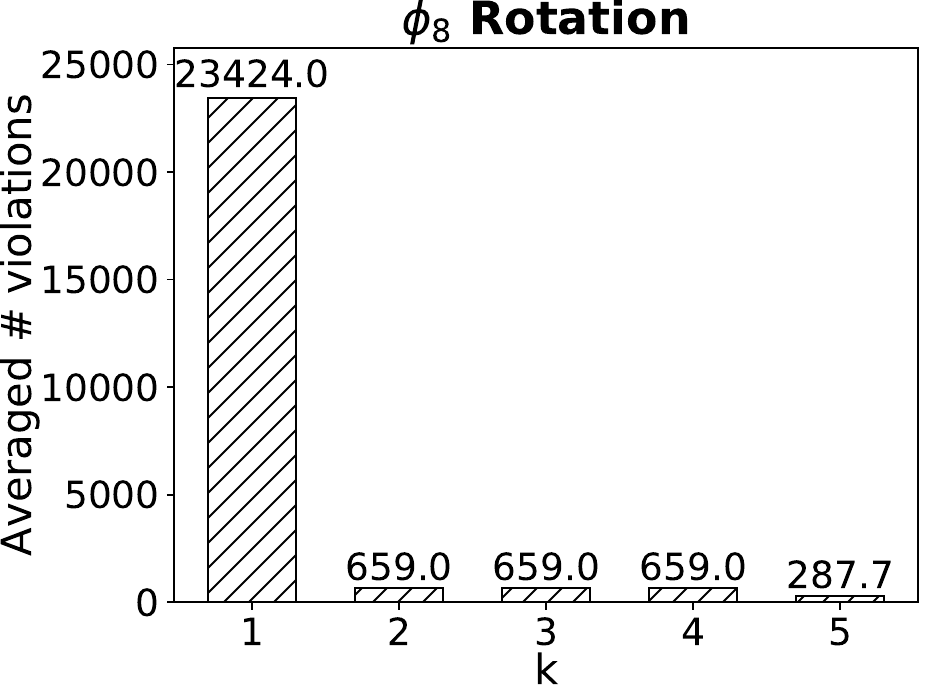}
\label{subfig:k_violation8}}
\hspace{-.7em}
\subfloat[]{
\includegraphics[width=0.16\textwidth, height=0.14\textwidth]{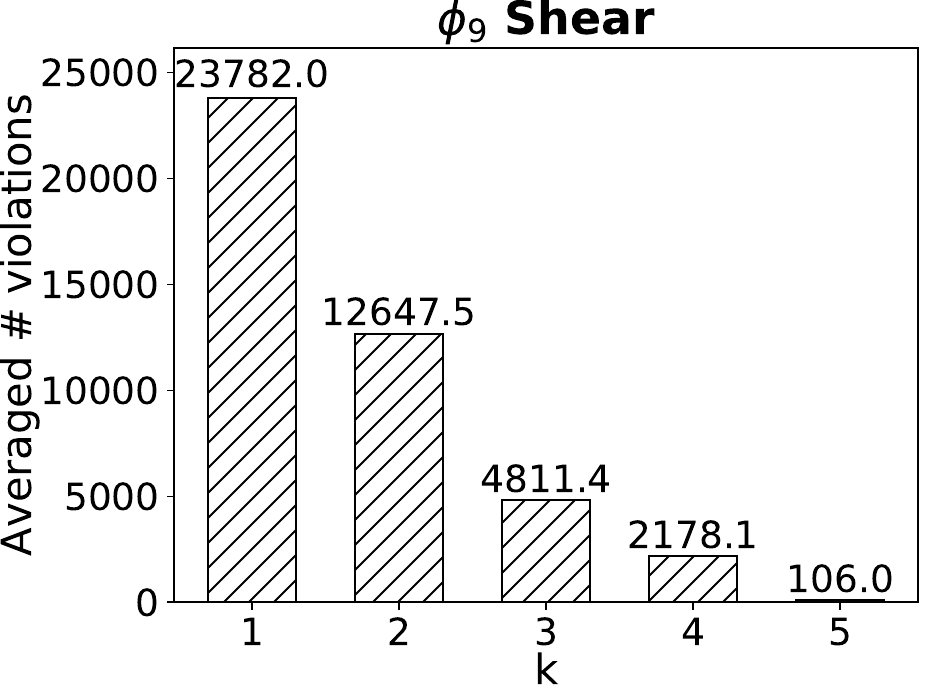}
\label{subfig:k_violation9}}
\hspace{-.7em}
\subfloat[]{
\includegraphics[width=0.16\textwidth, height=0.14\textwidth]{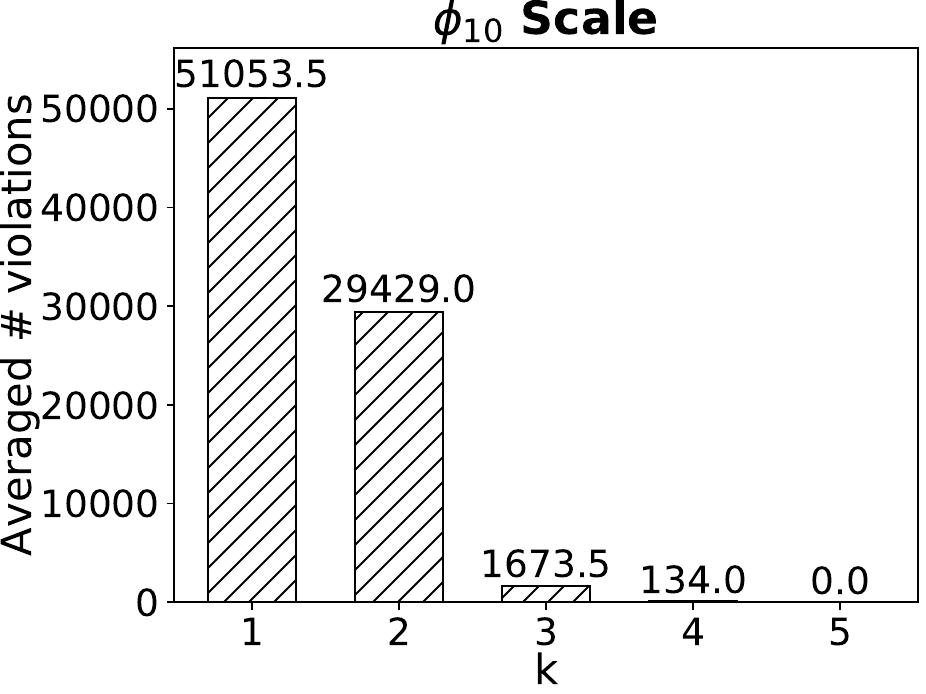}
\label{subfig:k_violation10}}
\hspace{-.7em}
\subfloat[]{
\includegraphics[width=0.16\textwidth, height=0.14\textwidth]{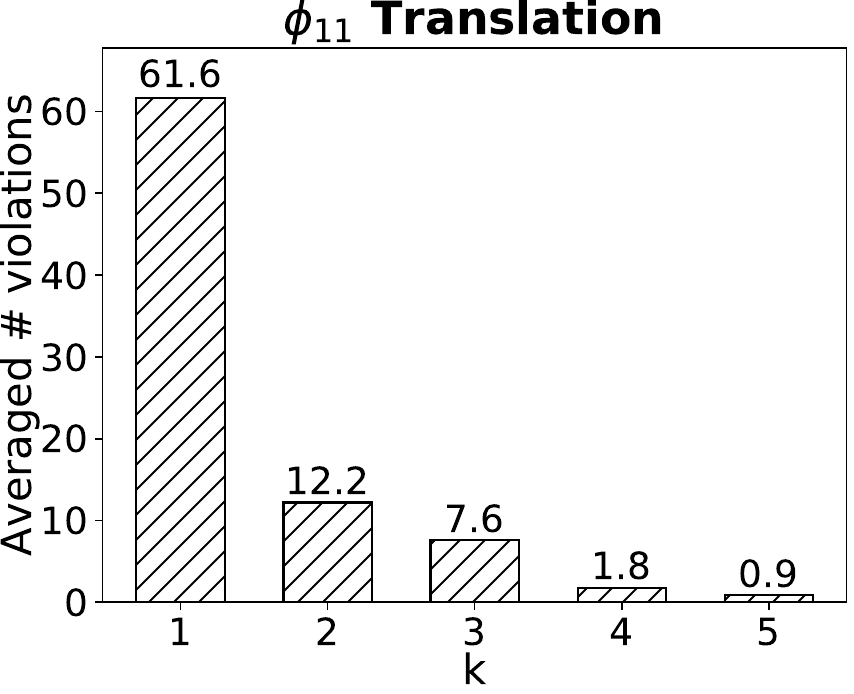}
\label{subfig:k_violation11}}
\hspace{-.7em}
\subfloat[]{
\includegraphics[width=0.16\textwidth, height=0.14\textwidth]{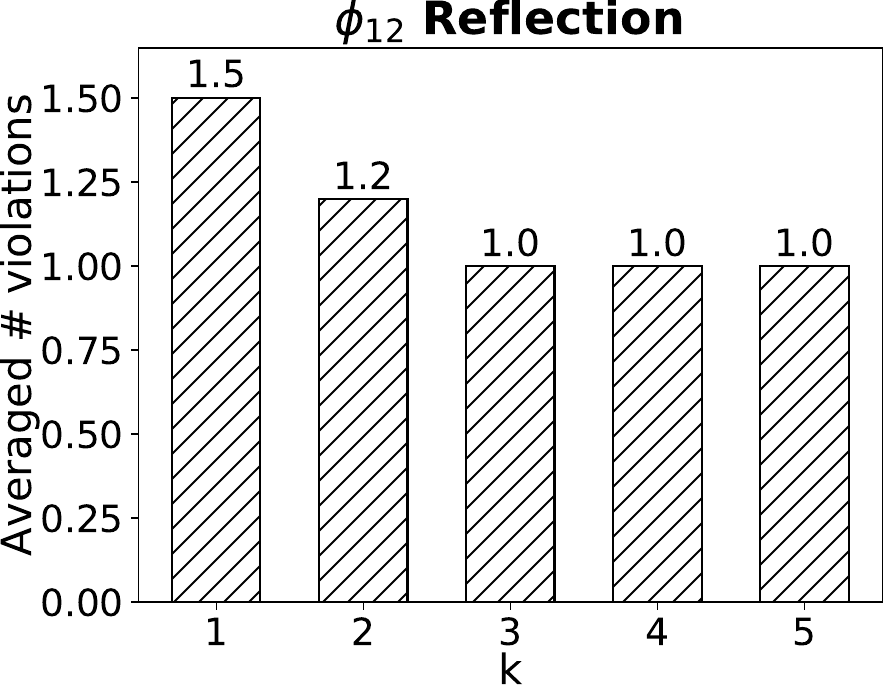}
\label{subfig:k_violation12}}

{\bf \scriptsize \hspace{0.025in} Altering $t$:
}

\subfloat[]{
\includegraphics[width=0.16\textwidth, height=0.14\textwidth]{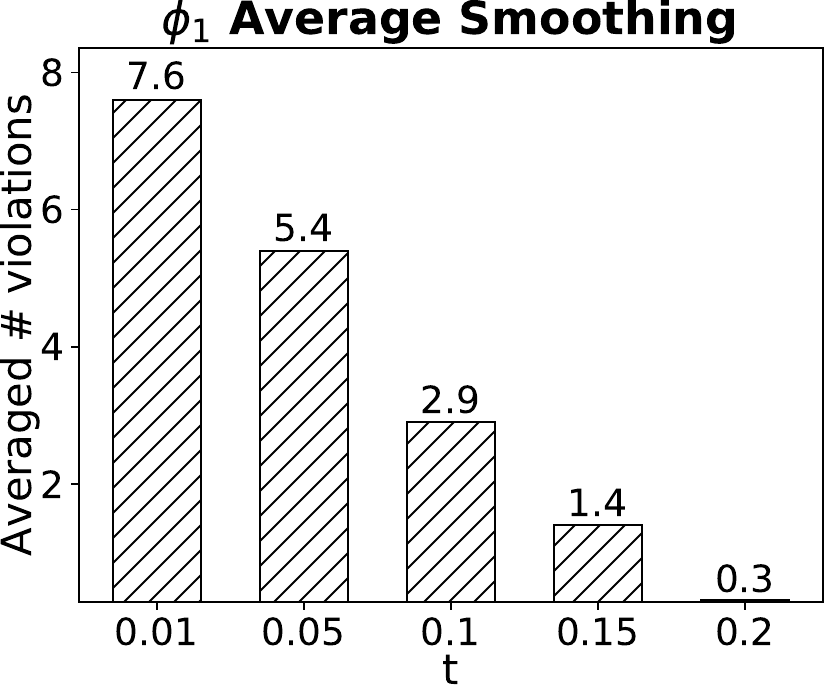}
\label{subfig:t_violation1}}
\hspace{-.7em}
\subfloat[]{
\includegraphics[width=0.16\textwidth, height=0.14\textwidth]{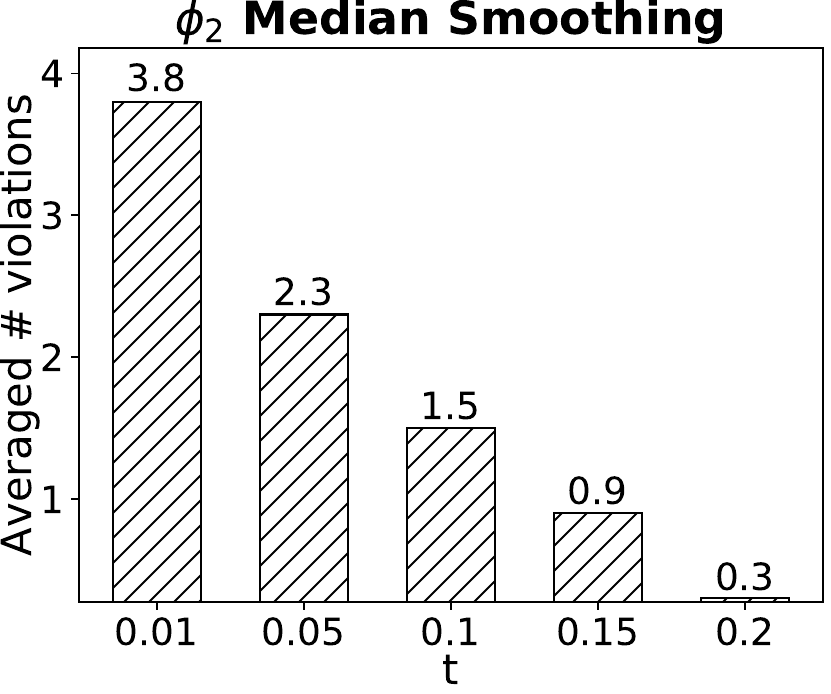}
\label{subfig:t_violation2}}
\hspace{-.7em}
\subfloat[]{
\includegraphics[width=0.16\textwidth, height=0.14\textwidth]{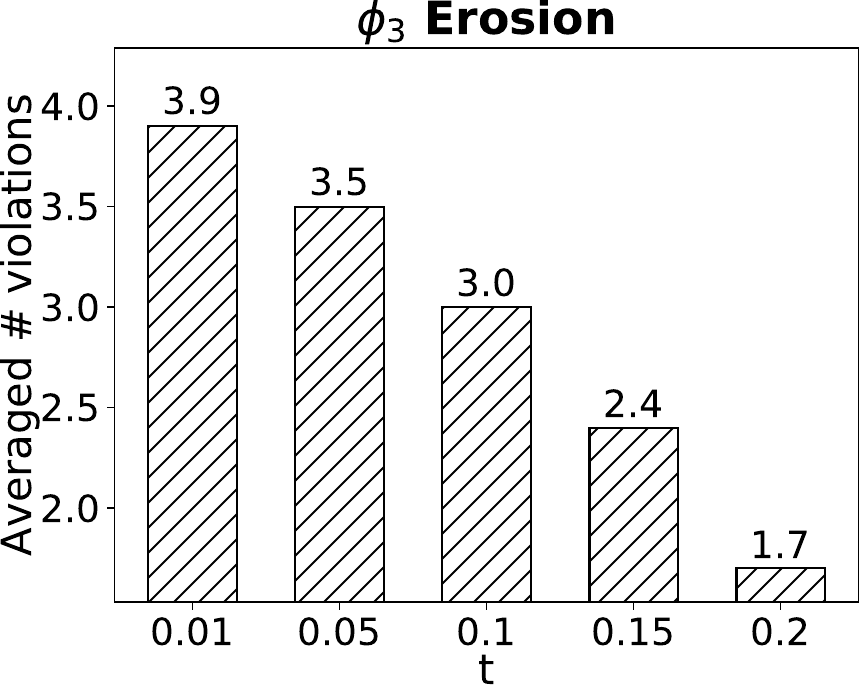}
\label{subfig:t_violation3}}
\hspace{-.7em}
\subfloat[]{
\includegraphics[width=0.16\textwidth, height=0.14\textwidth]{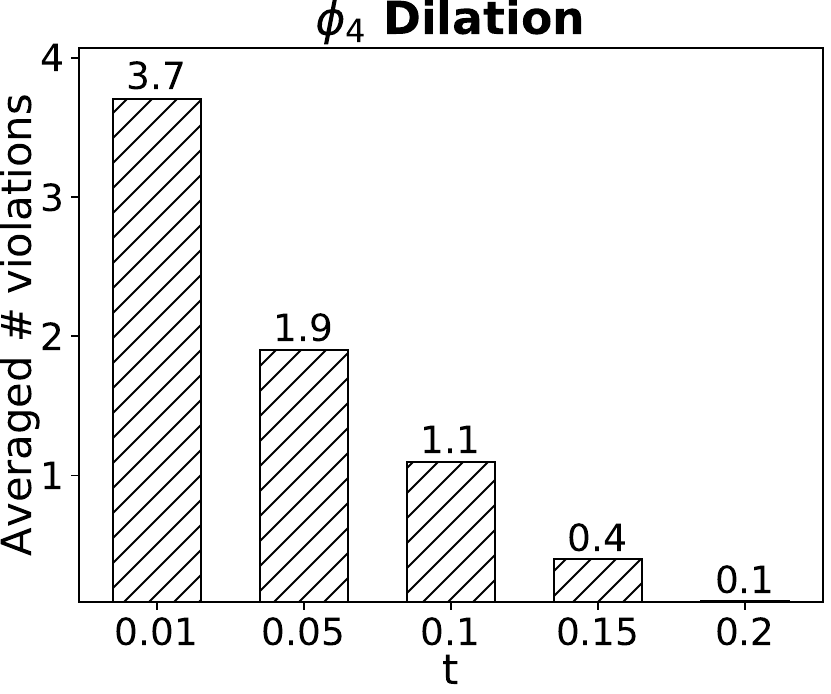}
\label{subfig:t_violation4}}
\hspace{-.7em}
\subfloat[]{
\includegraphics[width=0.16\textwidth, height=0.14\textwidth]{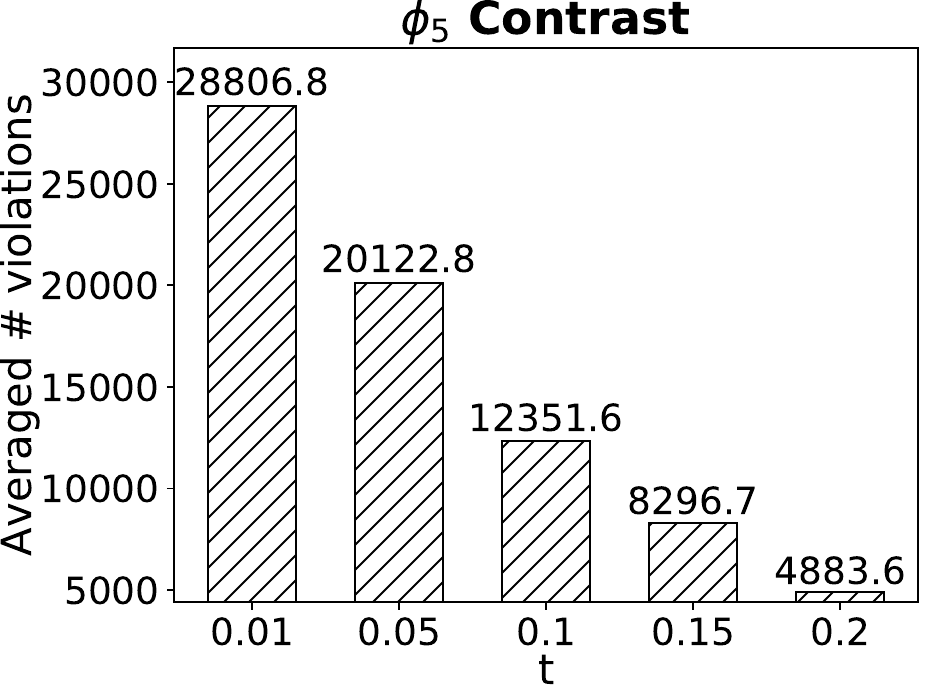}
\label{subfig:t_violation5}}
\hspace{-.7em}
\subfloat[]{
\includegraphics[width=0.16\textwidth, height=0.14\textwidth]{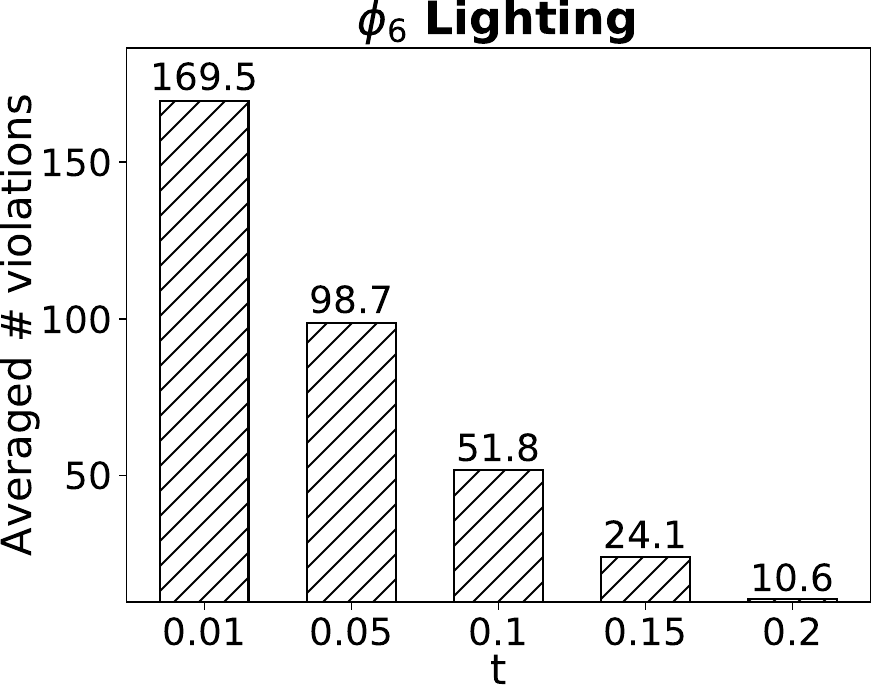}
\label{subfig:t_violation6}}

\subfloat[]{
\includegraphics[width=0.16\textwidth, height=0.14\textwidth]{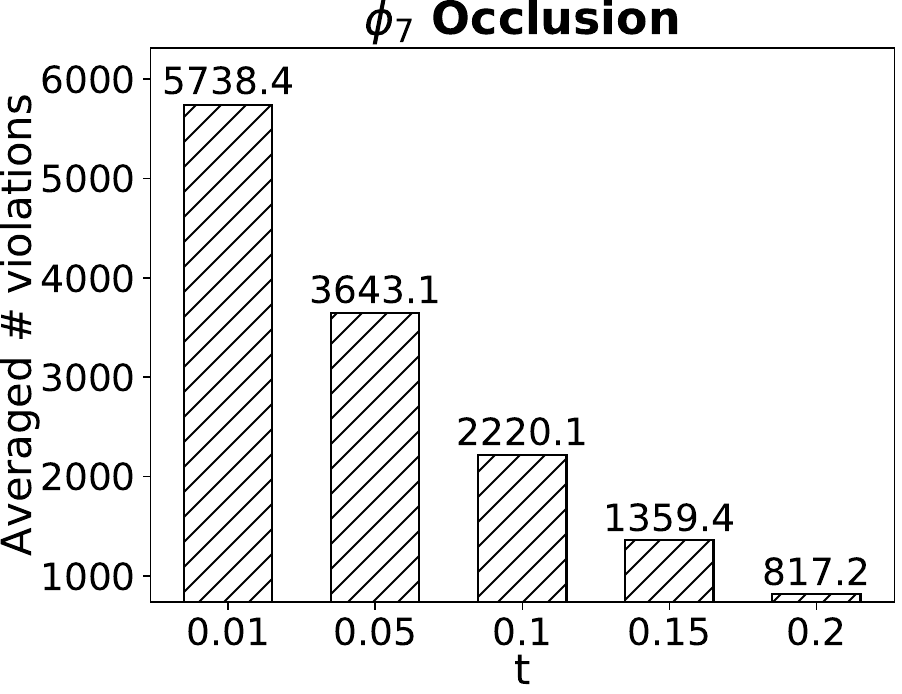}
\label{subfig:t_violation7}}
\hspace{-.7em}
\subfloat[]{
\includegraphics[width=0.16\textwidth, height=0.14\textwidth]{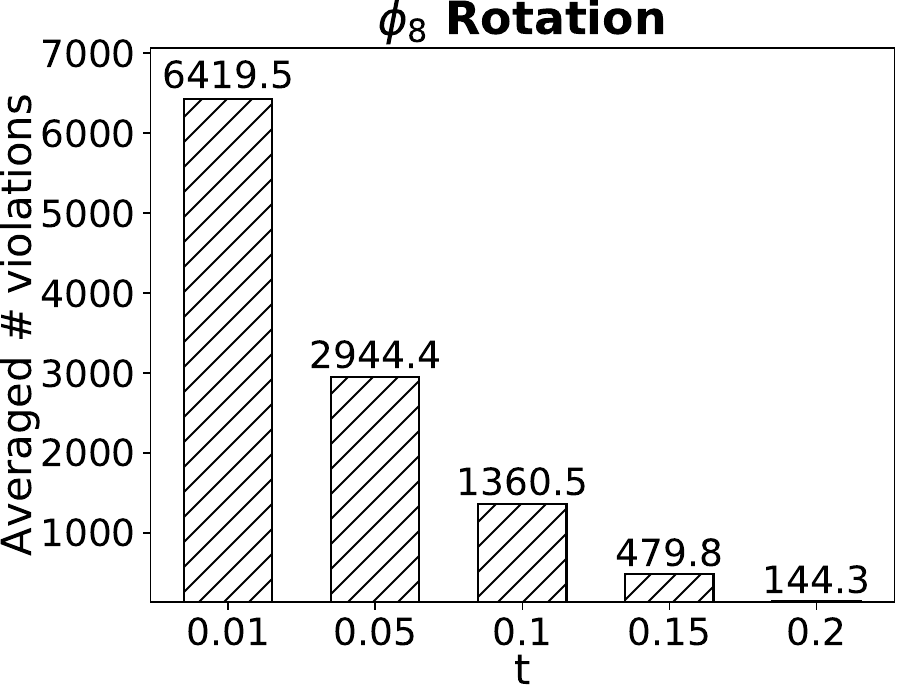}
\label{subfig:t_violation8}}
\hspace{-.7em}
\subfloat[]{
\includegraphics[width=0.16\textwidth, height=0.14\textwidth]{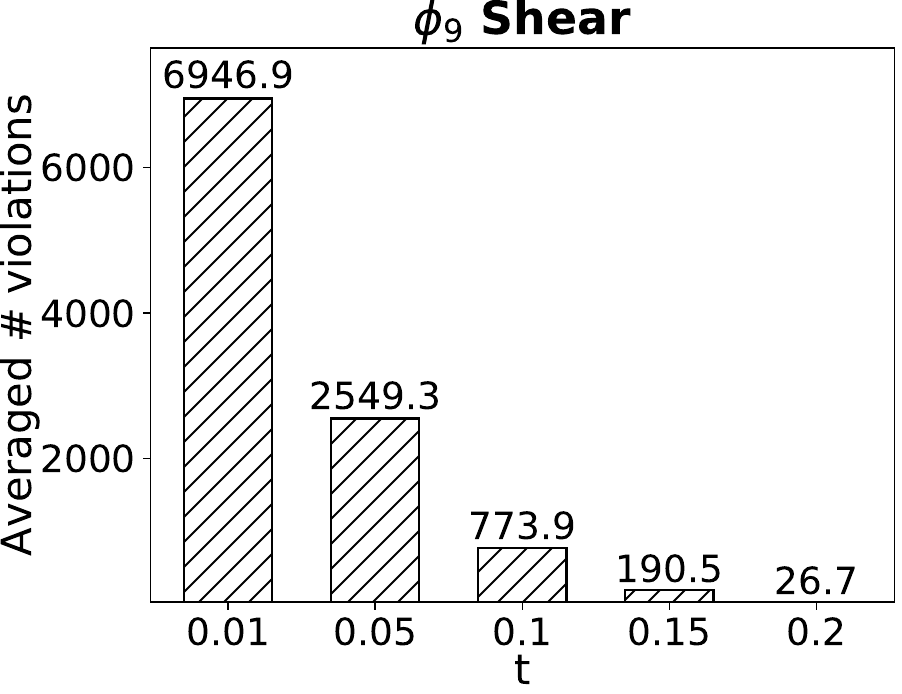}
\label{subfig:t_violation9}}
\hspace{-.7em}
\subfloat[]{
\includegraphics[width=0.16\textwidth, height=0.14\textwidth]{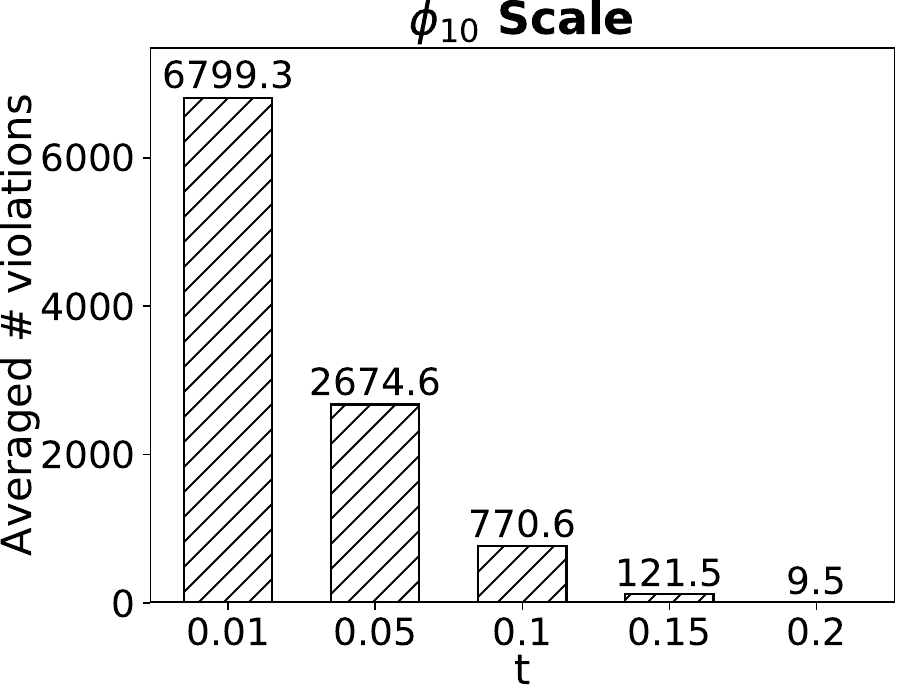}
\label{subfig:t_violation10}}
\hspace{-.7em}
\subfloat[]{
\includegraphics[width=0.16\textwidth, height=0.14\textwidth]{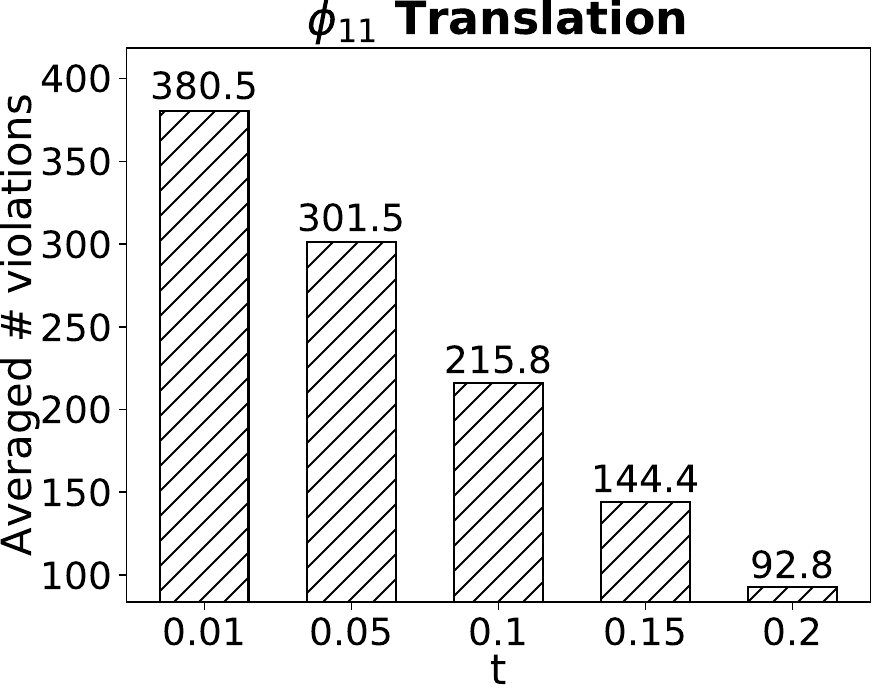}
\label{subfig:t_violation11}}
\hspace{-.7em}
\subfloat[]{
\includegraphics[width=0.16\textwidth, height=0.14\textwidth]{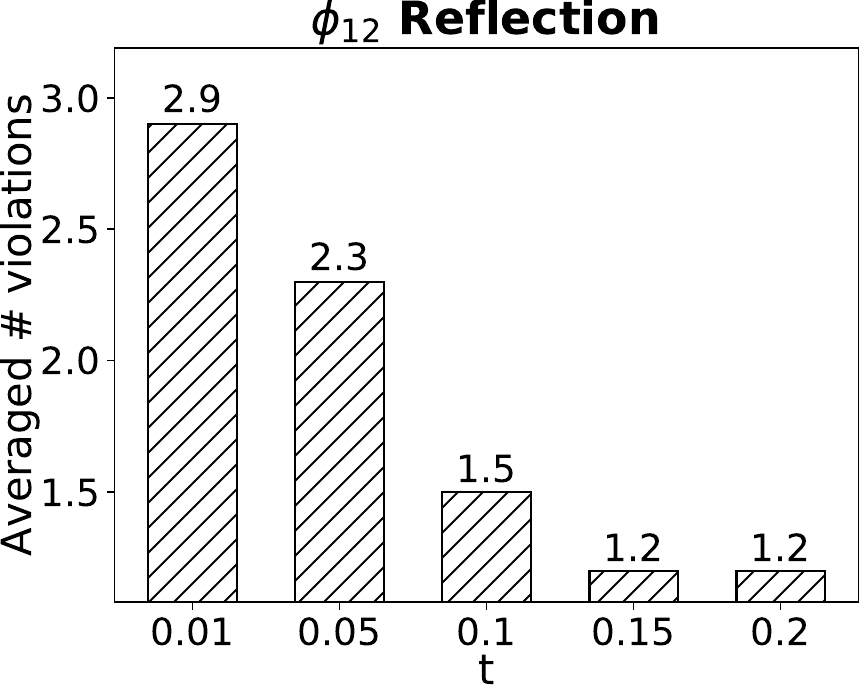}
\label{subfig:t_violation12}}

\caption{The change in the average number of violations of different safety properties as we increase $k$ (upper two rows) and $t$ (lower two rows) as defined in Section~\ref{sec:framework}. The number are averages over 10 images. The number of violations tend to decrease with increasing $k$ or $t$. The number above each bar shows the actual number of violations found for each $k$ or $t$ in IMG\_C3 and DRV\_C2 respectively.}
\label{fig:kt_violation}
\end{figure*}

\vspace{.1cm}\noindent\textbf{Effects of $k$ and $t$ on the number of violations.}
We present how the thresholds $k$ and $t$ of local safety properties defined in Section~\ref{sec:framework} influence the number of violations found by \sys. 
As shown in Figure~\ref{fig:kt_violation}, the number of violations decreases with increases in $k$ and $t$. This is intuitive as increasing $k$ and $t$ essentially increase the allowed margin of error for the vision systems. One interesting fact is that although the number of violations drops significantly when $k$ increases from 1 to 2, the changes in number of violations tend to be smaller when $k$ increases further. BY contrast, the decrease in the number of violations for different increasing values of t seems to be more uniform. 

\vspace{.1cm}\noindent\textbf{Violations for composition of transformations.}
We also explore the efficacy of \sys to verify safety properties involving composition of multiple transformation (e.g., $\phi_1$ and $\phi_6$).  For such cases, \sys computes $\mathbb{C}_{critical}$ for the new composite transformation by calculating the Cartesian product of the critical parameter values of each of the individual transformations.  Specifically, we run \sys for different compositions of $\phi_1$ (average smoothing), $\phi_6$ (lighting), and $\phi_{11}$ (translation).  

\begin{figure}[!t]
\centering
\includegraphics[width=0.8\columnwidth]{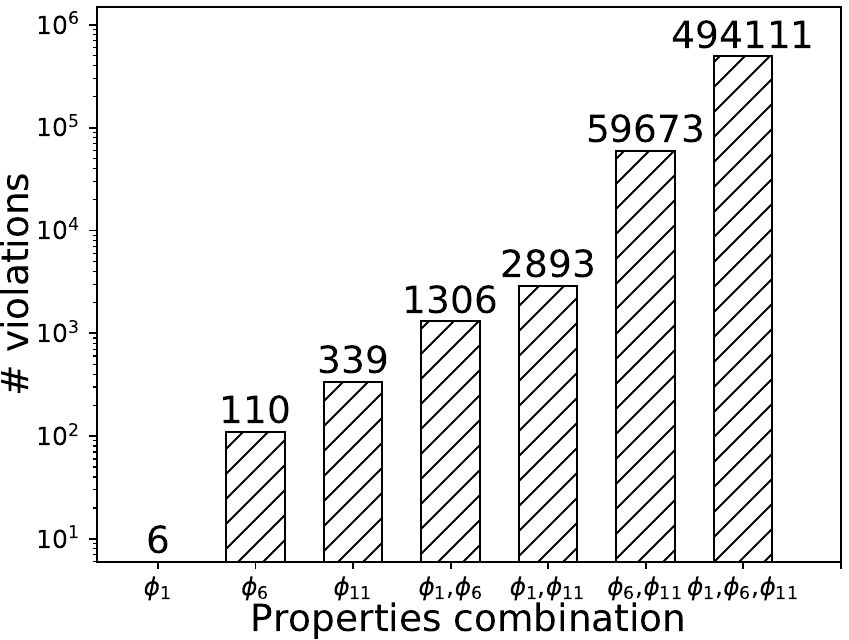}
\caption{Average number of violations found for properties $\phi_1$, $\phi_6$, and $\phi_{11}$ and their different compositions. We use  IMG\_C3 (MobileNet) as the target for verification.}
\label{fig:combination}
\end{figure}

Figure~\ref{fig:combination} shows the results averaged for ten random input images with IMG\_C3 as the verification target.  We find that the number of violations for composite transformations is larger than simply multiplying the number of violations for individual transformations. This indicates that the composition of different transformations result in new violations than combinations of the existing ones for each individual transformation. Therefore, verifying safety properties with composite transformations, besides individual transformations, is critical for safety- and security-critical vision systems.


\vspace{.1cm}\noindent\textbf{Distribution of violating parameter values.}
We also investigate how the violating parameter values are distributed in the parameter space. For example, we check whether the gain values for brightness transformation causing safety property violations follow some obvious patterns. Our results indicate that while some violating parameter values for transformations like average smoothing follow simple patterns (e.g., parameter values that are higher than a threshold cause violations), most of the transformations do not display any obvious patterns. 

\begin{figure}
\centering
\captionsetup[subfloat]{labelformat=empty}
\subfloat[$\phi_1$ (average smoothing) violation distribution (left) and the input (right)]{
\includegraphics[width=0.9\columnwidth, height=0.3\columnwidth]{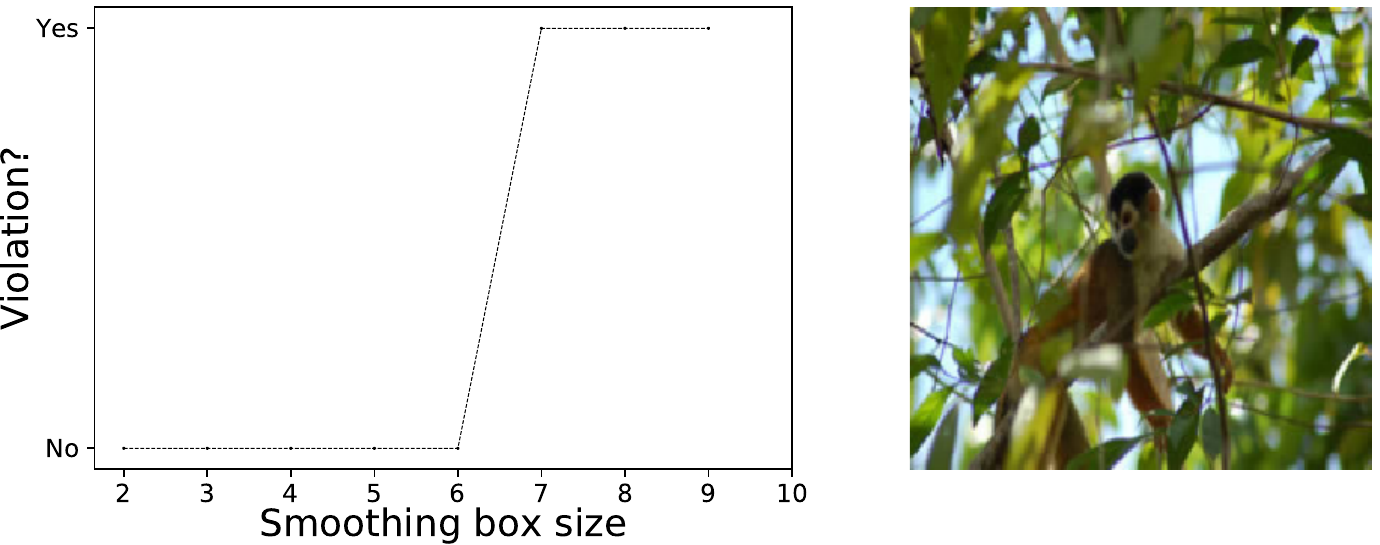}
\label{subfig:dist1}}

\subfloat[$\phi_6$ (lighting) violation distribution (left) and the input (right)]{
\includegraphics[width=0.9\columnwidth, height=0.3\columnwidth]{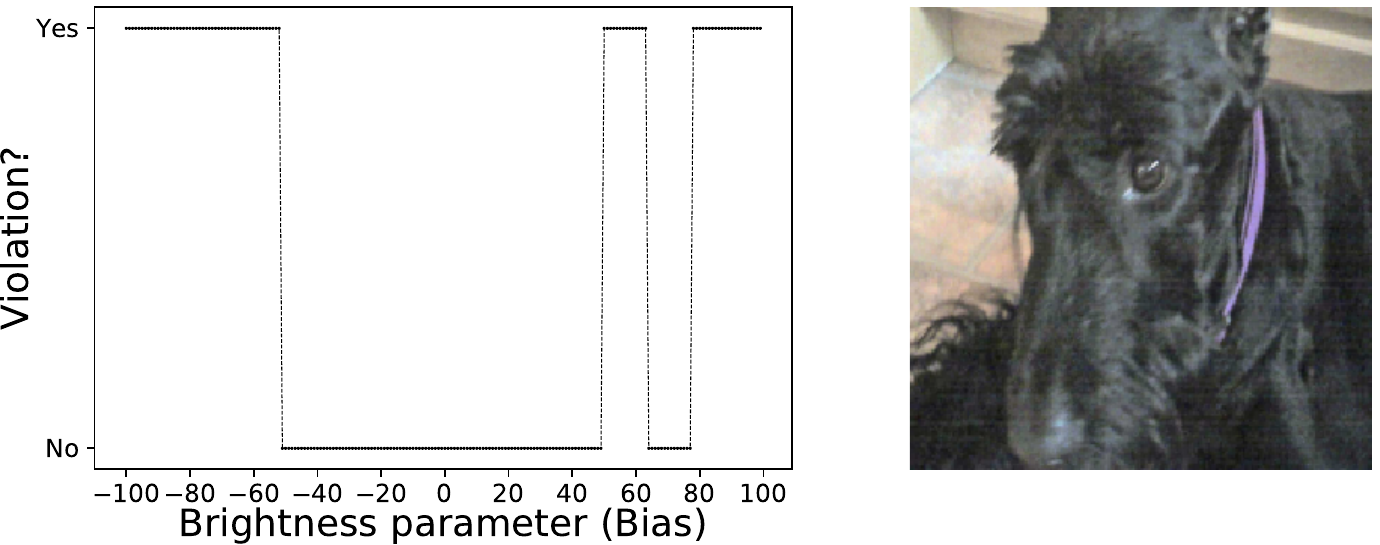}
\label{subfig:dist2}}

\caption{Distribution of the violations with increasing transformation parameter values for average smoothing ($\phi_1$) and lighting effect ($\phi_6$). ``Yes'' and ``No'' indicate the image transformed by this particular parameter violates and satisfy the safety property, respectively.}
\label{fig:distribution}
\end{figure}

Figure~\ref{fig:distribution} presents the results for IMG\_C3 (MobileNet) and two different transformations: average smoothing ($\phi_1$) and changing brightness ($\phi_6$) with the same experimental setting as those described in Section~\ref{subsec:violations_found}. We pick these two transformations as their parameter space in one-dimensional and relatively small enough for clear demonstration in a two-dimensional graph.  Figure~\ref{fig:distribution} shows that, for $\phi_1$, when the kernel size exceeds a threshold (\ie 6) for a specific input, the smoothing tends to always induce violations. However, by contrast, we do not find any obvious pattern among the  brightness parameter values for $\phi_6$ that result in violations.  

\vspace{.1cm}\noindent\textbf{Effect of increasing parameter space $\bm{\mathbb{C}_{\phi}}$ on violations.}
We show the number of violations found by \sys as the range of transformation space increases using Dave-orig (DRV\_C2) as the target vision system.  Figure~\ref{fig:bound_time_violation} shows that the number of violations increases as we increase the range due to increase in the number of critical parameter values that need to be checked. For example, the average number of violations for rotation ($\phi_8$) increases from $1360.5$ to $3573.1$ when we change the range of rotation degrees from $[-2,2]$ to $[-3,3]$.

\begin{figure*}
\centering
\captionsetup[subfloat]{captionskip=-.15cm,labelformat=empty}

\subfloat[]{
\includegraphics[width=0.16\textwidth, height=0.14\textwidth]{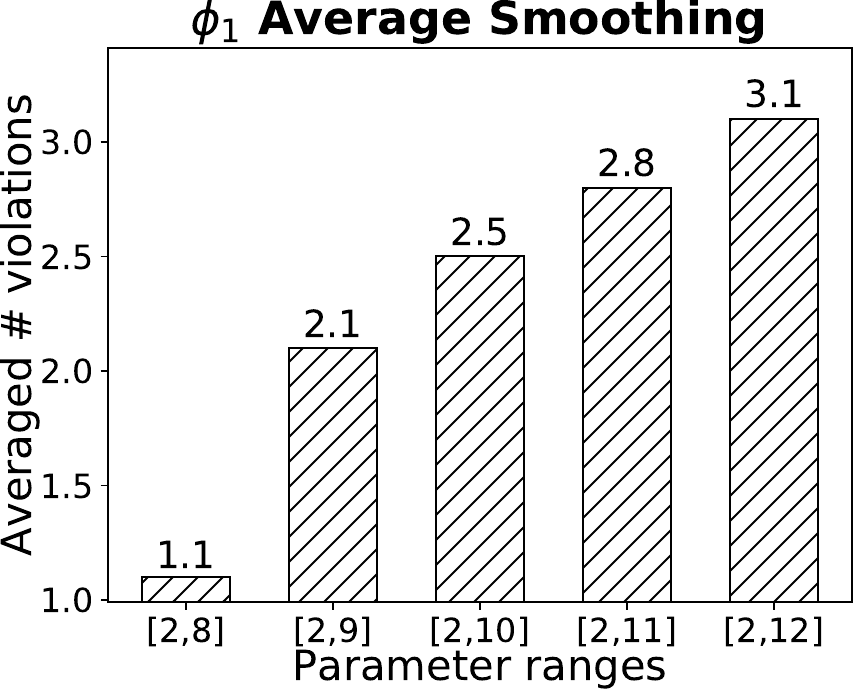}
\label{subfig:bound_violation1}}
\hspace{-.8em}
\subfloat[]{
\includegraphics[width=0.16\textwidth, height=0.14\textwidth]{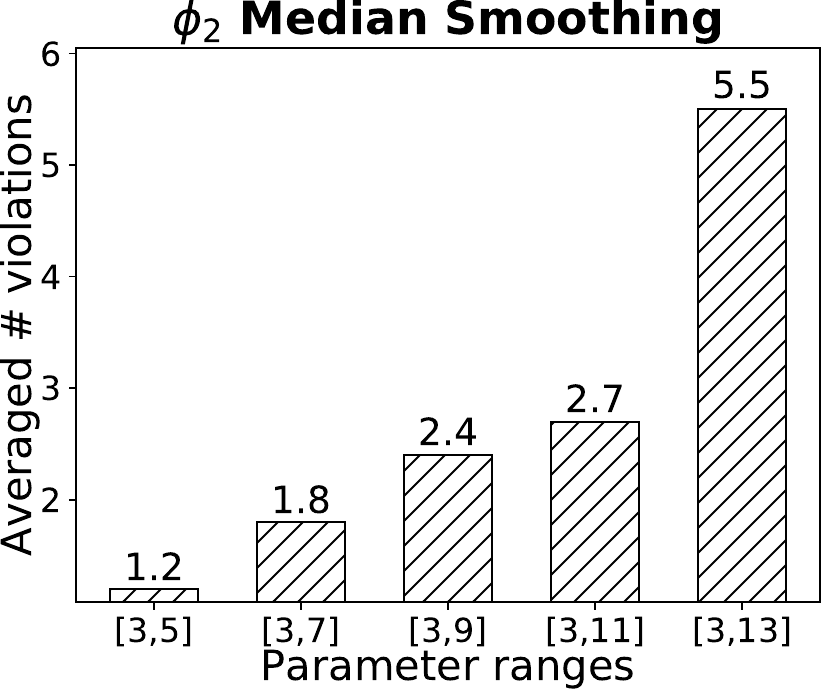}
\label{subfig:bound_violation2}}
\hspace{-.8em}
\subfloat[]{
\includegraphics[width=0.16\textwidth, height=0.14\textwidth]{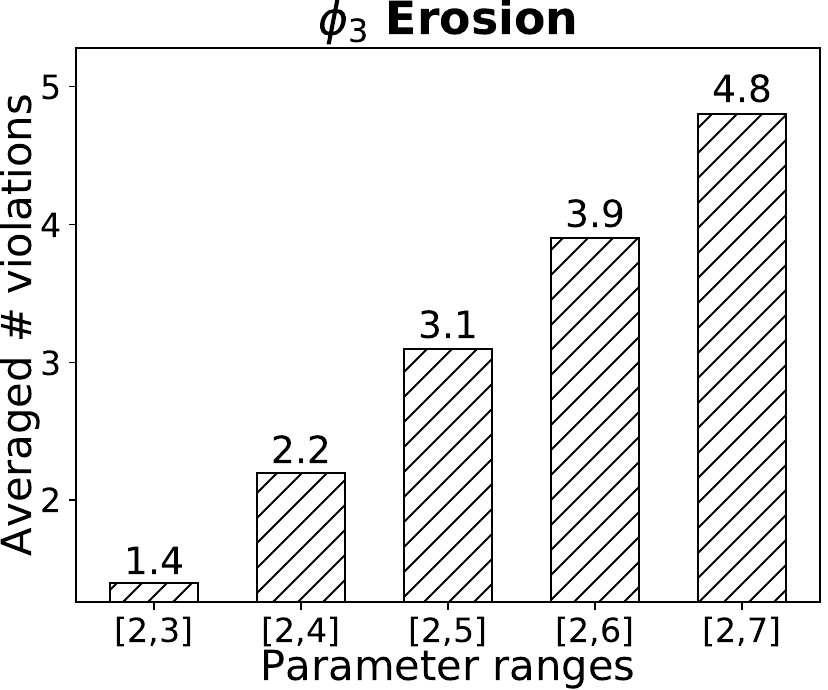}
\label{subfig:bound_violation3}}
\hspace{-.8em}
\subfloat[]{
\includegraphics[width=0.16\textwidth, height=0.14\textwidth]{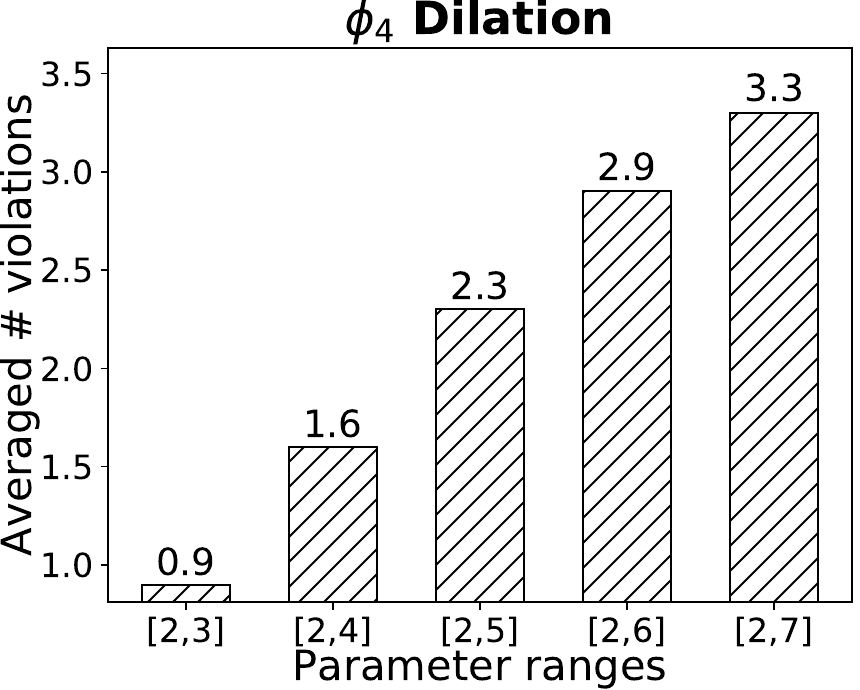}
\label{subfig:bound_violation4}}
\hspace{-.8em}
\subfloat[]{
\includegraphics[width=0.16\textwidth, height=0.14\textwidth]{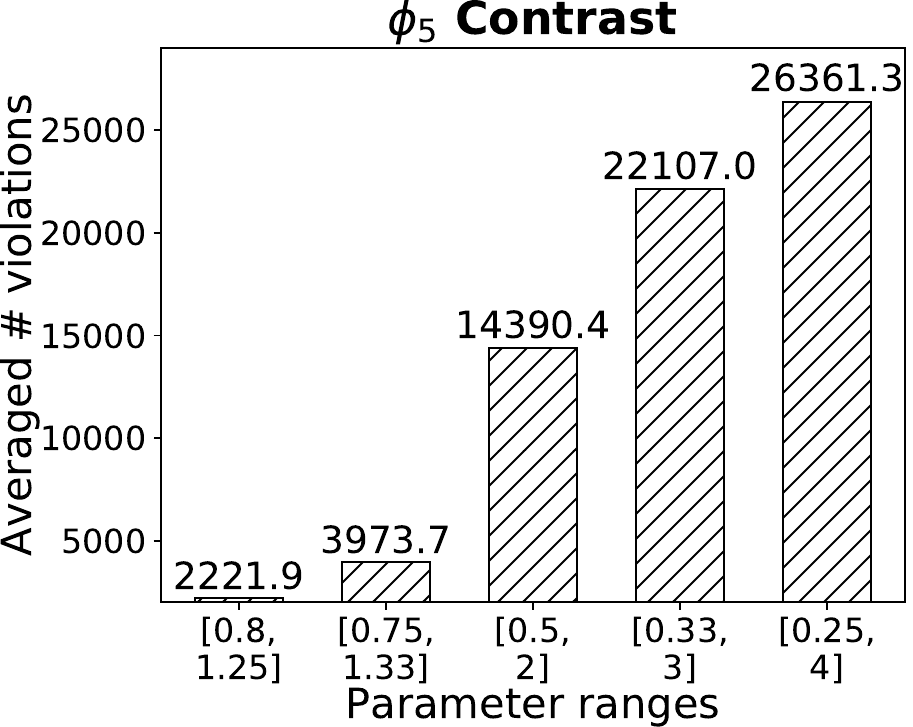}
\label{subfig:bound_violation5}}
\hspace{-.8em}
\subfloat[]{
\includegraphics[width=0.16\textwidth, height=0.14\textwidth]{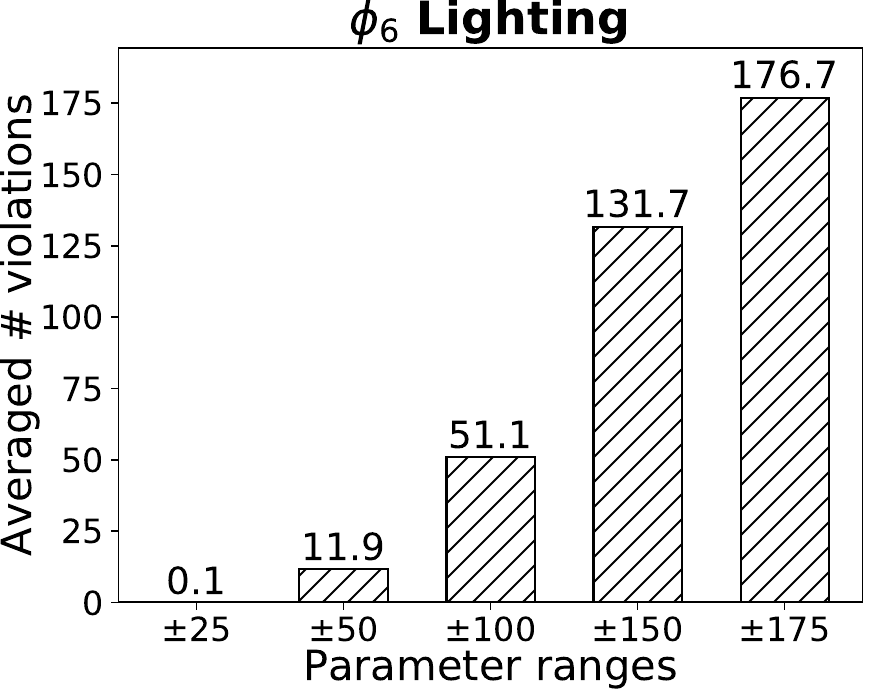}
\label{subfig:bound_violation6}}

\subfloat[]{
\includegraphics[width=0.16\textwidth, height=0.14\textwidth]{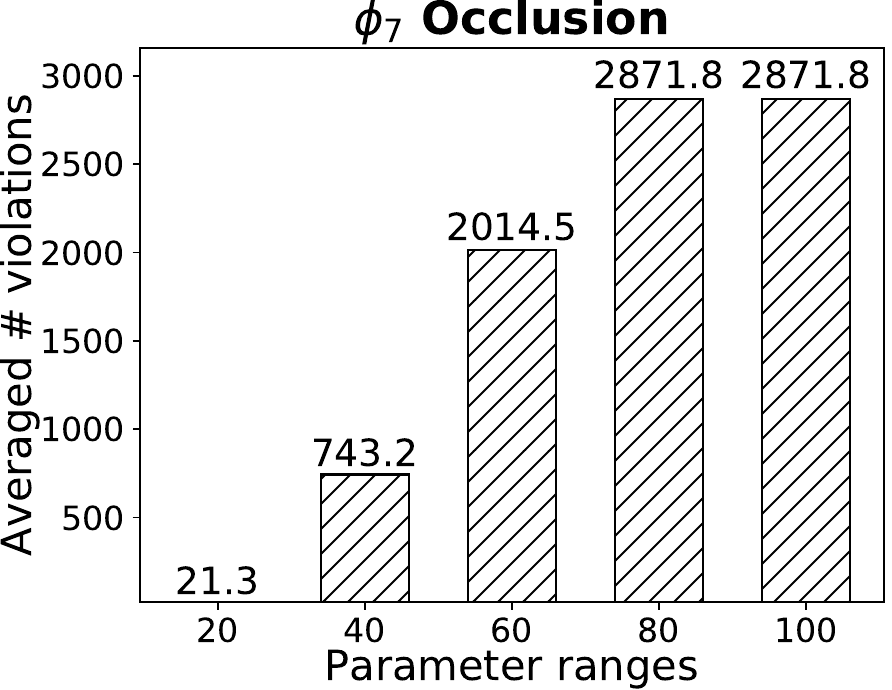}
\label{subfig:bound_violation7}}
\hspace{-.8em}
\subfloat[]{
\includegraphics[width=0.16\textwidth, height=0.14\textwidth]{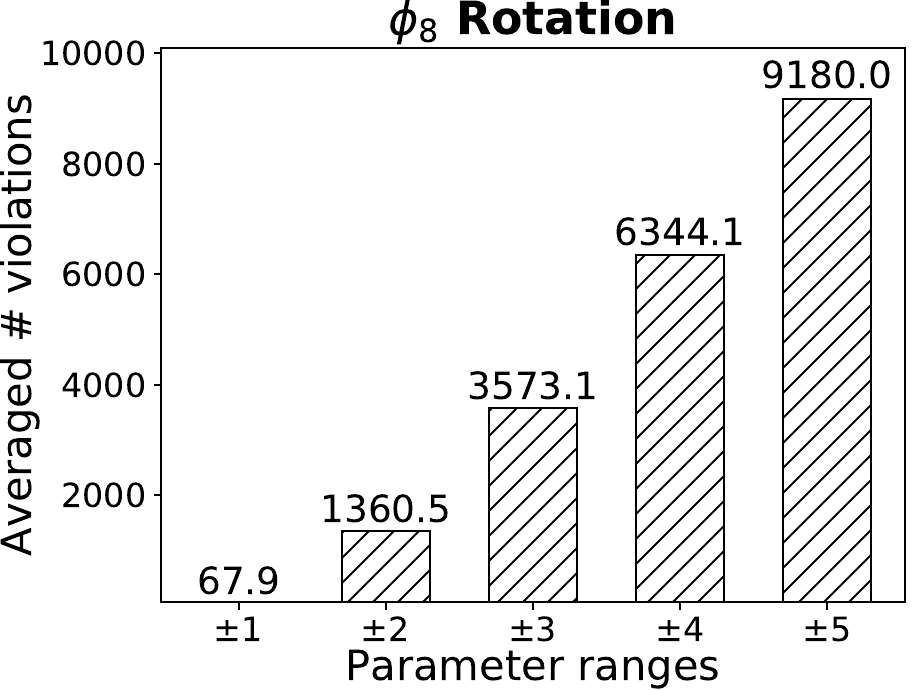}
\label{subfig:bound_violation8}}
\hspace{-.8em}
\subfloat[]{
\includegraphics[width=0.16\textwidth, height=0.14\textwidth]{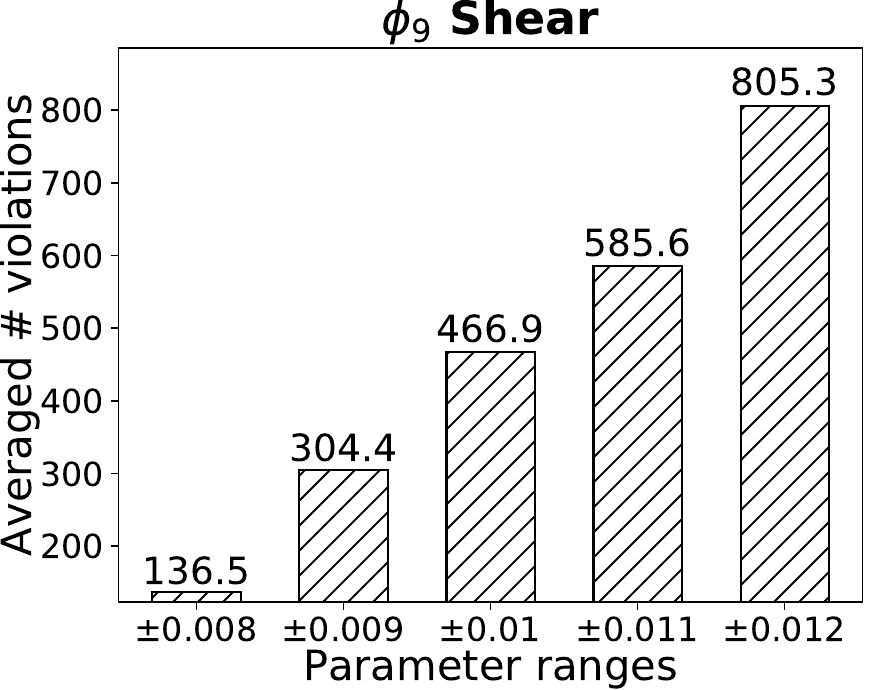}
\label{subfig:bound_violation9}}
\hspace{-.8em}
\subfloat[]{
\includegraphics[width=0.16\textwidth, height=0.14\textwidth]{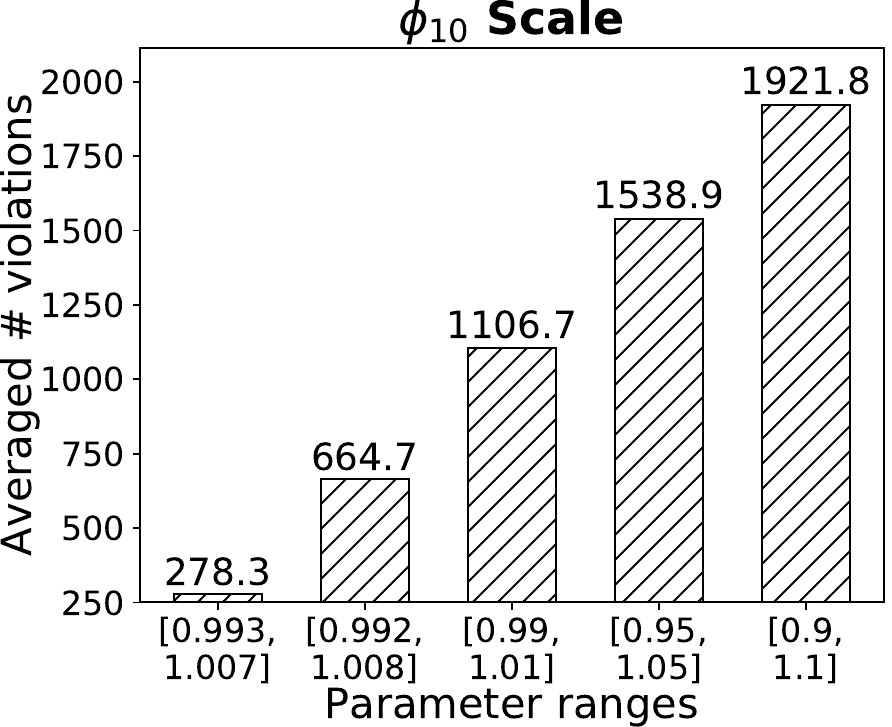}
\label{subfig:bound_violation10}}
\hspace{-.8em}
\subfloat[]{
\includegraphics[width=0.16\textwidth, height=0.14\textwidth]{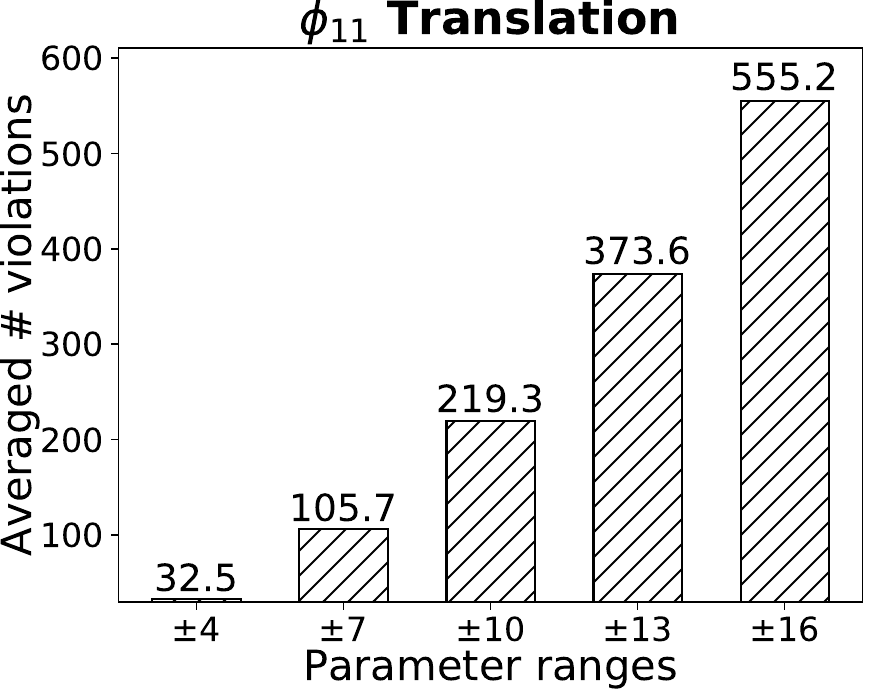}
\label{subfig:bound_violation11}}
\hspace{-.8em}
\subfloat[]{
\includegraphics[width=0.16\textwidth, height=0.14\textwidth]{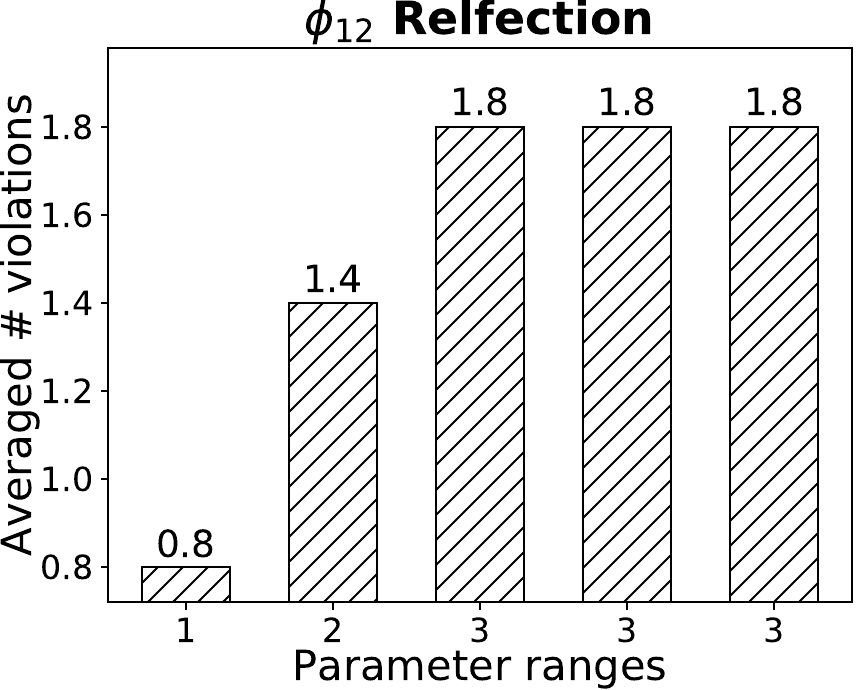}
\label{subfig:bound_violation12}}

\caption{The average numbers of violations found increase as we increase the range of the parameter space ($\mathbb{C}_{\phi}$). The number above each bar shows the exact number of violations with the corresponding bounds.}
\label{fig:bound_time_violation}
\end{figure*}


\vspace{.1cm}\noindent\textbf{Violations for complex transformations.}
Several real-world phenomena (e.g., fog, rain, etc.) that may affect input images are hard to replicate using the simple transformations described in the paper. However, one design custom transformations to mimic such effects and use \sys to check safety properties with these transformations. As an example of this approch, we demonstrate how a simple parameterized fog-simulating transformation can be designed and verified with \sys. For this transformation, we start with a fog mask, apply average smoothing on the mask, and apply the mask to the input image. By controlling the smoothing kernel size, we simulate different amounts of fog.

We use \sys to enumerate and check all critical parameter values for the fog transform described above,  i.e., different box sizes for average smoothing. \sys were able to find hundreds of violations in MobileNet (IMG C3), Google vision API (API C1), and dave-orig (DRV C2). Figure 9 shows three sample violations found by \sys. 

\begin{figure}
\centering
\captionsetup[subfloat]{labelformat=empty}
\subfloat[cougar]{
\includegraphics[width=0.3\columnwidth]{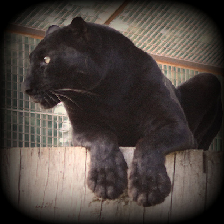}
\label{subfig:fog1}}
\subfloat[honeycomb]{
\includegraphics[width=0.3\columnwidth]{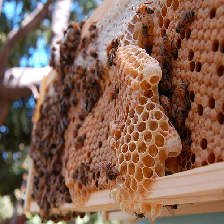}
\label{subfig:fog2}}
\subfloat[turn right]{
\includegraphics[width=0.3\columnwidth]{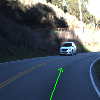}
\label{subfig:fog3}}

\subfloat[elephant]{
\includegraphics[width=0.3\columnwidth]{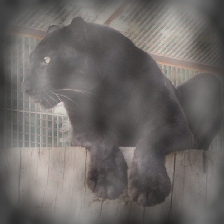}
\label{subfig:fog4}}
\subfloat[invertebrate]{
\includegraphics[width=0.3\columnwidth]{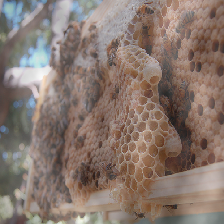}
\label{subfig:fog5}}
\subfloat[go straight]{
\includegraphics[width=0.3\columnwidth]{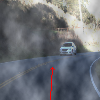}
\label{subfig:fog6}}

\caption{Violations found by \sys for a fog-simulating transformation in MobileNet (IMG\_C3), Google vision API (API\_C1), and dave-orig (DRV\_C2). The first and second rows show the original images and the foggy images that result in violations, respectively.}
\label{fig:fog}
\end{figure}

\subsection{Performance}
\label{subsec:performances}
In this section, we evaluate the performances of \sys in terms of the time it takes to verify different safety properties. We report the numbers for VGG-19 (IMG\_C2) and Rambo (DRV\_C1) as they have the slowest inference time and therefore illustrate the worst case behavior of \sys. All the test settings are the same as the ones described in Section~\ref{subsec:violations_found} and Table~\ref{tab:bounds} unless mentioned otherwise.

\vspace{.1cm}\noindent\textbf{Summary.}
Table~\ref{tab:total_time} shows the overall verification time required to verify each property for VGG-19 (IMG\_C2)  and Rambo (DRV\_C1) after adopting batch prediction. The total verification time of enumerating all possible critical parameter values per image varies from $0.3$ to $1863.5$ seconds.

\begin{table}[!htb]
\setlength{\tabcolsep}{2.1pt}
\scriptsize
\centering
\renewcommand{\arraystretch}{1.4}
\caption{The average verification time (in seconds) for verifying different properties for IMG\_C2 and DRV\_C1.}
\label{tab:total_time}
\begin{tabular}{|c|c|c|c|c|c|c|c|c|c|c|c|c|}
\hline
 & $\bm{\phi_1}$ & $\bm{\phi_2}$ & $\bm{\phi_3}$ & $\bm{\phi_4}$ & $\bm{\phi_5}$ & $\bm{\phi_6}$ & $\bm{\phi_7}$ & $\bm{\phi_8}$ & $\bm{\phi_9}$ & $\bm{\phi_{10}}$ & $\bm{\phi_{11}}$ & $\bm{\phi_{12}}$ \\ \hline
\textbf{IMG\_C2} & 1.4 & 1.1 & 1.1 & 1.1 & 225.3 & 5 & 238.4 & 714.5 & 1863.5 & 1798.9 & 8.6 & 1.1 \\ \hline
\textbf{DRV\_C1} & 0.3 & 0.8 & 0.8 & 0.8 & 75.1 & 1.6 & 81.2 & 242.1 & 596.6 & 596.1 & 3 & 0.7 \\ \hline
\end{tabular}
\end{table}

We find that the verification time primarily depends on the number of critical parameter values for a transformation as the image transformation operation is significantly cheaper than the testing time for each transformed image.  For example, verification $\phi_1$  (9 critical parameter values) takes only 0.3 seconds but $\phi_5$ (32,512 critical parameter values) takes around 75.1 seconds for DRV\_C1 as shown in Table~\ref{tab:total_time}. 

\vspace{.1cm}\noindent\textbf{Performance improvement with batch prediction.}
As described in Section~\ref{sec:impl}, \sys uses batch prediction~\cite{chollet2015keras}, using both GPU and CPU, to speed up the verification process by allowing vision systems to predict a batch of images in parallel.  Table~\ref{tab:dataset_dnn} shows batch prediction can speed up verification by upto 17.6$\times$ times for VGG-19 (IMG\_C2), and 197.6$\times$ times for Rambo (DRV\_C1). 

\begin{table}[!htb]
\setlength{\tabcolsep}{10pt}
\footnotesize
\centering
\renewcommand{\arraystretch}{1.2}
\caption{The average running time (millisecond per image) of \sys with and without batch prediction on the VGG-19 (IMG\_C2) and Rambo (DRV\_C1). The speedup is shown in the last column.}
\label{tab:speedup}
\begin{tabular}{|c|c|c|c|}
\hline
 & \textbf{Baseline} & \textbf{Batch pred.} & \textbf{Speed-up}\\ \hline
\textbf{IMG\_C2} & 103.9 & 5.9 & 17.6$\times$ \\ \hline
\textbf{DRV\_C1} & 33.6 & 0.17 & 197.6$\times$ \\ \hline
\end{tabular}
\end{table}


\subsection{Improving robustness with retraining on safety-violating images}
In this subsection, we investigate whether the robustness of the tested computer vision systems against these transformations can be improved by retraining the affected systems using the violations found by \sys. In particular, we borrow the idea of adversarial retraining as a data augmentation technique introduced by Goodfellow et al.~\cite{goodfellow2014explaining} to retrain the ML models on transformed images that induce safety violations. We pick VGG-16 for this experiment due to its large number of safety violations (the largest Top-1 loss as shown in Table~\ref{tab:dataset_dnn}).
 
\begin{figure}
\centering
\captionsetup[subfloat]{captionskip=-.15cm,labelformat=empty}
\subfloat[]{
\includegraphics[width=0.32\columnwidth]{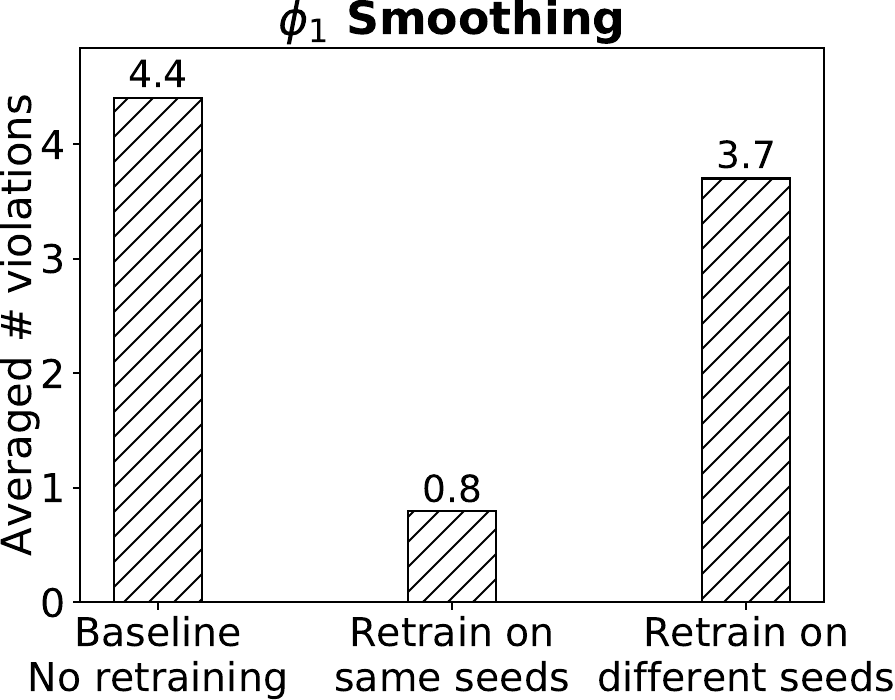}
\label{subfig:retrain1}}
\hspace{-.9em}
\subfloat[]{
\includegraphics[width=0.32\columnwidth]{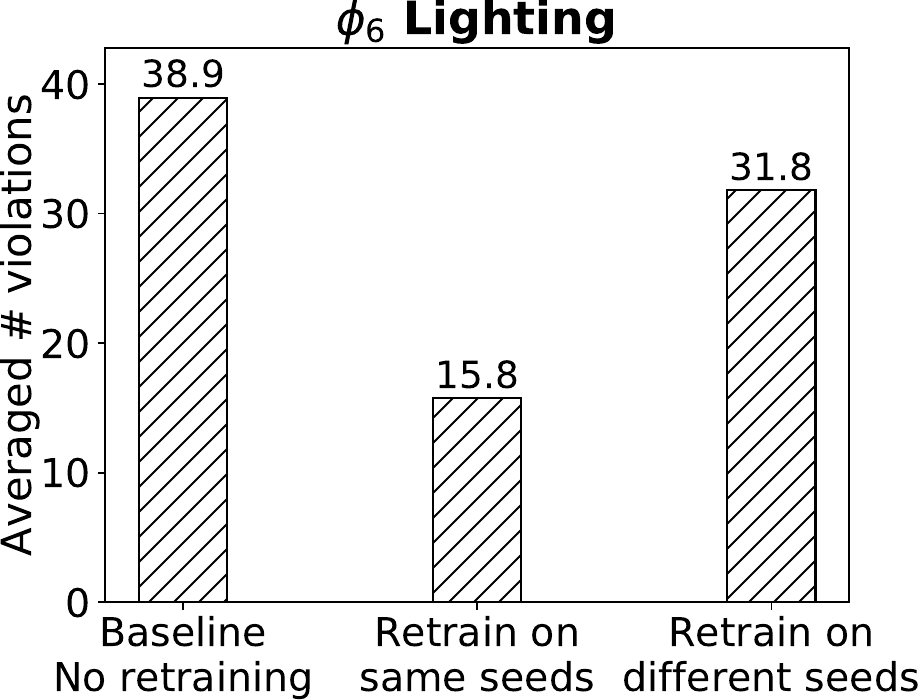}
\label{subfig:retrain2}}
\hspace{-.9em}
\subfloat[]{
\includegraphics[width=0.32\columnwidth]{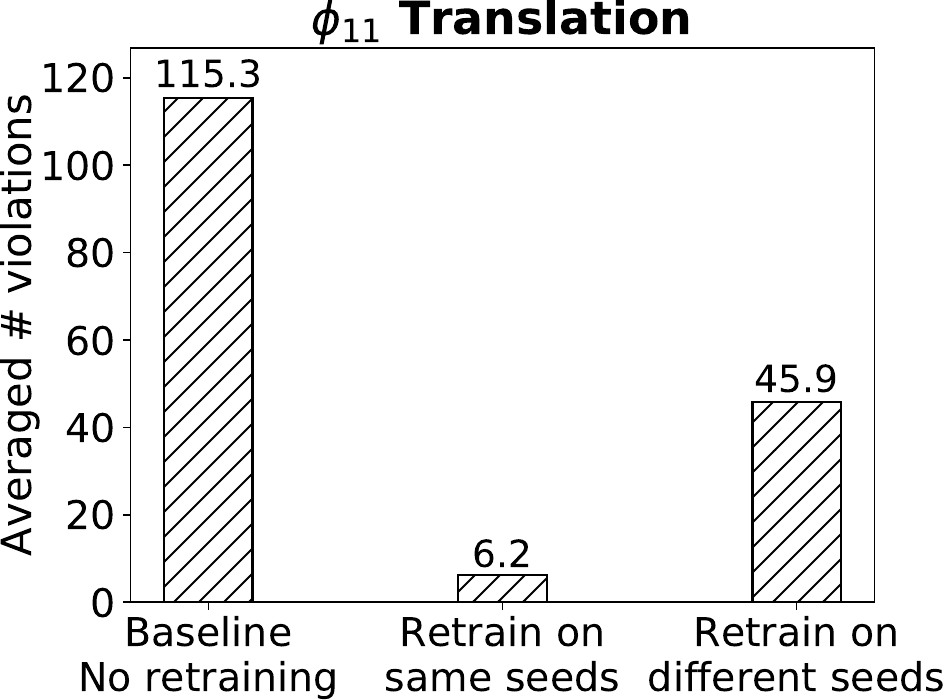}
\label{subfig:retrain3}}

\caption{The number of average violations found before and after retraining of IMG\_C1 on violations of safety properties $\phi_1$, $\phi_6$, and $\phi_{11}$. The left bar shows the baseline without retraining. The middle and right bars show after retraining on (1) violations generated from same images as those ten test images, and (2) violations generated from different images, respectively. Even in the latter case, the number of violations can drop by up to 60\%.}
\label{fig:retrain}
\end{figure}

We compare the violations found by \sys for the retrained models and the original ones using ten randomly selected images from the ILSVRC test set. We use two different training data augmentation strategies: (1) retraining on the safety-violating images generated from \textit{the test images} and (2) retraining on the safety-violating images generated from \textit{different images} than the test images. Note that the the second approach is less likely to overfit than the first approach. Figure~\ref{fig:retrain} shows the number of violations found for randomly-drawn 10 testing images before and after retraining the VGG-16 with our two tactics. The results show that both approaches of retraining described above reduce the number of violations and therefore improve the robustness of the ML model. In addition to apparent drop of number of violations for retraining on same test images, the number of violations are also reduced by up to 60.2\% for retraining on different images.



%


\section{Related Work}
\label{sec:related}
\noindent{\bf Testing and verification of machine learning.}
\label{subsec:robust_ml}
Given the deployment of ML systems in security- and safety-critical settings, some recent studies have focused on generating diverse and realistic corner-case test inputs for testing ML systems and have found numerous incorrect behavior in state-of-the-art systems~\cite{pei2017deepxplore,tian2017deeptest}. However, none of these systems, unlike \sys, can provide any guarantee about non-existence of safety violations.

Several researchers have also recently started exploring the possibility of whitebox formal verification of DNNs and ensure correct behavior under different settings~\cite{pulina2010abstraction, huang2017safety, katz2017reluplex}. Unlike \sys, these techniques either fail to provide strong guarantees on state-of-the-art ML systems, or they cannot scale to larger ML systems (\ie neural network with thousands of neurons). For example, Pulina et al.~\cite{pulina2010abstraction} was able to demonstrate that the output class of DNNs with only one hidden layer remains constant for different neighboring inputs.
Huang et al.~\cite{huang2017safety} extended this approach by discretizing the input space and then propagating and tracking output through each layer of a DNN. While not restricted to small neural net, the discretization assumption cannot provide strong guarantee about non-existence of erroneous inputs. 
Katz et al.~\cite{katz2017reluplex} leverage linear programming solvers and extend simplex algorithm to verify some properties of simple DNNs using ReLUs~\cite{nair2010rectified} as activation functions. However, they can only support a specific type of neural network with only few input features and neurons.

\noindent{\bf Adversarial machine learning.}
Adversarial machine learning is a popular research area that focuses on finding vulnerabilities of the ML models by generating error-inducing test inputs by adding minimal perturbations to an existing input~\cite{szegedy2013intriguing,laskov2014practical,goodfellow2014explaining,nguyen2015deep,xu2016automatically,wilber2016can, sharif2016accessorize,papernot2016crafting,liu2016delving} and studying how to improve the robustness of the ML systems against such attacks~\cite{gu2014towards, wang2014man, papernot2016distillation, bastani2016measuring, zheng2016improving, huang2017safety, carlini2017towards, cisse2017parseval, metzen2017detecting}. We refer the interested readers to the survey by Papernot et al.~\cite{papernot2016towards} for more details on these works.

The key difference between \sys and this line of work is twofold: (1) adversarial inputs only focus \textit{generating one type of test case} by adding adversarial noises/perturbations) and check if target ML models behave unexpectedly. By contrast, \sys is a general safety property verification framework that supports a broad set of safety properties with more realistic transformations that even weak attackers might be able to induce; and (2) unlike any of the adversarial machine learning projects, \sys can ensure non-existence of inputs that violate a given safety property for state-of-the-art ML systems.




\section{Conclusion}
\label{sec:conclusion}

In this paper, we have formulated a general framework for verifying the robustness of ML systems regarding different real-world safety properties which can model different attacker capabilities. 
We have designed, implemented, and extensively evaluated \sys, a scalable verification system that can verify a diverse set of safety properties for state-of-the-art computer vision systems with only blackbox access.
By defining formal decomposition framework for image transformation, \sys is able to reduce the potentially continuous input space into finite, discrete, and polynomial number of critical states.
This state reduction enabled efficient model-checking the critical states such that \sys found thousands of safety violations in fifteen vision systems including ten state-of-the-art DNNs and five commercial third-party vision APIs for twelve different safety properties. 
In addition, \sys showed substantial improvement over existing testing and verification of ML systems in terms of both formal guarantees and the scope of modeling attacker capabilities.
Future work includes extending \sys to other application domains and incorporating whitebox access (\ie model parameters) to further reduce the size of verification state space for some safety properties that are super-polynomial or exponential.

\begin{table*}[!b]
\setlength{\tabcolsep}{10pt}
\footnotesize
\centering
\renewcommand{\arraystretch}{0.5}
\caption{Violations of transformation-invariant properties on Udacity driving dataset for different models and transformations.}
\label{tab:violations_car}
\begin{tabular}{c|cc|cc|c}
\multirow{2}{*}{\textbf{Model ID}} & \multicolumn{2}{c|}{\textbf{Original}} & \multicolumn{2}{c|}{\textbf{Transformed}} & \multicolumn{1}{c}{\multirow{2}{*}{\textbf{Violated}}} \\
 & \textit{Image} & \textit{Steering angle prediction} & \textit{Image} & \textit{Steering angle prediction} & \multicolumn{1}{c}{} \\ \hline
DRV\_C1 & \vspace{.01cm} \includegraphics[width=0.2\columnwidth,height=0.2\columnwidth,valign=c]{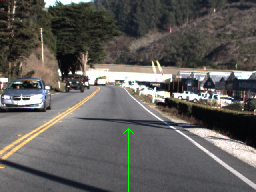}\vspace{.01cm} 
& \begin{tabular}[c]{@{}c@{}}near straight (left 0.1$^\circ$)\end{tabular} & \vspace{.01cm} \includegraphics[width=0.2\columnwidth,height=0.2\columnwidth,valign=c]{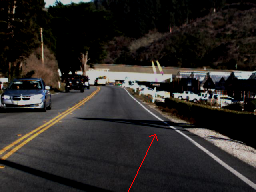}\vspace{.01cm} 
& turn right (right $25^\circ$) & $\phi_6$ \\ \hline
DRV\_C2 & \vspace{.01cm} \includegraphics[width=0.2\columnwidth,height=0.2\columnwidth,valign=c]{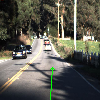}\vspace{.01cm} & \begin{tabular}[c]{@{}c@{}}near straight (right $4^\circ$)\end{tabular} & \vspace{.01cm} \includegraphics[width=0.2\columnwidth,height=0.2\columnwidth,valign=c]{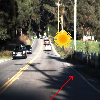}\vspace{.01cm} & turn right (right $44^\circ$) & $\phi_7$ \\ \hline
DRV\_C3 & \vspace{.01cm} \includegraphics[width=0.2\columnwidth,height=0.2\columnwidth,valign=c]{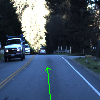}\vspace{.01cm} & \begin{tabular}[c]{@{}c@{}}near straight (left $3.7^\circ$)\end{tabular} & \vspace{.01cm} \includegraphics[width=0.2\columnwidth,,height=0.2\columnwidth,valign=c]{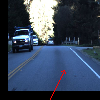}\vspace{.01cm} & turn right (right $27^\circ$) & $\phi_{11}$ \\ \hline
DRV\_C4 & \vspace{.01cm} \includegraphics[width=0.2\columnwidth,height=0.2\columnwidth,valign=c]{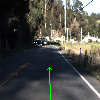}\vspace{.01cm} & \begin{tabular}[c]{@{}c@{}}near straight (left $0.03^\circ$)\end{tabular} & \vspace{.01cm} \includegraphics[width=0.2\columnwidth,height=0.2\columnwidth,valign=c]{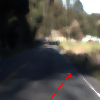}\vspace{.01cm} & turn right (right $38^\circ$) & $\phi_2$  \\ \hline
\end{tabular}
\end{table*}

\bibliographystyle{abbrv}
\bibliography{paper}

\begin{appendix}
\section{Evaluation and Results}

\subsection{Sample errors found by \sys}

Table~\ref{tab:violations_imagenet} and \ref{tab:violations_car} illustrates some of sample inputs found by \sys that induce errors of invariance properties of the vision systems we have tested.
It clearly shows that a simple transformation, which human should not have difficulty in producing correct decisions (\ie recognition or steering) on the transformed inputs, can mislead the state-of-the-art models to output (obvious) incorrect decisions.

\begin{table*}[!t]
\setlength{\tabcolsep}{12pt}
\scriptsize
\centering
\renewcommand{\arraystretch}{1.1}
\caption{Sample errors of safety properties found by \sys in different vision systems trained on ImageNet}
\label{tab:violations_imagenet}
\begin{tabular}{c|cc|cc|c}
\multirow{2}{*}{\textbf{Model ID}} & \multicolumn{2}{c|}{\textbf{Original}} & \multicolumn{2}{c|}{\textbf{Transformed}} & \multicolumn{1}{c}{\multirow{2}{*}{\textbf{Violated}}} \\
 & \textit{Image} & \textit{Labels (Top-5)} & \textit{Image} & \textit{Label (Top-1)} & \multicolumn{1}{c}{} \\ \hline
IMG\_C1 & \vspace{.01cm} \includegraphics[width=0.2\columnwidth,height=0.2\columnwidth,valign=c]{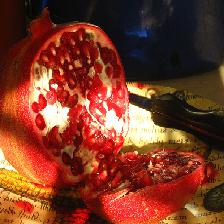}\vspace{.01cm} 
& \begin{tabular}[c]{@{}c@{}}pomegranate\\ bell\_pepper\\ acorn\_squash\\ pizza\\ trifle\end{tabular} & \vspace{.01cm} \includegraphics[width=0.2\columnwidth,height=0.2\columnwidth,valign=c]{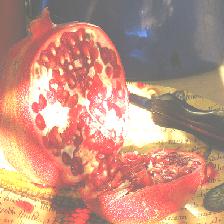}\vspace{.01cm} 
& starfish & $\phi_6$ \\ \hline
IMG\_C2 & \vspace{.01cm} \includegraphics[width=0.2\columnwidth,height=0.2\columnwidth,valign=c]{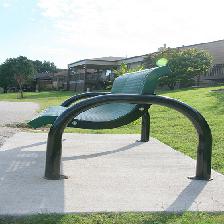}\vspace{.01cm} & \begin{tabular}[c]{@{}c@{}}park\_bench\\ sundial\\ cannon\\ bannister\\ plow\end{tabular} & \vspace{.01cm} \includegraphics[width=0.2\columnwidth,height=0.2\columnwidth,valign=c]{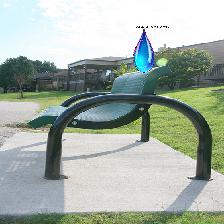}\vspace{.01cm} & peacock & $\phi_7$ \\ \hline
IMG\_C3 & \vspace{.01cm} \includegraphics[width=0.2\columnwidth,height=0.2\columnwidth,valign=c]{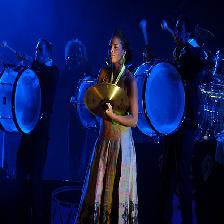}\vspace{.01cm} & \begin{tabular}[c]{@{}c@{}}stage\\ cornet\\ trombone\\ cello\\ violin\\ bathtub\end{tabular} & \vspace{.01cm} \includegraphics[width=0.2\columnwidth,,height=0.2\columnwidth,valign=c]{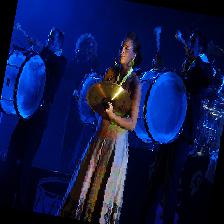}\vspace{.01cm} & jellyfish & $\phi_8$ \\ \hline
IMG\_C4 & \vspace{.01cm} \includegraphics[width=0.2\columnwidth,height=0.2\columnwidth,valign=c]{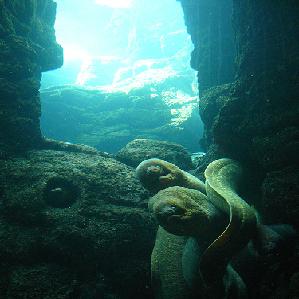}\vspace{.01cm} & \begin{tabular}[c]{@{}c@{}}sea\_snake\\ scuba\_diver\\ sea\_lion\\ coral\_reef\\ brain\_coral\end{tabular} & \vspace{.01cm} \includegraphics[width=0.2\columnwidth,height=0.2\columnwidth,valign=c]{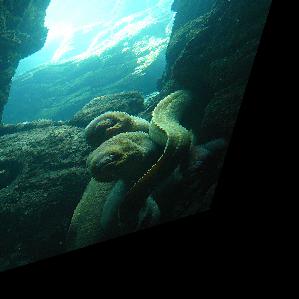}\vspace{.01cm} & wreck & $\phi_9$  \\ \hline
IMG\_C5 & \vspace{.01cm}  \includegraphics[width=0.2\columnwidth,height=0.2\columnwidth,valign=c]{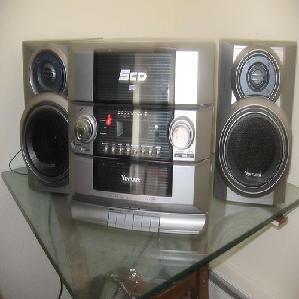}\vspace{.01cm} & \begin{tabular}[c]{@{}c@{}}cassette\_player\\ tape\_player\\ radio\\ CD\_player\\ loudspeaker\end{tabular} & \vspace{.01cm} \includegraphics[width=0.2\columnwidth,height=0.2\columnwidth,valign=c]{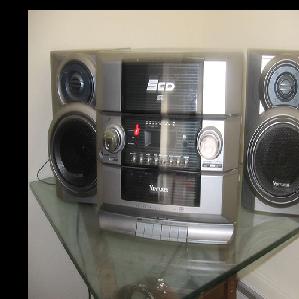}\vspace{.01cm} & Polaroid\_camera & $\phi_{11}$  \\ \hline
IMG\_C6 & \vspace{.01cm}  \includegraphics[width=0.2\columnwidth,height=0.2\columnwidth,valign=c]{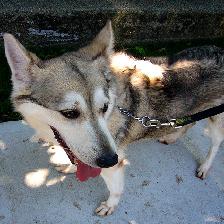}\vspace{.01cm} & \begin{tabular}[c]{@{}c@{}}Eskimo\_dog\\ Siberian\_husky\\ malamute\\ Norwegian\_elkhoun\\ Pembroke\end{tabular} & \vspace{.01cm} \includegraphics[width=0.2\columnwidth,height=0.2\columnwidth,valign=c]{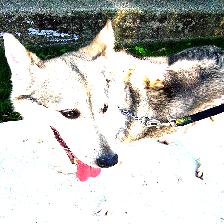}\vspace{.01cm} & Cardigan & $\phi_5$ \\ \hline
API\_C1 & \vspace{.01cm}  \includegraphics[width=0.2\columnwidth,height=0.2\columnwidth,valign=c]{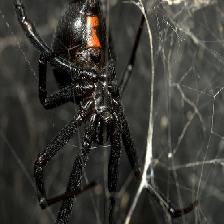}\vspace{.01cm} & \begin{tabular}[c]{@{}c@{}}spider\\ arachnid\\ invertebrate\\ tangle\_web\_spider\\ arthropod\end{tabular} & \vspace{.01cm} \includegraphics[width=0.2\columnwidth,height=0.2\columnwidth,valign=c]{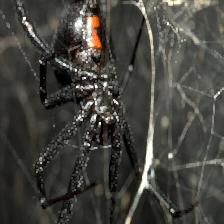}\vspace{.01cm} & water & $\phi_{3}$ \\ \hline
API\_C2 & \vspace{.01cm}  \includegraphics[width=0.2\columnwidth,height=0.2\columnwidth,valign=c]{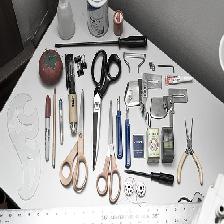}\vspace{.01cm} & \begin{tabular}[c]{@{}c@{}}scissors\\ equipment\\ tool\\ steel\\ work\end{tabular} & \vspace{.01cm} \includegraphics[width=0.2\columnwidth,height=0.2\columnwidth,valign=c]{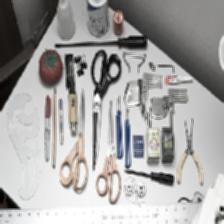}\vspace{.01cm} & business & $\phi_{1}$ \\ \hline
API\_C3 & \vspace{.01cm}  \includegraphics[width=0.2\columnwidth,height=0.2\columnwidth,valign=c]{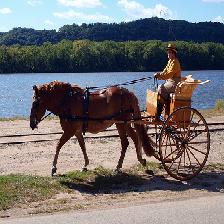}\vspace{.01cm} & \begin{tabular}[c]{@{}c@{}}carthorse\\ odd-toed\_ungulate\\ mammal\\ animal\\ buggy\end{tabular} & \vspace{.01cm} \includegraphics[width=0.2\columnwidth,height=0.2\columnwidth,valign=c]{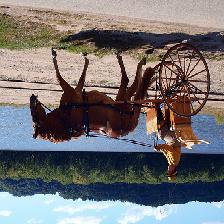}\vspace{.01cm} & elk & $\phi_{12}$ \\ \hline
API\_C4 & \vspace{.01cm}  \includegraphics[width=0.2\columnwidth,height=0.2\columnwidth,valign=c]{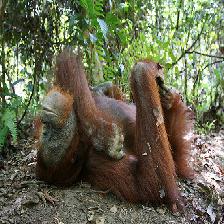}\vspace{.01cm} & \begin{tabular}[c]{@{}c@{}}tree\\ outdoor\\ animal\\ mammal\\ ape\end{tabular} & \vspace{.01cm} \includegraphics[width=0.2\columnwidth,height=0.2\columnwidth,valign=c]{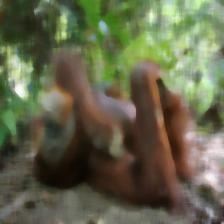}\vspace{.01cm} & fungus & $\phi_2$ \\ \hline
API\_C5 &  \includegraphics[width=0.2\columnwidth,height=0.2\columnwidth,valign=c]{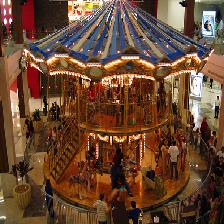}\vspace{.05cm} & \begin{tabular}[c]{@{}c@{}}Belt\\ Carousel\\ Indoors\\ Lobby\\ Reception\end{tabular} & \includegraphics[width=0.2\columnwidth,height=0.2\columnwidth,valign=c]{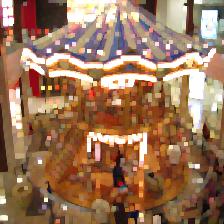}\vspace{.05cm} & architecture & $\phi_4$ \\
\end{tabular}
\end{table*}

\begin{table*}[t!]
\setlength{\tabcolsep}{12pt}
\scriptsize
\centering
\renewcommand{\arraystretch}{1.1}
\caption{Sample errors of safety properties found by \sys in different self-driving systems}
\label{tab:violations_car}
\begin{tabular}{c|cc|cc|c}
\multirow{2}{*}{\textbf{Model ID}} & \multicolumn{2}{c|}{\textbf{Original}} & \multicolumn{2}{c|}{\textbf{Transformed}} & \multicolumn{1}{c}{\multirow{2}{*}{\textbf{Violated}}} \\
 & \textit{Image} & \textit{Steering angle prediction} & \textit{Image} & \textit{Steering angle prediction} & \multicolumn{1}{c}{} \\ \hline
DRV\_C1 & \vspace{.01cm} \includegraphics[width=0.2\columnwidth,height=0.2\columnwidth,valign=c]{figs/car/illu/rambo_-0_00160922610667_to_-0_440718531609_p_-61_78_200_orig.png}\vspace{.01cm} 
& \begin{tabular}[c]{@{}c@{}}near straight (left 0.1$^\circ$)\end{tabular} & \vspace{.01cm} \includegraphics[width=0.2\columnwidth,height=0.2\columnwidth,valign=c]{figs/car/illu/rambo_-0_00160922610667_to_-0_440718531609_p_-61_78_200_gen.png}\vspace{.01cm} 
& turn right (right $25^\circ$) & $\phi_6$ \\ \hline
DRV\_C2 & \vspace{.01cm} \includegraphics[width=0.2\columnwidth,height=0.2\columnwidth,valign=c]{figs/car/occl/dave-orig_-0_069287955761_to_-0_770116508007_p__29__53__26_6400_orig.png}\vspace{.01cm} & \begin{tabular}[c]{@{}c@{}}near straight (right $4^\circ$)\end{tabular} & \vspace{.01cm} \includegraphics[width=0.2\columnwidth,height=0.2\columnwidth,valign=c]{figs/car/occl/dave-orig_-0_069287955761_to_-0_770116508007_p__29__53__26_6400_gen.png}\vspace{.01cm} & turn right (right $44^\circ$) & $\phi_7$ \\ \hline
DRV\_C3 & \vspace{.01cm} \includegraphics[width=0.2\columnwidth,height=0.2\columnwidth,valign=c]{figs/car/shft/dave-orig_0_0642189010978_to_-0_46838209033_p__8__-9__1_400_orig.png}\vspace{.01cm} & \begin{tabular}[c]{@{}c@{}}near straight (left $3.7^\circ$)\end{tabular} & \vspace{.01cm} \includegraphics[width=0.2\columnwidth,,height=0.2\columnwidth,valign=c]{figs/car/shft/dave-orig_0_0642189010978_to_-0_46838209033_p__8__-9__1_400_gen.png}\vspace{.01cm} & turn right (right $27^\circ$) & $\phi_{11}$ \\ \hline
DRV\_C4 & \vspace{.01cm} \includegraphics[width=0.2\columnwidth,height=0.2\columnwidth,valign=c]{figs/car/mblur/dave-dropout_0_000613341107965_to_-0_658129930496_p_5_2_4_orig.png}\vspace{.01cm} & \begin{tabular}[c]{@{}c@{}}near straight (left $0.03^\circ$)\end{tabular} & \vspace{.01cm} \includegraphics[width=0.2\columnwidth,height=0.2\columnwidth,valign=c]{figs/car/mblur/dave-dropout_0_000613341107965_to_-0_658129930496_p_5_2_4_gen.png}\vspace{.01cm} & turn right (right $38^\circ$) & $\phi_2$  \\ \hline
\end{tabular}
\end{table*}

\subsection{Distribution of error-inducing parameter values}

\begin{figure*}
\captionsetup[subfloat]{captionskip=-.15cm, labelformat=empty}

\vspace{-.15cm}{\bf \scriptsize \hspace{0.025in} Average smoothing:\vspace{-.3cm}}%

\subfloat[]{
\includegraphics[width=0.155\textwidth]{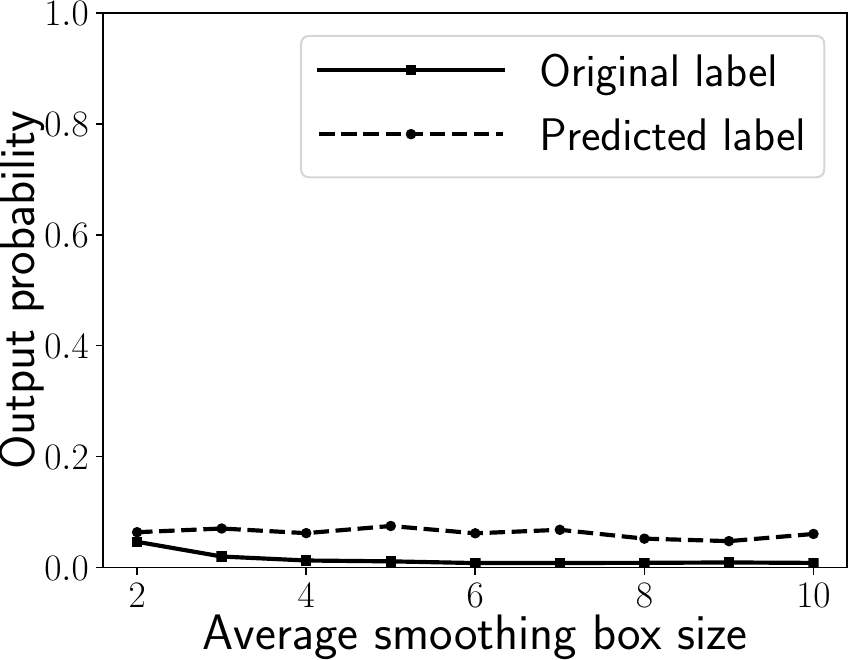}}%
\hfill
\subfloat[]{
\includegraphics[width=0.155\textwidth]{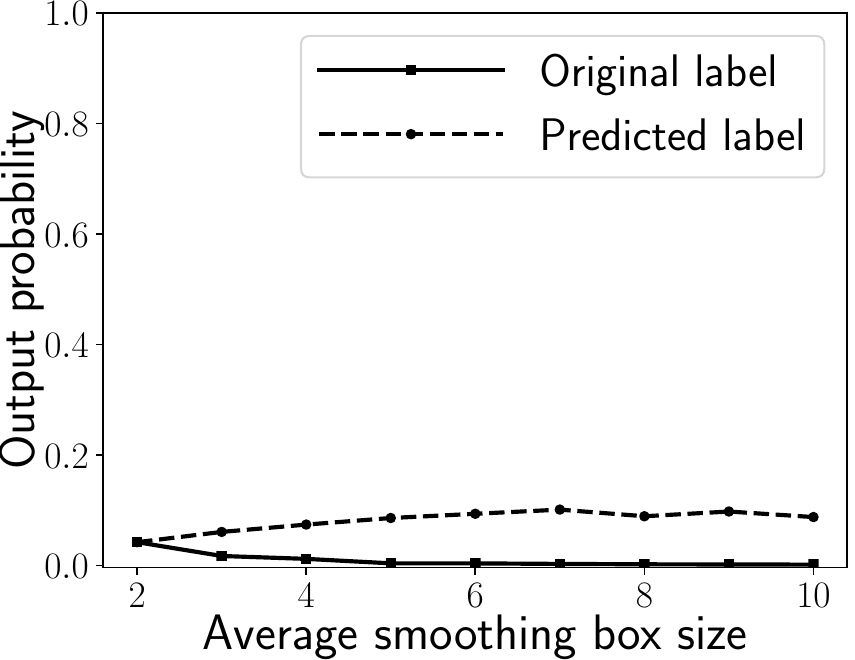}}%
\hfill
\subfloat[]{
\includegraphics[width=0.155\textwidth]{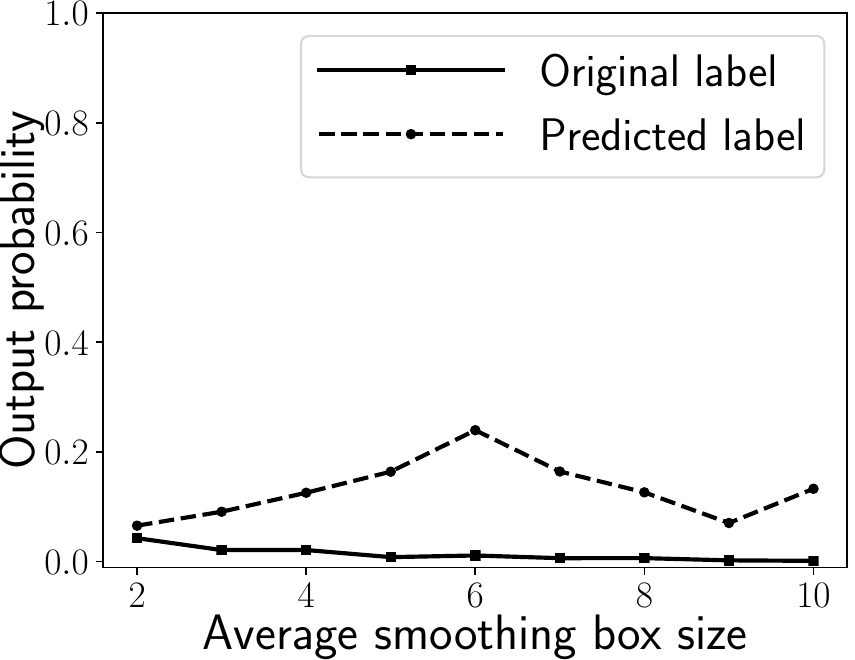}}%
\hfill
\subfloat[]{
\includegraphics[width=0.155\textwidth]{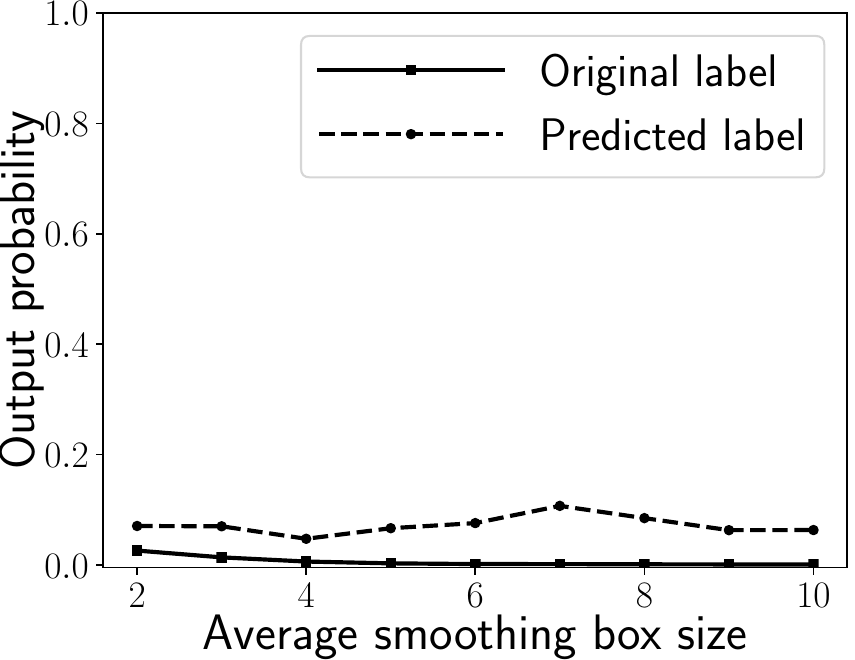}}%
\hfill
\subfloat[]{
\includegraphics[width=0.155\textwidth]{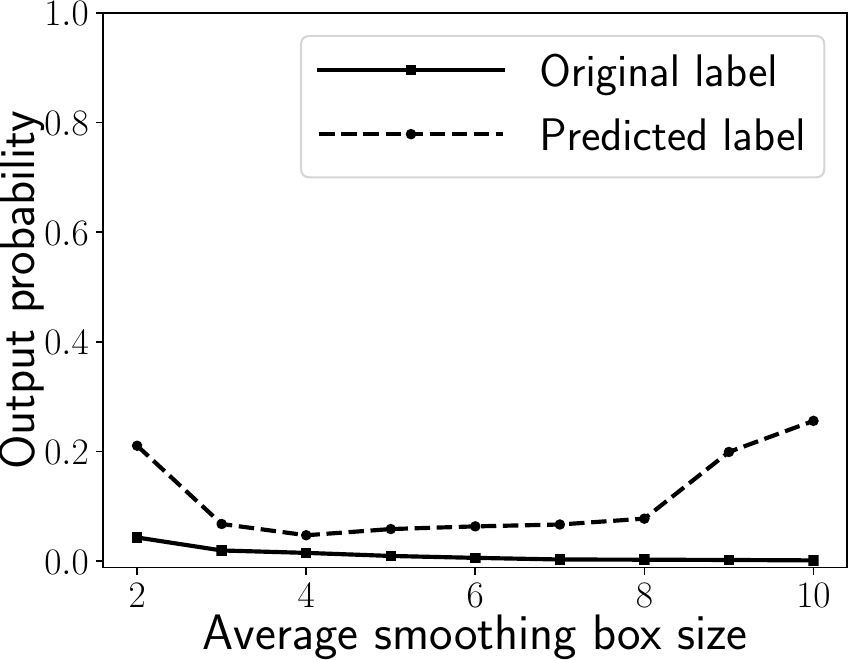}}%
\hfill
\subfloat[]{
\includegraphics[width=0.155\textwidth]{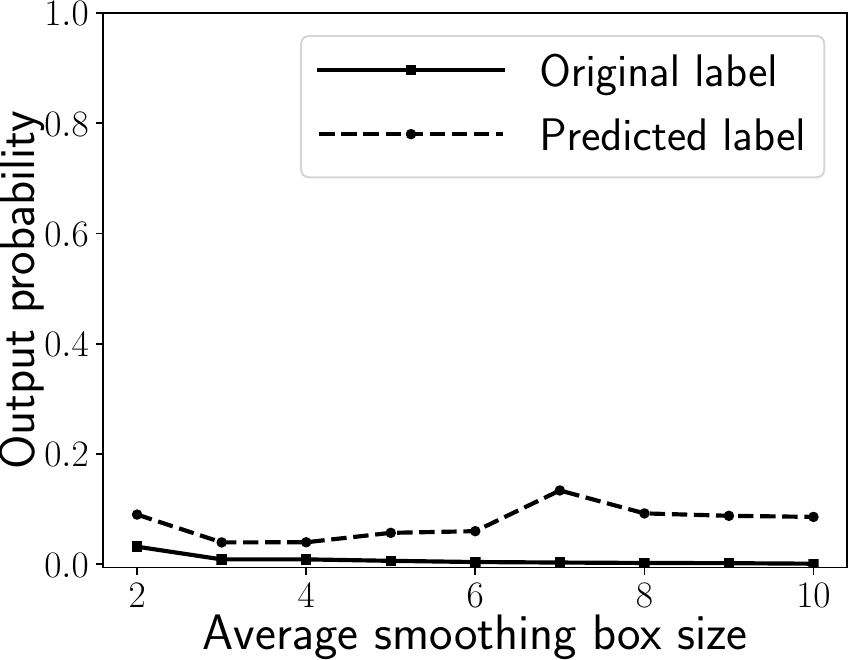}}%
\vspace{-.5cm}
\subfloat[]{
\includegraphics[width=0.155\textwidth]{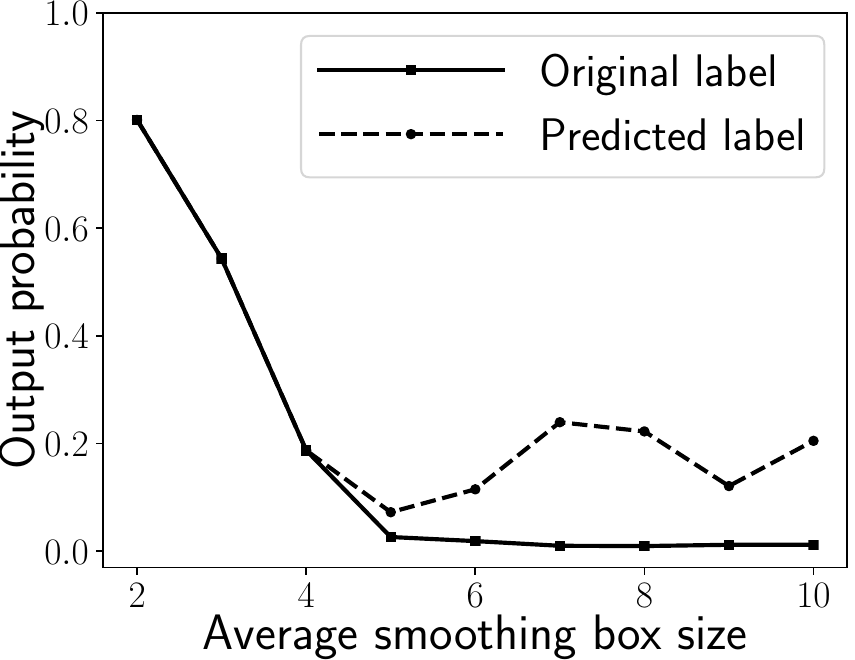}}%
\hfill
\subfloat[]{
\includegraphics[width=0.155\textwidth]{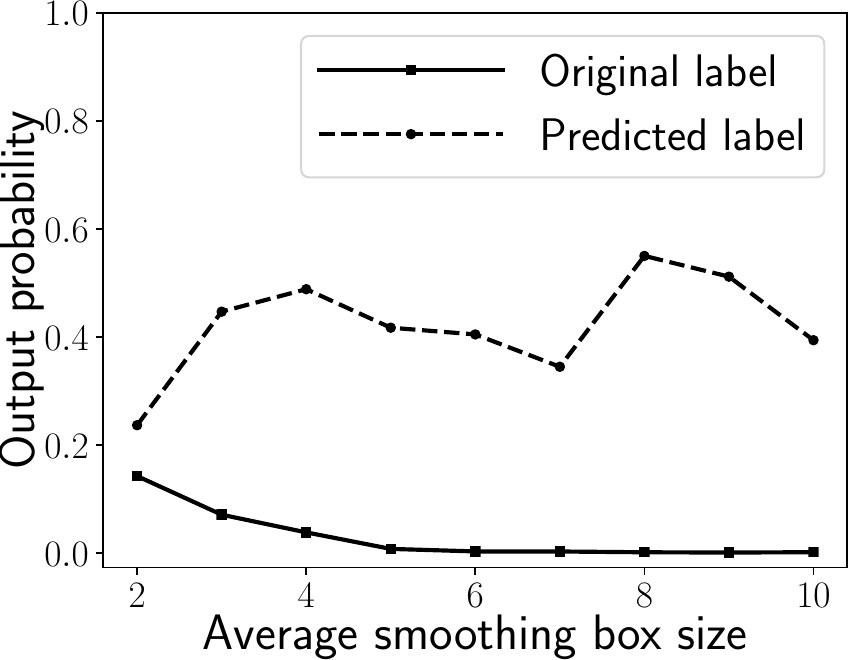}}%
\hfill
\subfloat[]{
\includegraphics[width=0.155\textwidth]{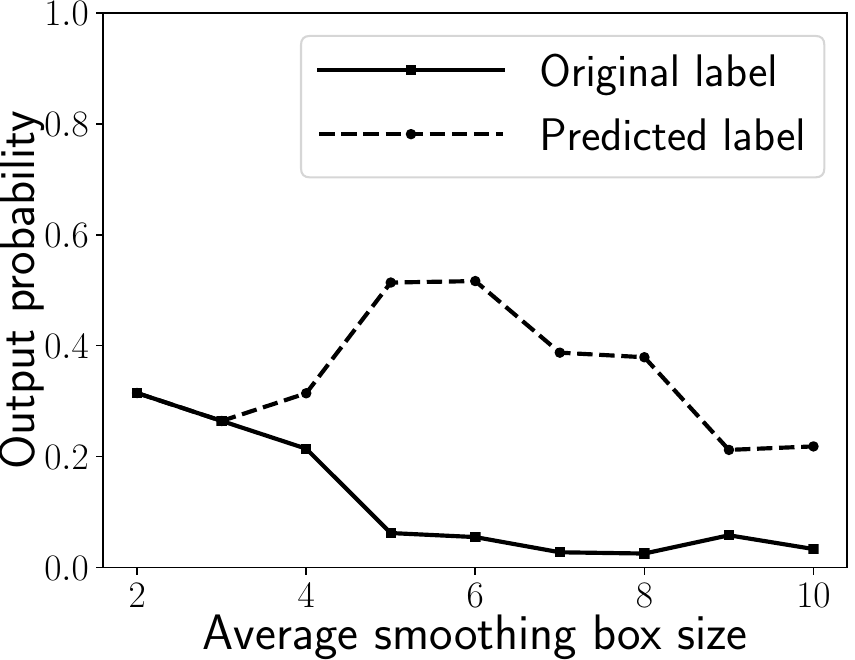}}%
\hfill
\subfloat[]{
\includegraphics[width=0.155\textwidth]{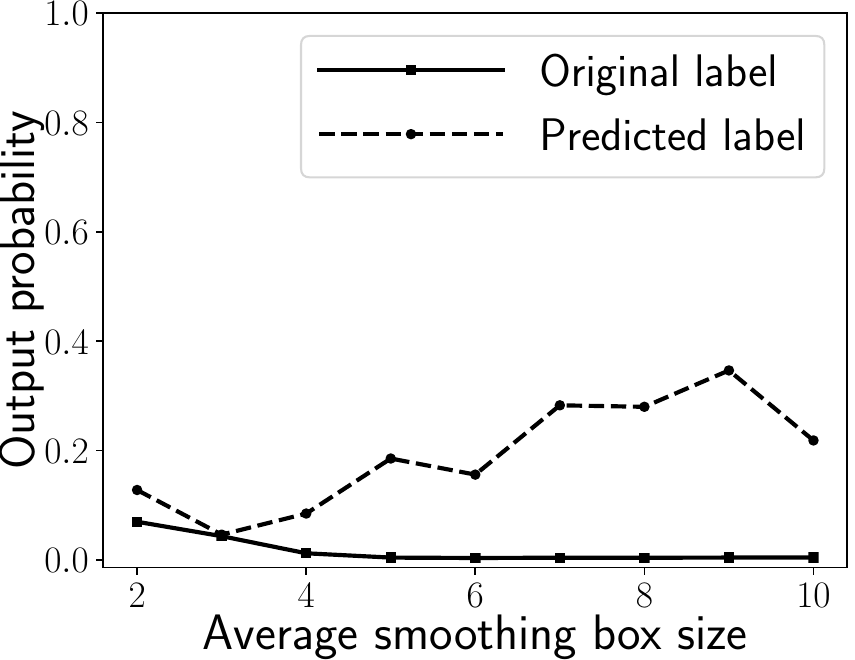}}%
\hfill
\subfloat[]{
\includegraphics[width=0.155\textwidth]{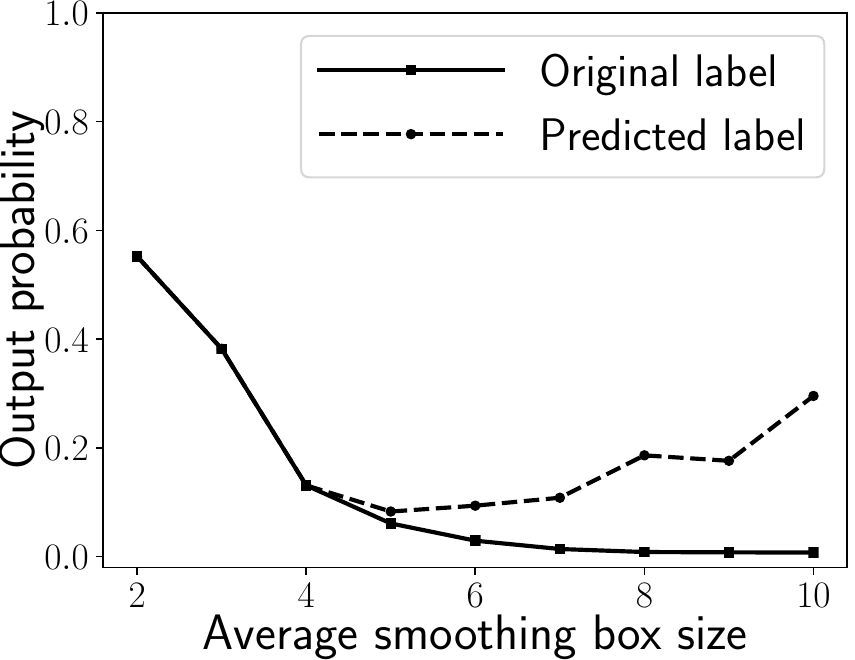}}%
\hfill
\subfloat[]{
\includegraphics[width=0.155\textwidth]{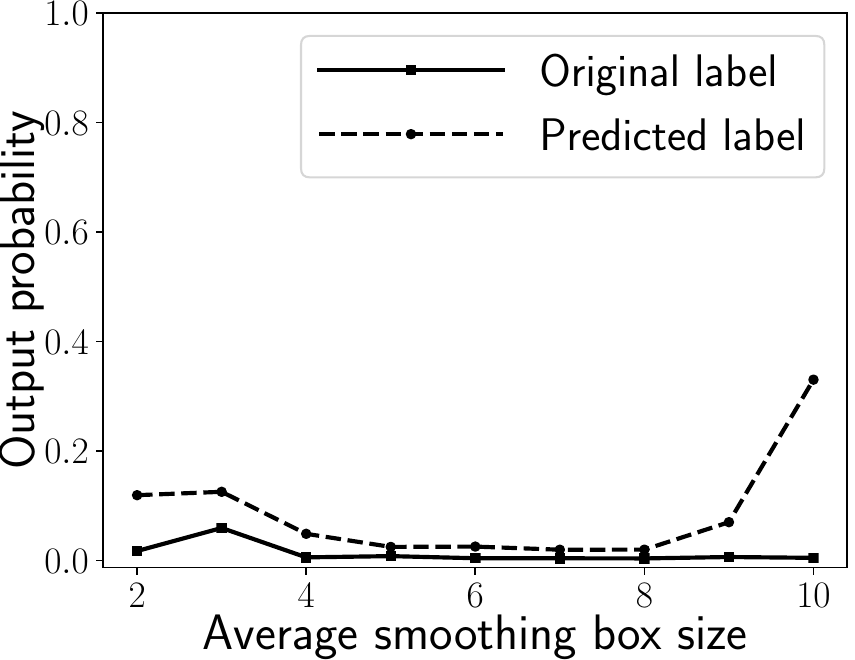}}%

\vspace{-.15cm}{\bf \scriptsize \hspace{0.025in} Median smoothing:\vspace{-.3cm}}%

\subfloat[]{
\includegraphics[width=0.155\textwidth]{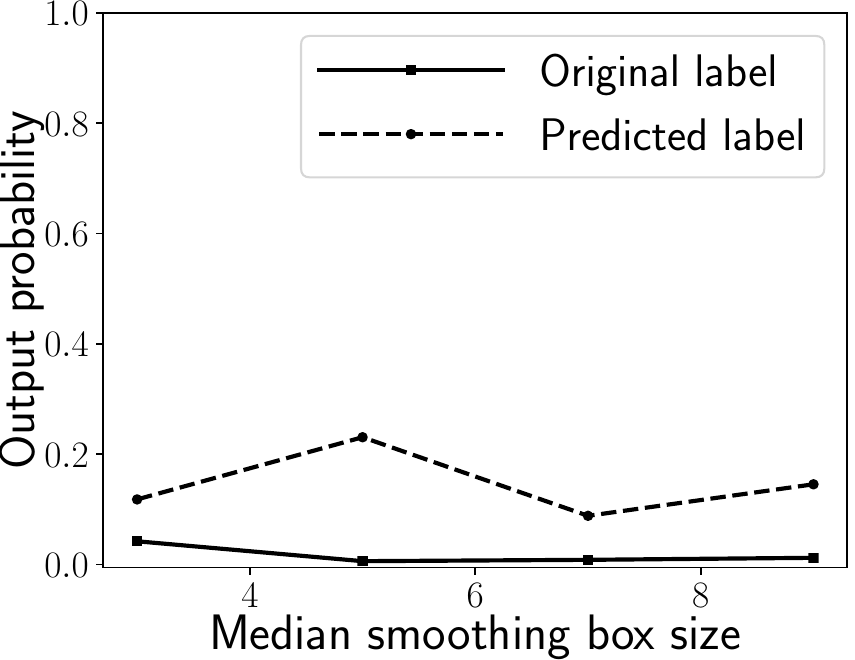}}%
\hfill
\subfloat[]{
\includegraphics[width=0.155\textwidth]{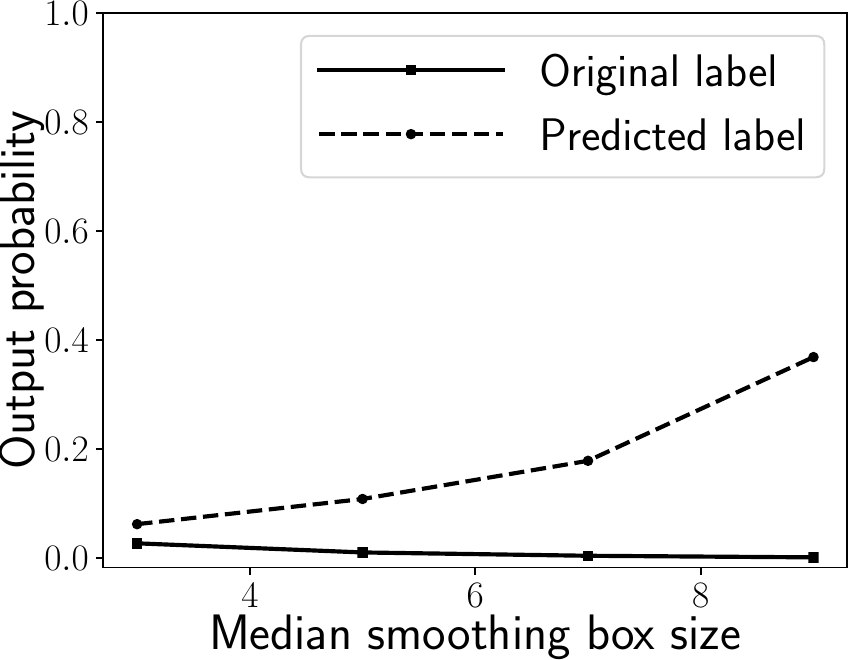}}%
\hfill
\subfloat[]{
\includegraphics[width=0.155\textwidth]{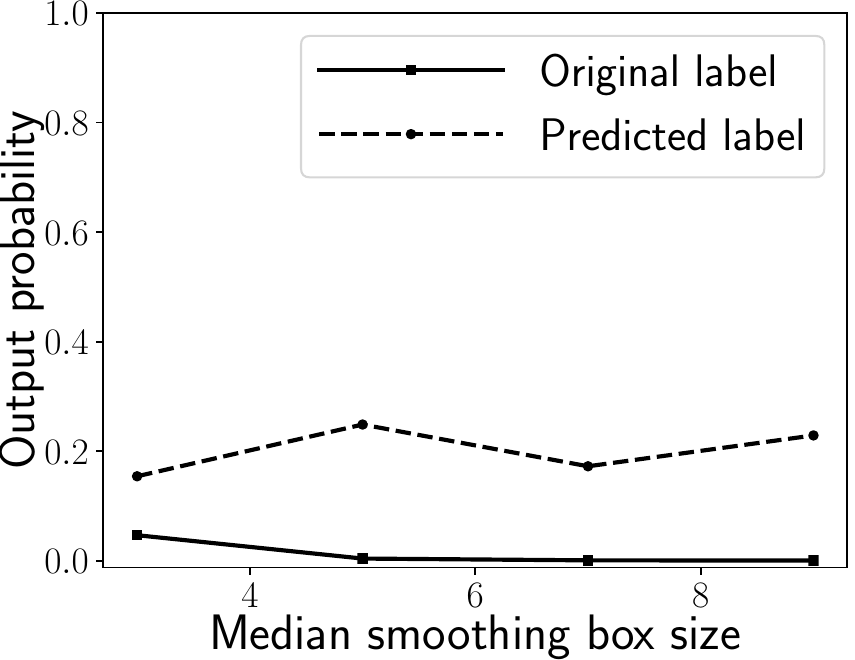}}%
\hfill
\subfloat[]{
\includegraphics[width=0.155\textwidth]{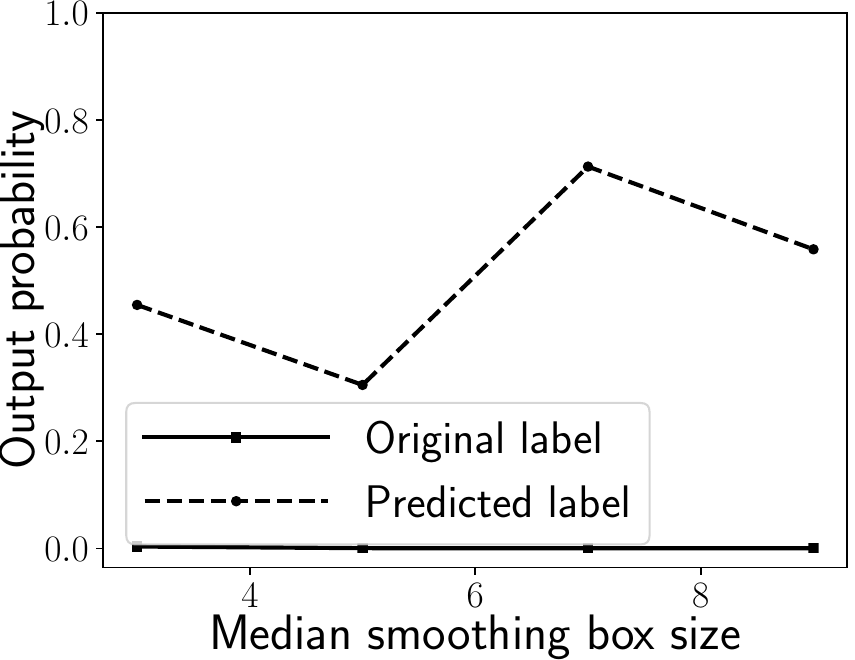}}%
\hfill
\subfloat[]{
\includegraphics[width=0.155\textwidth]{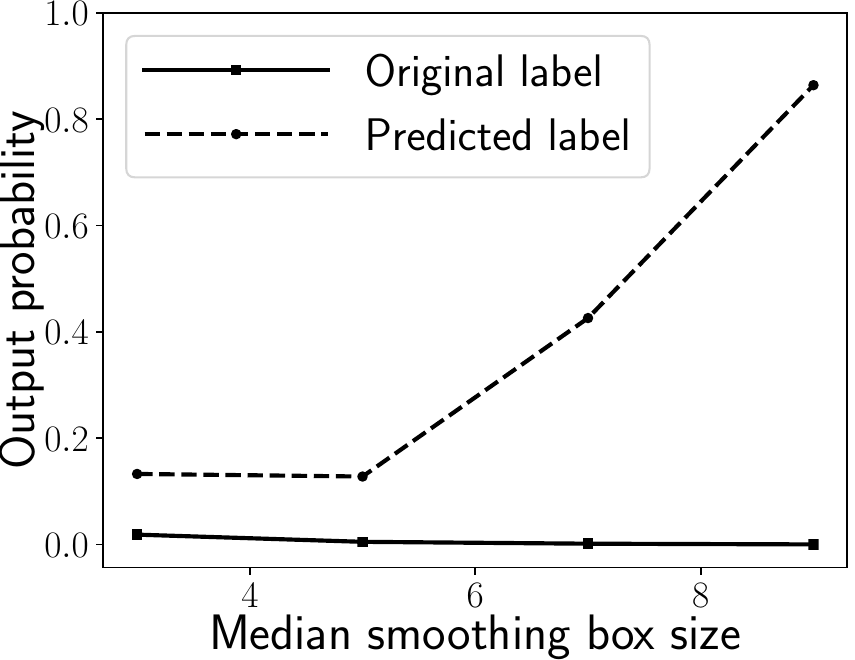}}%
\hfill
\subfloat[]{
\includegraphics[width=0.155\textwidth]{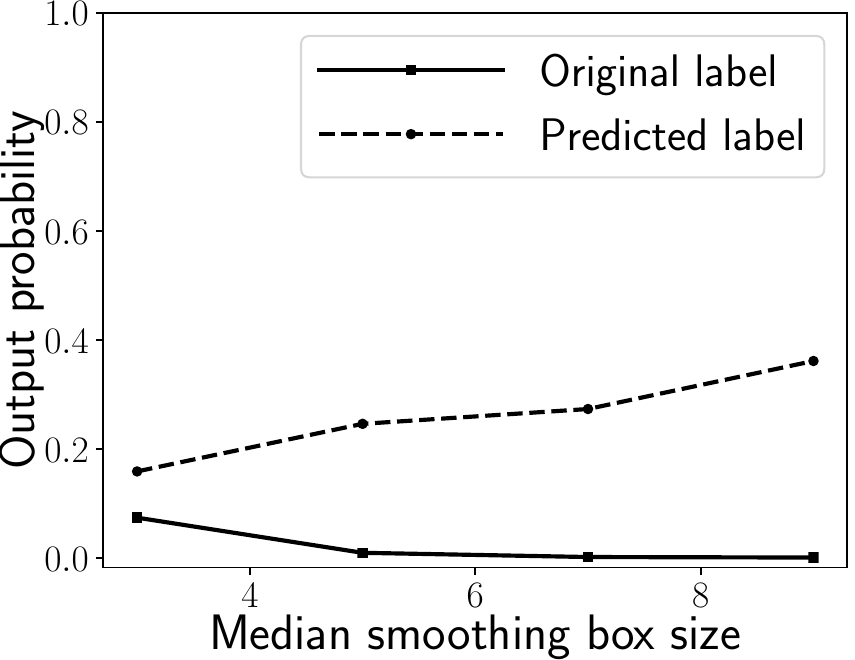}}%
\vspace{-.5cm}
\subfloat[]{
\includegraphics[width=0.155\textwidth]{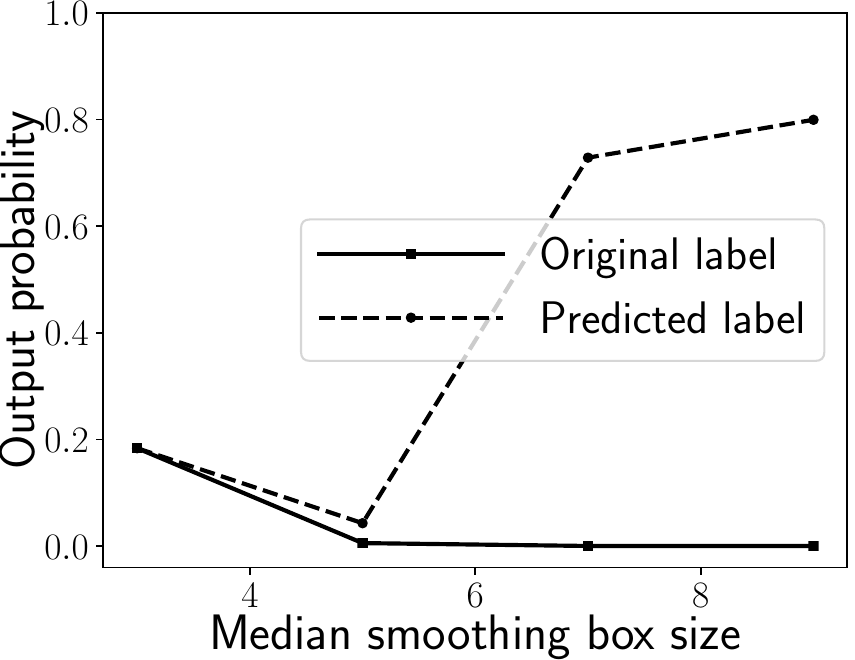}}%
\hfill
\subfloat[]{
\includegraphics[width=0.155\textwidth]{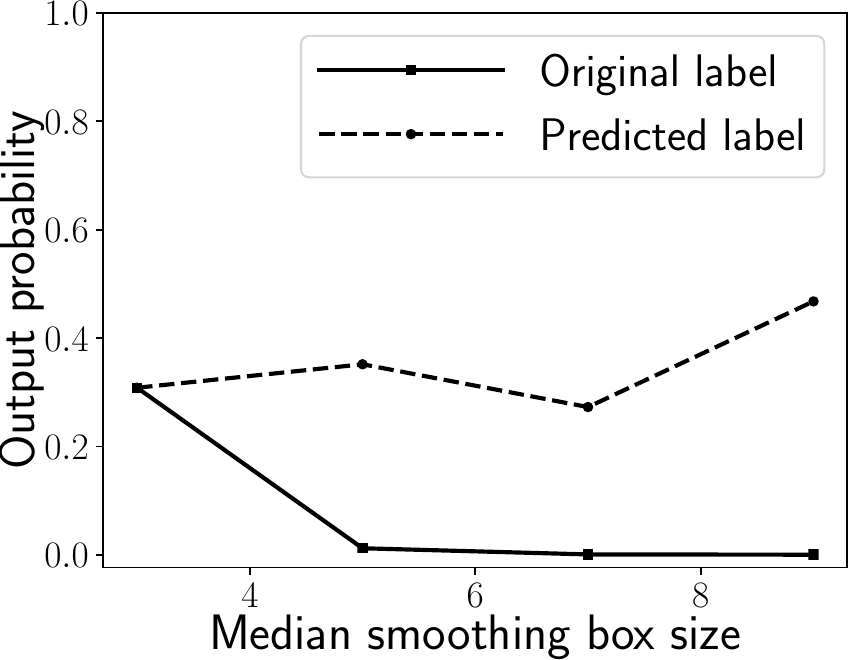}}%
\hfill
\subfloat[]{
\includegraphics[width=0.155\textwidth]{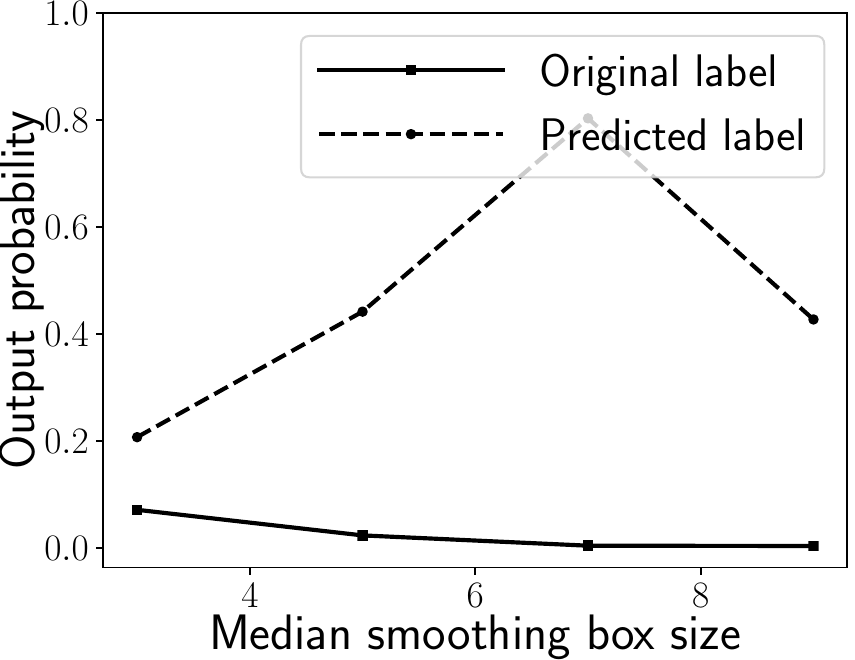}}%
\hfill
\subfloat[]{
\includegraphics[width=0.155\textwidth]{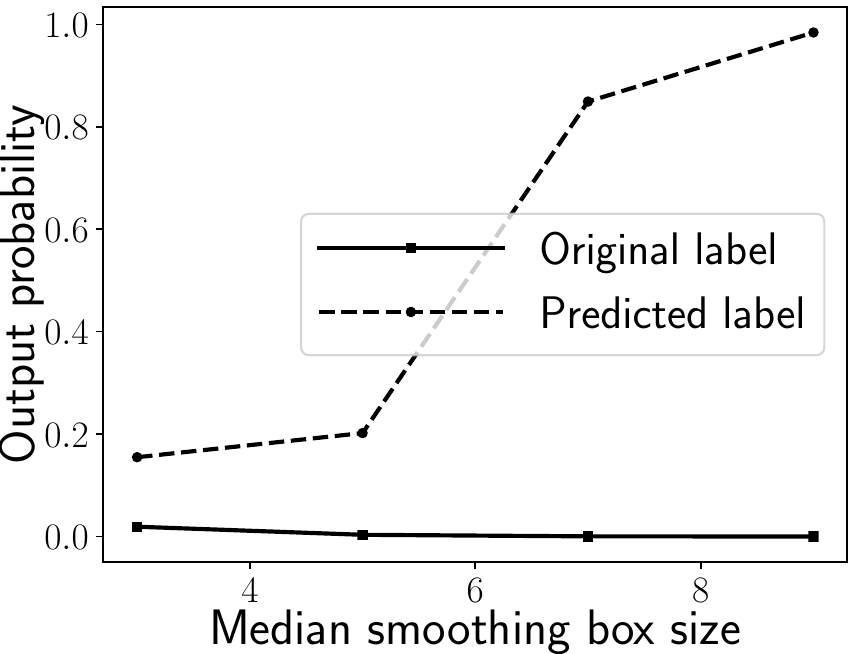}}%
\hfill
\subfloat[]{
\includegraphics[width=0.155\textwidth]{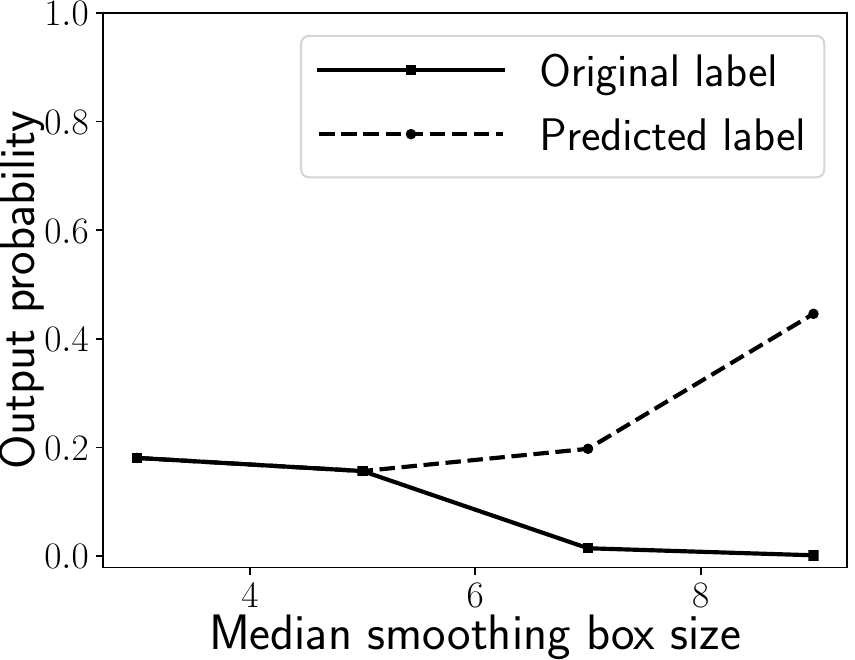}}%
\hfill
\subfloat[]{
\includegraphics[width=0.155\textwidth]{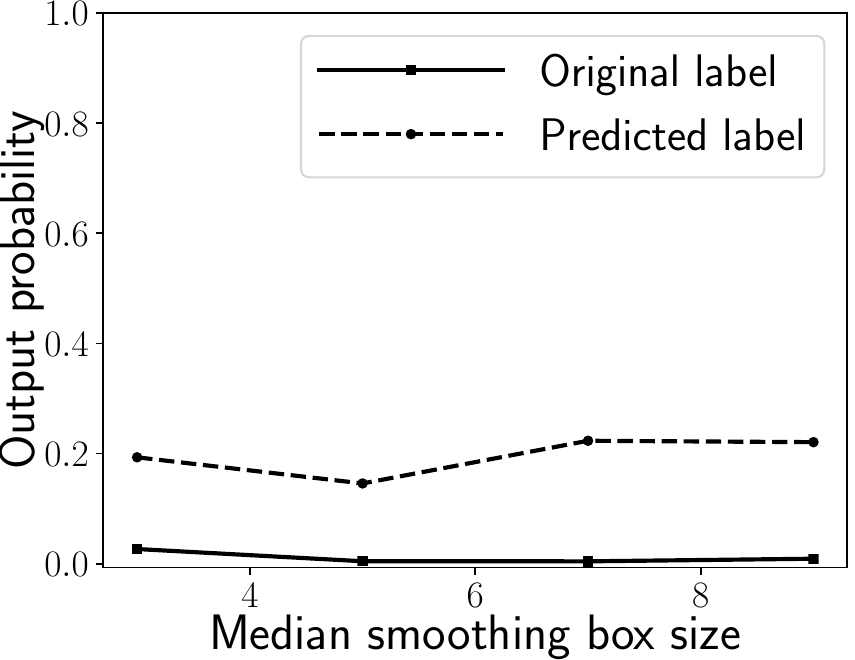}}%

\vspace{-.15cm}{\bf \scriptsize \hspace{0.025in} Erosion:\vspace{-.3cm}}%

\subfloat[]{
\includegraphics[width=0.155\textwidth]{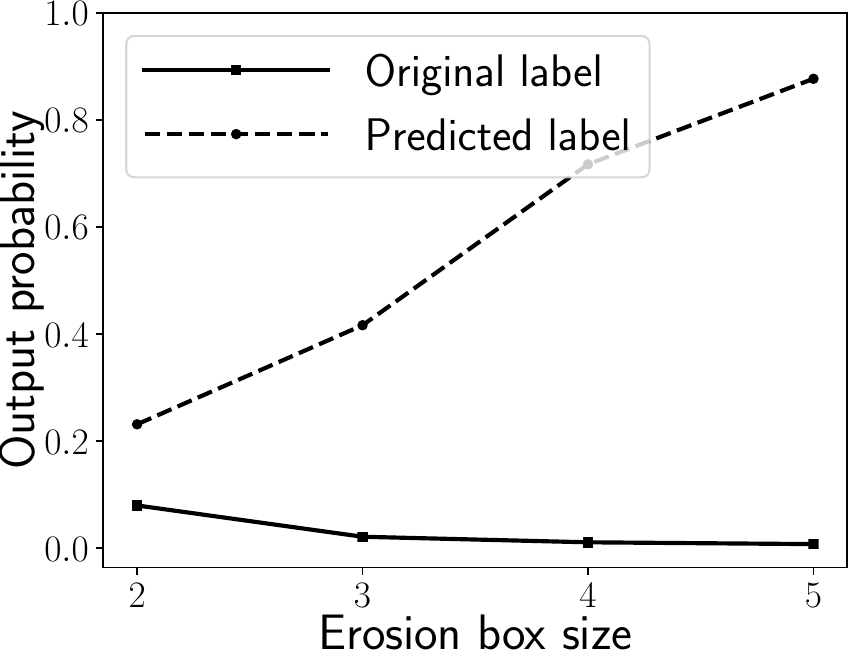}}%
\hfill
\subfloat[]{
\includegraphics[width=0.155\textwidth]{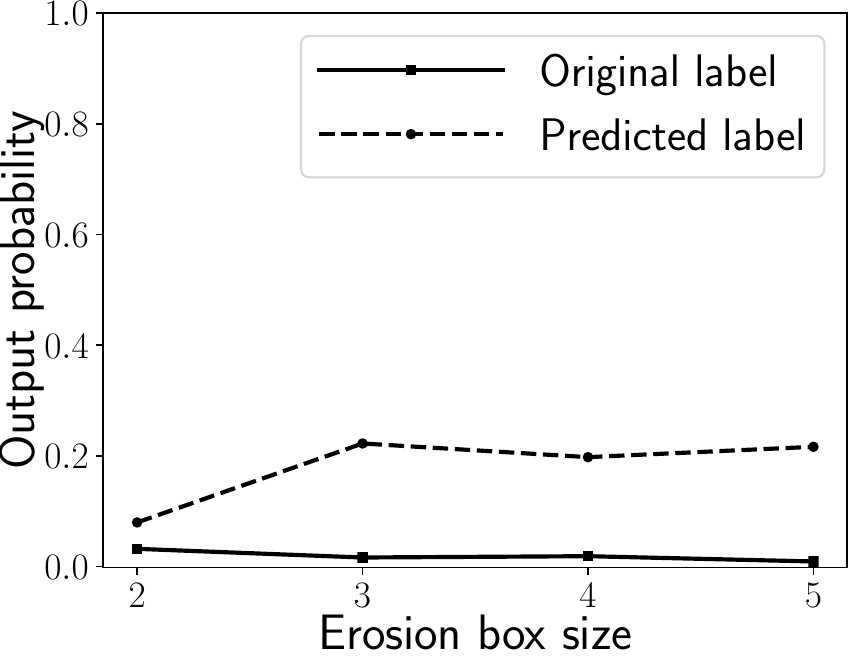}}%
\hfill
\subfloat[]{
\includegraphics[width=0.155\textwidth]{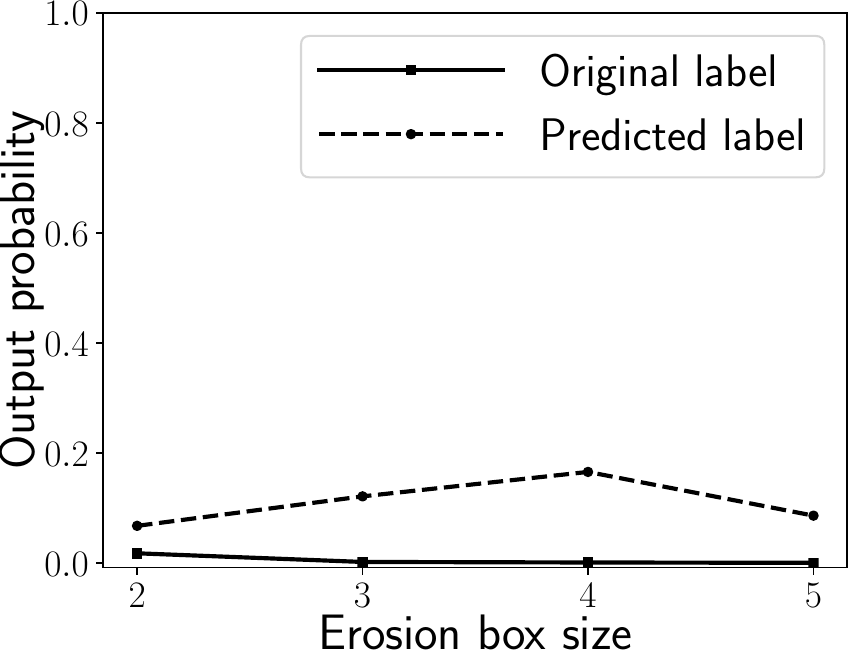}}%
\hfill
\subfloat[]{
\includegraphics[width=0.155\textwidth]{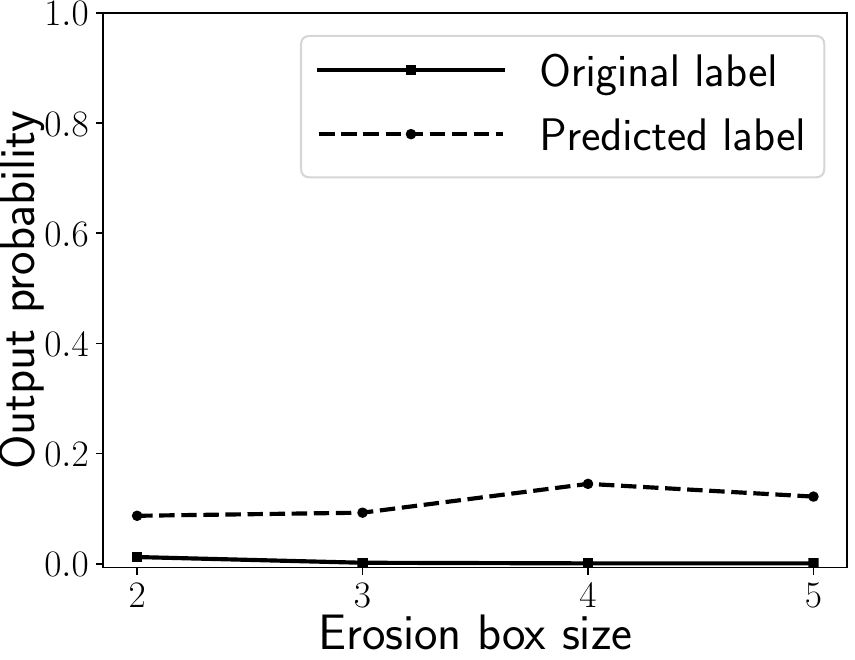}}%
\hfill
\subfloat[]{
\includegraphics[width=0.155\textwidth]{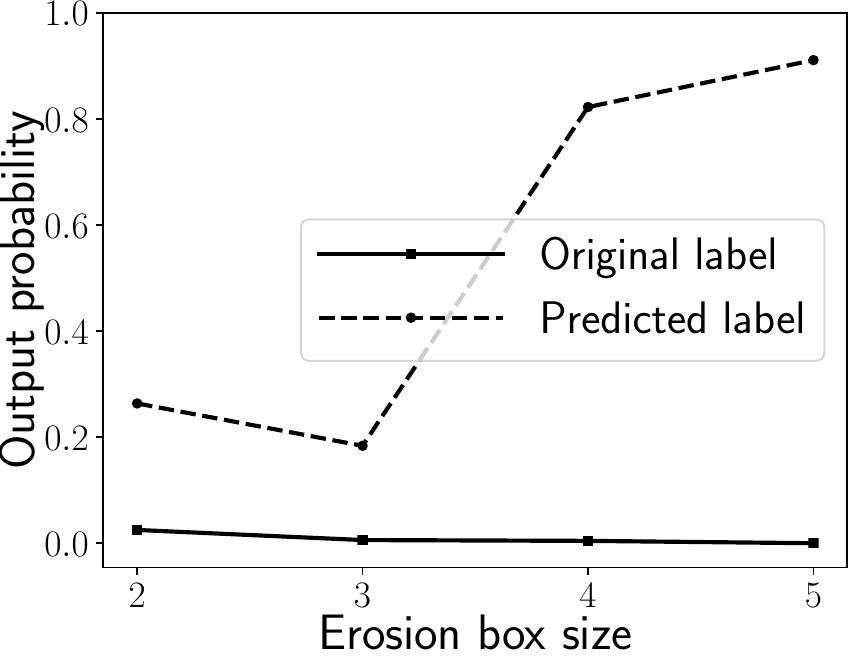}}%
\hfill
\subfloat[]{
\includegraphics[width=0.155\textwidth]{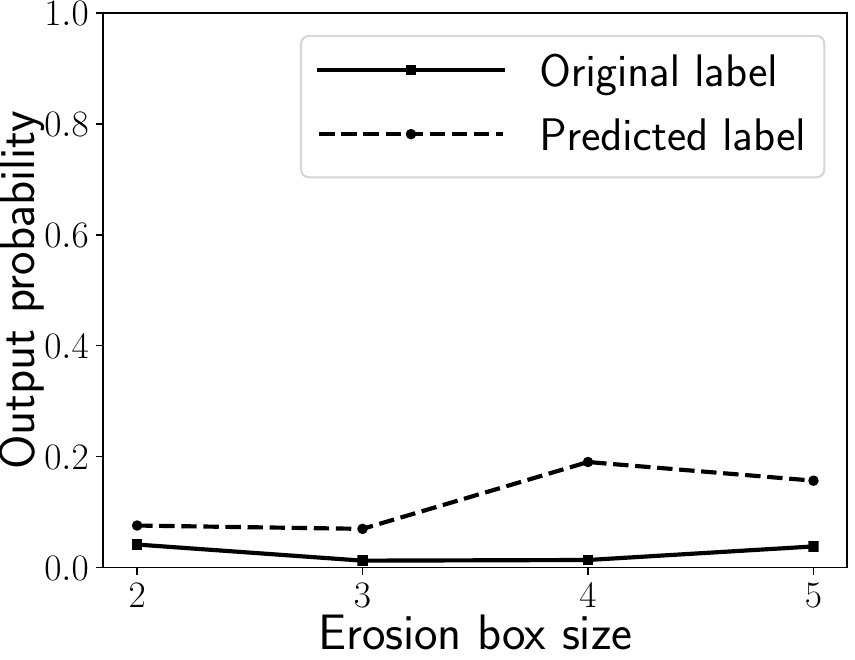}}%
\vspace{-.5cm}
\subfloat[]{
\includegraphics[width=0.155\textwidth]{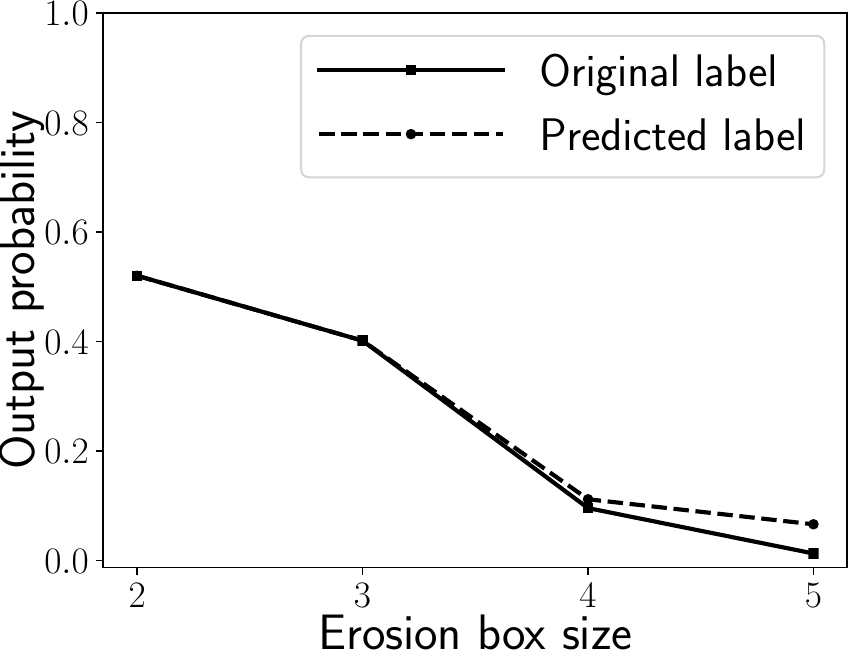}}%
\hfill
\subfloat[]{
\includegraphics[width=0.155\textwidth]{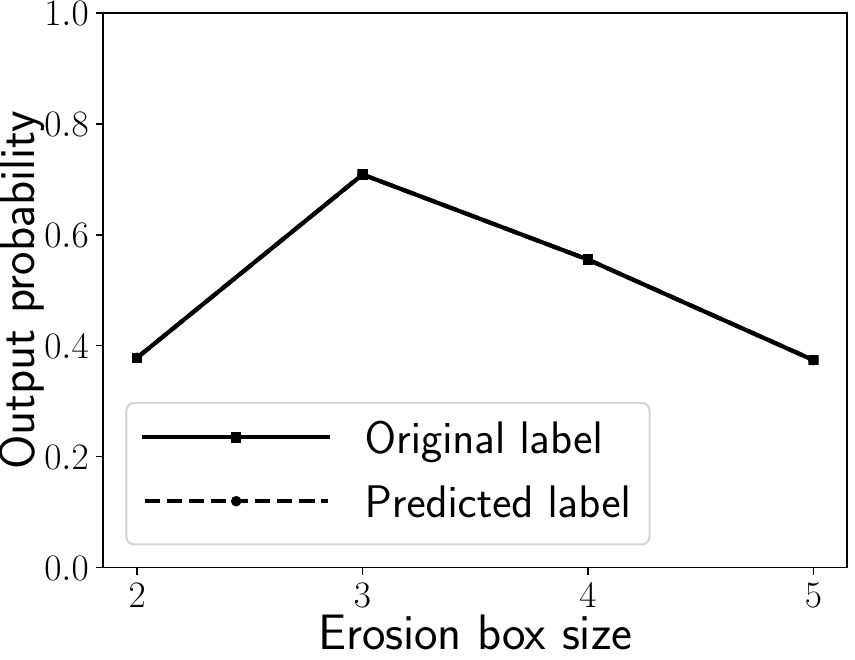}}%
\hfill
\subfloat[]{
\includegraphics[width=0.155\textwidth]{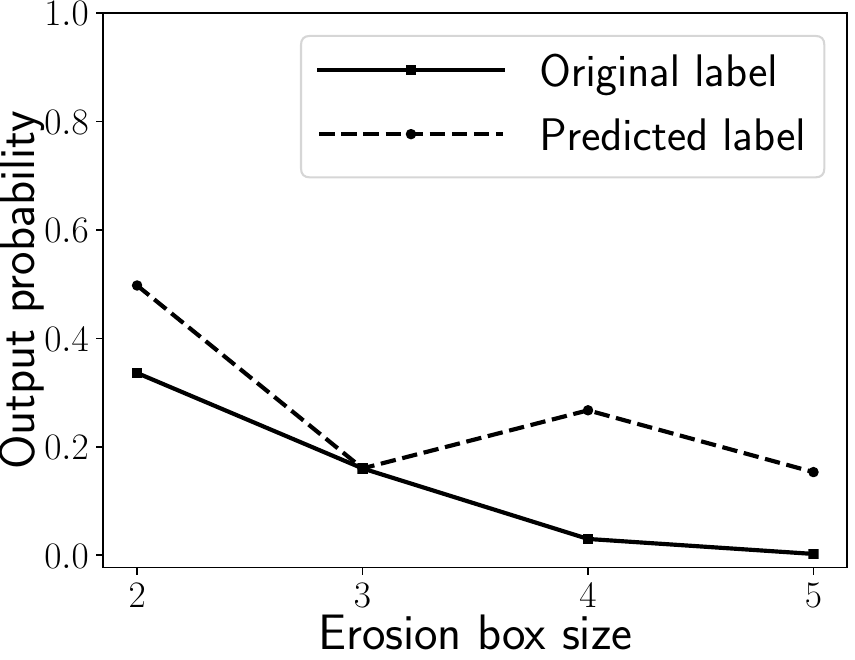}}%
\hfill
\subfloat[]{
\includegraphics[width=0.155\textwidth]{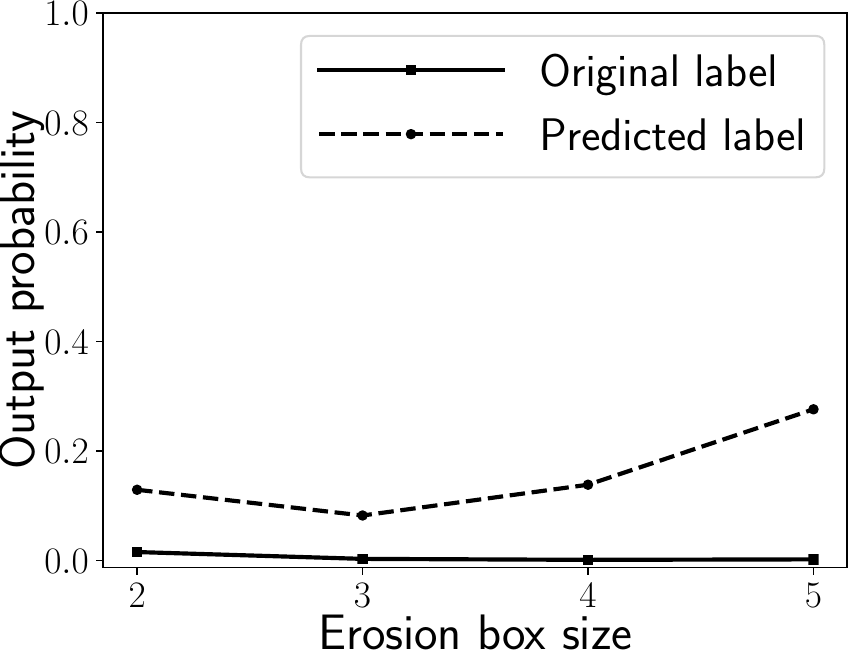}}%
\hfill
\subfloat[]{
\includegraphics[width=0.155\textwidth]{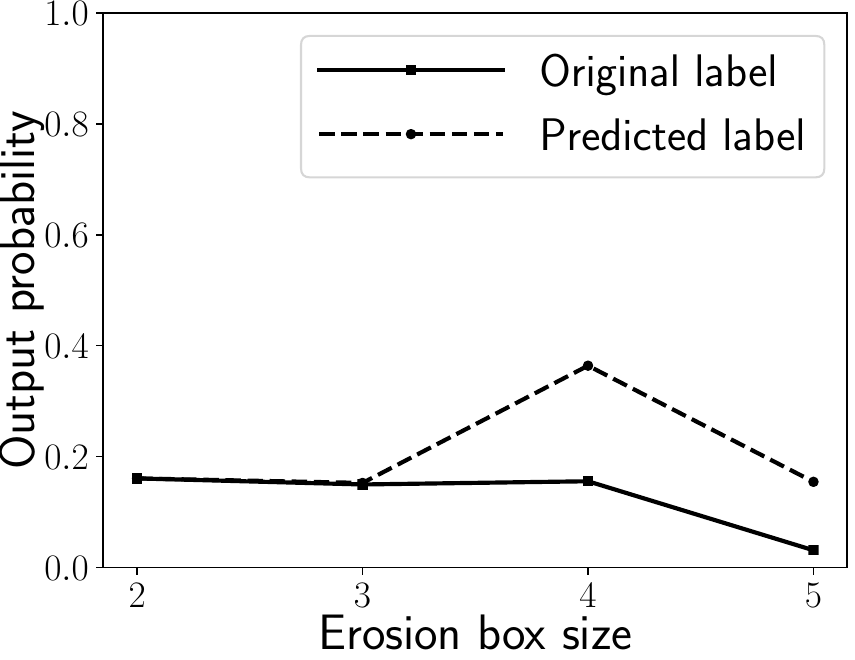}}%
\hfill
\subfloat[]{
\includegraphics[width=0.155\textwidth]{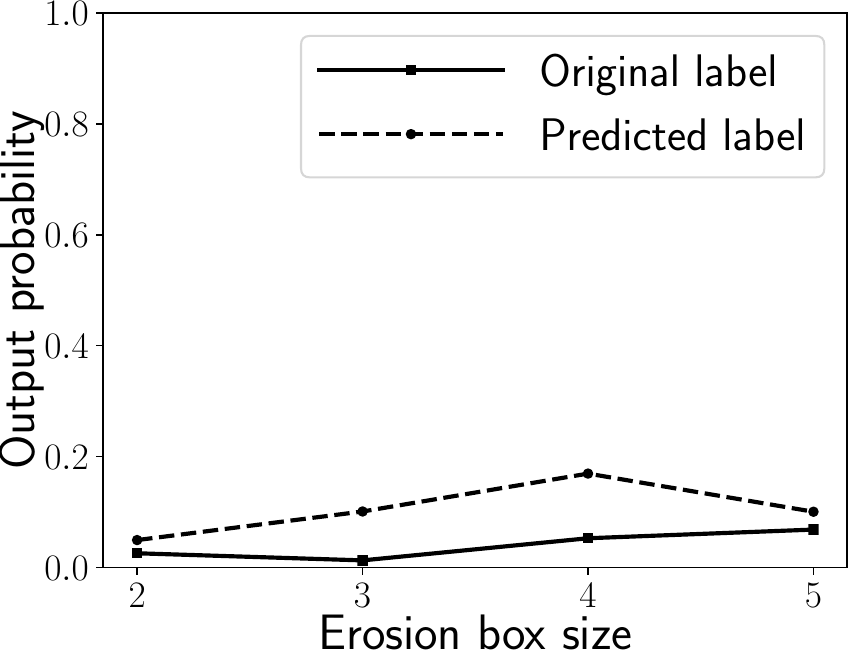}}%

\vspace{-.15cm}{\bf \scriptsize \hspace{0.025in} Dilation:\vspace{-.3cm}}%

\subfloat[]{
\includegraphics[width=0.155\textwidth]{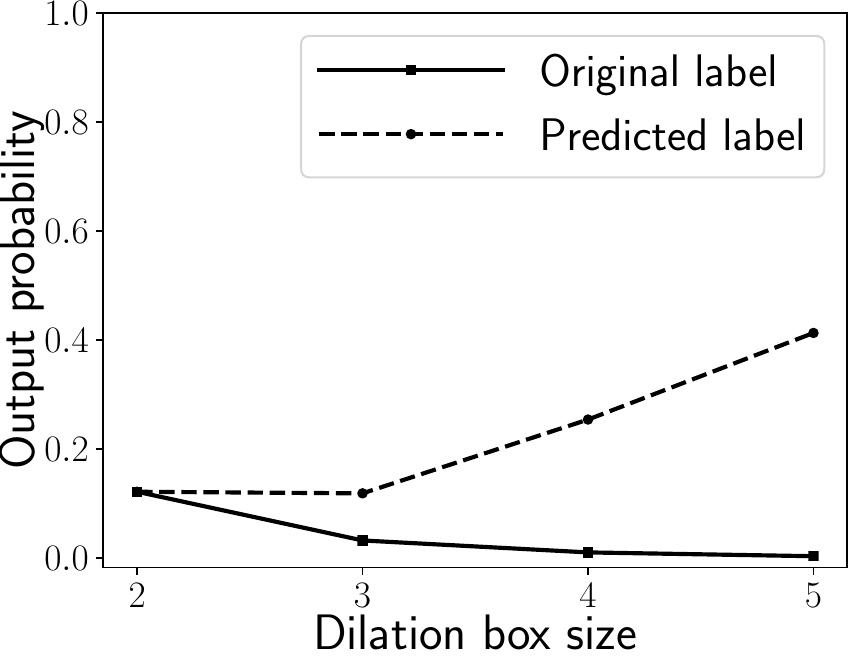}}%
\hfill
\subfloat[]{
\includegraphics[width=0.155\textwidth]{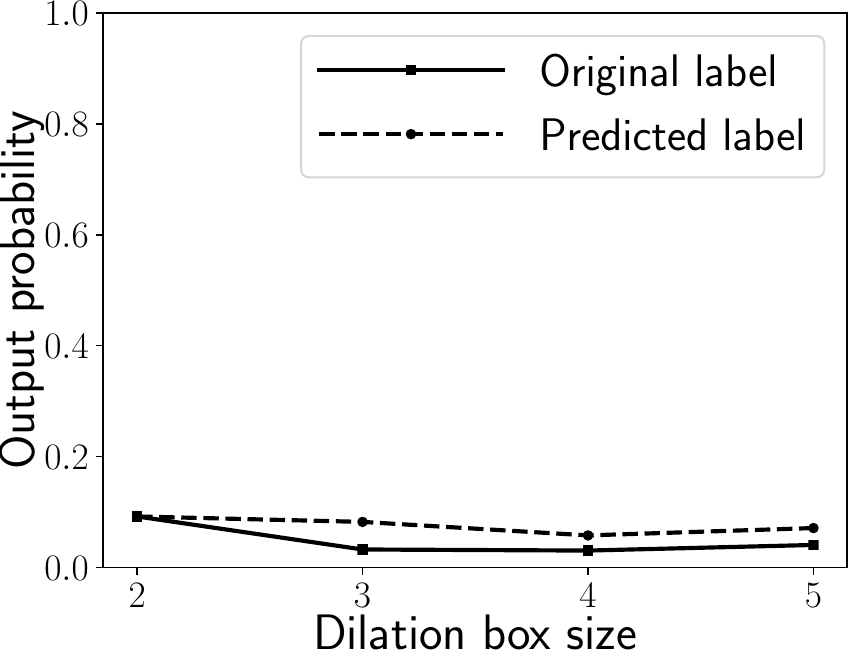}}%
\hfill
\subfloat[]{
\includegraphics[width=0.155\textwidth]{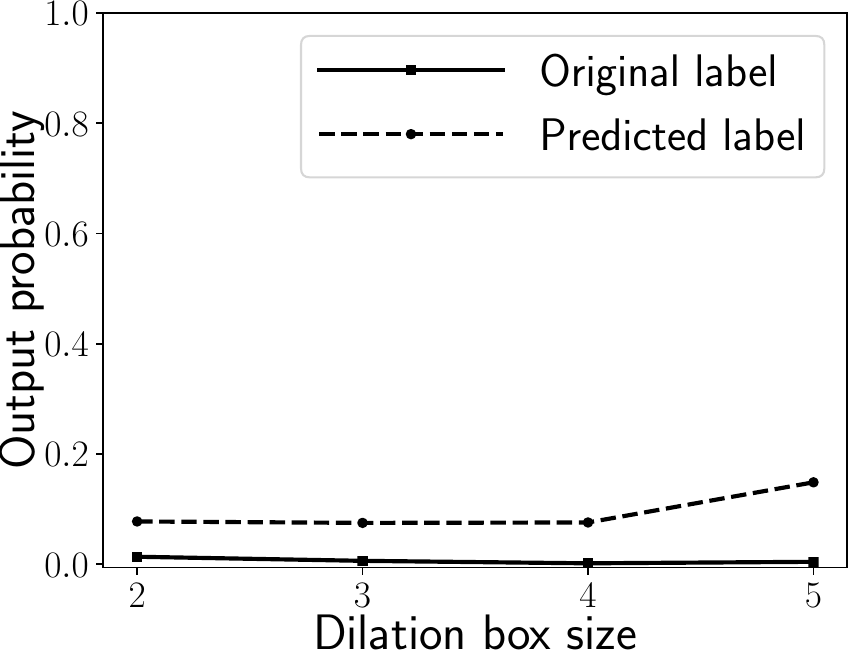}}
\hfill
\subfloat[]{
\includegraphics[width=0.155\textwidth]{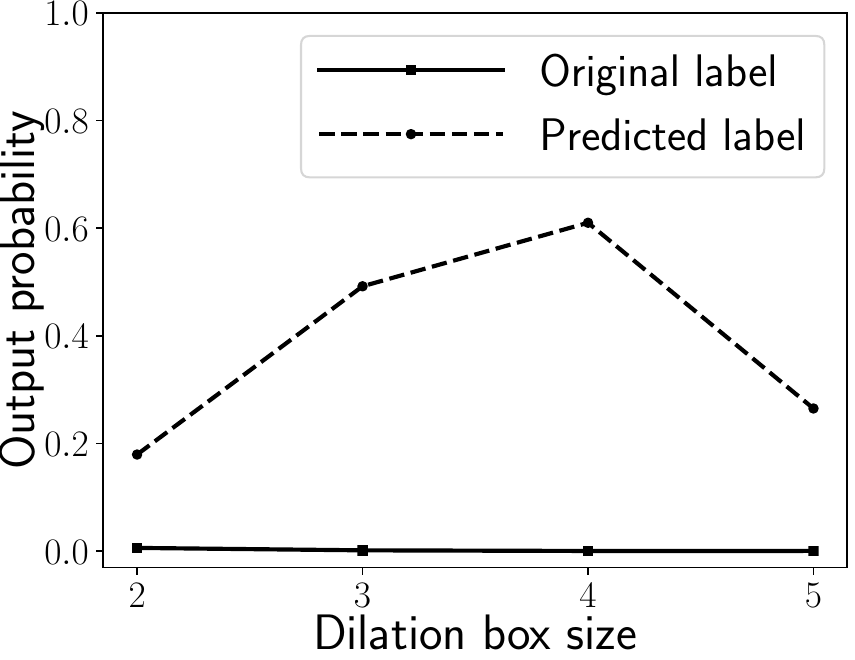}}%
\hfill
\subfloat[]{
\includegraphics[width=0.155\textwidth]{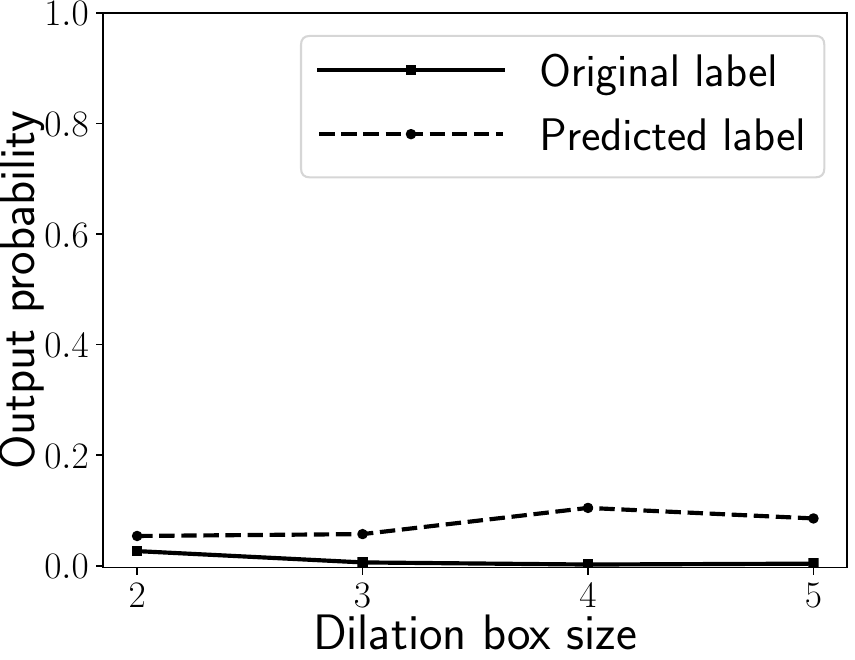}}%
\hfill
\subfloat[]{
\includegraphics[width=0.155\textwidth]{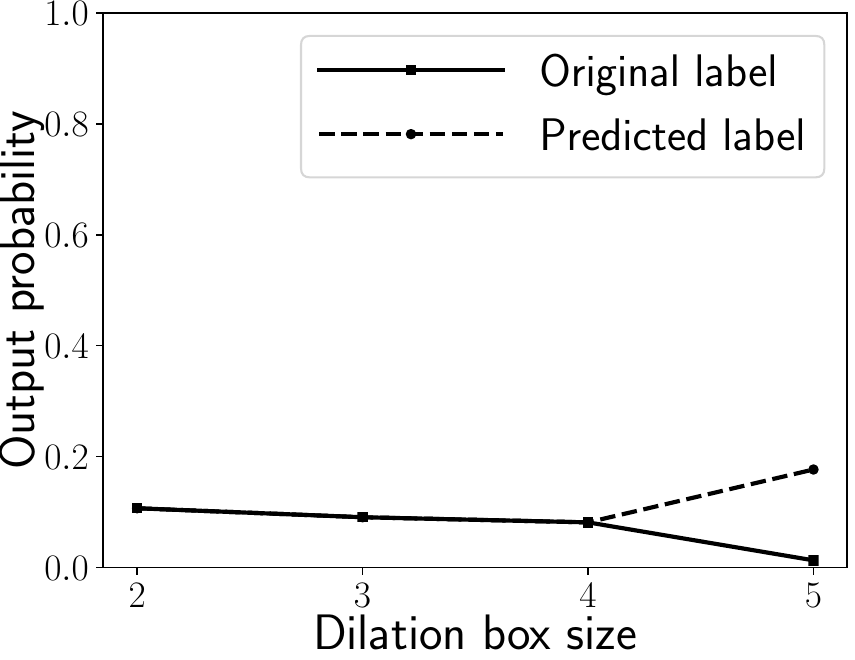}}%
\vspace{-.5cm}
\subfloat[]{
\includegraphics[width=0.155\textwidth]{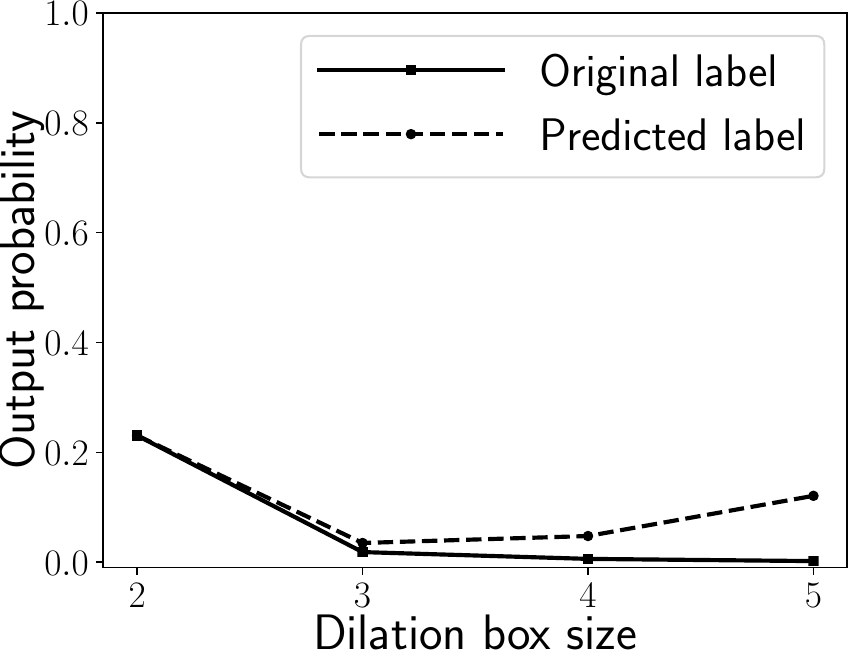}}%
\hfill
\subfloat[]{
\includegraphics[width=0.155\textwidth]{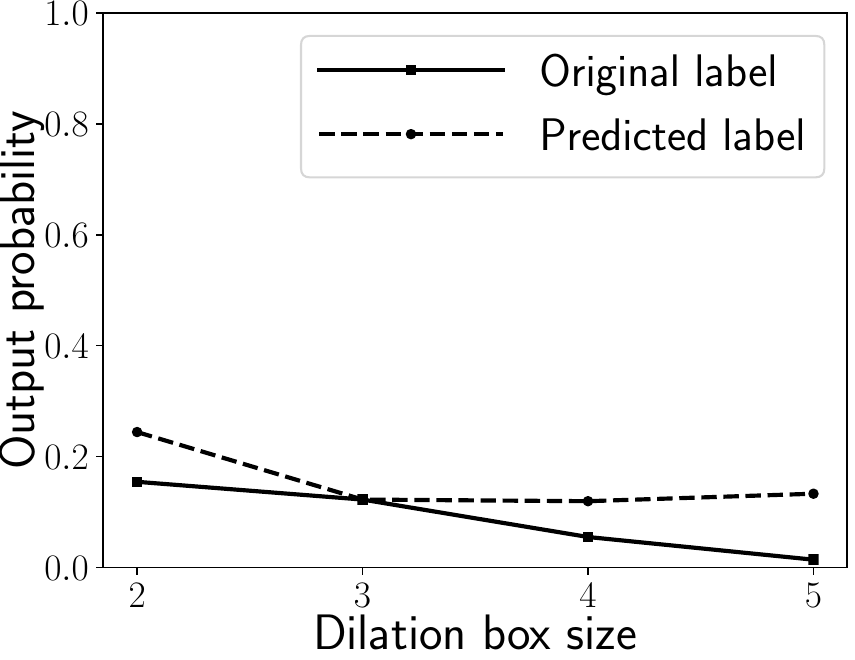}}%
\hfill
\subfloat[]{
\includegraphics[width=0.155\textwidth]{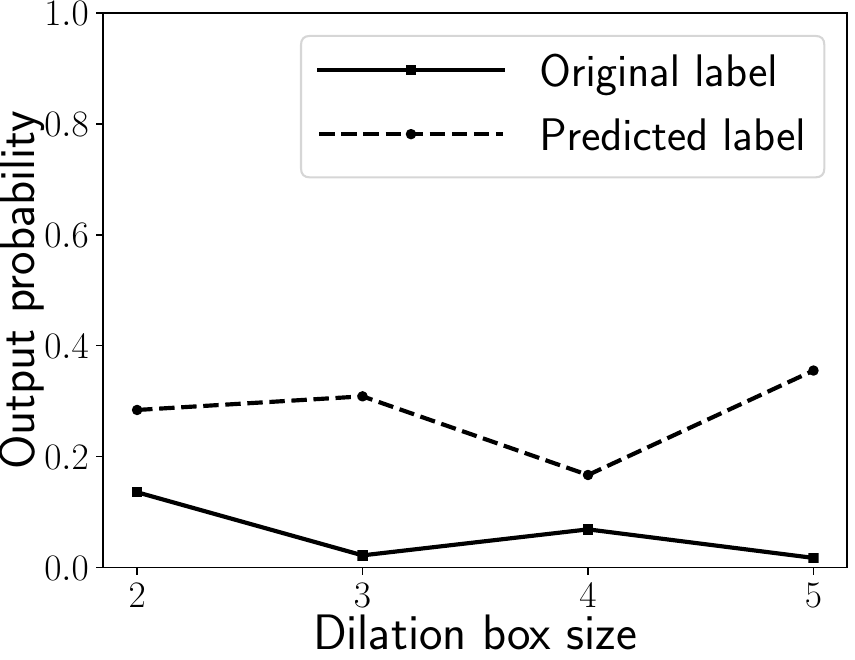}}%
\hfill
\subfloat[]{
\includegraphics[width=0.155\textwidth]{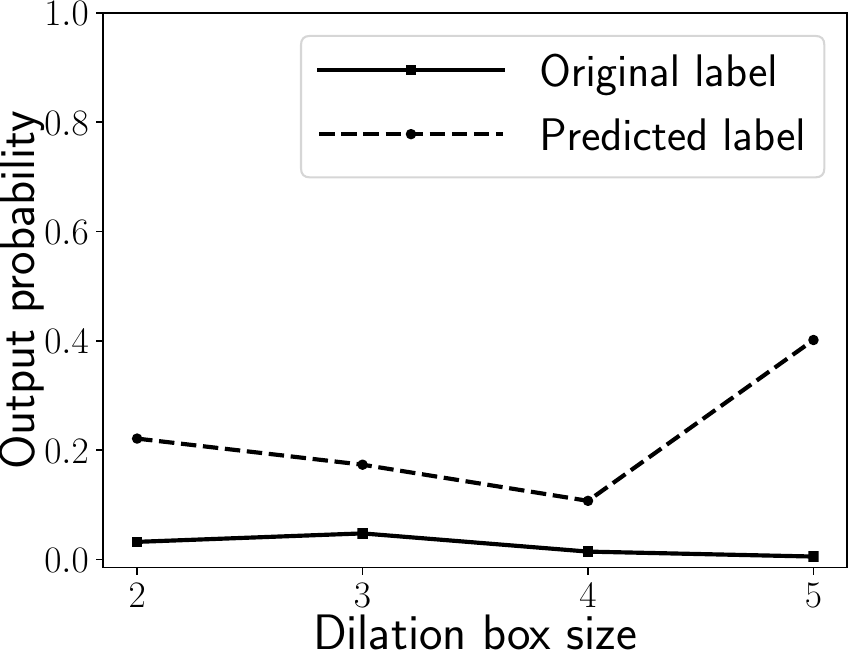}}%
\hfill
\subfloat[]{
\includegraphics[width=0.155\textwidth]{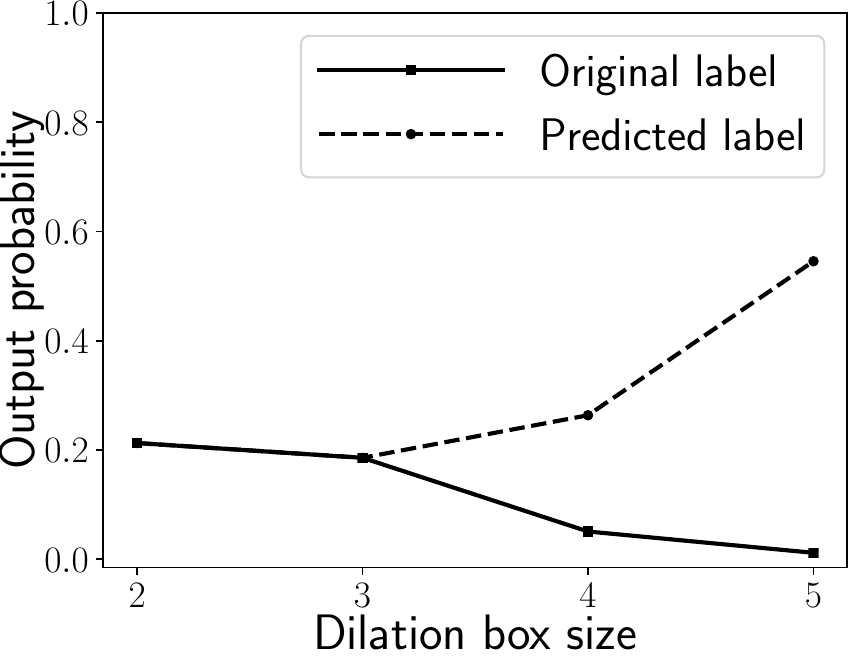}}%
\hfill
\subfloat[]{
\includegraphics[width=0.155\textwidth]{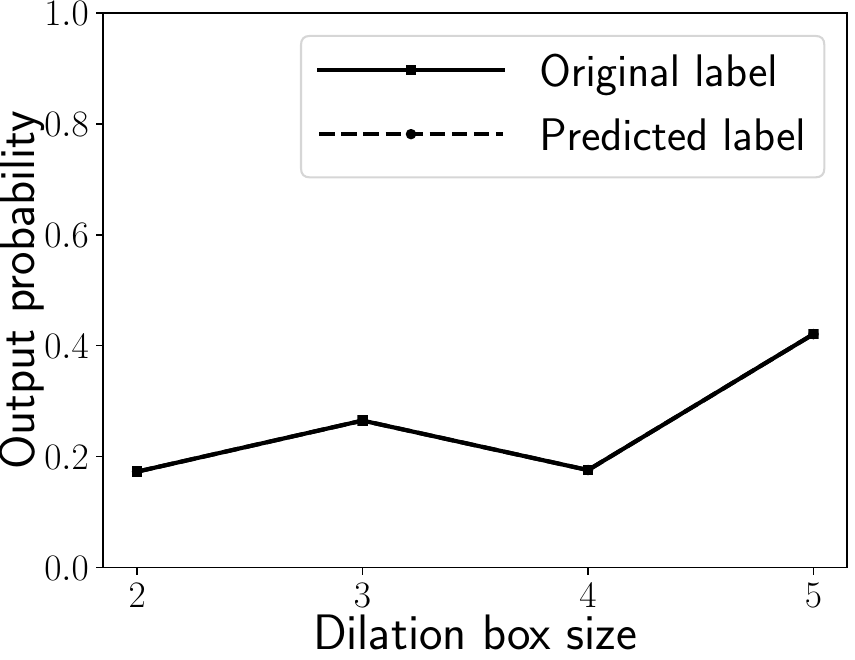}}%

\caption{Distribution of the prediction errors of convolution transforms in $\phi_1$ to $\phi_4$. The solid and dotted lines denote the output probability for the original and predicted labels respectively.
The higher the gap between the solid and dotted lines the higher the error (overlap means no error).}
\label{fig:prob_dist1}
\end{figure*}

\begin{figure*}
\captionsetup[subfloat]{captionskip=-.15cm, labelformat=empty}

\vspace{-.15cm}{\bf \scriptsize \hspace{0.025in} Contrast:\vspace{-.3cm}}

\subfloat[]{
\includegraphics[width=0.155\textwidth]{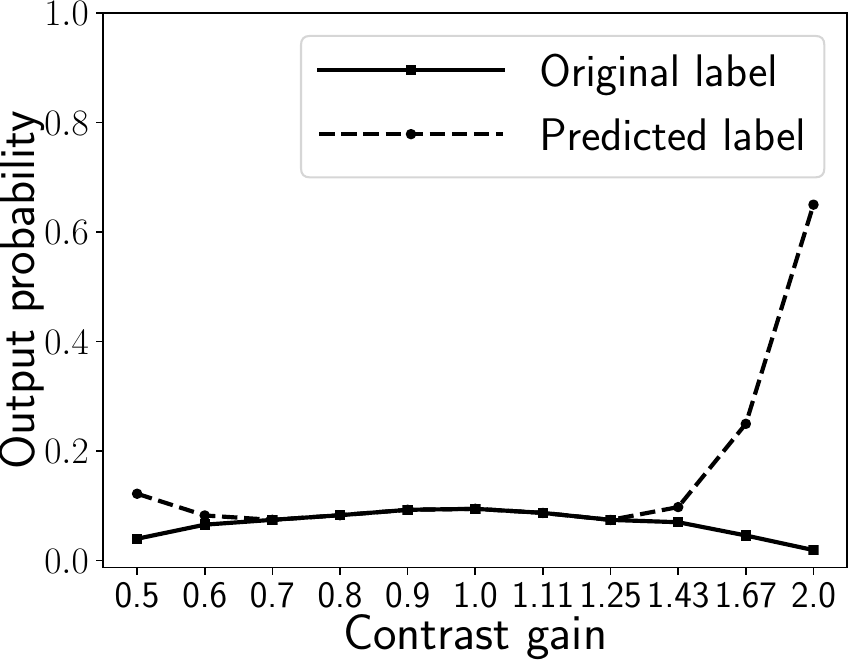}}%
\hfill
\subfloat[]{
\includegraphics[width=0.155\textwidth]{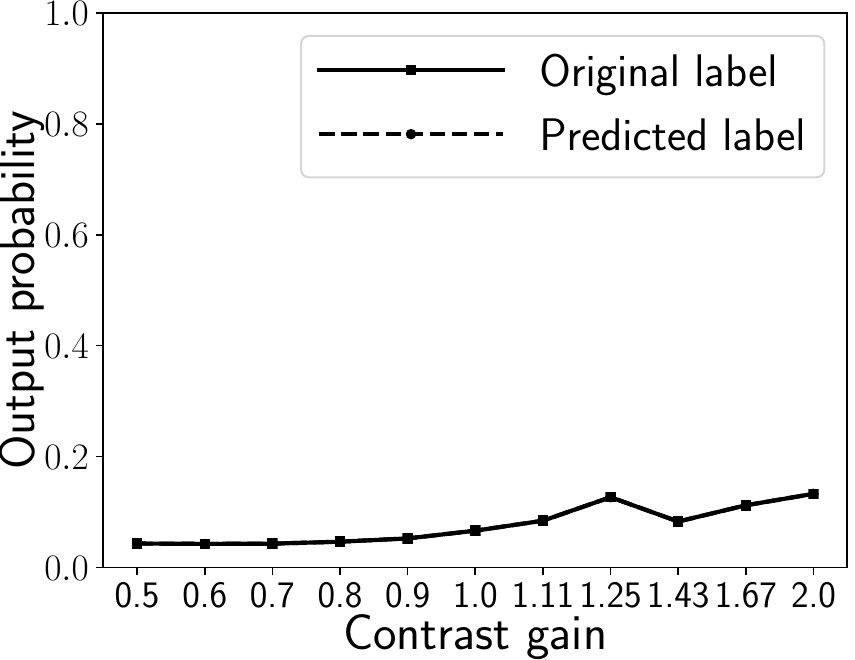}}%
\hfill
\subfloat[]{
\includegraphics[width=0.155\textwidth]{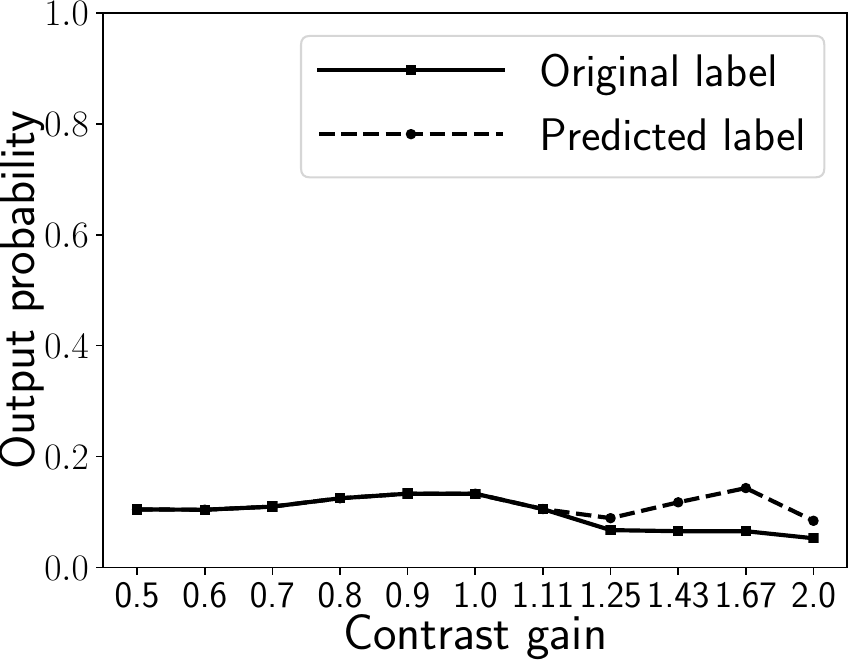}}%
\hfill
\subfloat[]{
\includegraphics[width=0.155\textwidth]{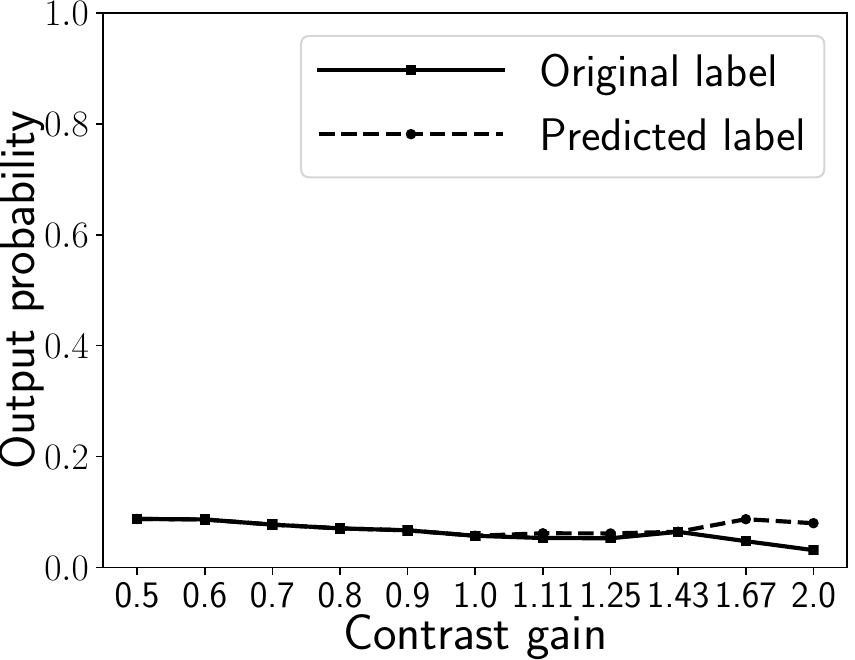}}%
\hfill
\subfloat[]{
\includegraphics[width=0.155\textwidth]{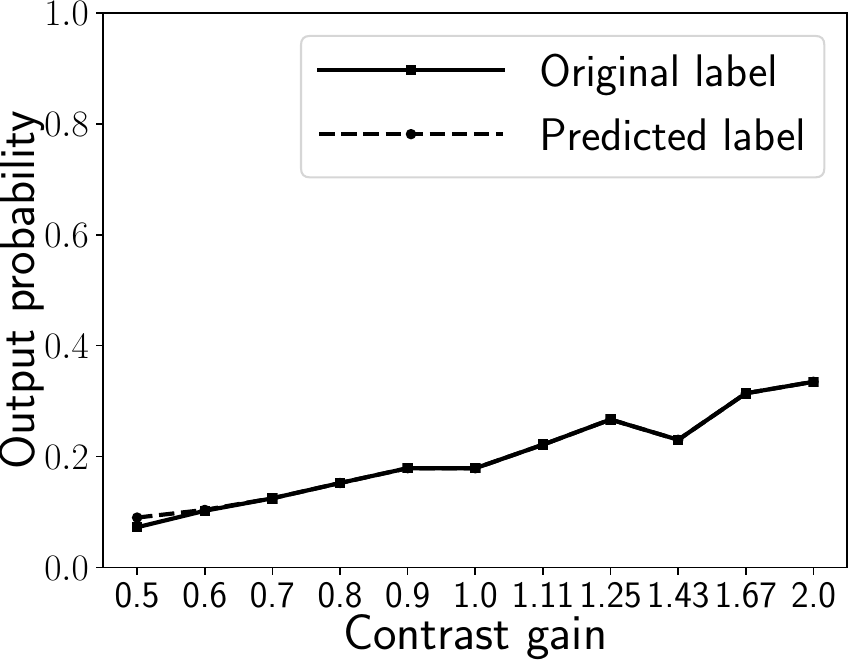}}%
\hfill
\subfloat[]{
\includegraphics[width=0.155\textwidth]{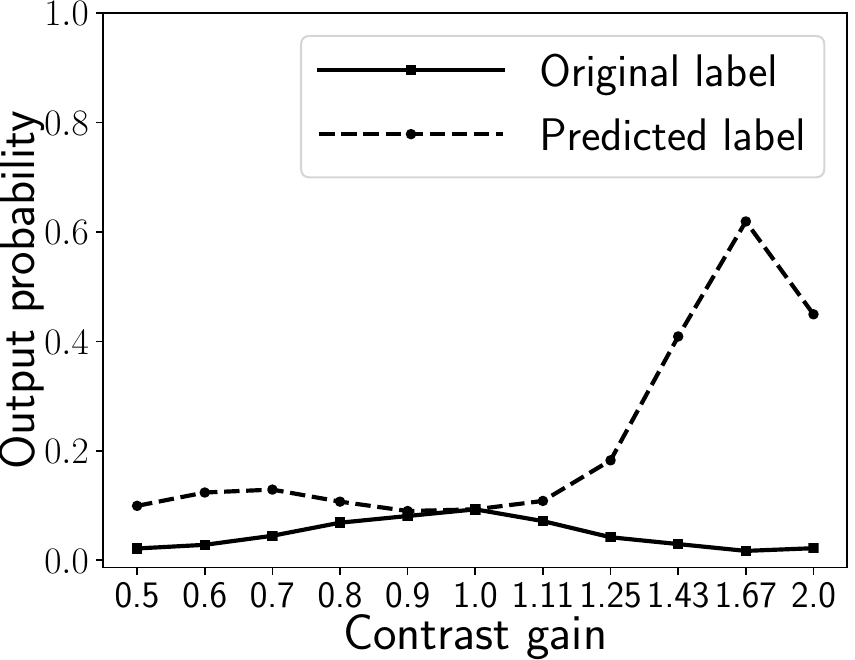}}%
\vspace{-.5cm}
\subfloat[]{
\includegraphics[width=0.155\textwidth]{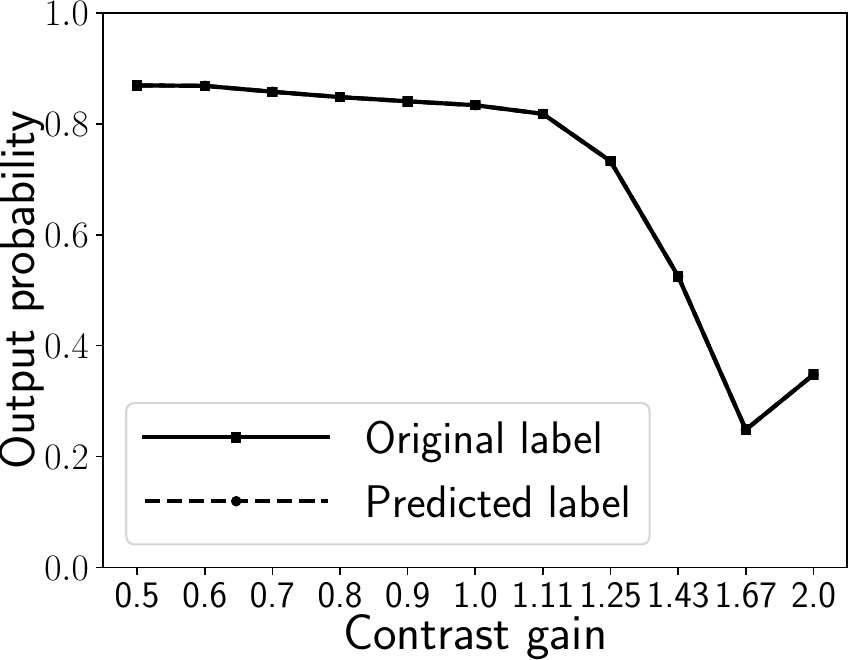}}%
\hfill
\subfloat[]{
\includegraphics[width=0.155\textwidth]{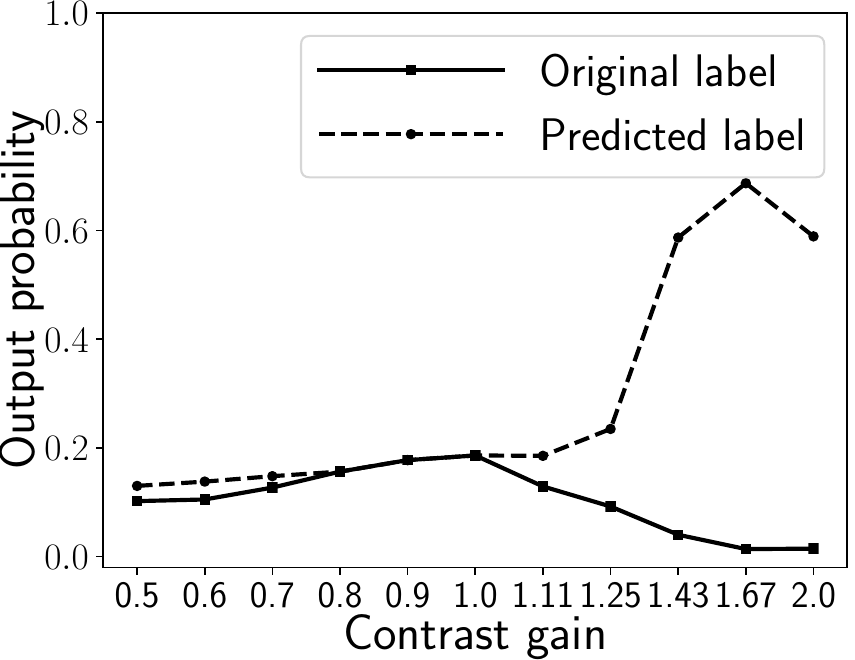}}%
\hfill
\subfloat[]{
\includegraphics[width=0.155\textwidth]{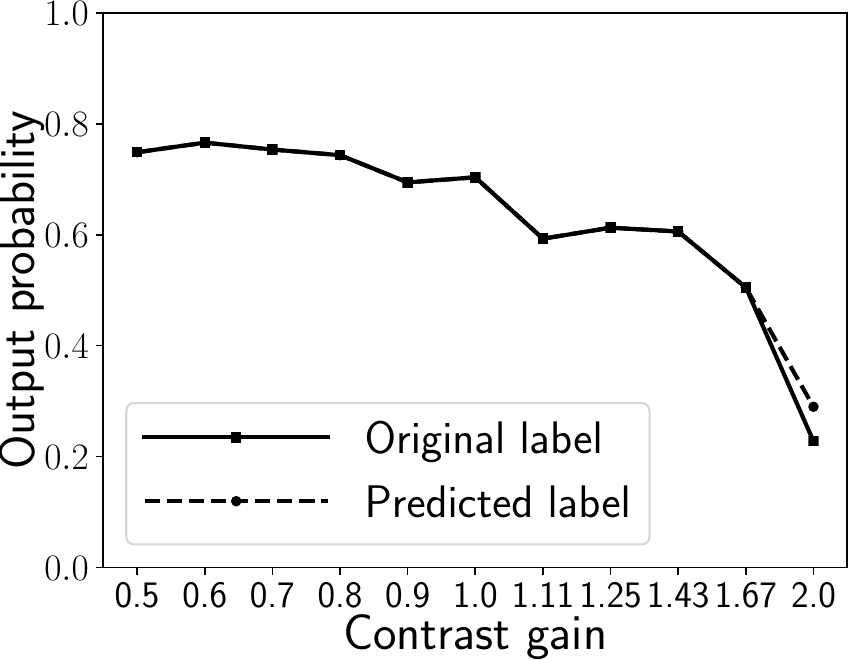}}%
\hfill
\subfloat[]{
\includegraphics[width=0.155\textwidth]{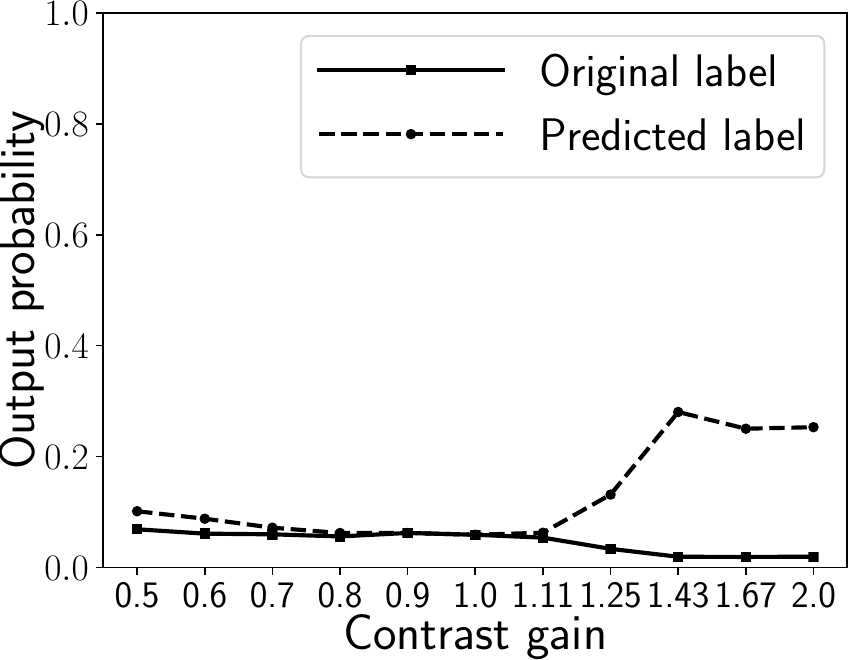}}%
\hfill
\subfloat[]{
\includegraphics[width=0.155\textwidth]{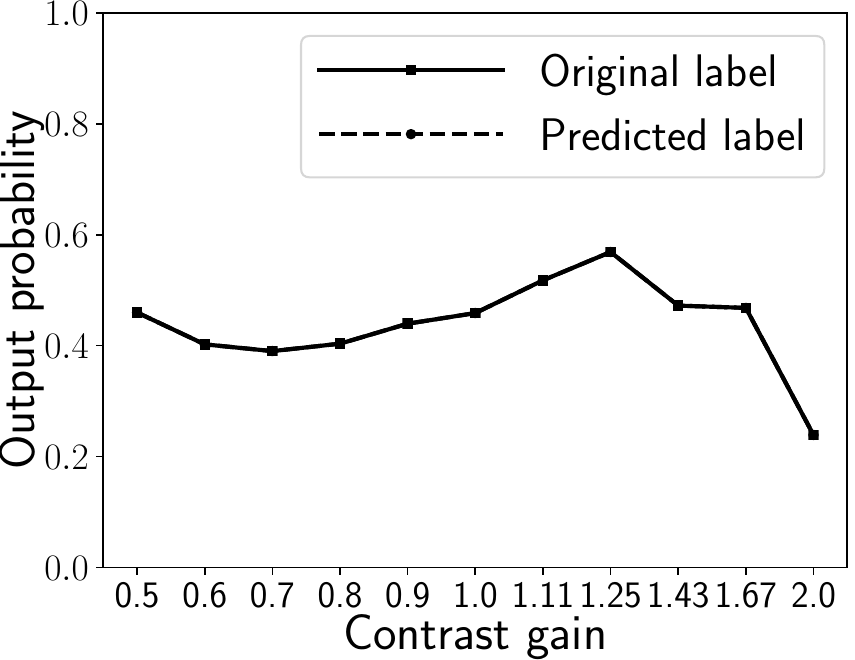}}%
\hfill
\subfloat[]{
\includegraphics[width=0.155\textwidth]{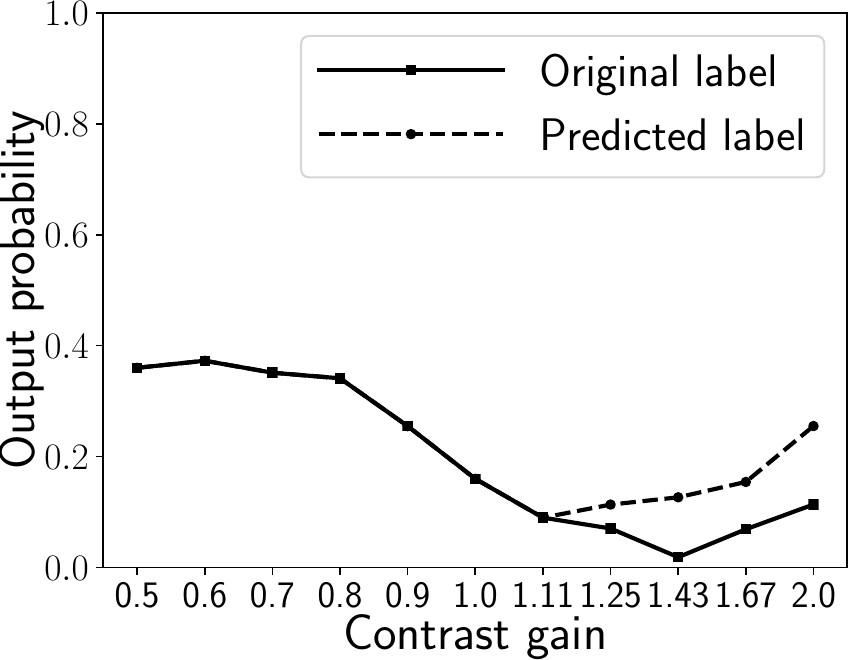}}%

\vspace{-.15cm}{\bf \scriptsize \hspace{0.025in} Brightness:\vspace{-.3cm}}%

\subfloat[]{
\includegraphics[width=0.155\textwidth]{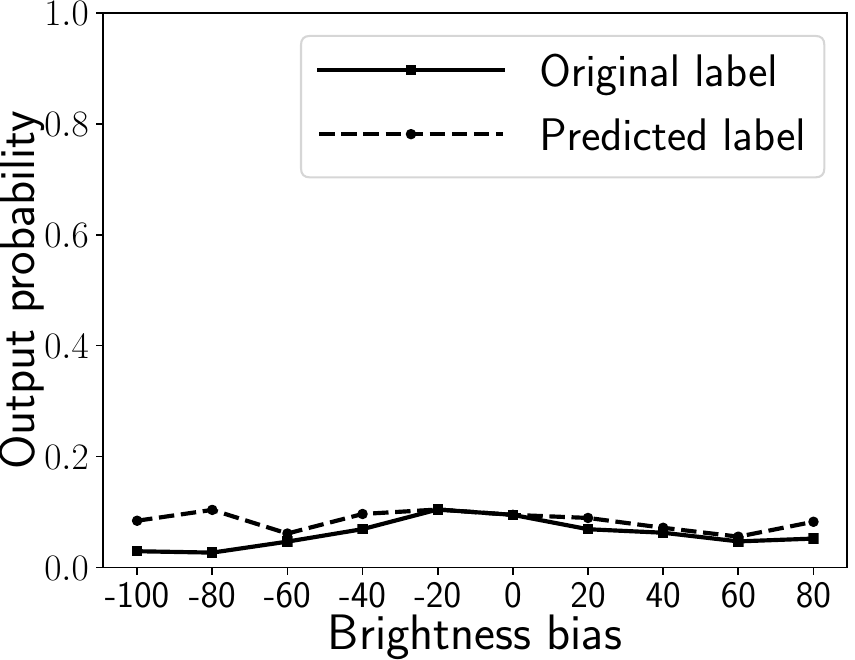}}%
\hfill
\subfloat[]{
\includegraphics[width=0.155\textwidth]{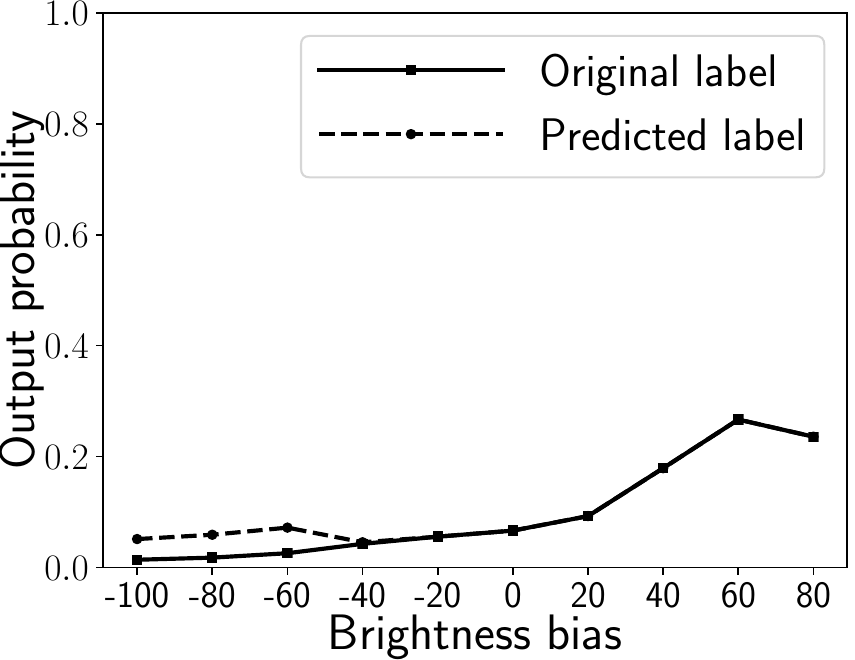}}%
\hfill
\subfloat[]{
\includegraphics[width=0.155\textwidth]{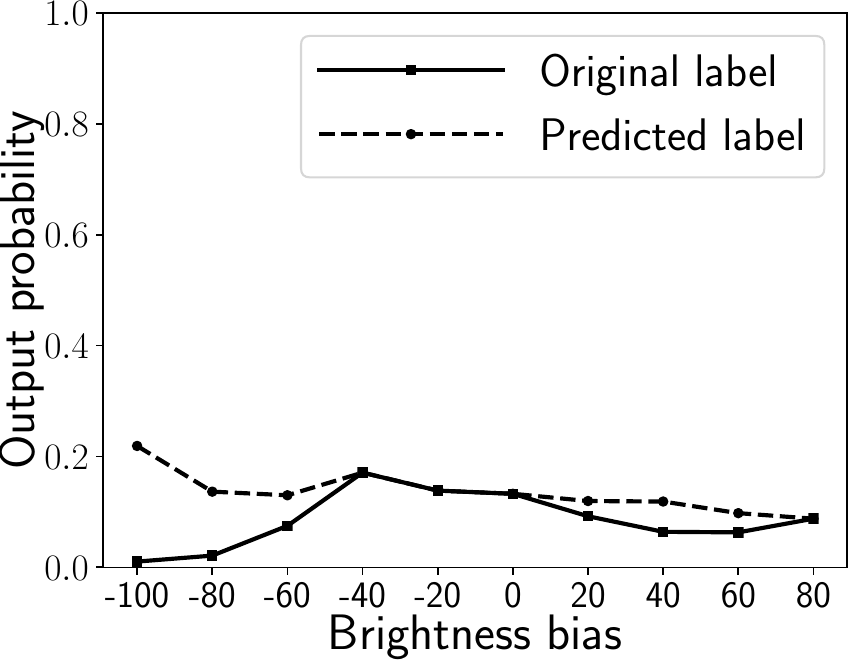}}%
\hfill
\subfloat[]{
\includegraphics[width=0.155\textwidth]{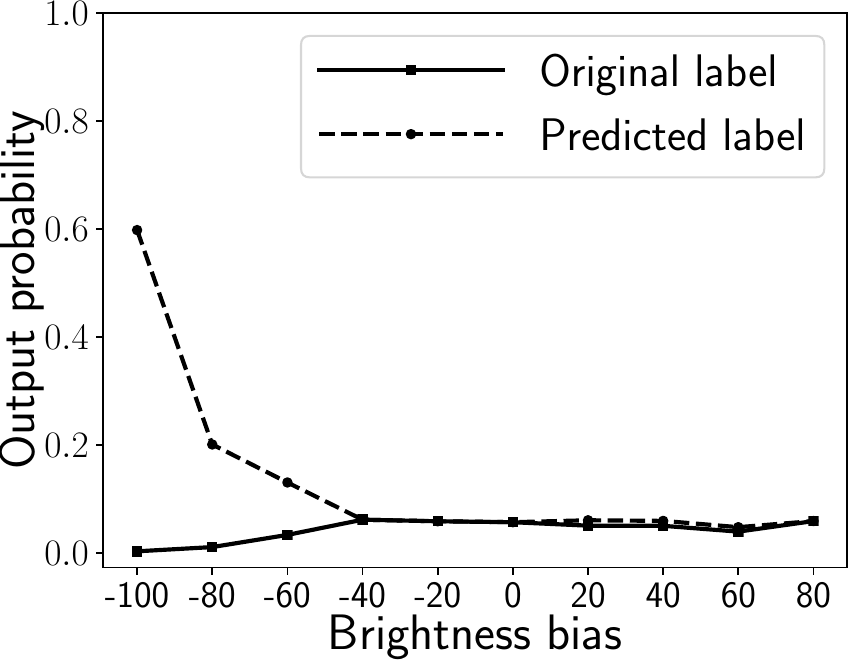}}%
\hfill
\subfloat[]{
\includegraphics[width=0.155\textwidth]{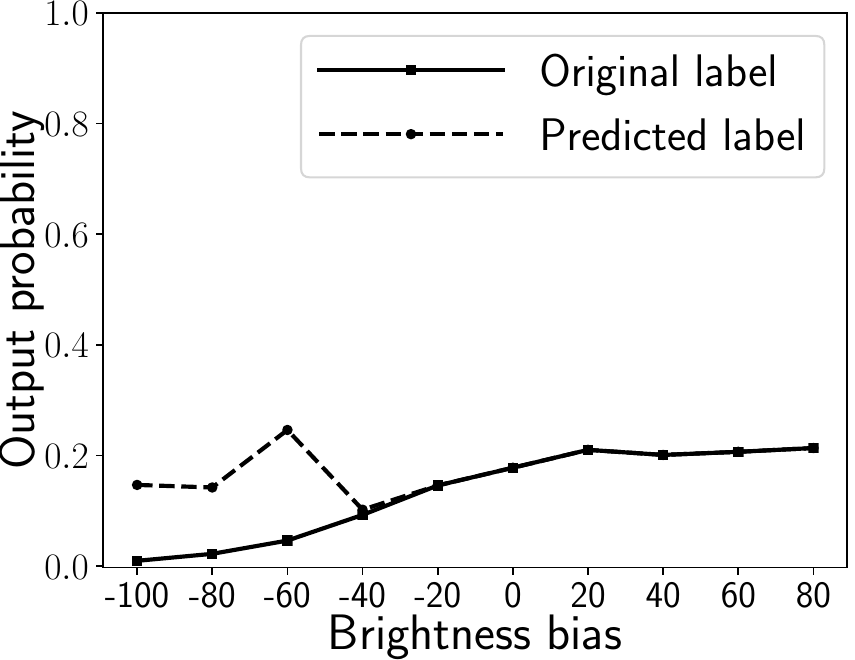}}%
\hfill
\subfloat[]{
\includegraphics[width=0.155\textwidth]{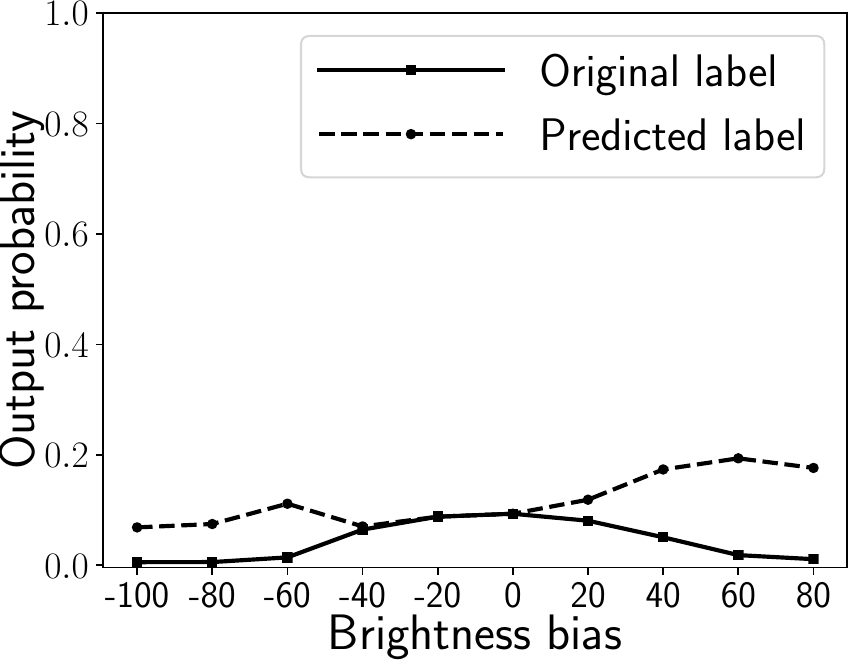}}%
\vspace{-.5cm}
\subfloat[]{
\includegraphics[width=0.155\textwidth]{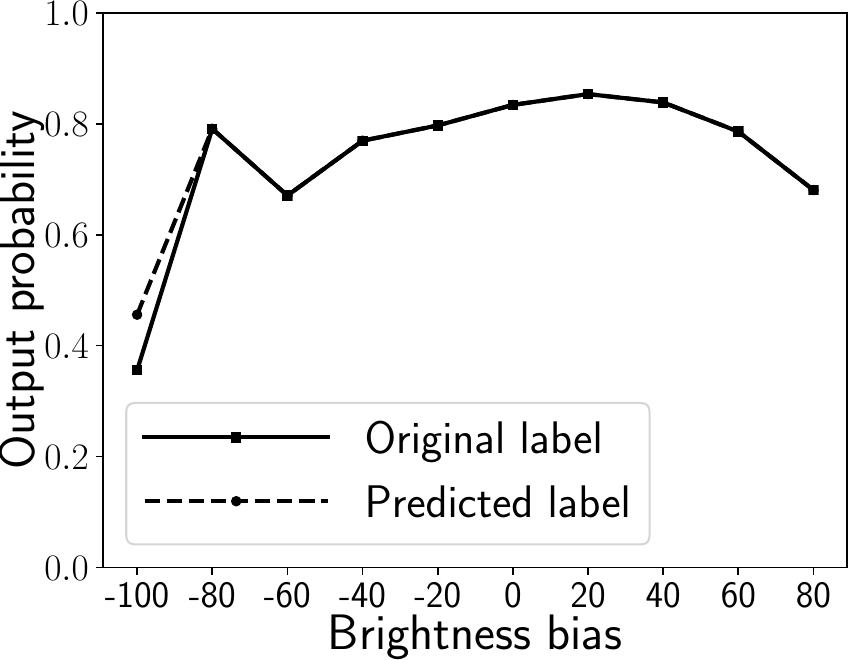}}%
\hfill
\subfloat[]{
\includegraphics[width=0.155\textwidth]{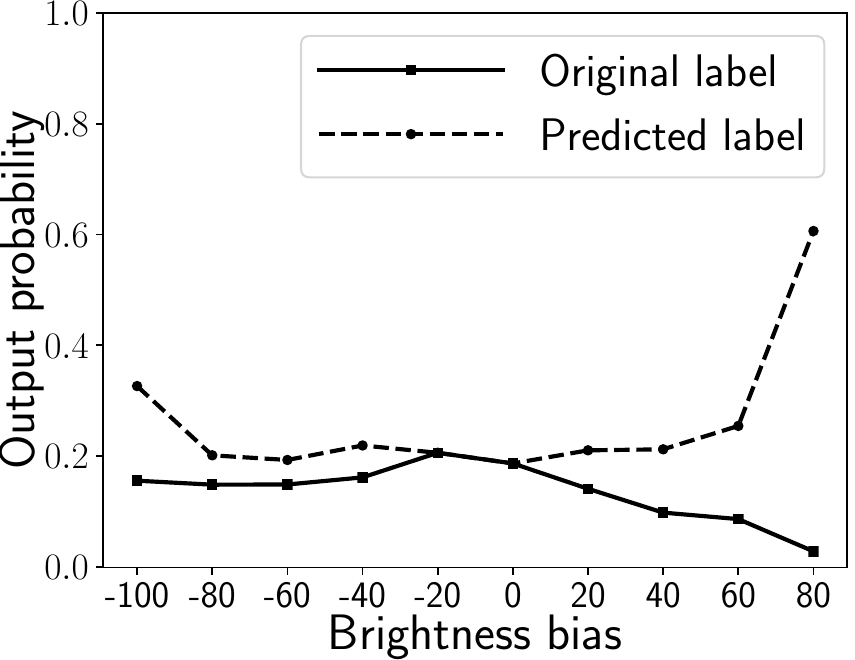}}%
\hfill
\subfloat[]{
\includegraphics[width=0.155\textwidth]{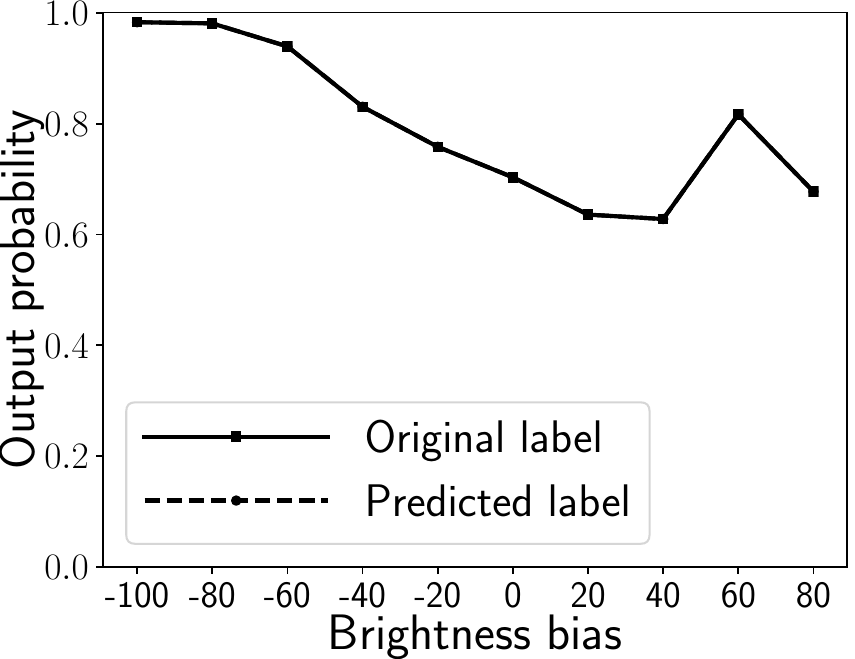}}%
\hfill
\subfloat[]{
\includegraphics[width=0.155\textwidth]{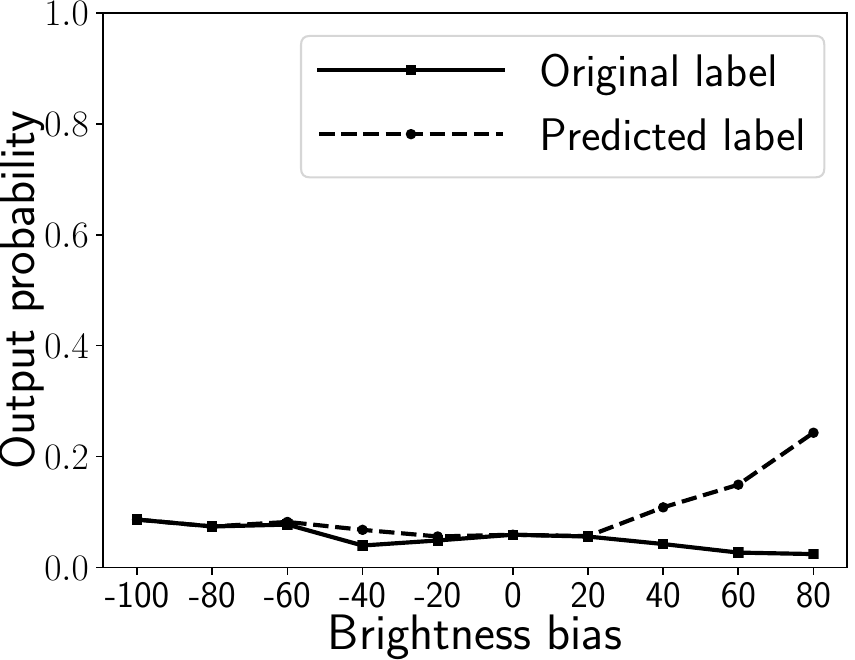}}%
\hfill
\subfloat[]{
\includegraphics[width=0.155\textwidth]{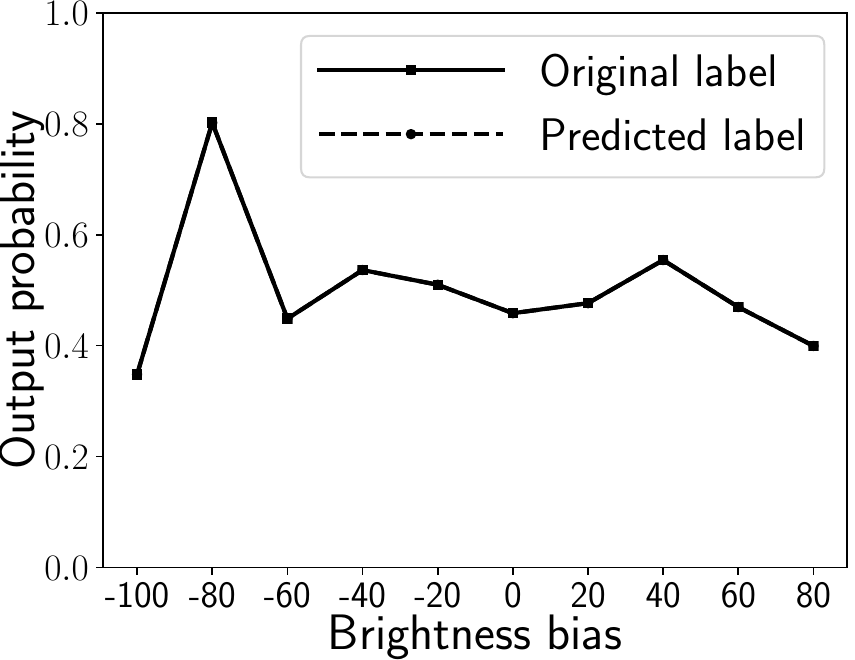}}%
\hfill
\subfloat[]{
\includegraphics[width=0.155\textwidth]{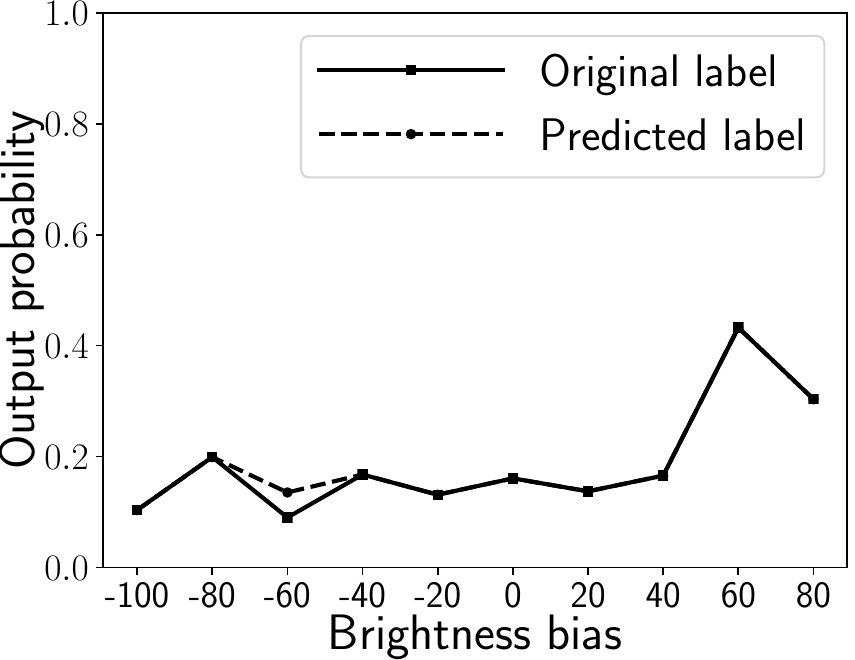}}%

\vspace{-.15cm}{\bf \scriptsize \hspace{0.025in} Occlusion:\vspace{-.3cm}}%

\subfloat[]{
\includegraphics[width=0.155\textwidth]{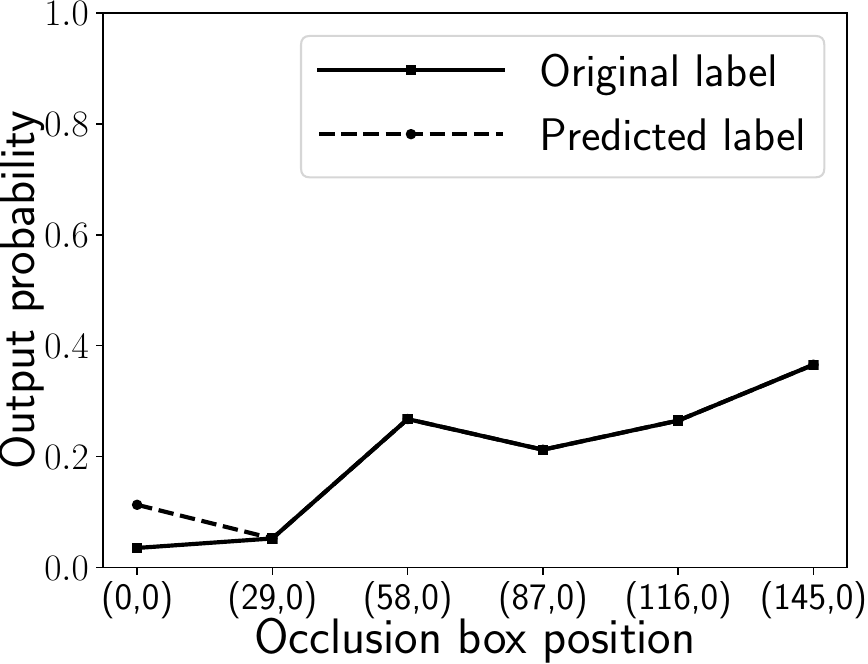}}%
\hfill
\subfloat[]{
\includegraphics[width=0.155\textwidth]{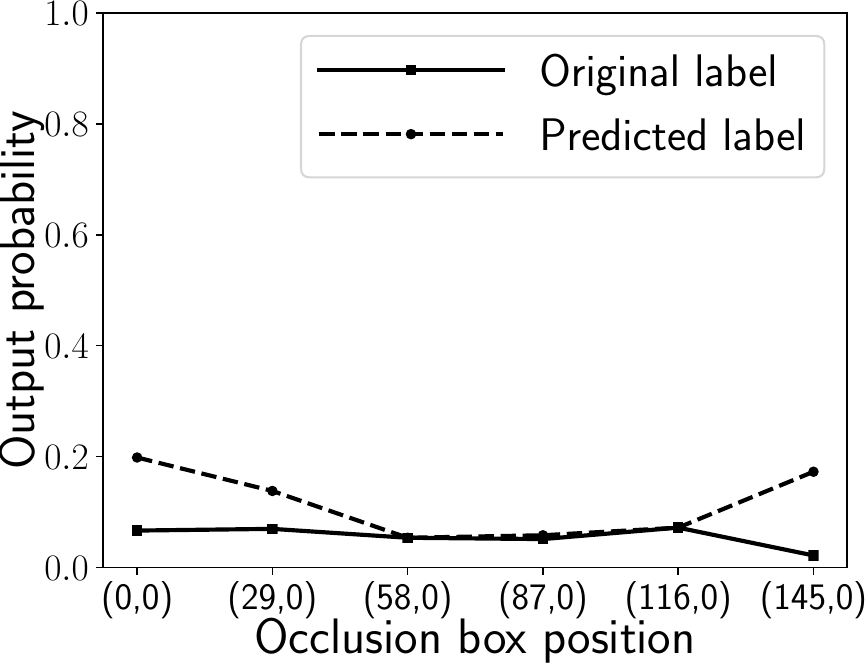}}%
\hfill
\subfloat[]{
\includegraphics[width=0.155\textwidth]{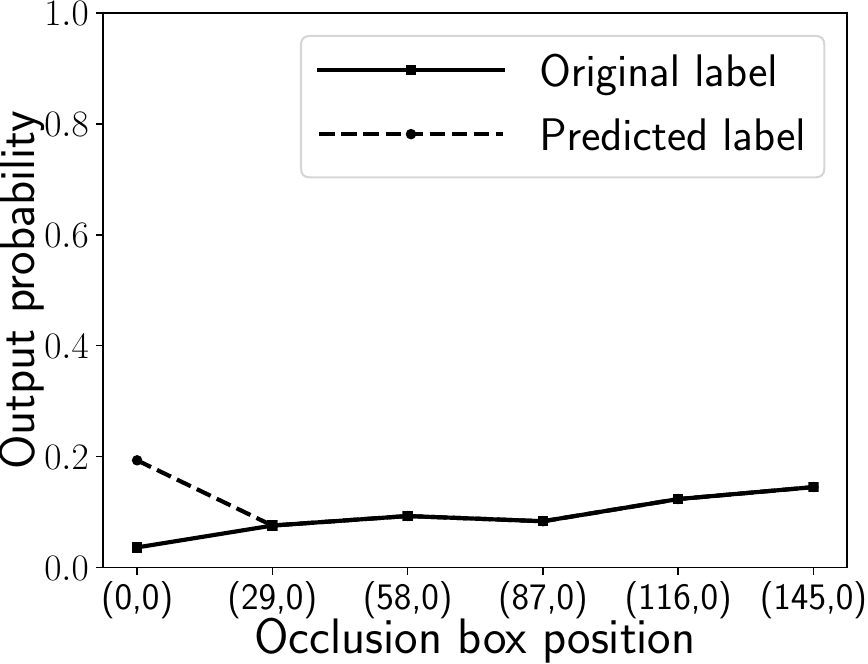}}%
\hfill
\subfloat[]{
\includegraphics[width=0.155\textwidth]{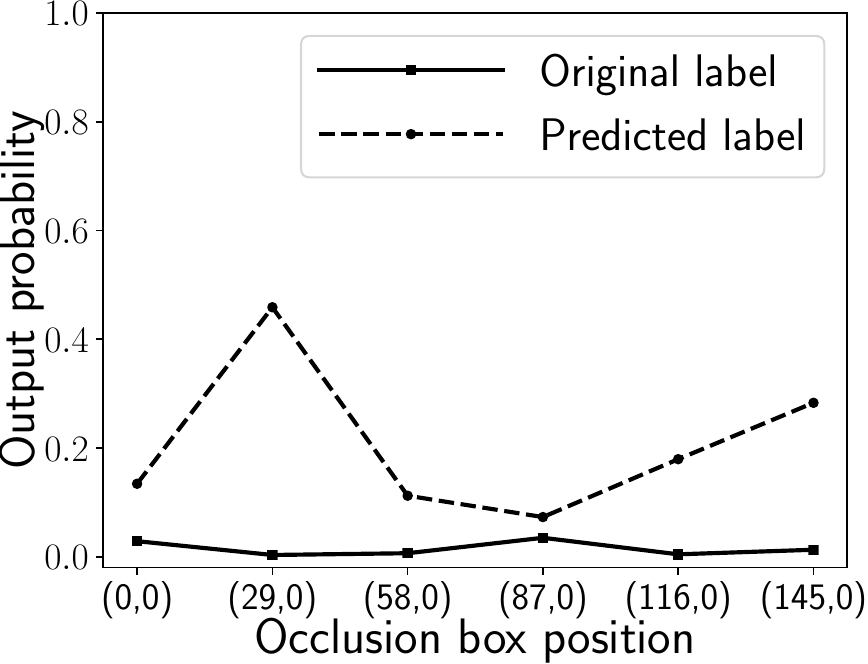}}%
\hfill
\subfloat[]{
\includegraphics[width=0.155\textwidth]{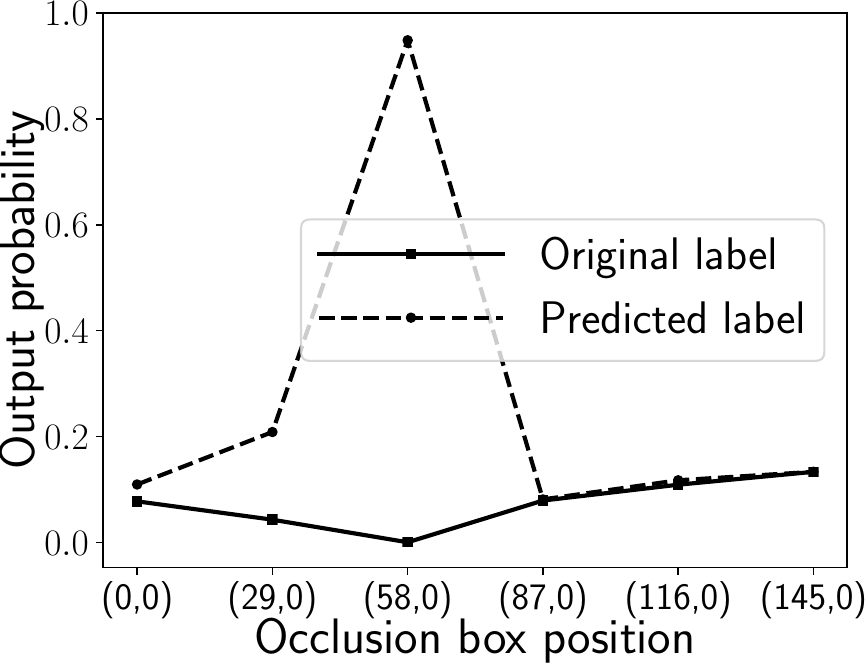}}%
\hfill
\subfloat[]{
\includegraphics[width=0.155\textwidth]{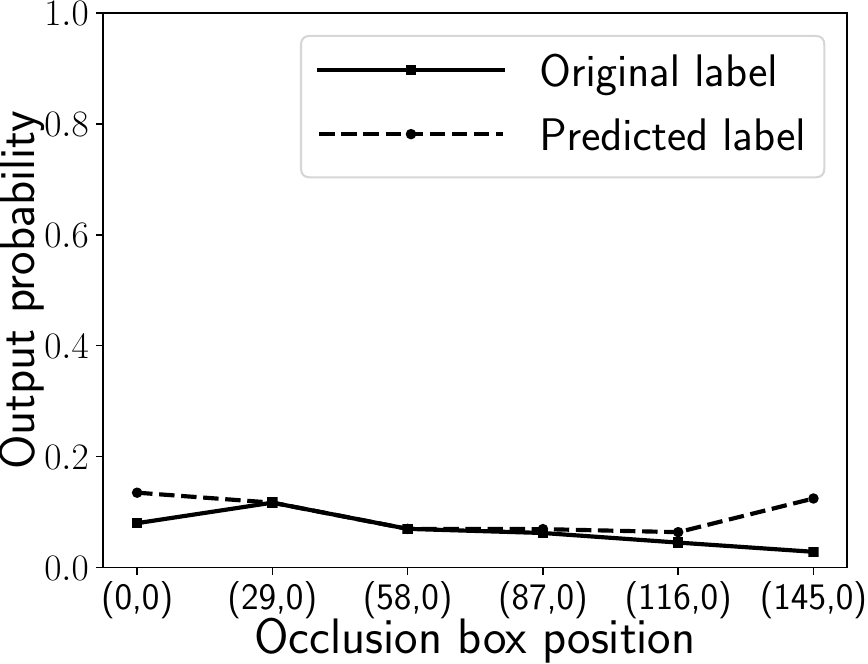}}%
\vspace{-.5cm}
\subfloat[]{
\includegraphics[width=0.155\textwidth]{./figs/prob_dist/probdist_illu_inceptionresnetv2.pdf}}%
\hfill
\subfloat[]{
\includegraphics[width=0.155\textwidth]{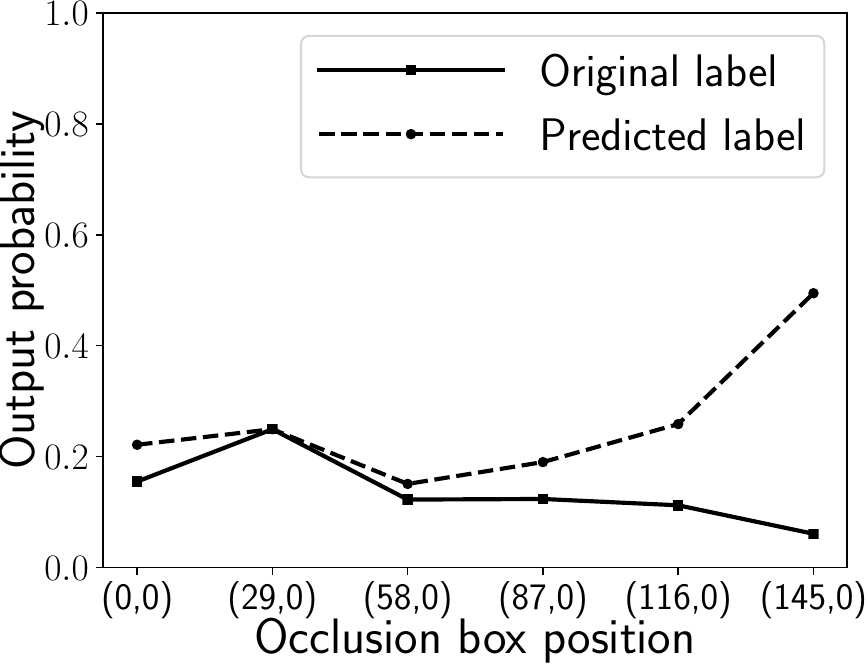}}%
\hfill
\subfloat[]{
\includegraphics[width=0.155\textwidth]{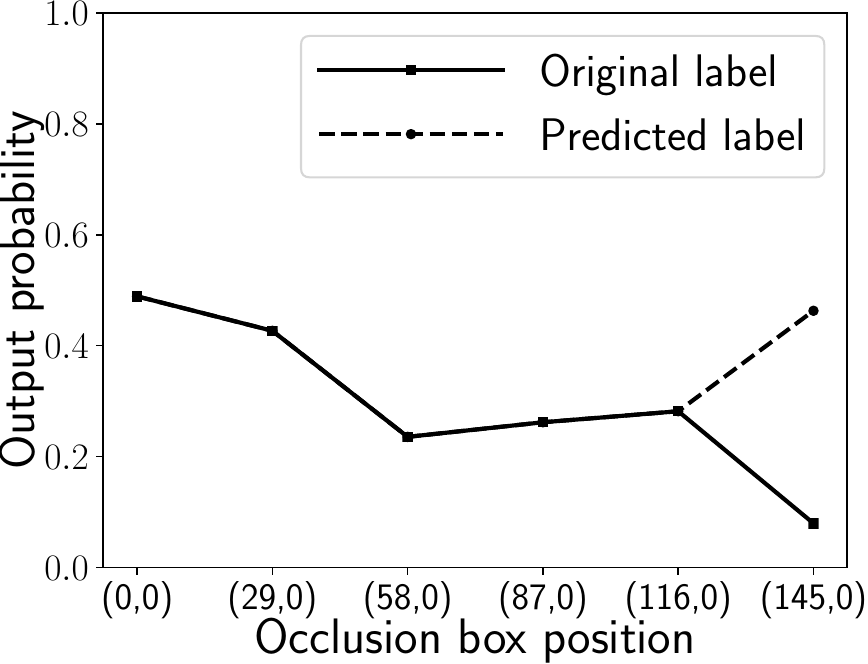}}%
\hfill
\subfloat[]{
\includegraphics[width=0.155\textwidth]{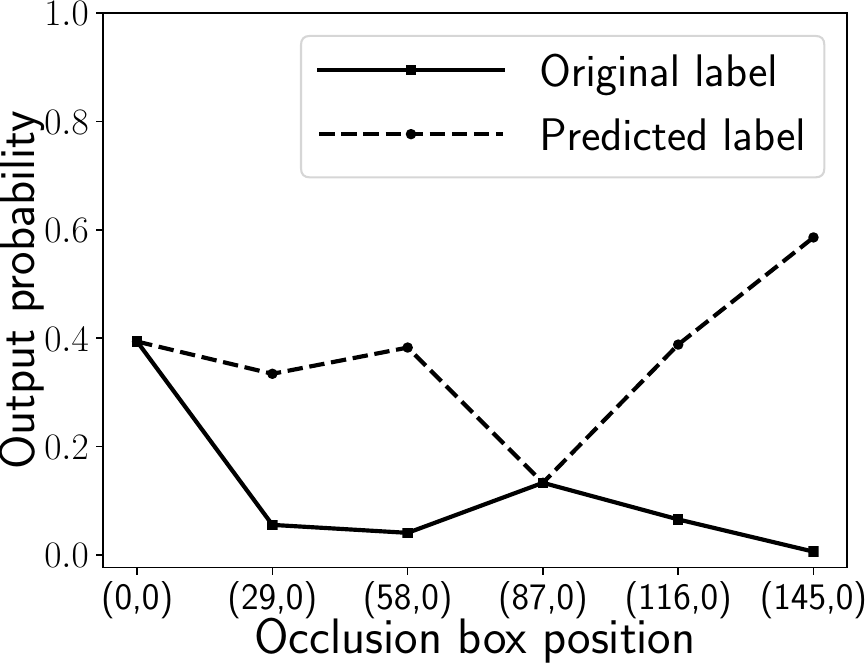}}%
\hfill
\subfloat[]{
\includegraphics[width=0.155\textwidth]{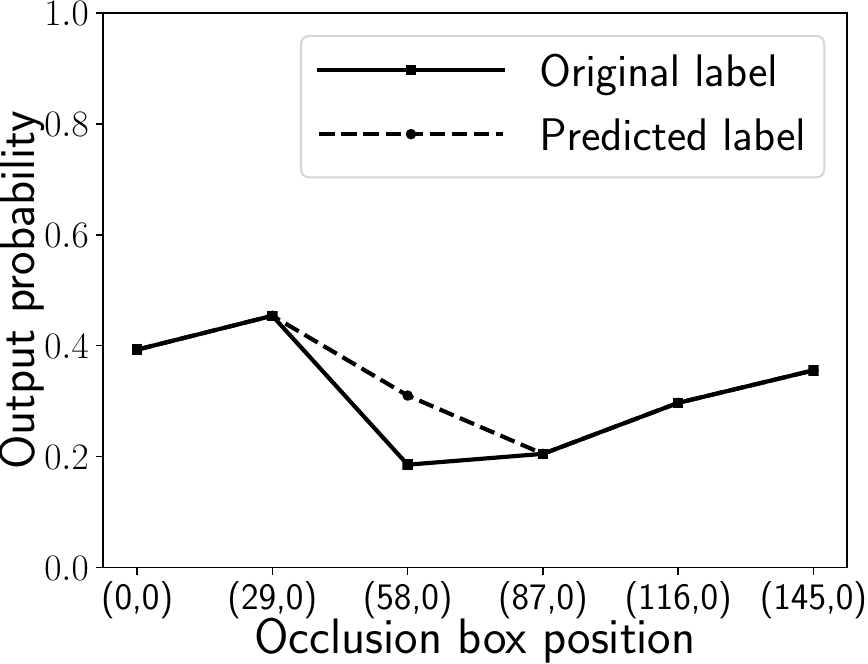}}%
\hfill
\subfloat[]{
\includegraphics[width=0.155\textwidth]{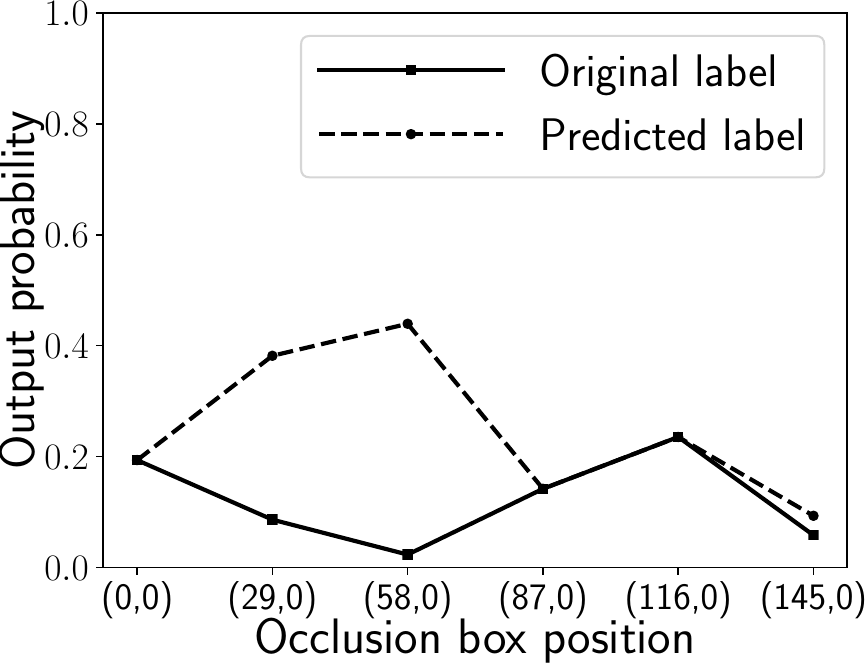}}%

\caption{Distribution of the prediction errors of contrast ($\phi_5$), brightening ($\phi_6$), and occlusion ($\phi_7$). The solid and dotted lines denote the output probability for the original and predicted labels respectively.
The higher the gap between the solid and dotted lines the higher the error (overlap means no error).}
\label{fig:prob_dist2}
\end{figure*}

\begin{figure*}
\captionsetup[subfloat]{captionskip=-.15cm, labelformat=empty}

\vspace{-.15cm}{\bf \scriptsize \hspace{0.025in} Rotation:\vspace{-.3cm}}%

\subfloat[]{
\includegraphics[width=0.155\textwidth]{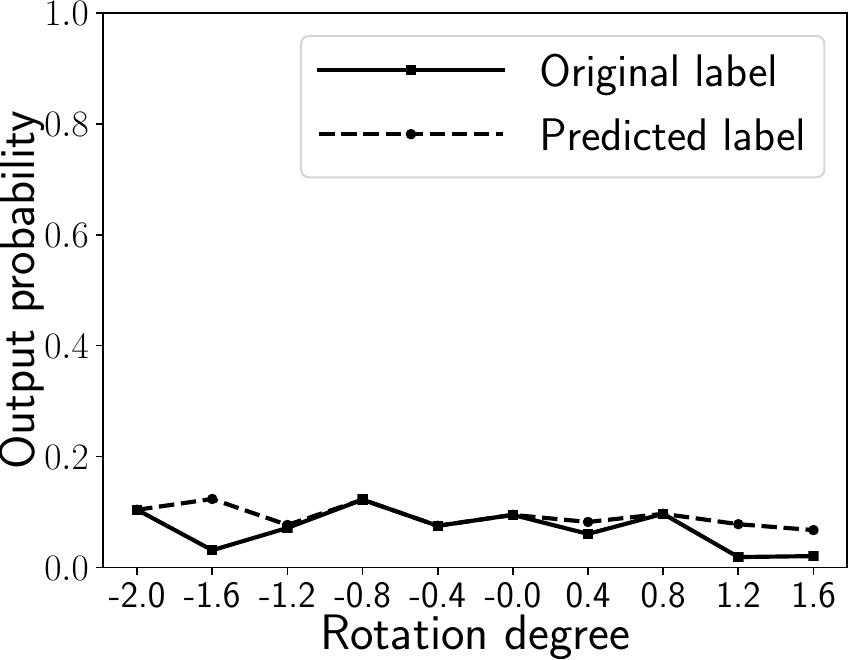}}%
\hfill
\subfloat[]{
\includegraphics[width=0.155\textwidth]{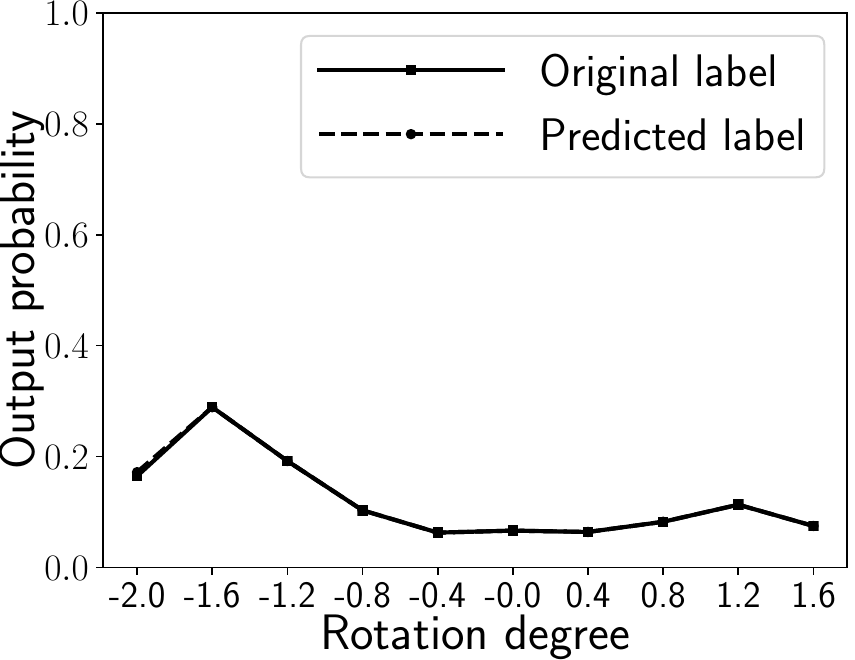}}%
\hfill
\subfloat[]{
\includegraphics[width=0.155\textwidth]{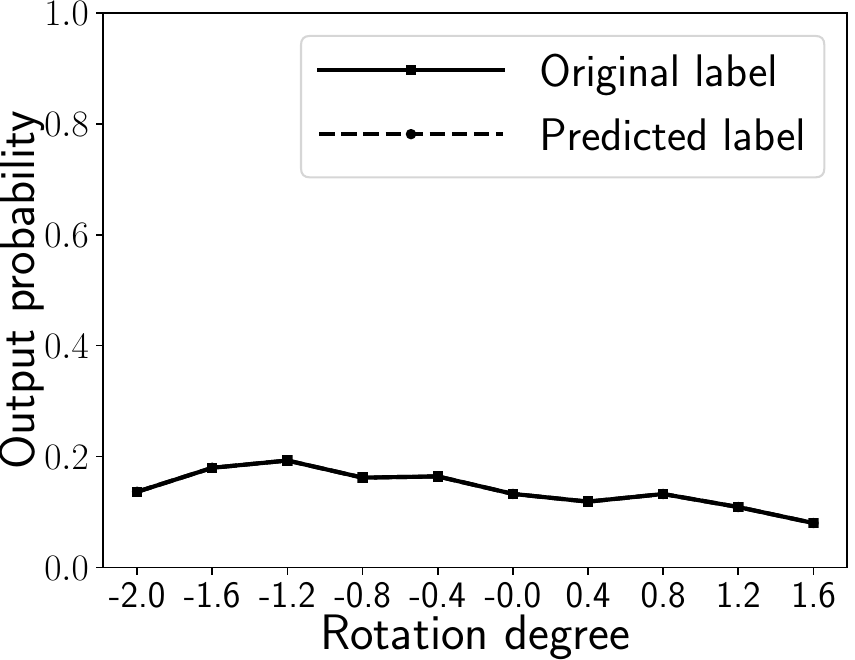}}%
\hfill
\subfloat[]{
\includegraphics[width=0.155\textwidth]{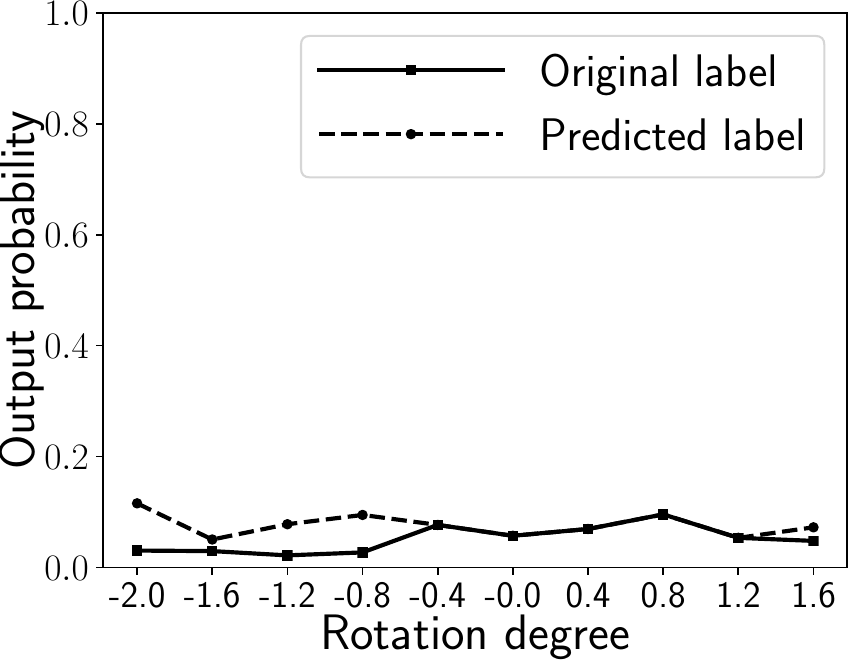}}%
\hfill
\subfloat[]{
\includegraphics[width=0.155\textwidth]{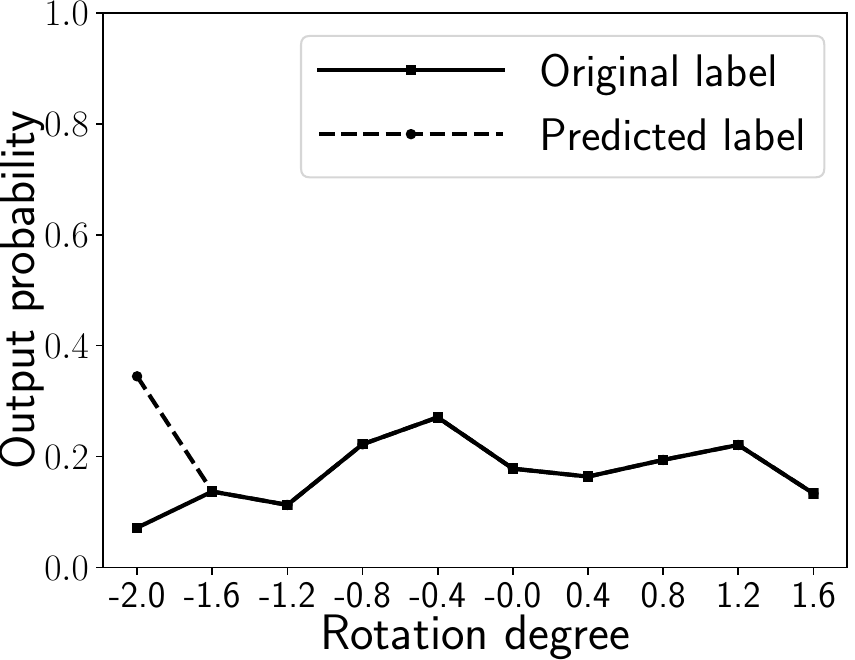}}%
\hfill
\subfloat[]{
\includegraphics[width=0.155\textwidth]{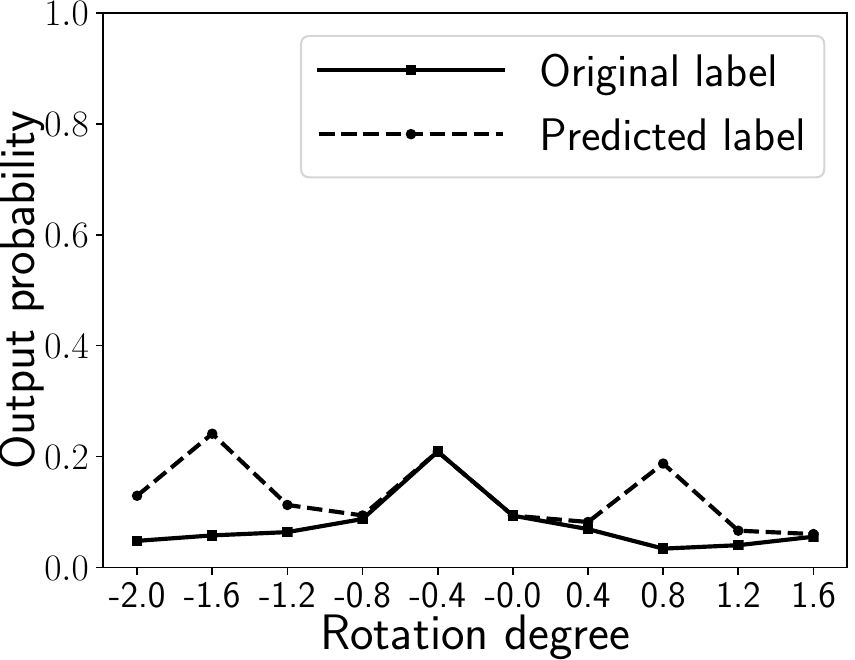}}%
\vspace{-.5cm}
\subfloat[]{
\includegraphics[width=0.155\textwidth]{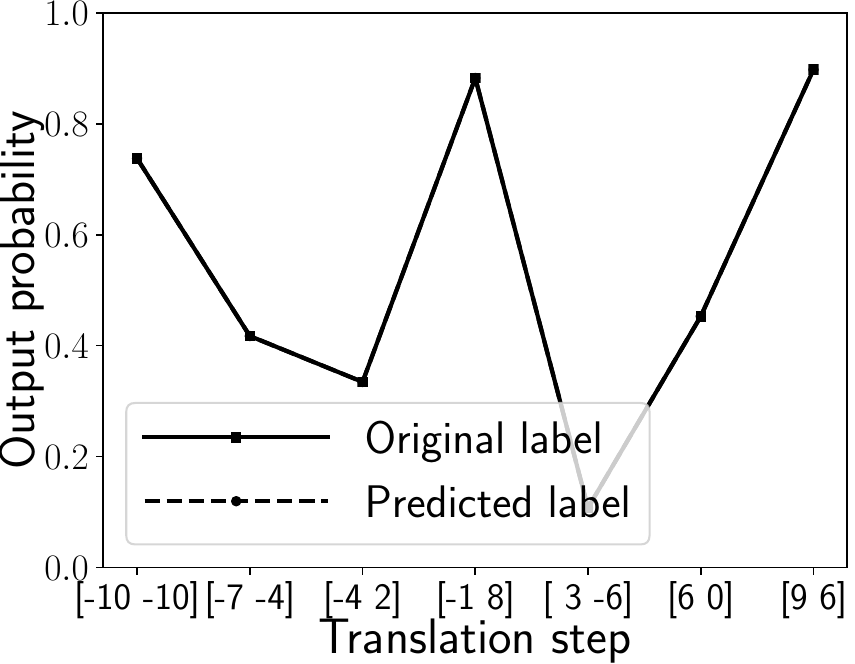}}%
\hfill
\subfloat[]{
\includegraphics[width=0.155\textwidth]{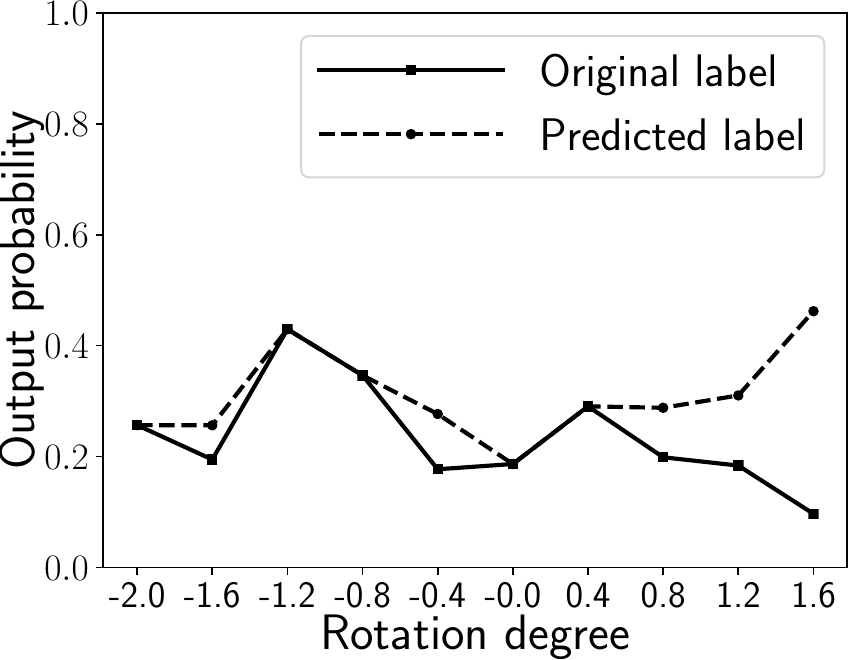}}%
\hfill
\subfloat[]{
\includegraphics[width=0.155\textwidth]{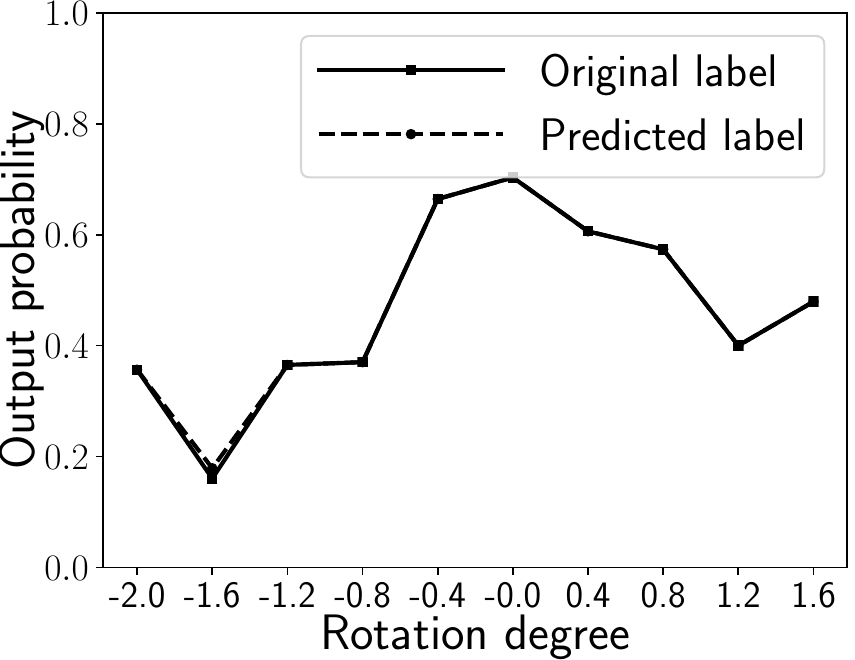}}%
\hfill
\subfloat[]{
\includegraphics[width=0.155\textwidth]{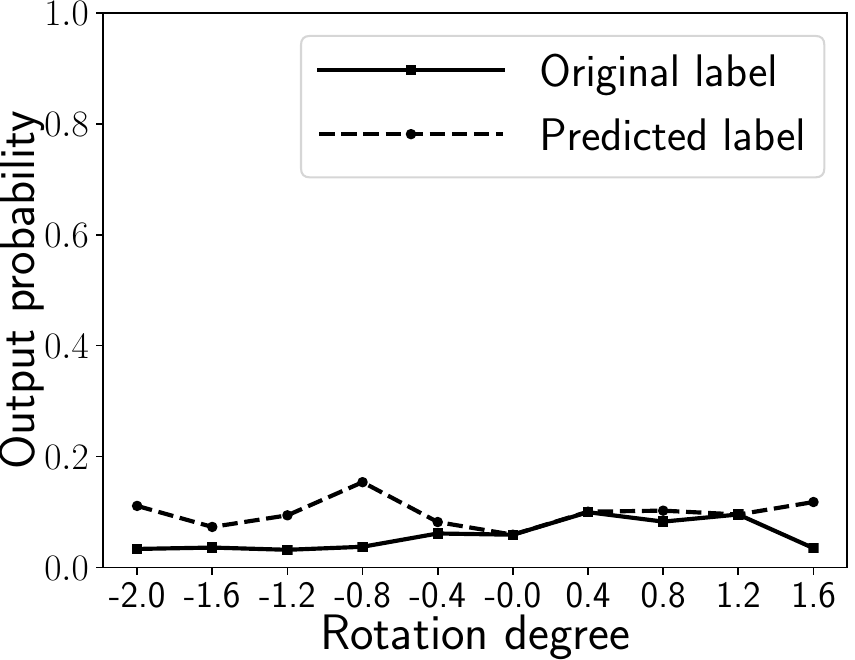}}%
\hfill
\subfloat[]{
\includegraphics[width=0.155\textwidth]{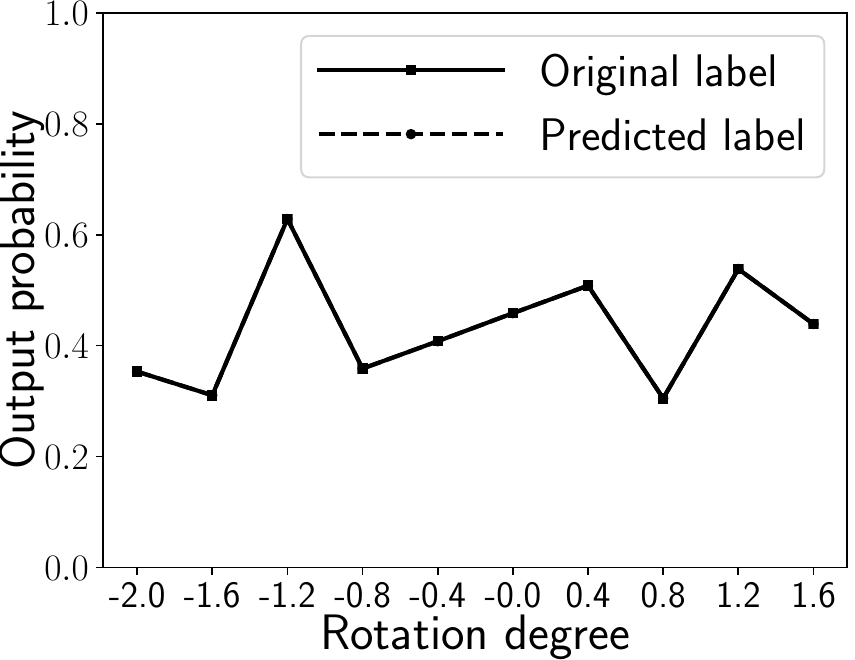}}%
\hfill
\subfloat[]{
\includegraphics[width=0.155\textwidth]{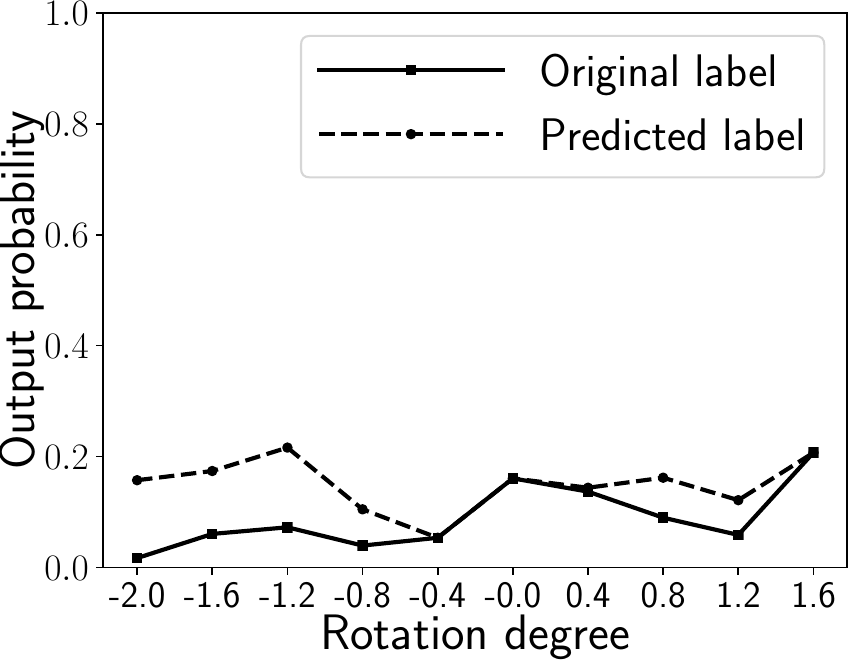}}%

\vspace{-.15cm}{\bf \scriptsize \hspace{0.025in} Shear:\vspace{-.3cm}}%

\subfloat[]{
\includegraphics[width=0.155\textwidth]{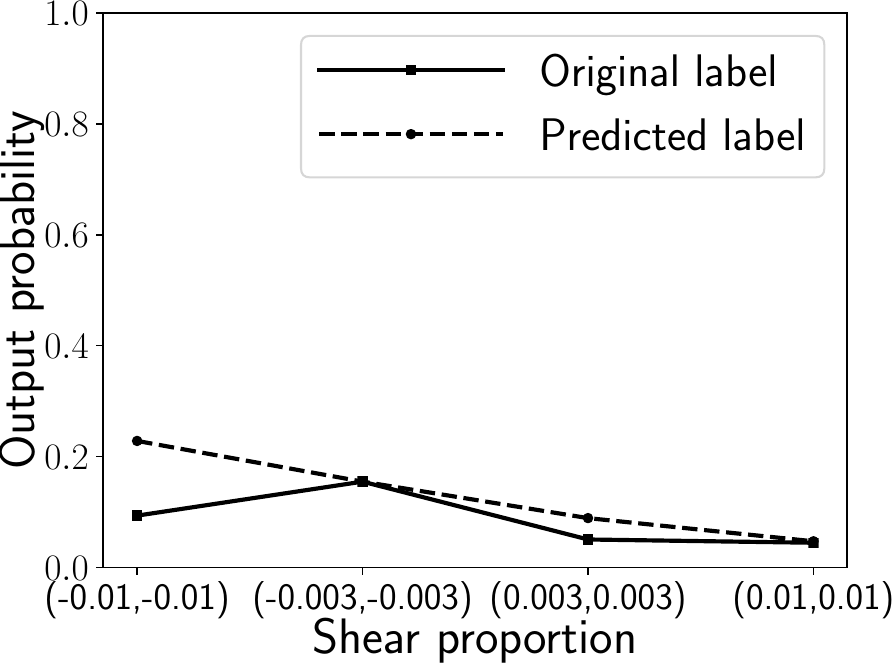}}%
\hfill
\subfloat[]{
\includegraphics[width=0.155\textwidth]{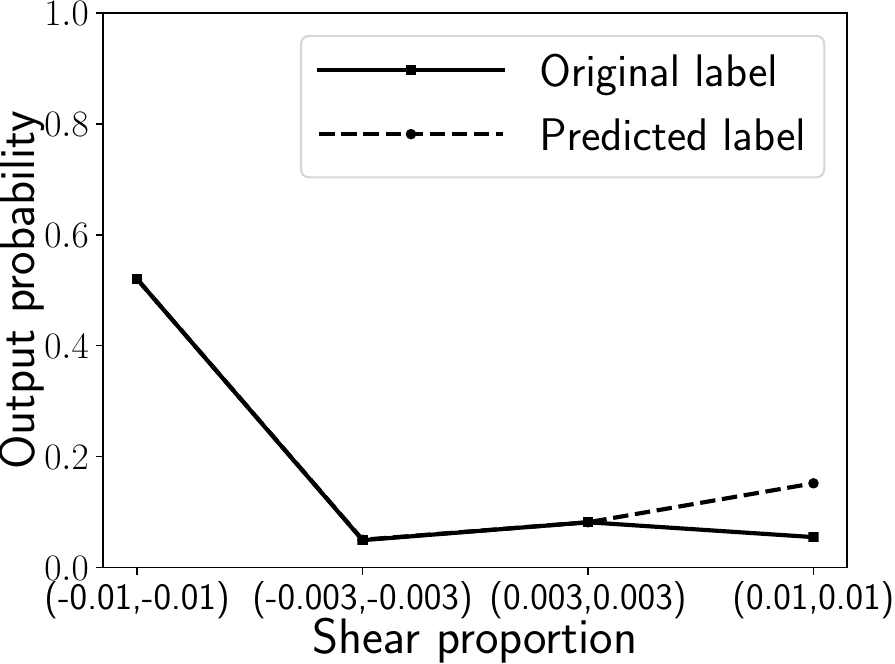}}%
\hfill
\subfloat[]{
\includegraphics[width=0.155\textwidth]{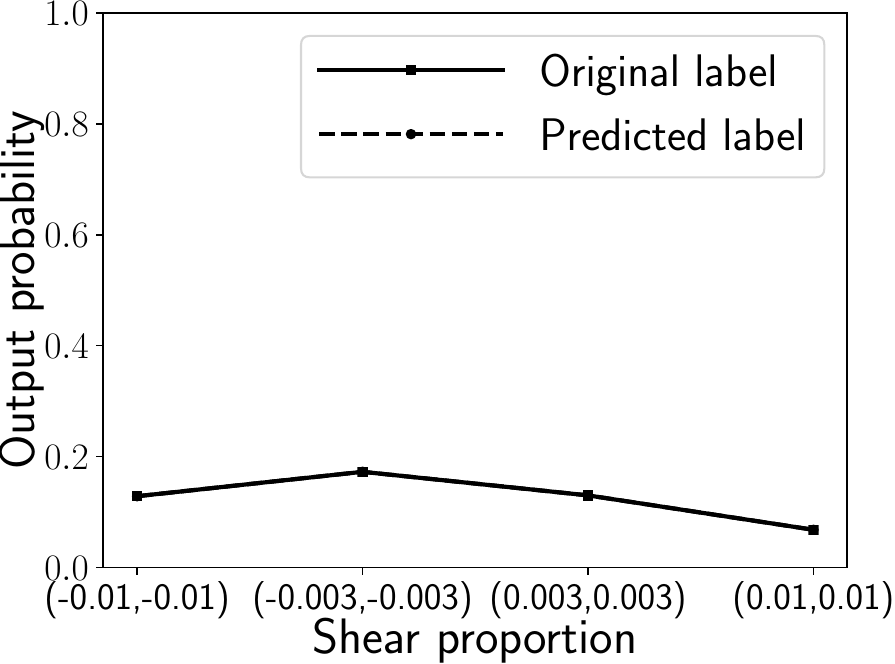}}%
\hfill
\subfloat[]{
\includegraphics[width=0.155\textwidth]{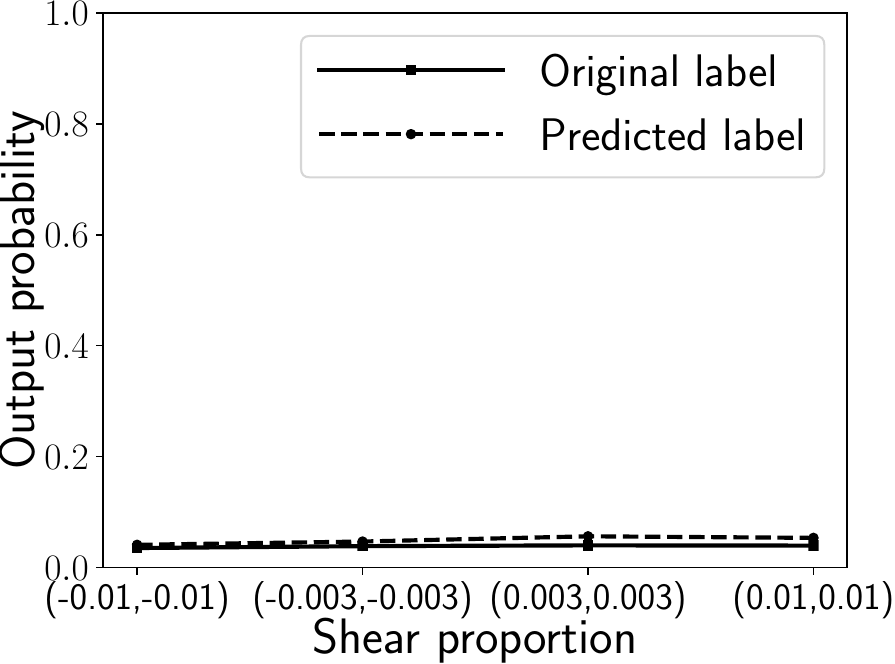}}%
\hfill
\subfloat[]{
\includegraphics[width=0.155\textwidth]{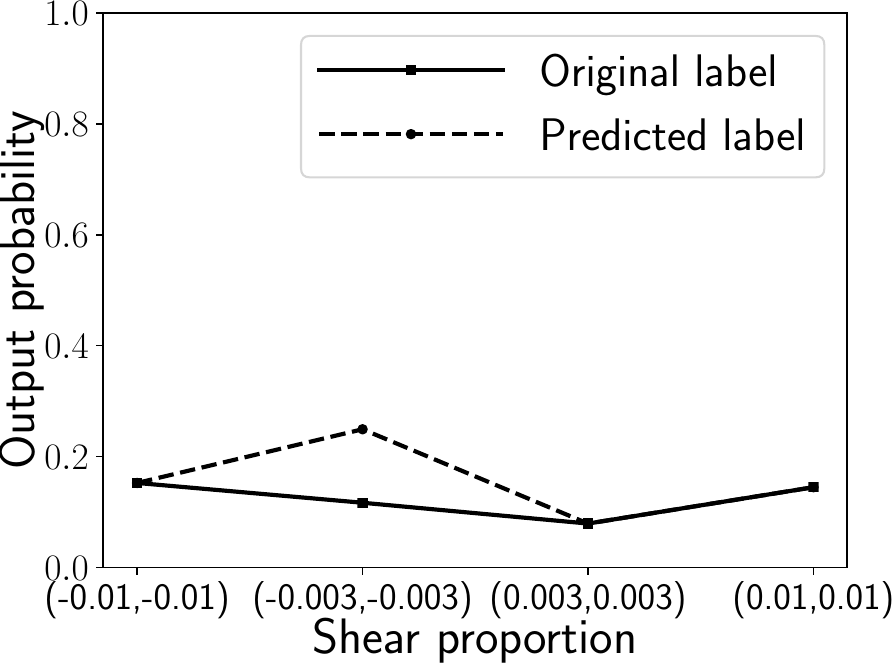}}%
\hfill
\subfloat[]{
\includegraphics[width=0.155\textwidth]{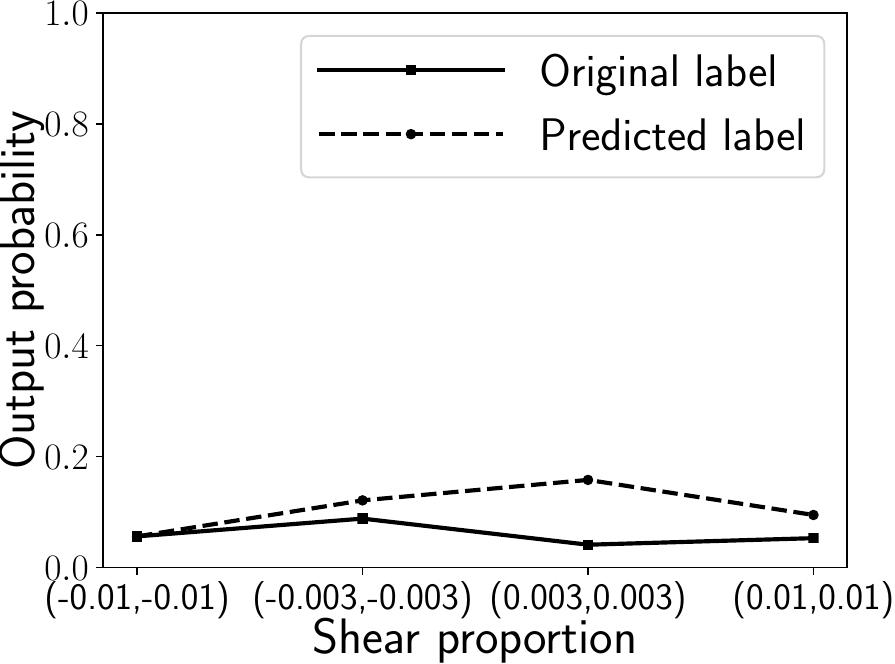}}%
\vspace{-.5cm}
\subfloat[]{
\includegraphics[width=0.155\textwidth]{./figs/prob_dist/probdist_shft_inceptionresnetv2.pdf}}%
\hfill
\subfloat[]{
\includegraphics[width=0.155\textwidth]{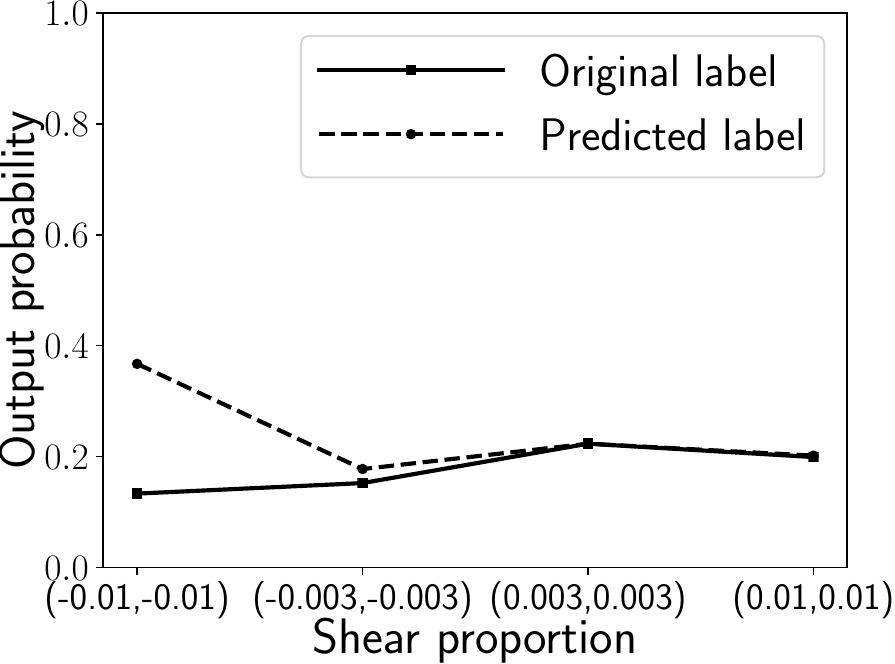}}%
\hfill
\subfloat[]{
\includegraphics[width=0.155\textwidth]{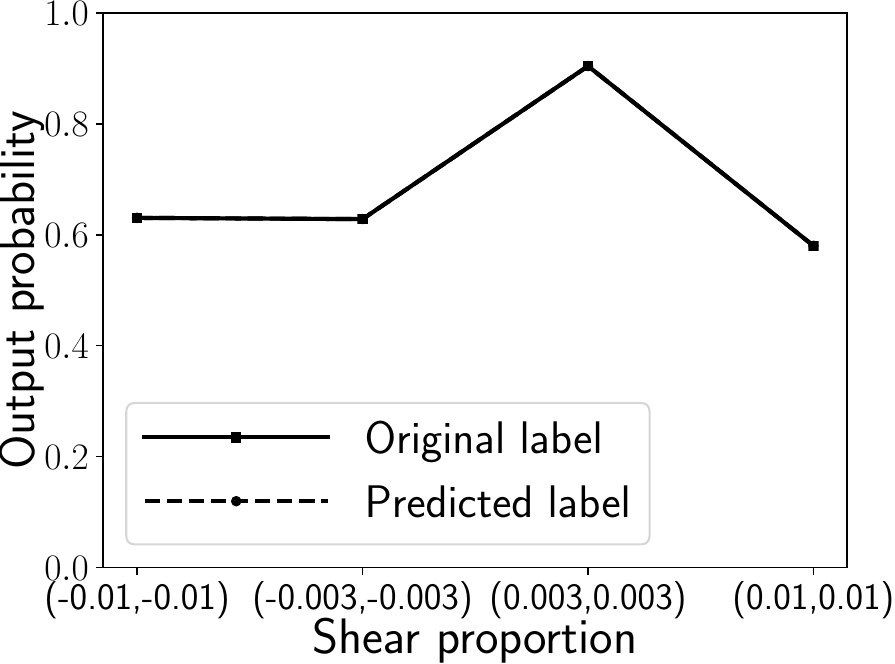}}%
\hfill
\subfloat[]{
\includegraphics[width=0.155\textwidth]{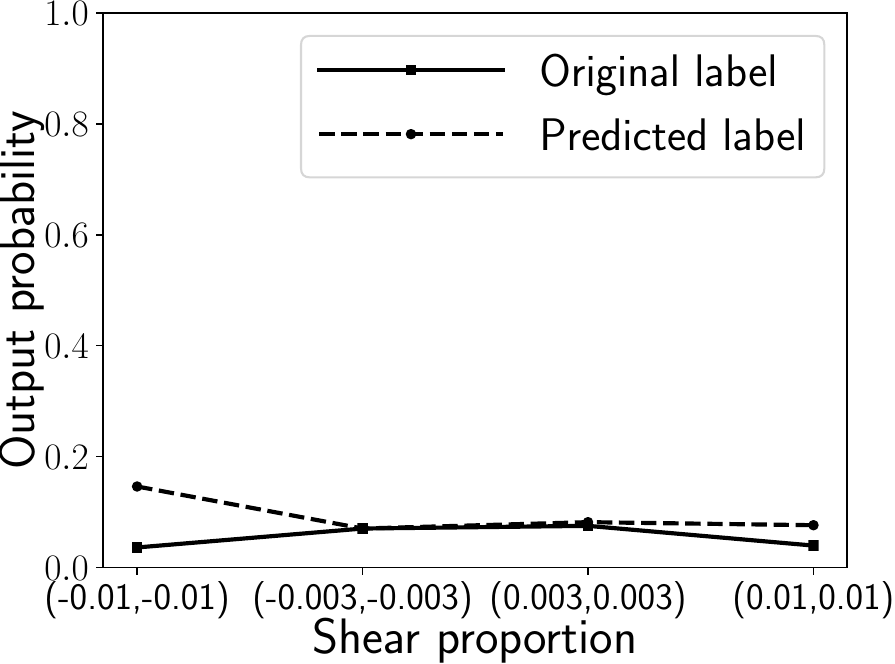}}%
\hfill
\subfloat[]{
\includegraphics[width=0.155\textwidth]{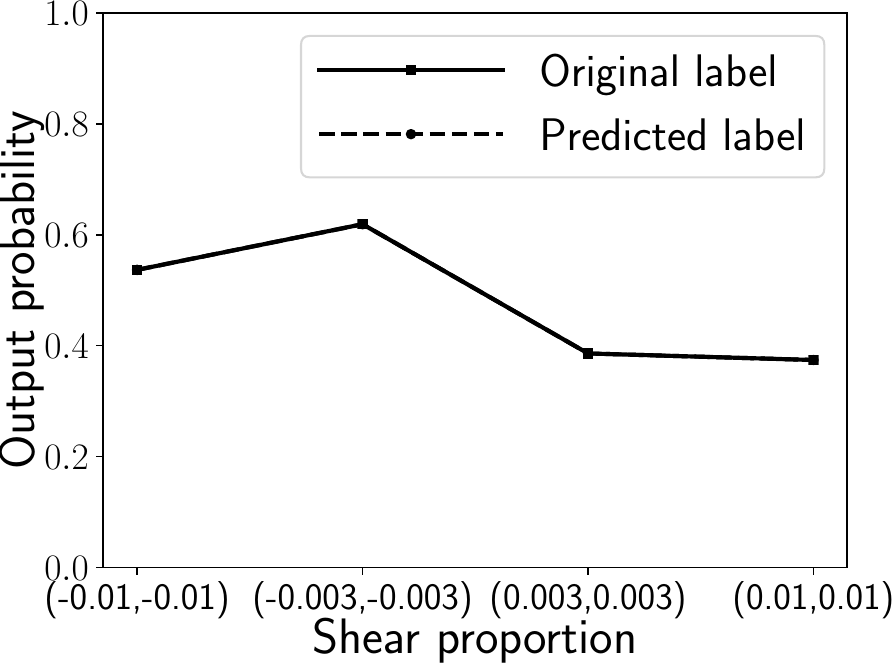}}%
\hfill
\subfloat[]{
\includegraphics[width=0.155\textwidth]{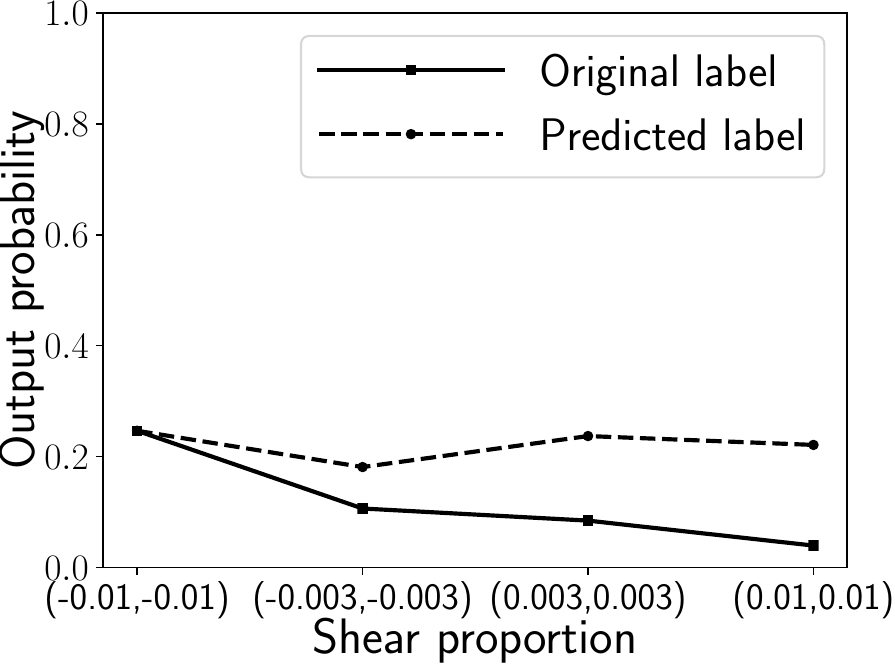}}%

\vspace{-.15cm}{\bf \scriptsize \hspace{0.025in} Scale:\vspace{-.3cm}}%

\subfloat[]{
\includegraphics[width=0.155\textwidth]{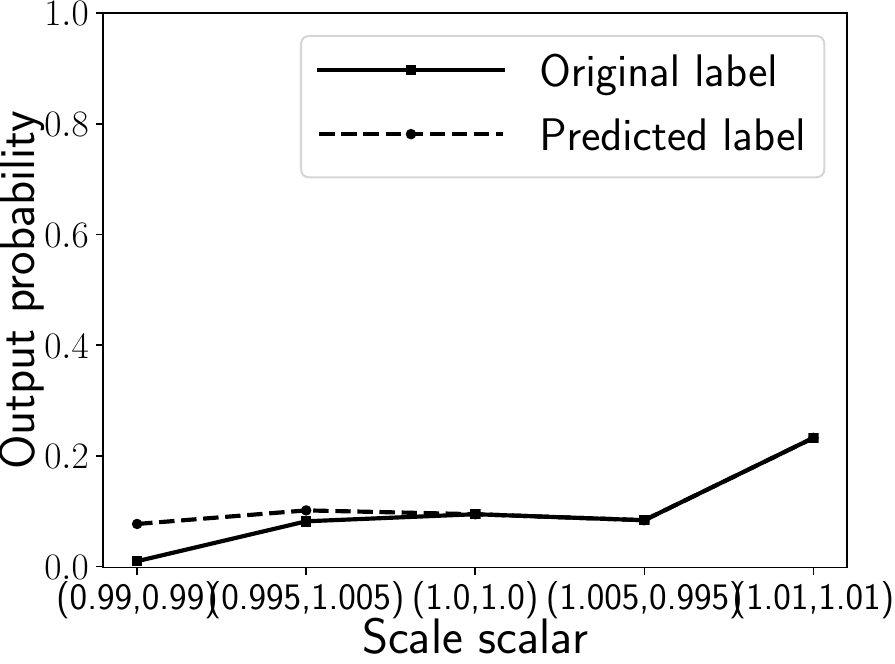}}%
\hfill
\subfloat[]{
\includegraphics[width=0.155\textwidth]{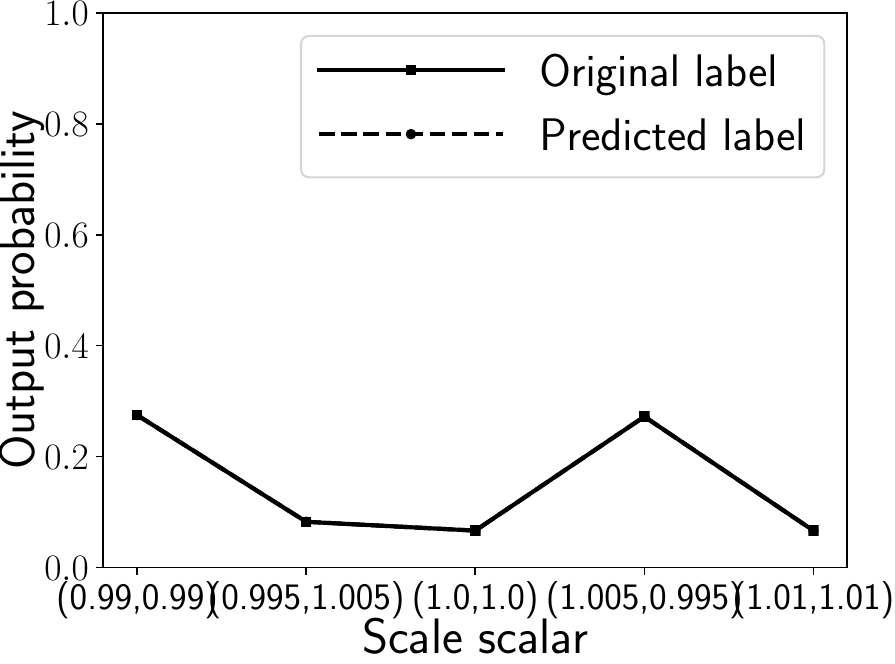}}%
\hfill
\subfloat[]{
\includegraphics[width=0.155\textwidth]{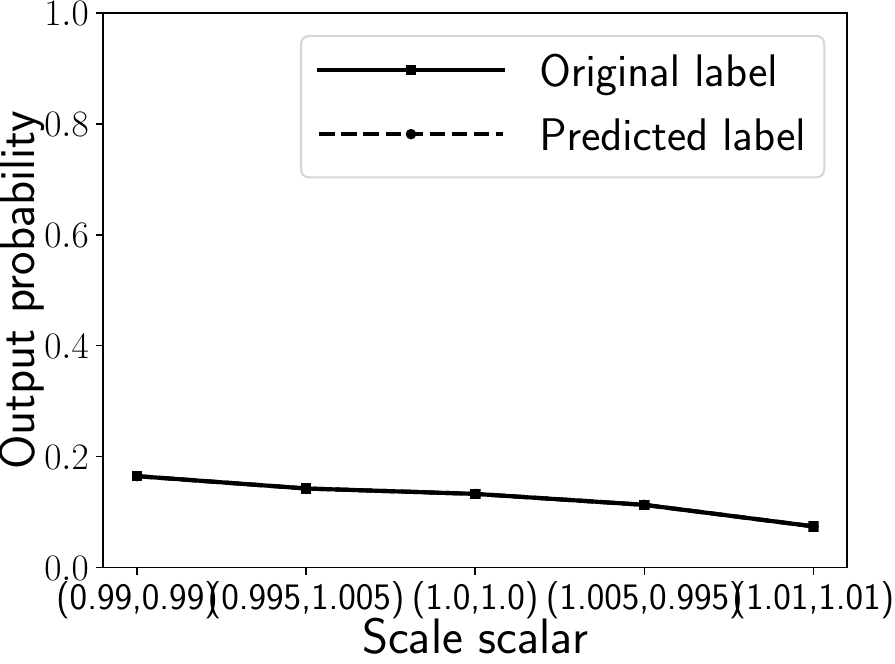}}%
\hfill
\subfloat[]{
\includegraphics[width=0.155\textwidth]{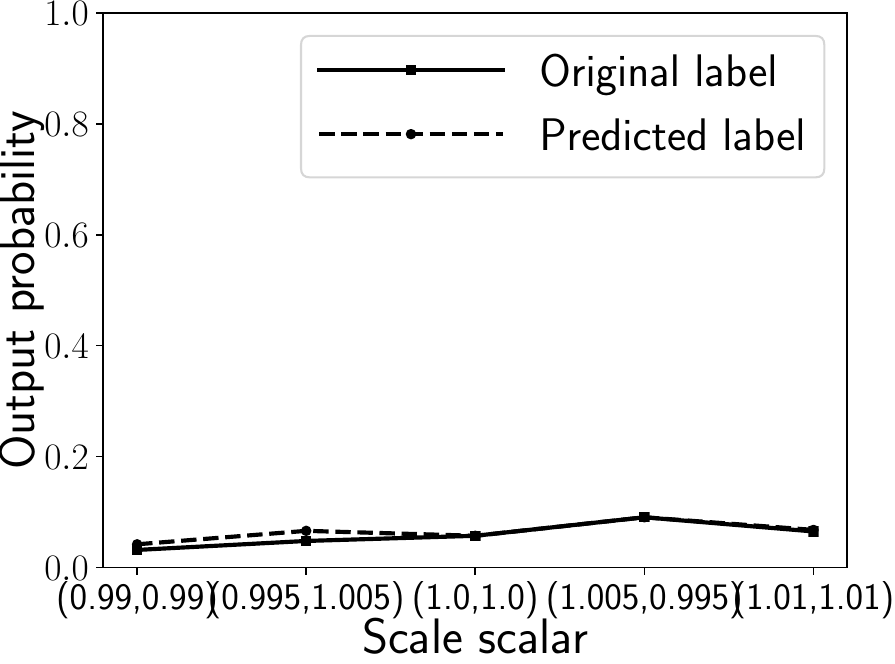}}%
\hfill
\subfloat[]{
\includegraphics[width=0.155\textwidth]{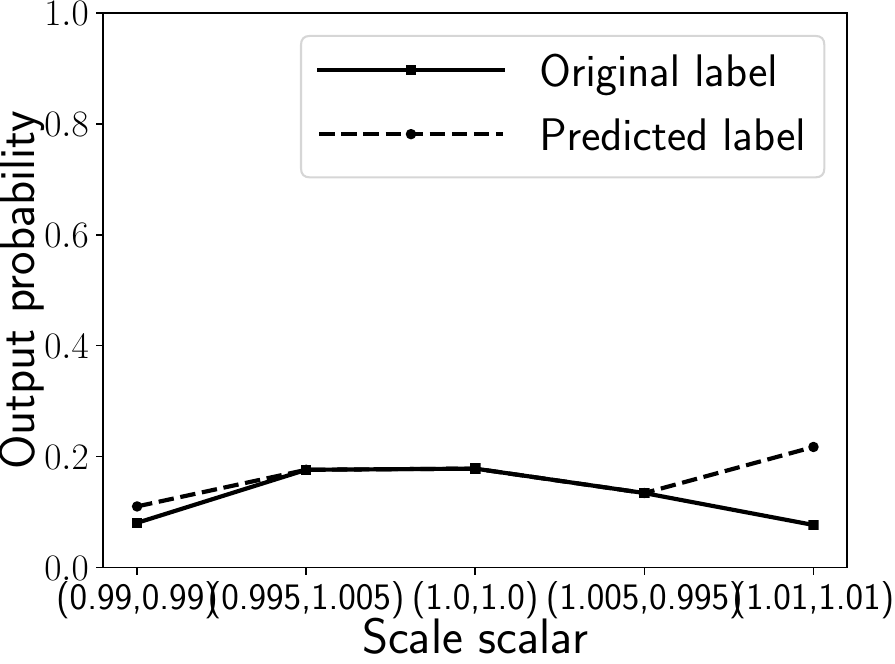}}%
\hfill
\subfloat[]{
\includegraphics[width=0.155\textwidth]{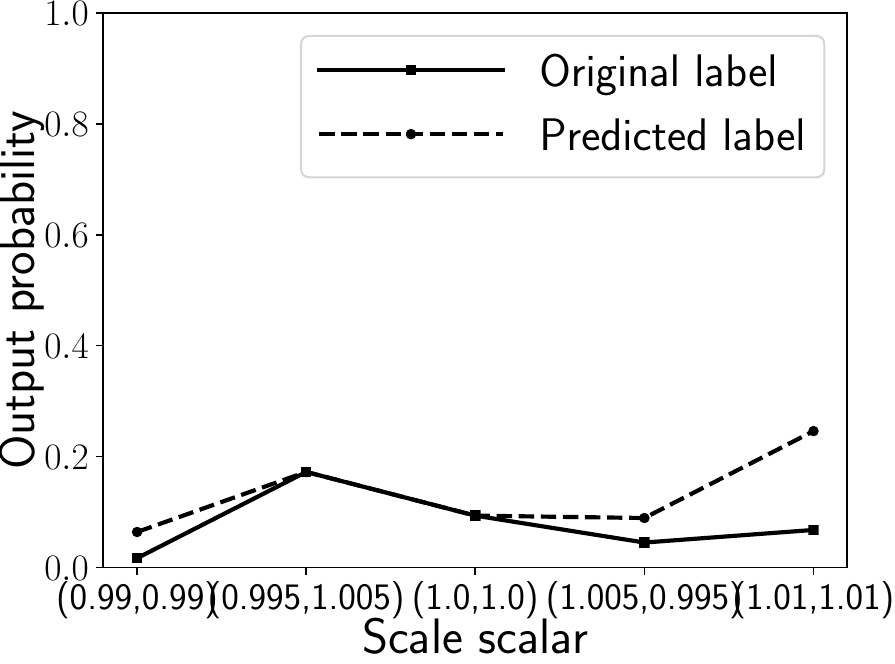}}%
\vspace{-.5cm}
\subfloat[]{
\includegraphics[width=0.155\textwidth]{./figs/prob_dist/probdist_shft_inceptionresnetv2.pdf}}%
\hfill
\subfloat[]{
\includegraphics[width=0.155\textwidth]{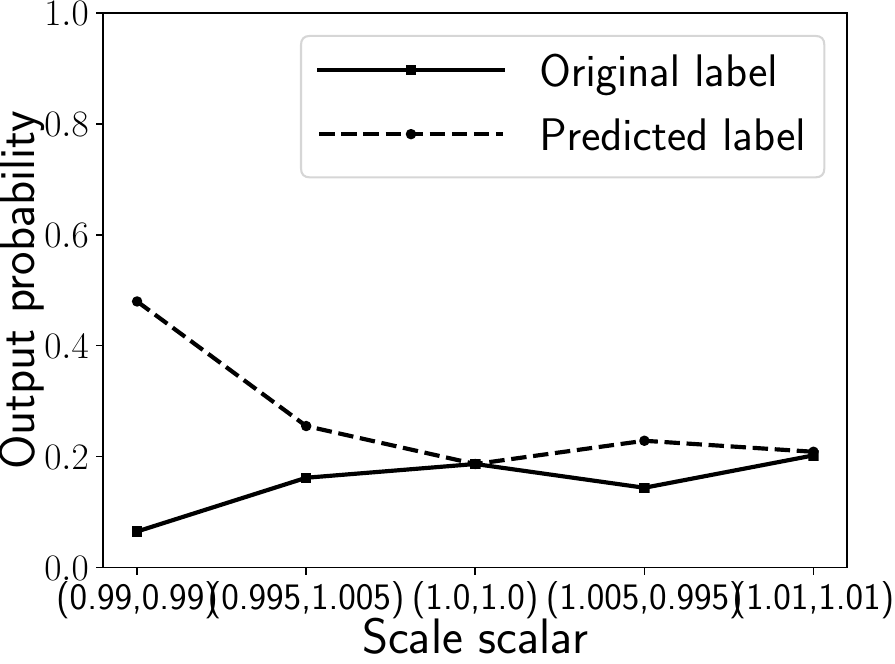}}%
\hfill
\subfloat[]{
\includegraphics[width=0.155\textwidth]{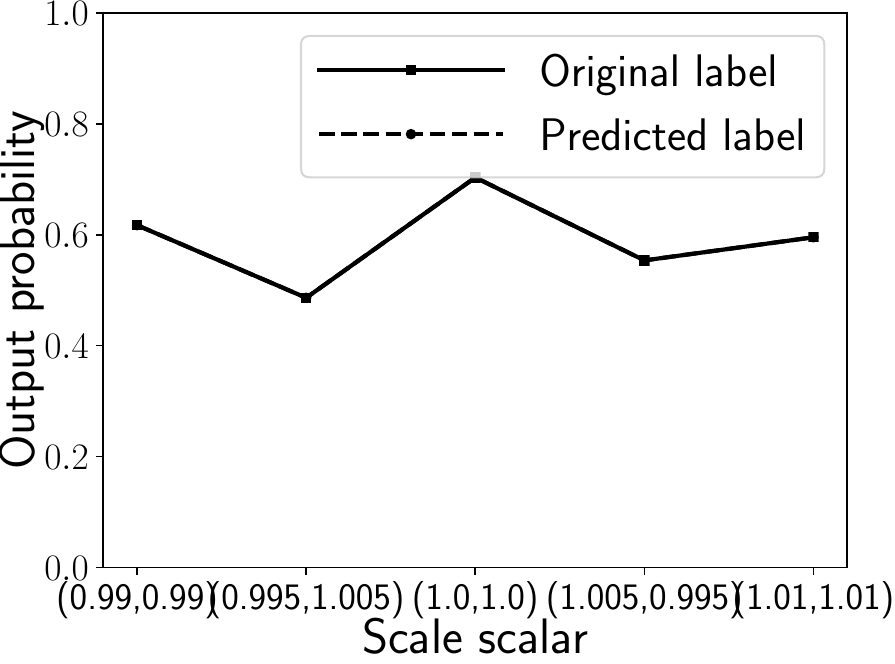}}%
\hfill
\subfloat[]{
\includegraphics[width=0.155\textwidth]{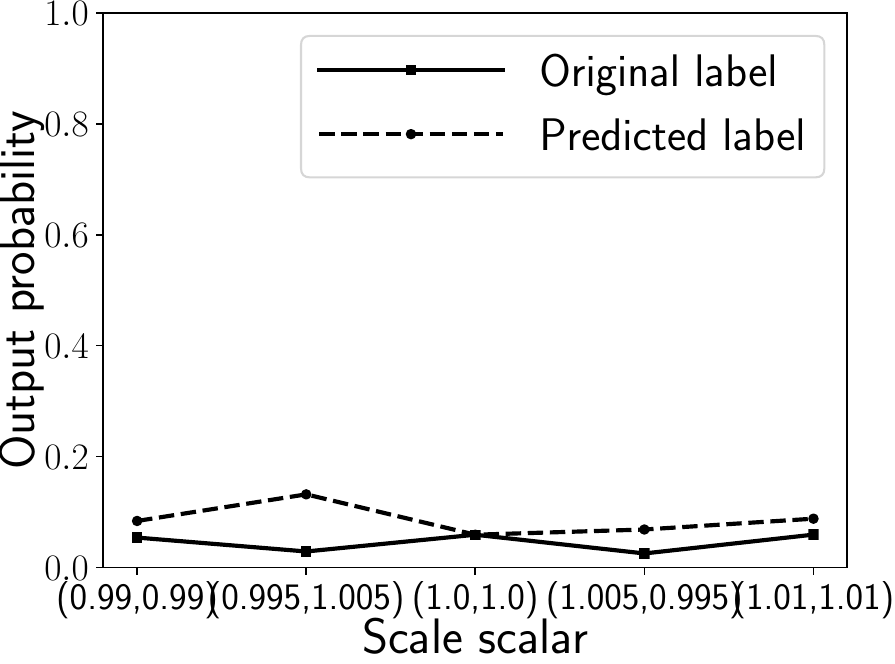}}%
\hfill
\subfloat[]{
\includegraphics[width=0.155\textwidth]{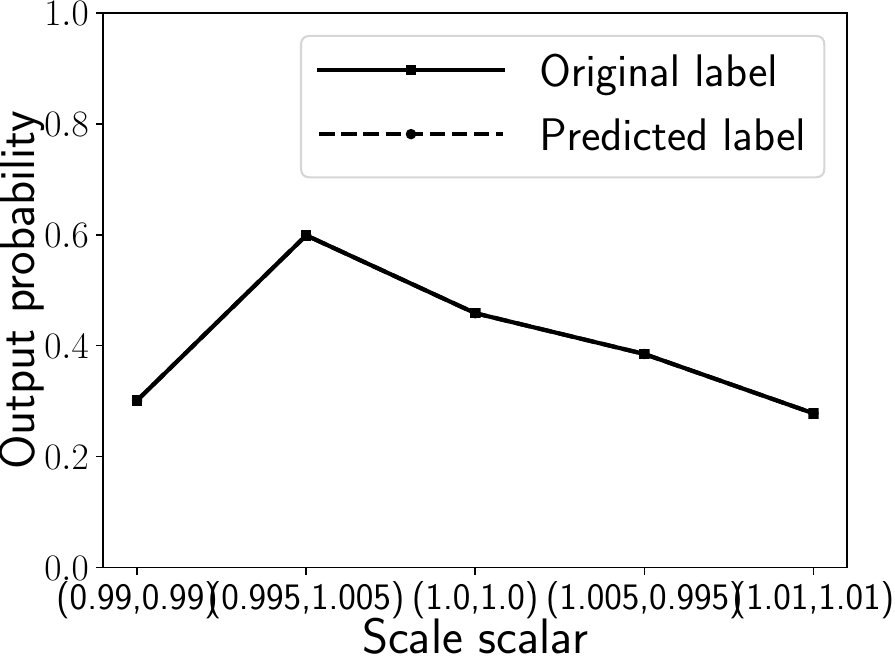}}%
\hfill
\subfloat[]{
\includegraphics[width=0.155\textwidth]{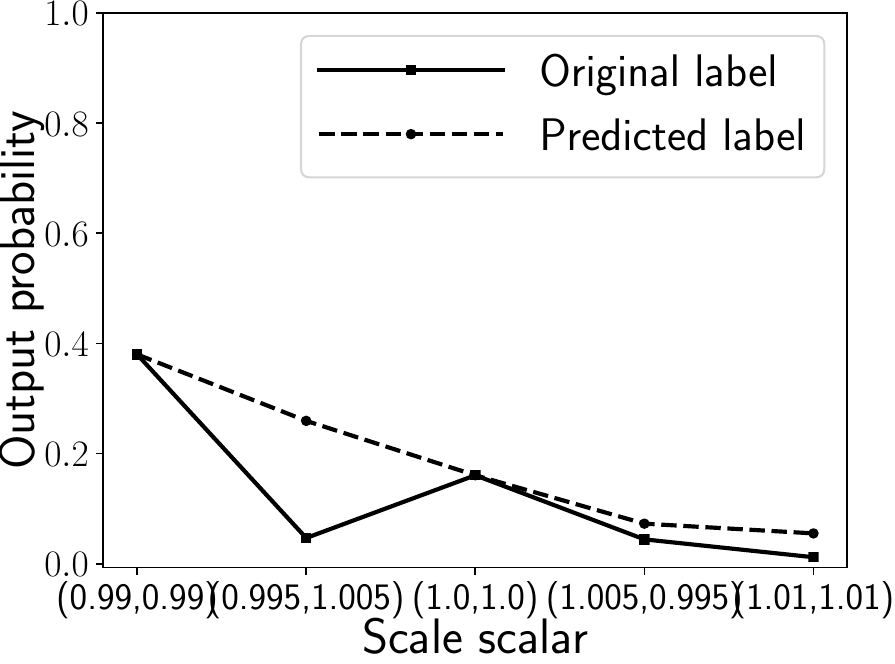}}%

\vspace{-.15cm}{\bf \scriptsize \hspace{0.025in} Translation:\vspace{-.3cm}}%

\subfloat[]{
\includegraphics[width=0.155\textwidth]{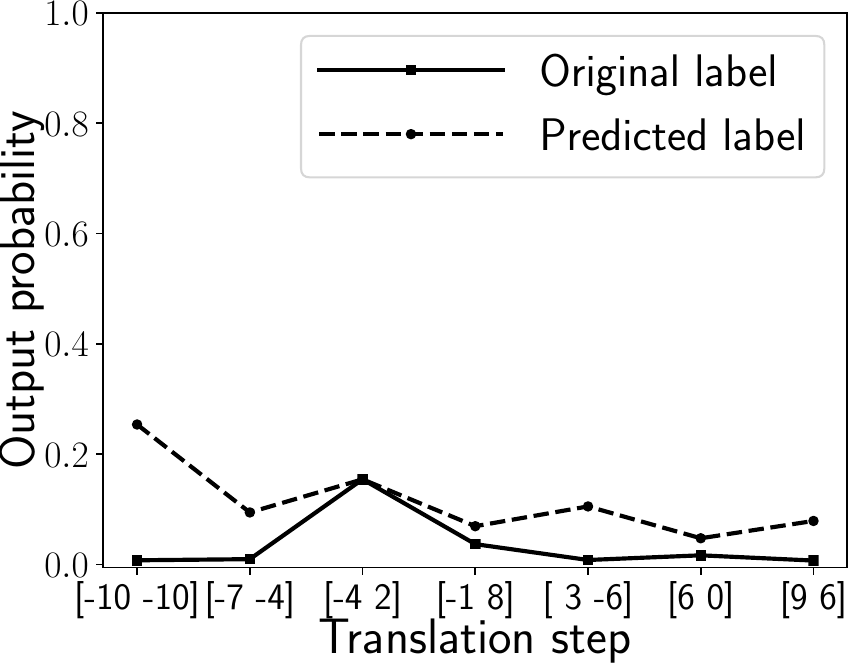}}%
\hfill
\subfloat[]{
\includegraphics[width=0.155\textwidth]{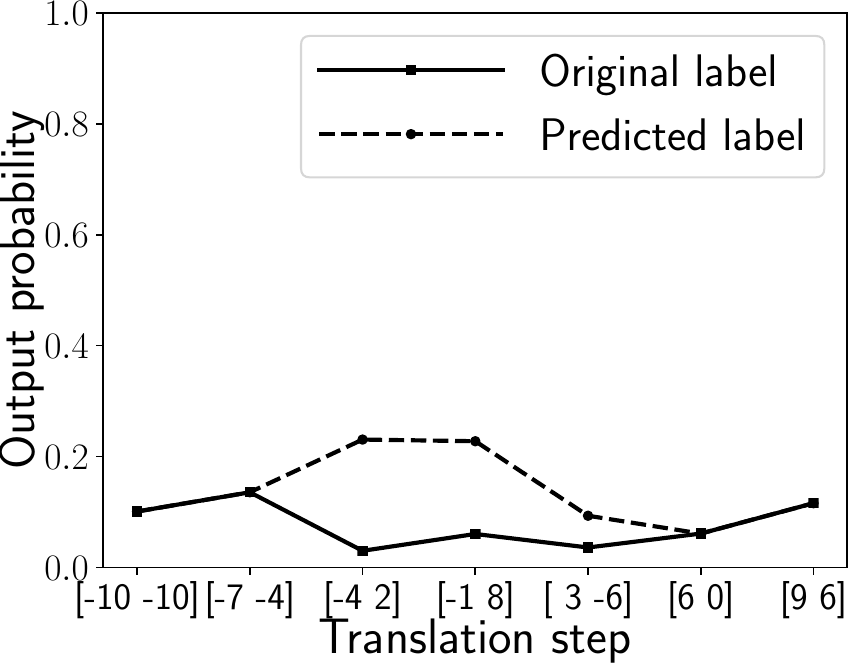}}%
\hfill
\subfloat[]{
\includegraphics[width=0.155\textwidth]{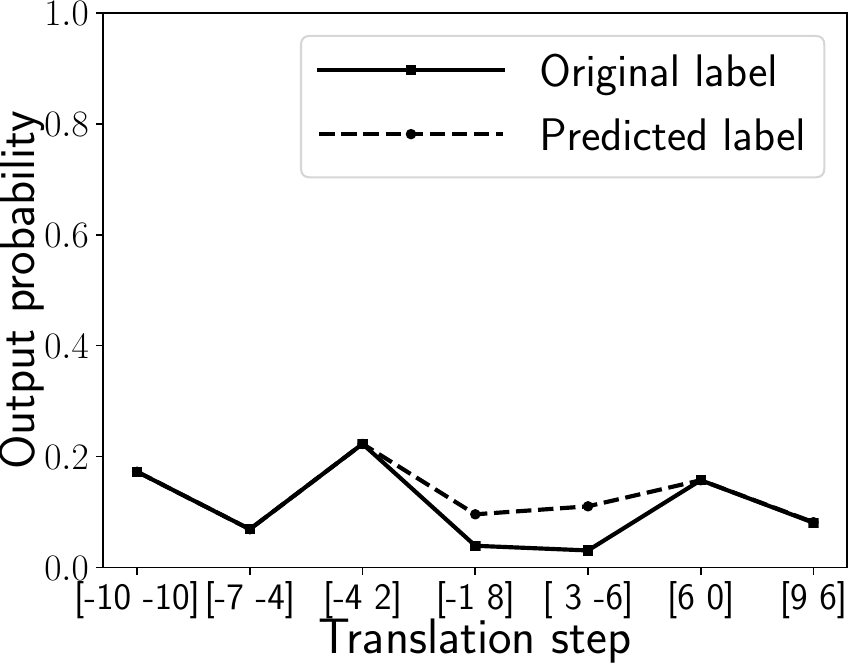}}%
\hfill
\subfloat[]{
\includegraphics[width=0.155\textwidth]{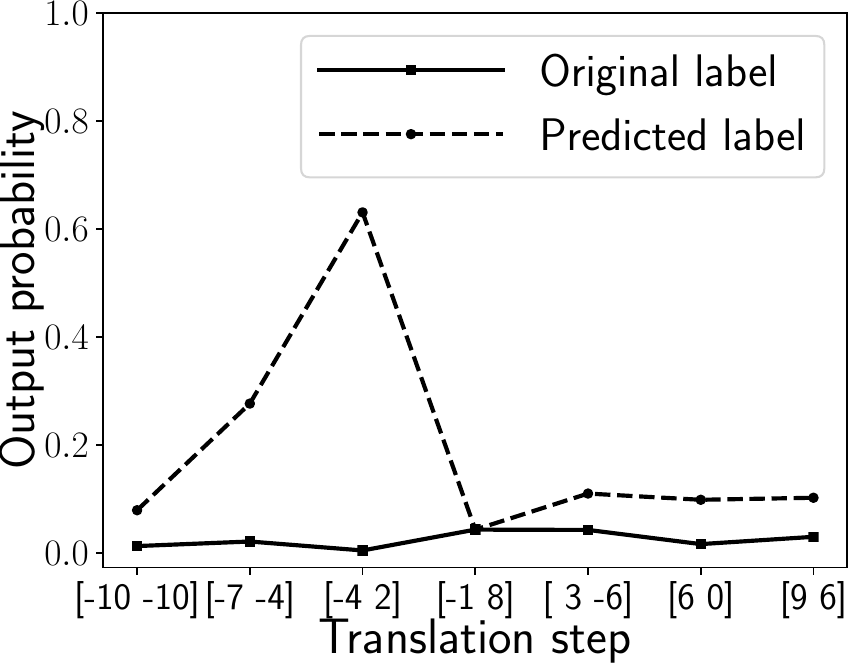}}%
\hfill
\subfloat[]{
\includegraphics[width=0.155\textwidth]{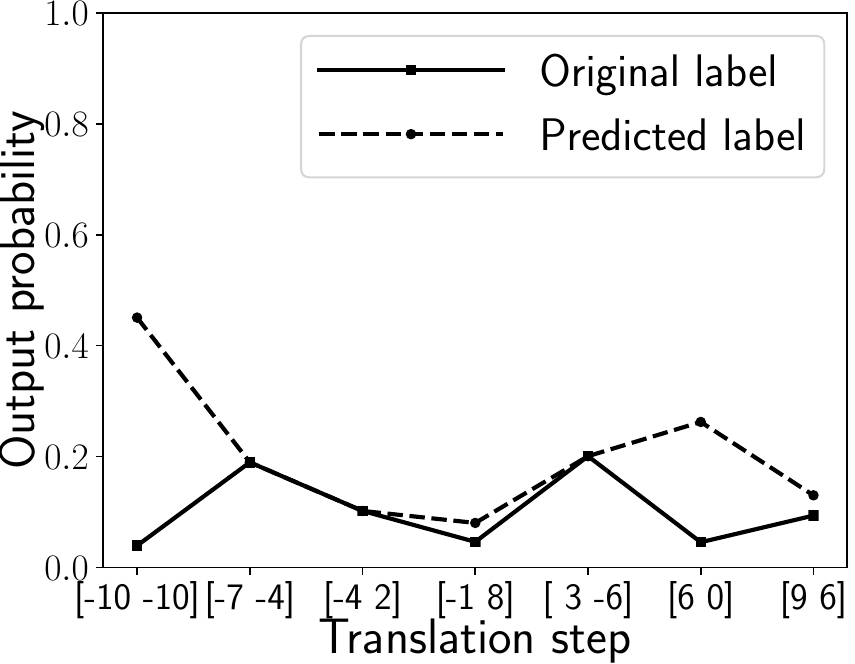}}%
\hfill
\subfloat[]{
\includegraphics[width=0.155\textwidth]{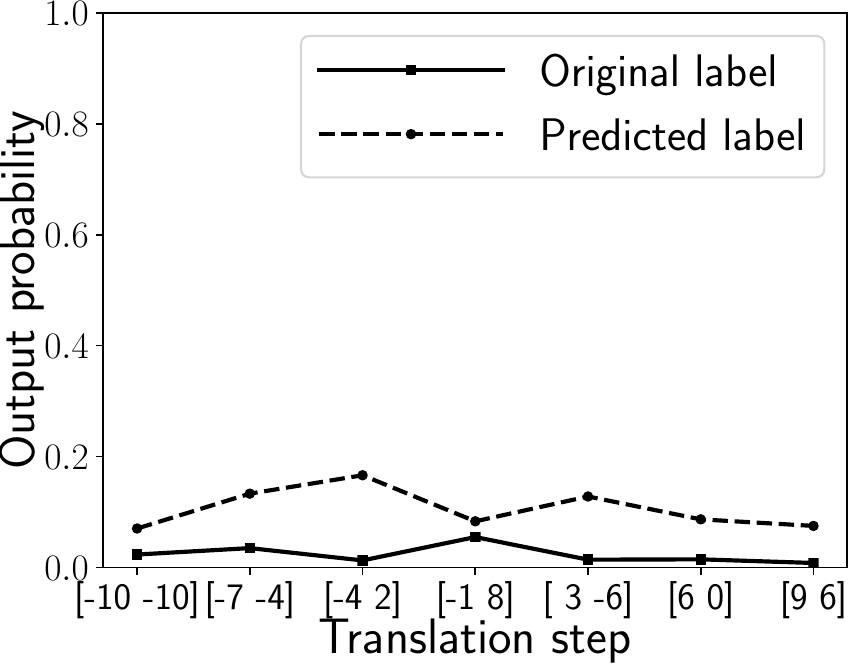}}%
\vspace{-.5cm}
\subfloat[]{
\includegraphics[width=0.155\textwidth]{./figs/prob_dist/probdist_shft_inceptionresnetv2.pdf}}%
\hfill
\subfloat[]{
\includegraphics[width=0.155\textwidth]{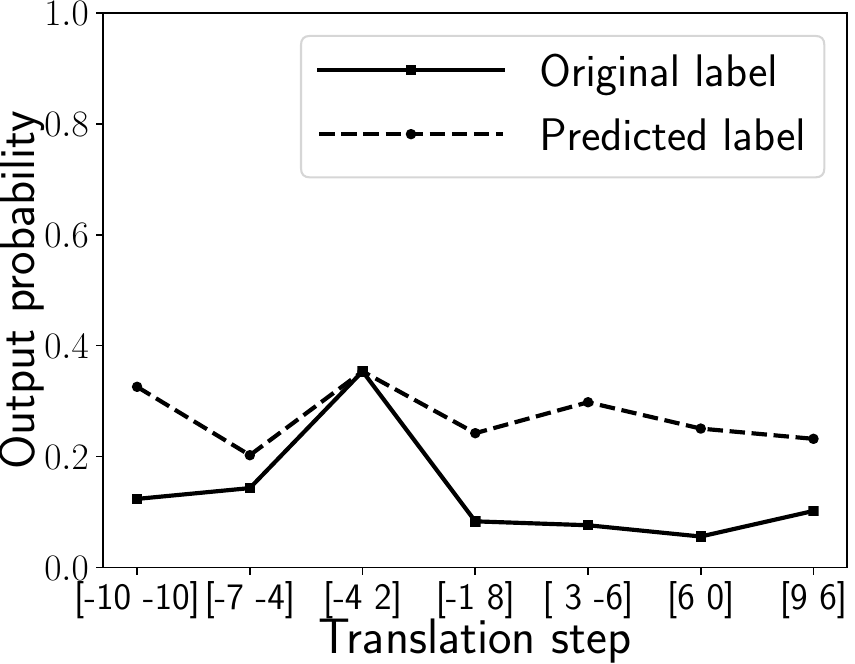}}%
\hfill
\subfloat[]{
\includegraphics[width=0.155\textwidth]{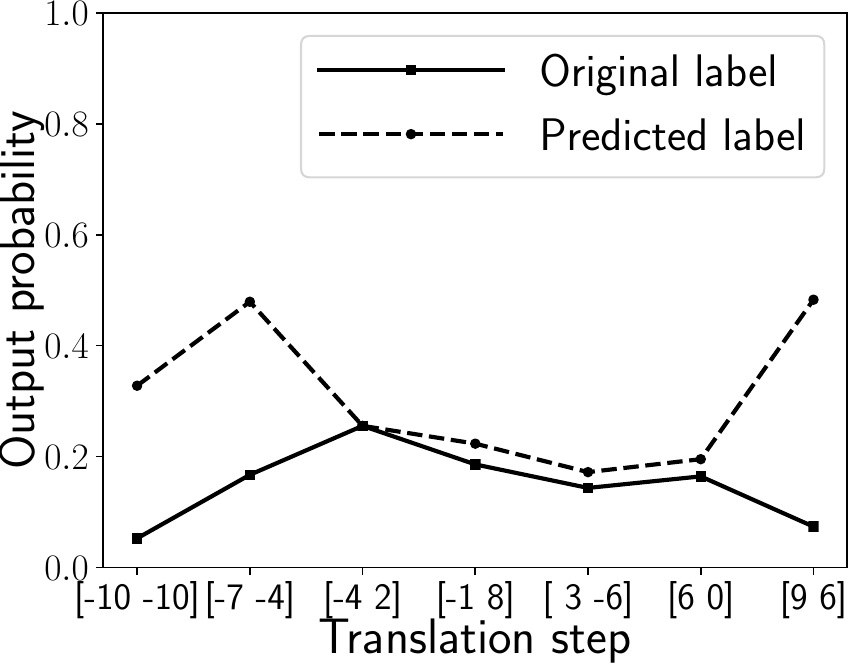}}%
\hfill
\subfloat[]{
\includegraphics[width=0.155\textwidth]{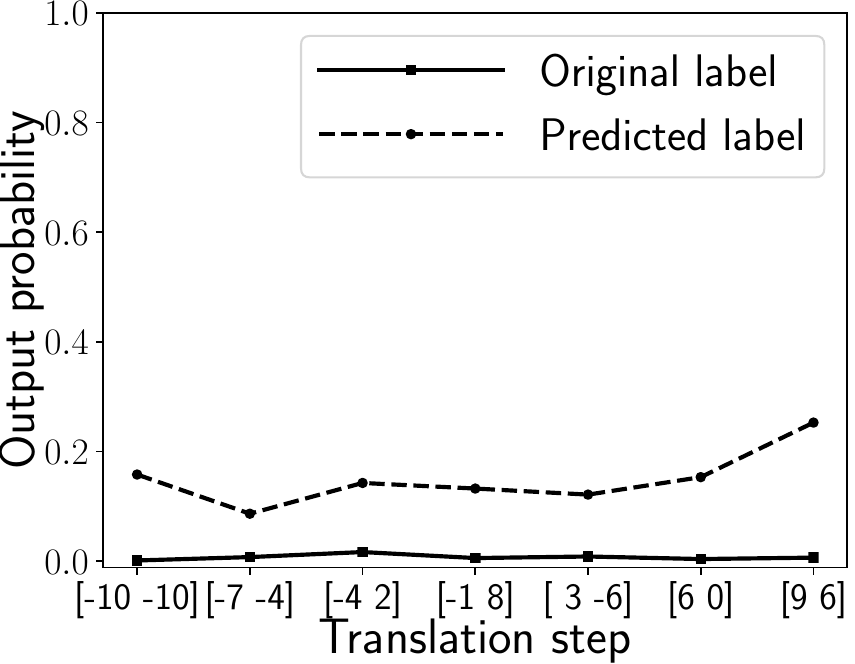}}%
\hfill
\subfloat[]{
\includegraphics[width=0.155\textwidth]{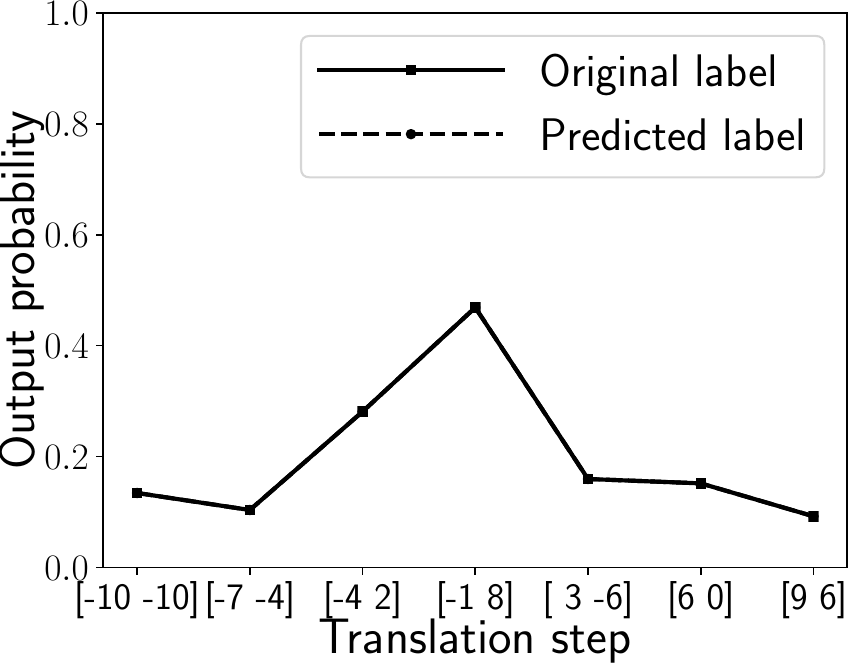}}%
\hfill
\subfloat[]{
\includegraphics[width=0.155\textwidth]{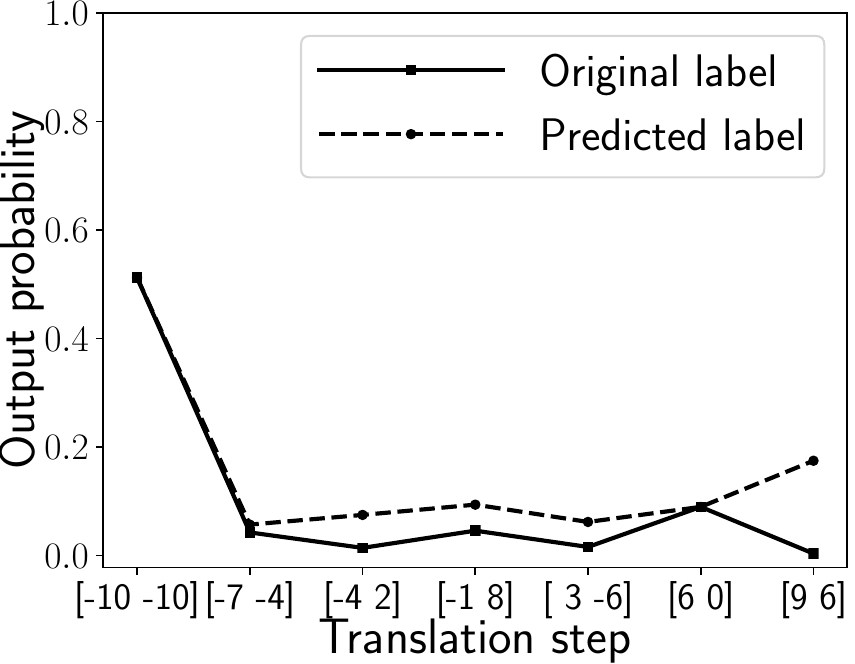}}%

\vspace{-.15cm}{\bf \scriptsize \hspace{0.025in} Reflection:\vspace{-.3cm}}

\subfloat[]{
\includegraphics[width=0.155\textwidth]{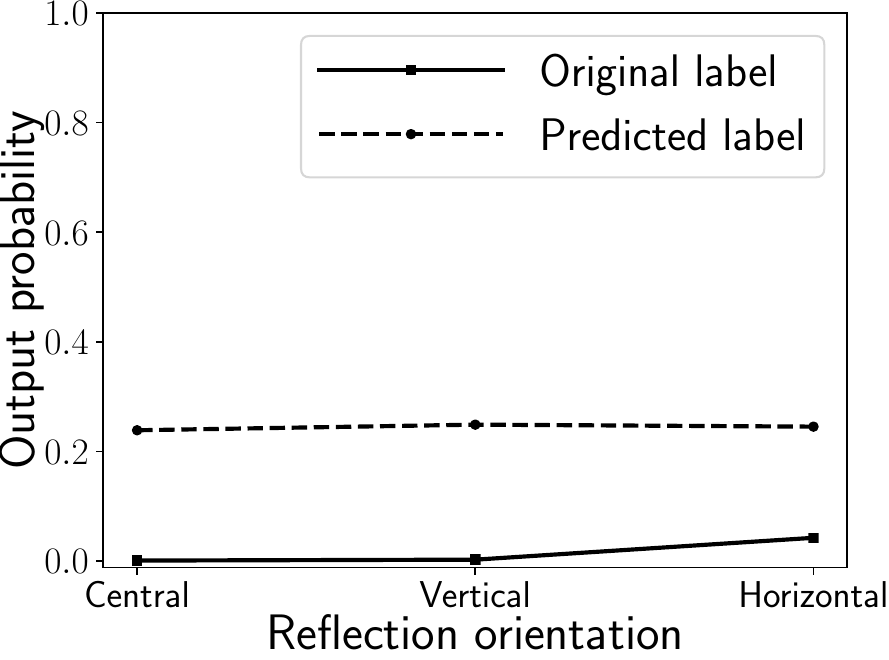}}%
\hfill
\subfloat[]{
\includegraphics[width=0.155\textwidth]{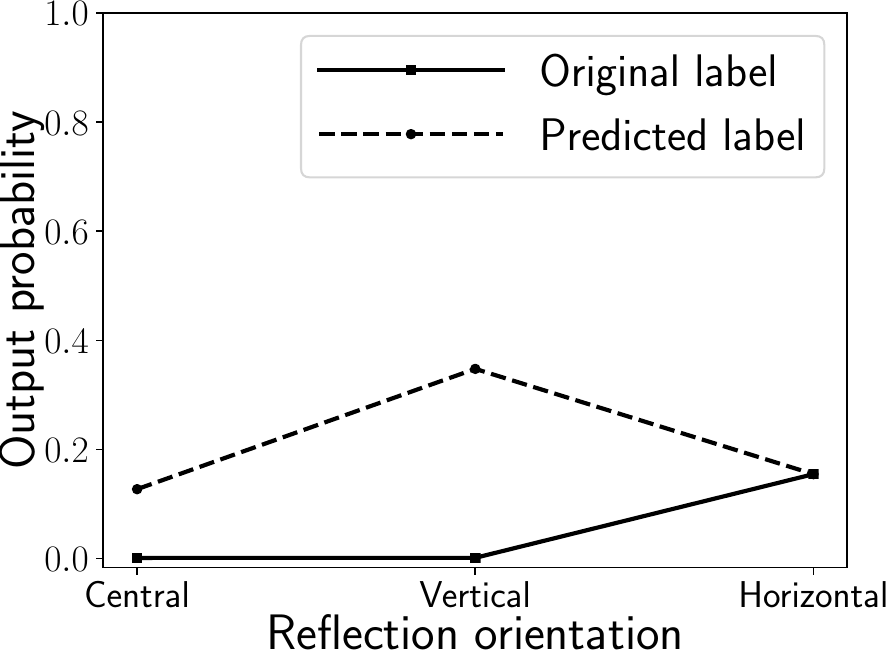}}%
\hfill
\subfloat[]{
\includegraphics[width=0.155\textwidth]{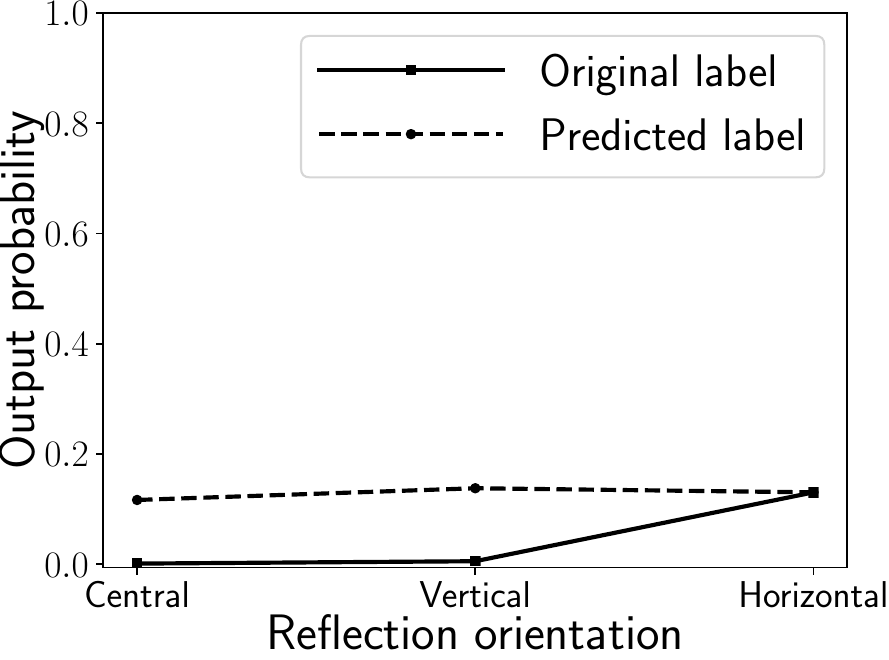}}%
\hfill
\subfloat[]{
\includegraphics[width=0.155\textwidth]{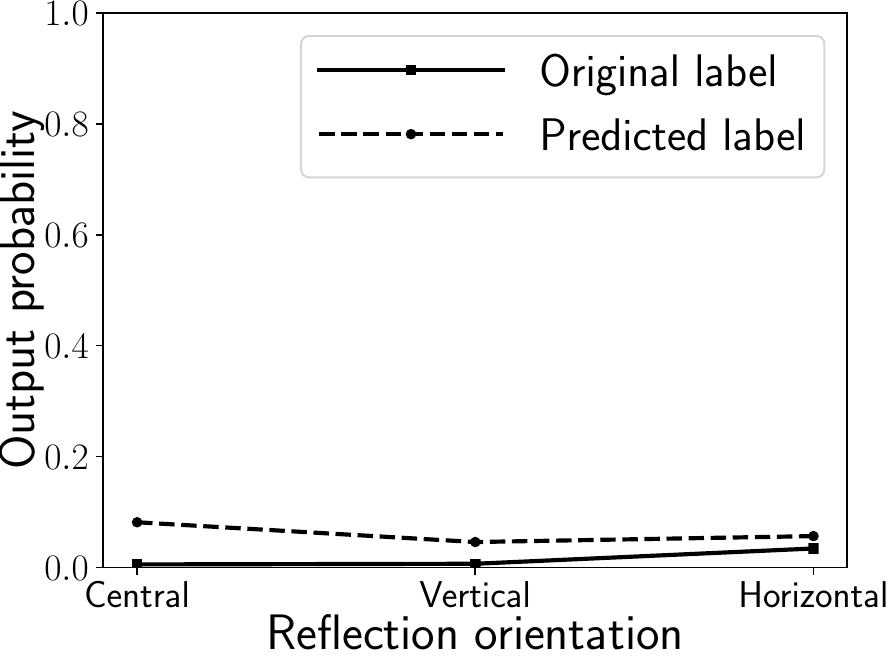}}%
\hfill
\subfloat[]{
\includegraphics[width=0.155\textwidth]{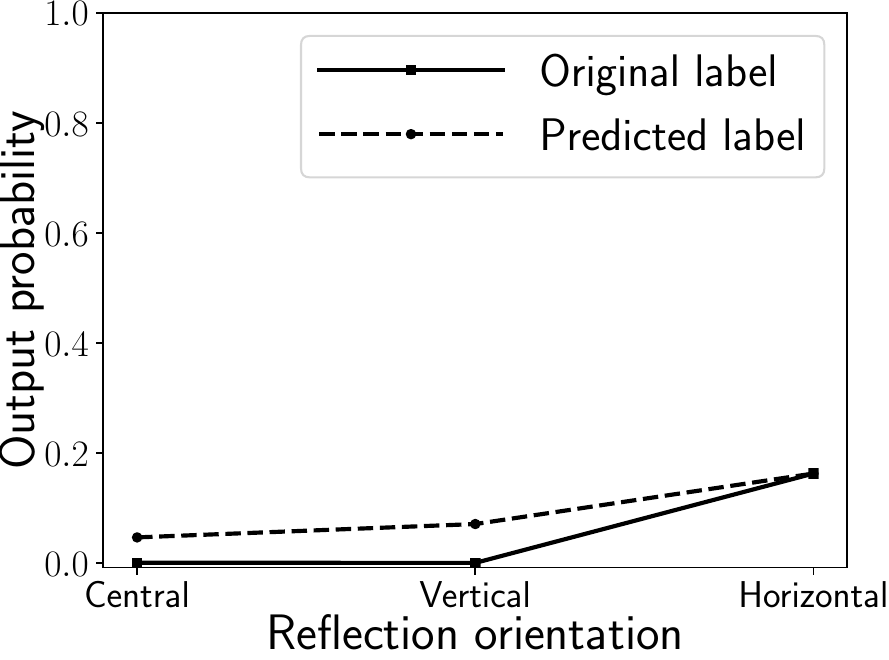}}%
\hfill
\subfloat[]{
\includegraphics[width=0.155\textwidth]{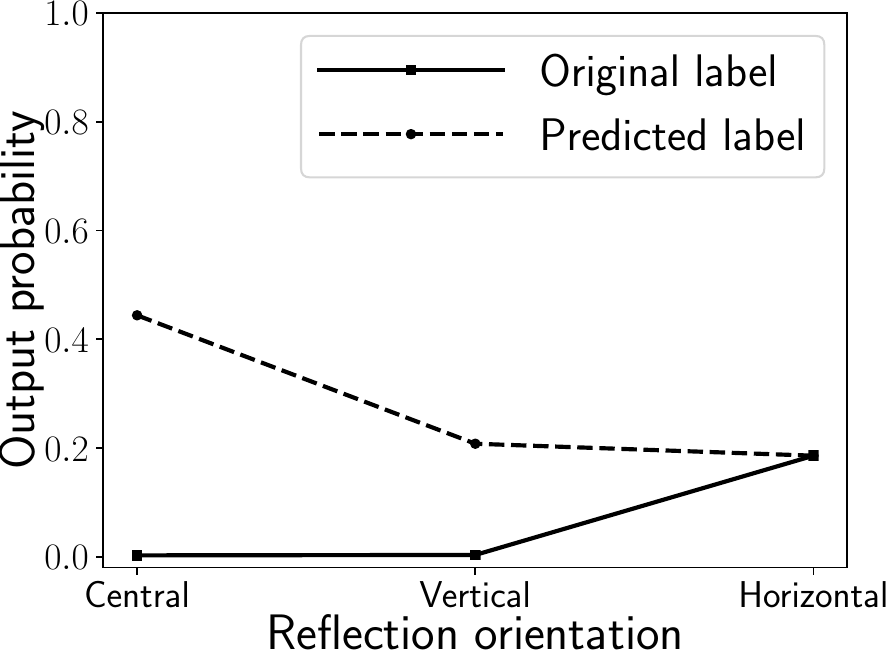}}%
\vspace{-.5cm}
\subfloat[]{
\includegraphics[width=0.155\textwidth]{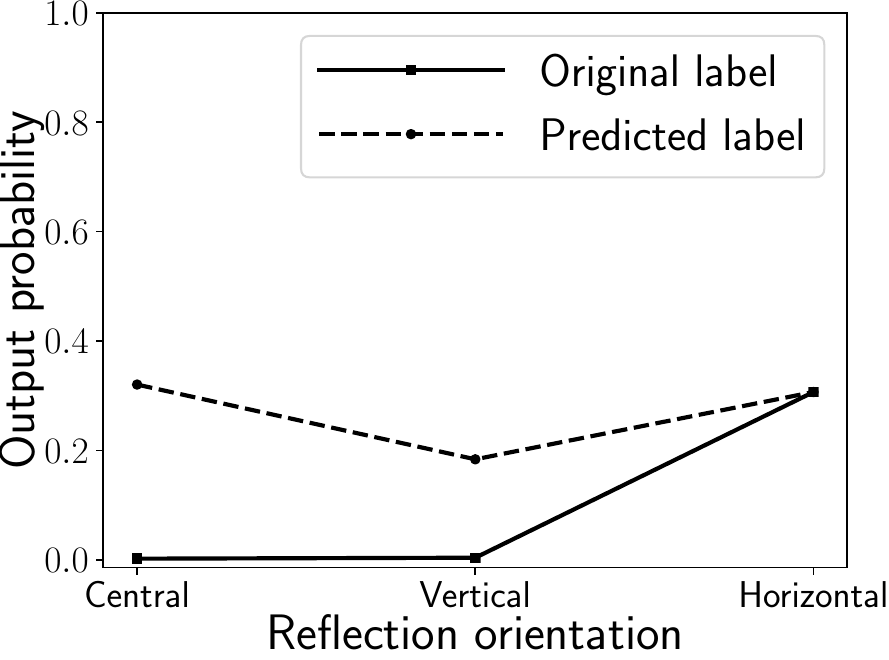}}%
\hfill
\subfloat[]{
\includegraphics[width=0.155\textwidth]{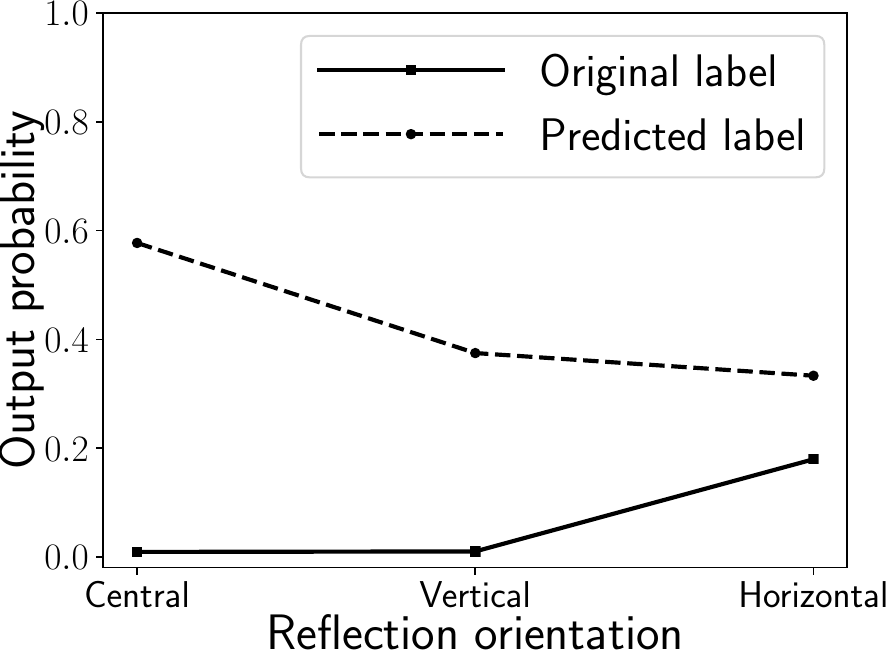}}%
\hfill
\subfloat[]{
\includegraphics[width=0.155\textwidth]{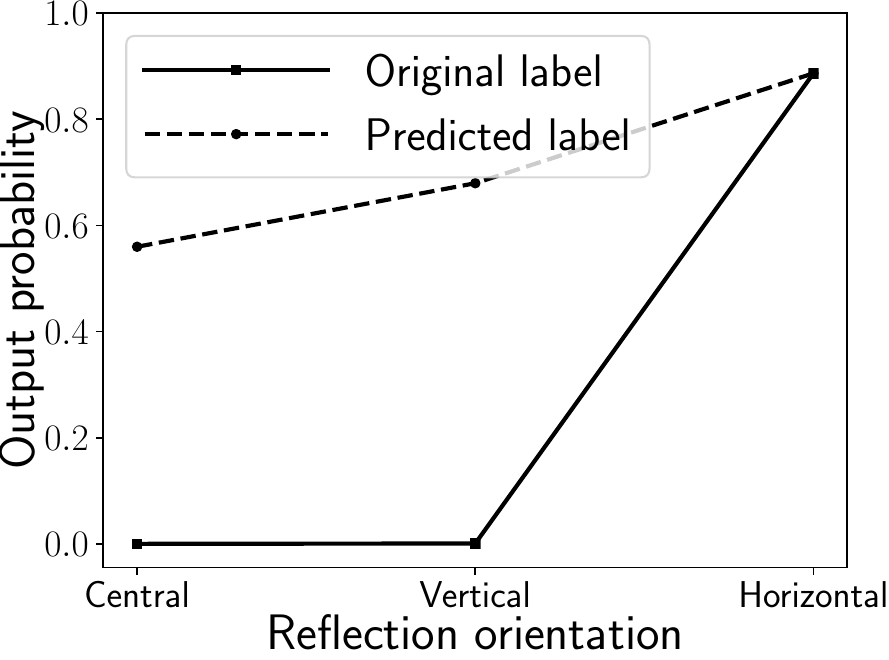}}%
\hfill
\subfloat[]{
\includegraphics[width=0.155\textwidth]{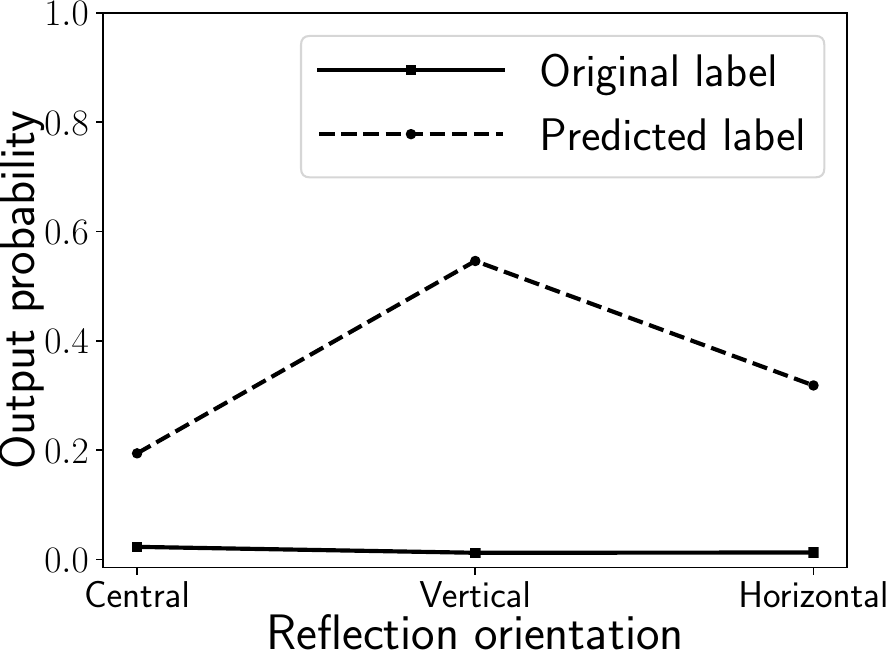}}%
\hfill
\subfloat[]{
\includegraphics[width=0.155\textwidth]{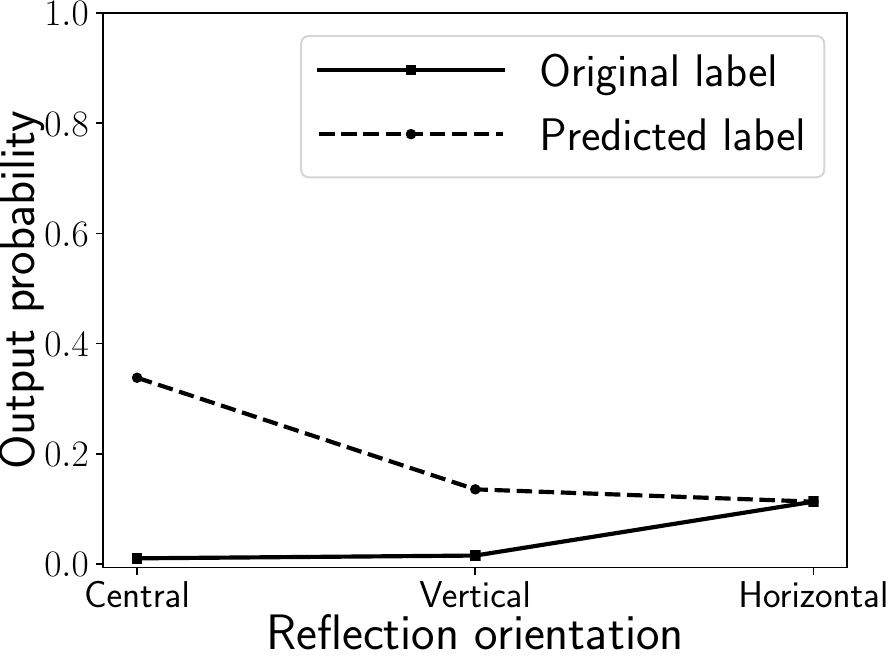}}%
\hfill
\subfloat[]{
\includegraphics[width=0.155\textwidth]{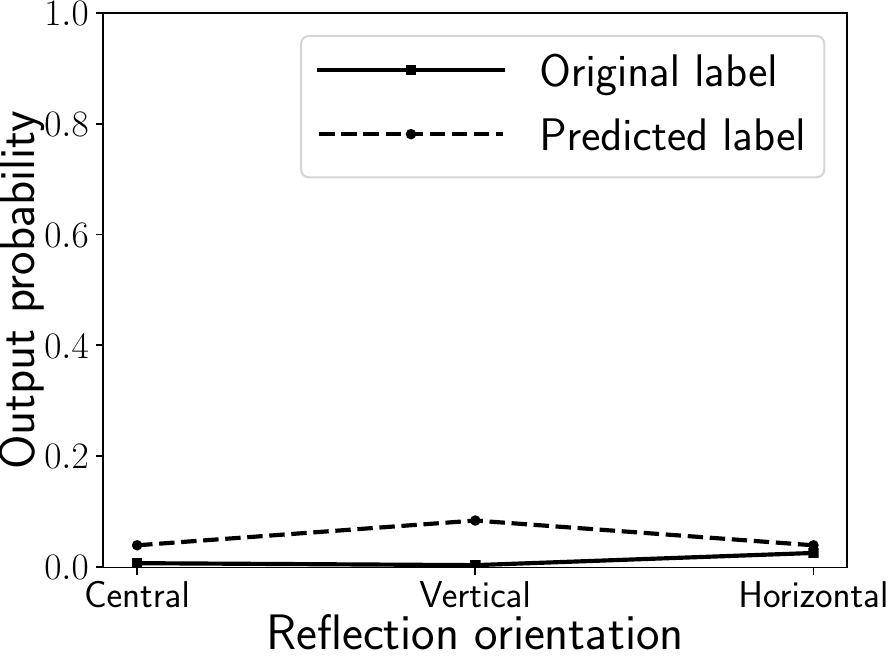}}%

\caption{Distribution of the prediction errors of affine transformations ($\phi_8$ to $\phi_{12}$). The solid and dotted lines denote the output probability for the original and predicted labels respectively.
The higher the gap between the solid and dotted lines the higher the error (overlap means no error).}
\label{fig:prob_dist3}
\end{figure*}

Figure~\ref{fig:prob_dist1}, \ref{fig:prob_dist2}, and \ref{fig:prob_dist3} present the prediction distributions of all twelve models deployed for ImageNet classification under all twelve transformations. 
Our results indicate that while some error-inducing parameter values for transformations like average smoothing follow simple patterns (\eg parameter values greater than a threshold will lead to errors), most of the transformations do not follow any obvious patterns (\ie does not normally distribute or fall into a certain range).
This further justifies the weakness of sampling-based testing approach, which may result in biased conclusion if it accidentally misses some critical error-inducing parameters.

\subsection{Effect of increasing parameter space on number of errors}

We show the number of errors found by \sys as the bound of transformation space increases using Dave-orig (DRV\_C2) as the target vision system.  
Figure~\ref{fig:bound_time_violation} shows that the number of errors increases as we increase the range due to increase in the number of critical parameter values that need to be checked. For example, the average number of errors for rotation ($\phi_8$) increases from $1360.5$ to $3573.1$ when we change the range of rotation degrees from $[-2,2]$ to $[-3,3]$.

\begin{figure*}
\centering
\captionsetup[subfloat]{captionskip=-.15cm,labelformat=empty}

\subfloat[]{
\includegraphics[width=0.152\textwidth]{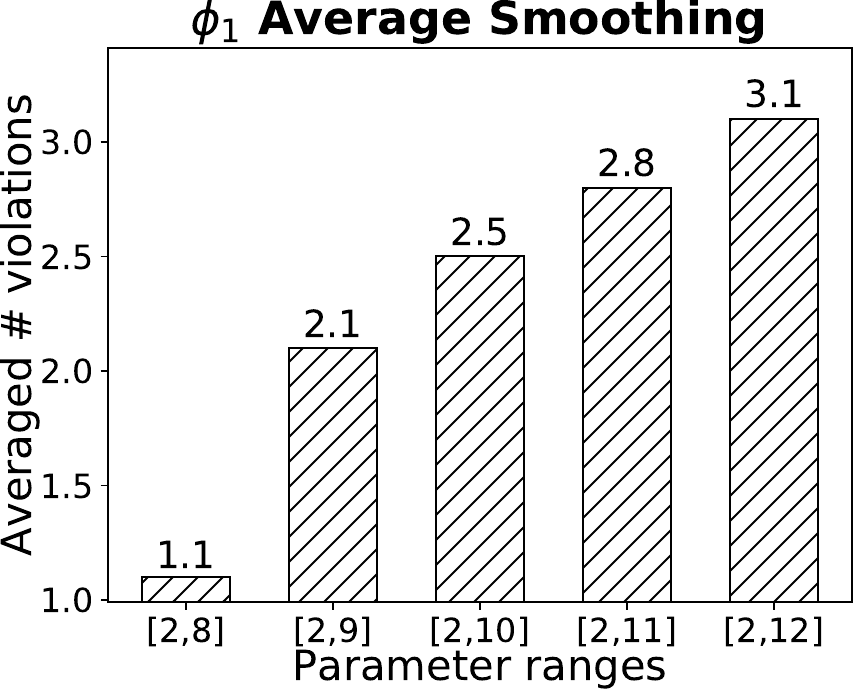}
\label{subfig:bound_violation1}}
\hfill
\subfloat[]{
\includegraphics[width=0.152\textwidth]{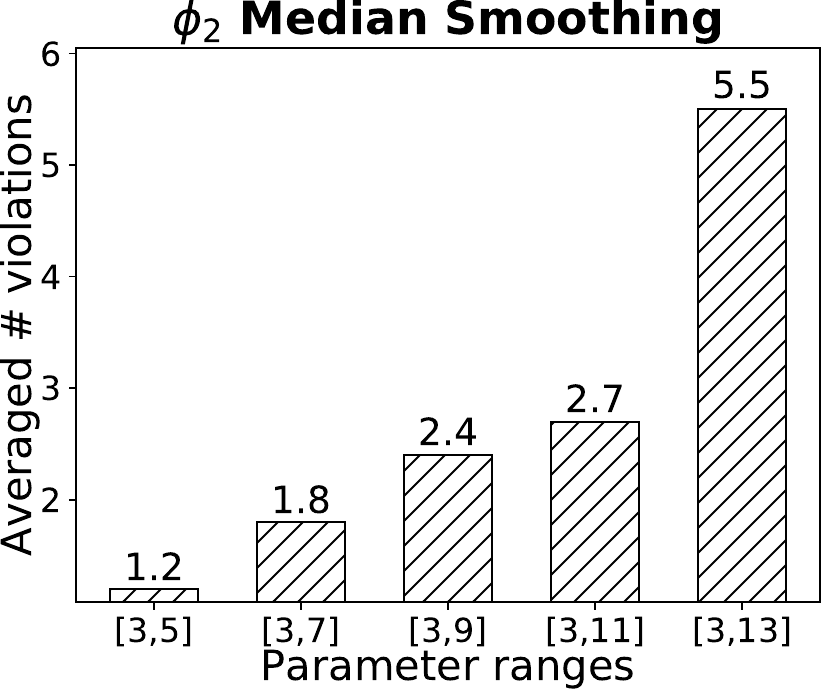}
\label{subfig:bound_violation2}}
\hfill
\subfloat[]{
\includegraphics[width=0.152\textwidth]{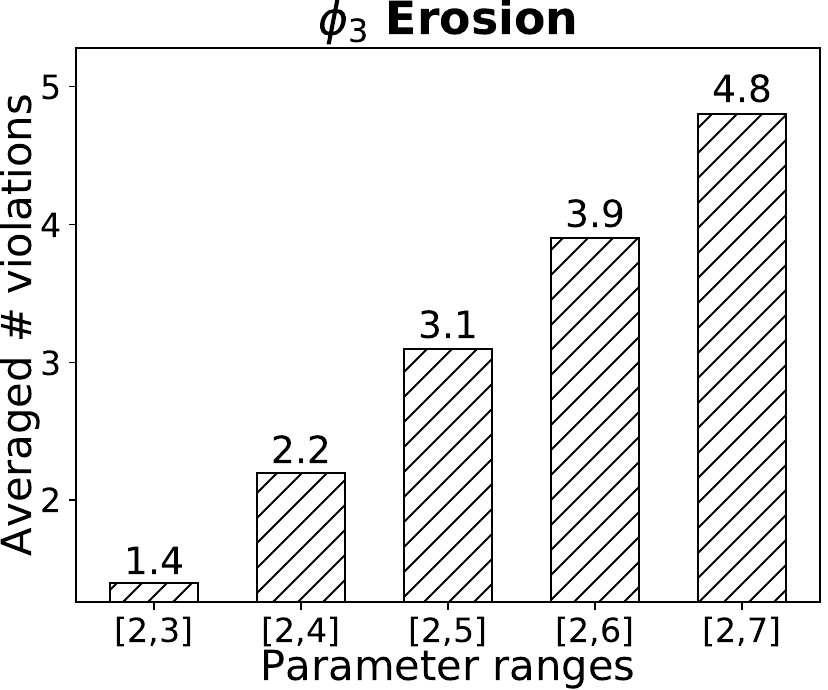}
\label{subfig:bound_violation3}}
\hfill
\subfloat[]{
\includegraphics[width=0.152\textwidth]{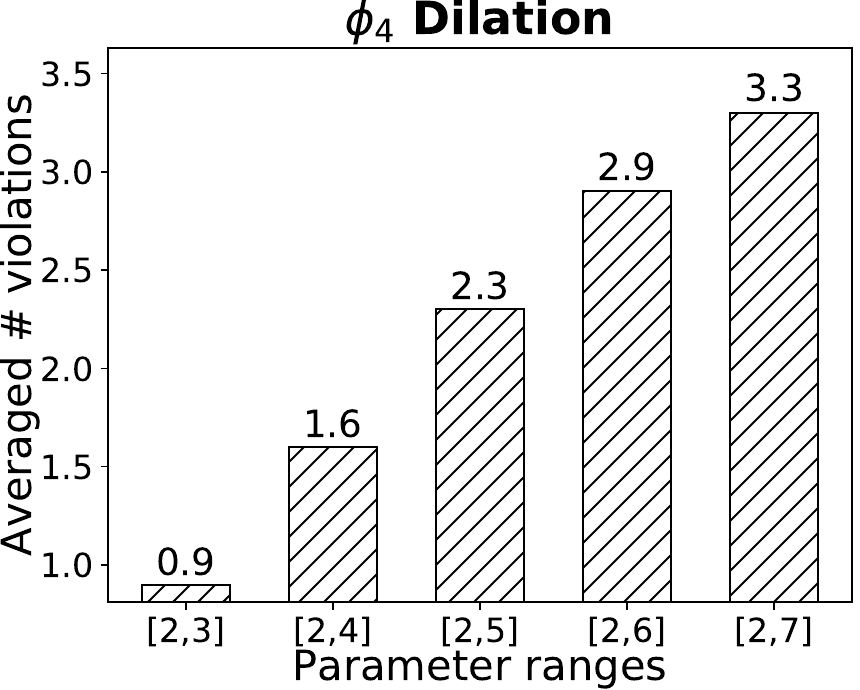}
\label{subfig:bound_violation4}}
\hfill
\subfloat[]{
\includegraphics[width=0.152\textwidth]{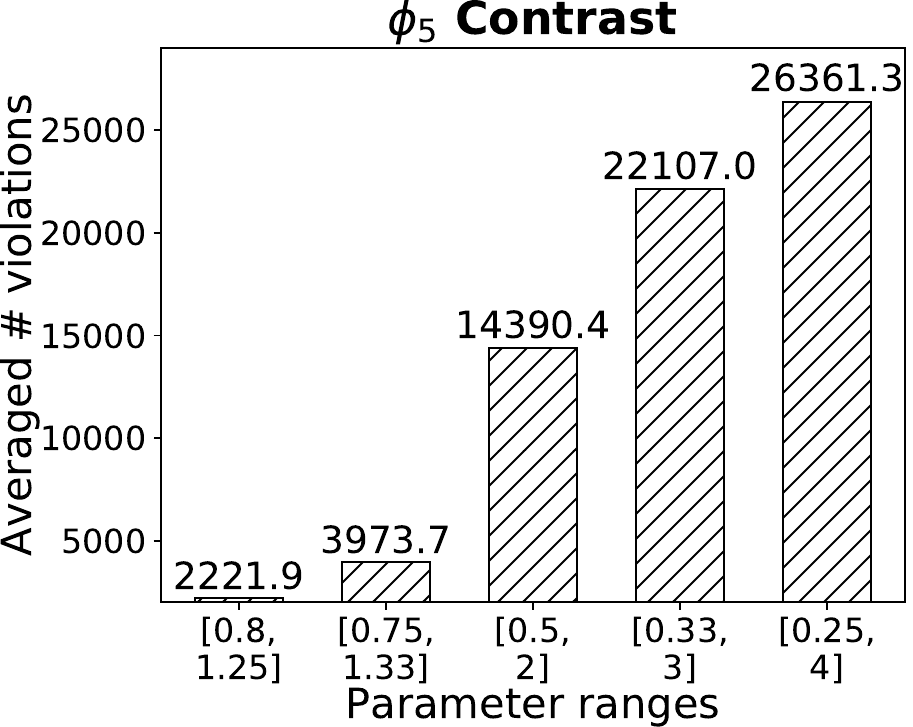}
\label{subfig:bound_violation5}}
\hfill
\subfloat[]{
\includegraphics[width=0.152\textwidth]{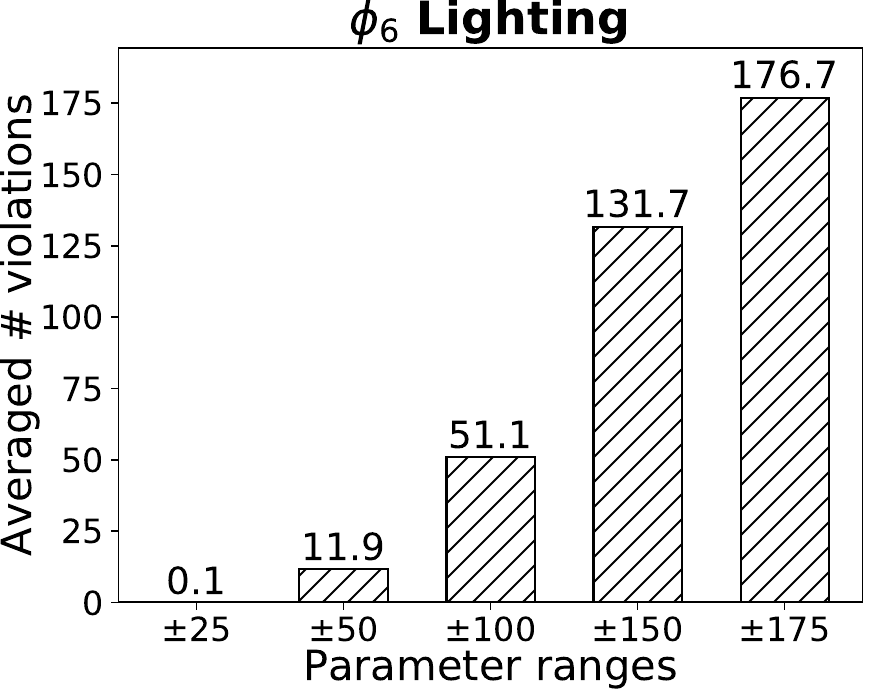}
\label{subfig:bound_violation6}}
\vspace{-.5cm}
\subfloat[]{
\includegraphics[width=0.152\textwidth]{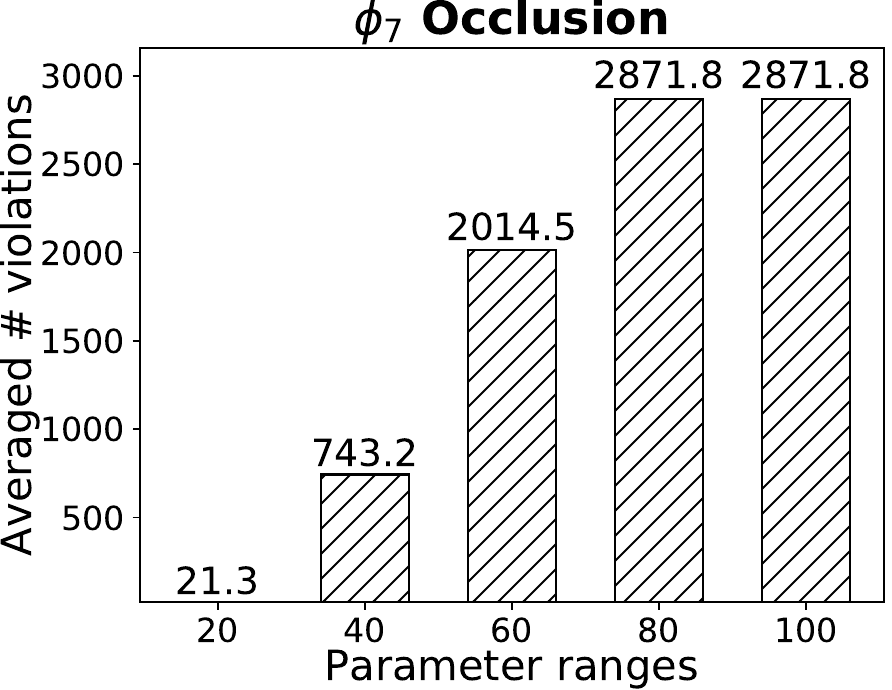}
\label{subfig:bound_violation7}}
\hfill
\subfloat[]{
\includegraphics[width=0.152\textwidth]{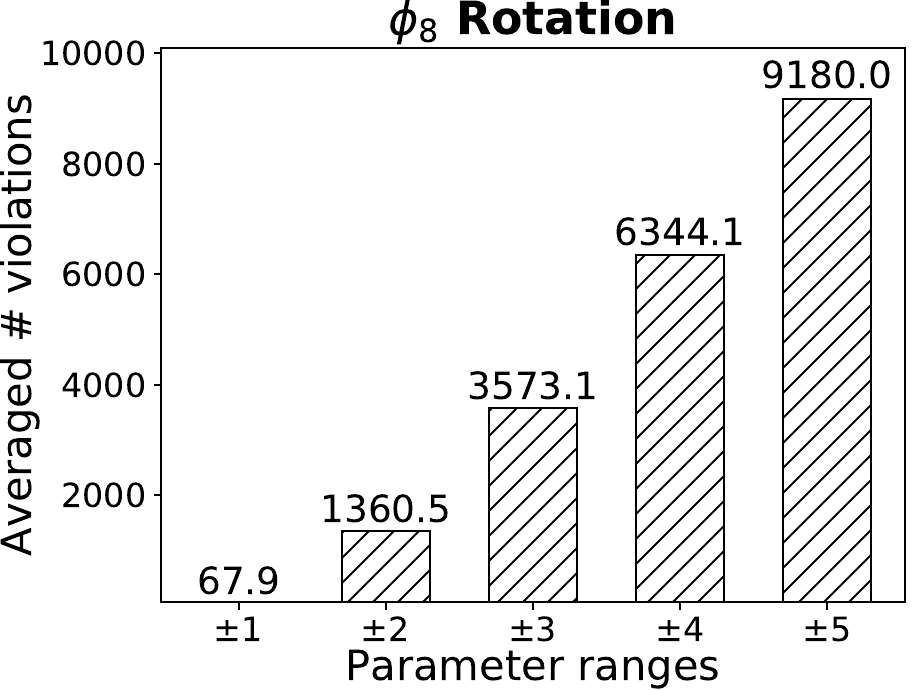}
\label{subfig:bound_violation8}}
\hfill
\subfloat[]{
\includegraphics[width=0.152\textwidth]{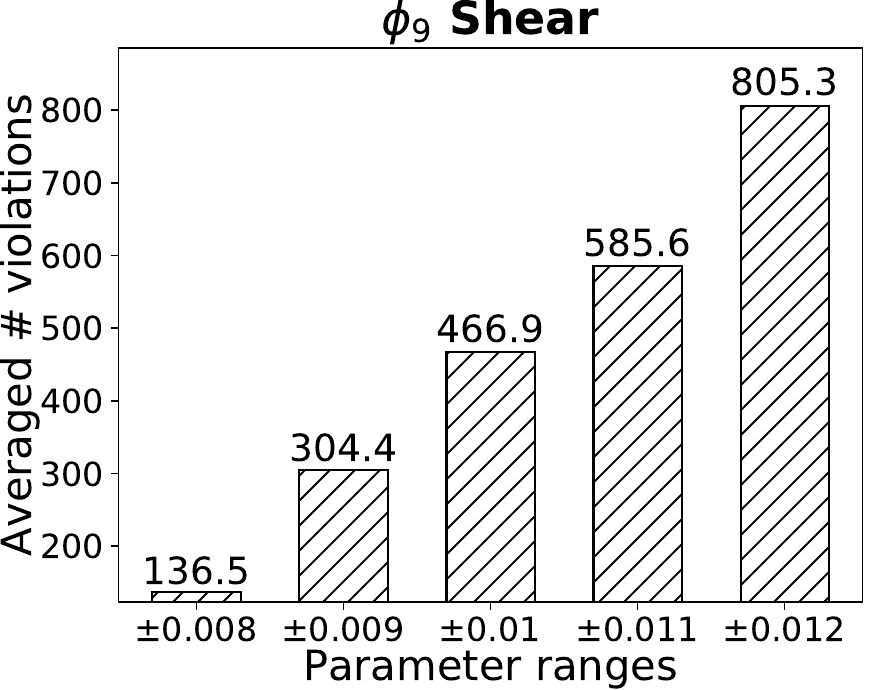}
\label{subfig:bound_violation9}}
\hfill
\subfloat[]{
\includegraphics[width=0.152\textwidth]{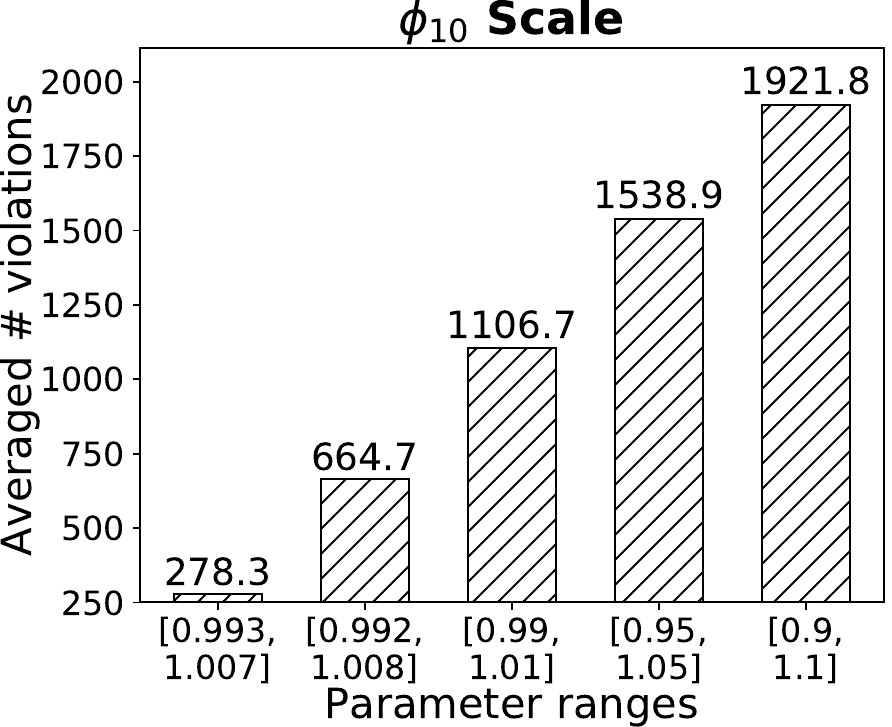}
\label{subfig:bound_violation10}}
\hfill
\subfloat[]{
\includegraphics[width=0.152\textwidth]{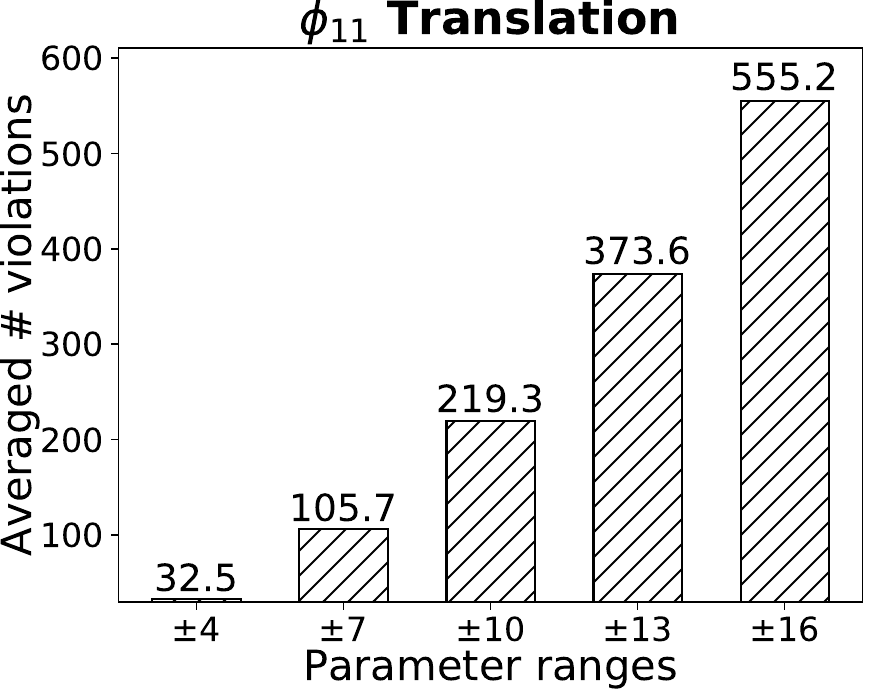}
\label{subfig:bound_violation11}}
\hfill
\subfloat[]{
\includegraphics[width=0.152\textwidth]{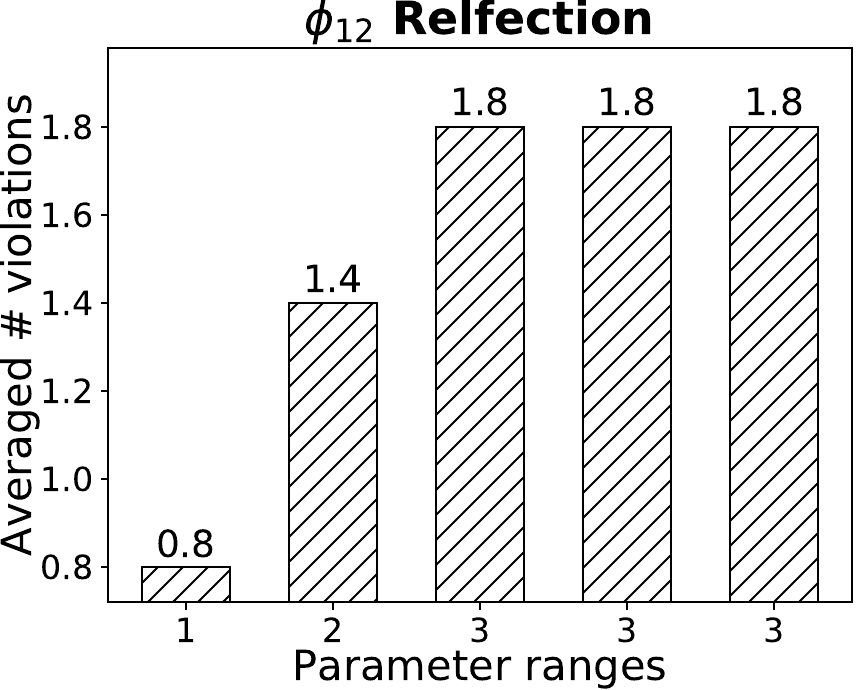}
\label{subfig:bound_violation12}}

\caption{The average numbers of errors found increase as we increase the bound of the parameter space ($\mathbb{C}_{\phi}$). The number above each bar shows the exact number of errors with the corresponding bounds.}

\label{fig:bound_time_violation}
\end{figure*}

\subsection{Effects of $k$ and $t$ on the number of errors}
Figure~\ref{subfig:k_violation1} and \ref{subfig:t_violation1} present how the thresholds $k$ and $t$ of invariance properties defined in the paper ($k$/$t$-invariance) influence the number of errors found by \sys. 
As shown, the number of errors decreases with increases in $k$ and $t$. This is intuitive as increasing $k$ and $t$ essentially increase the allowed margin of error for the vision systems. One interesting fact is that although the number of errors drops significantly when $k$ increases from 1 to 2, the changes in number of errors tend to be smaller when $k$ increases further. By contrast, the decrease in the number of errors for different increasing values of $t$ seems to be more uniform. 

\begin{figure*}
\captionsetup[subfloat]{captionskip=-.15cm, labelformat=empty}

{\bf \scriptsize \hspace{0.025in} Altering $k$:\vspace{-0.3cm}}

\subfloat[]{
\includegraphics[width=0.15\textwidth, height=0.14\textwidth]{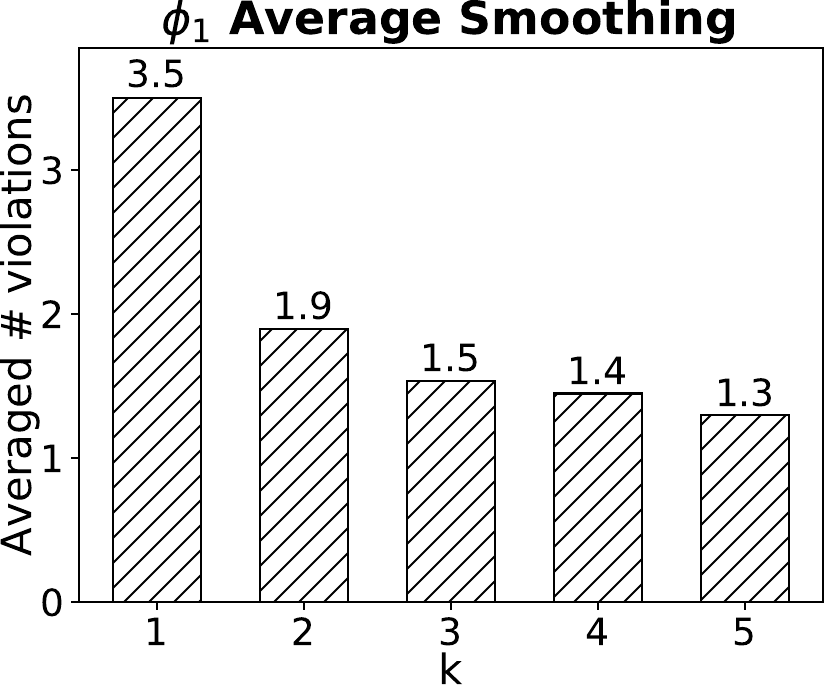}
\label{subfig:k_violation1}}
\hfill
\subfloat[]{
\includegraphics[width=0.15\textwidth, height=0.14\textwidth]{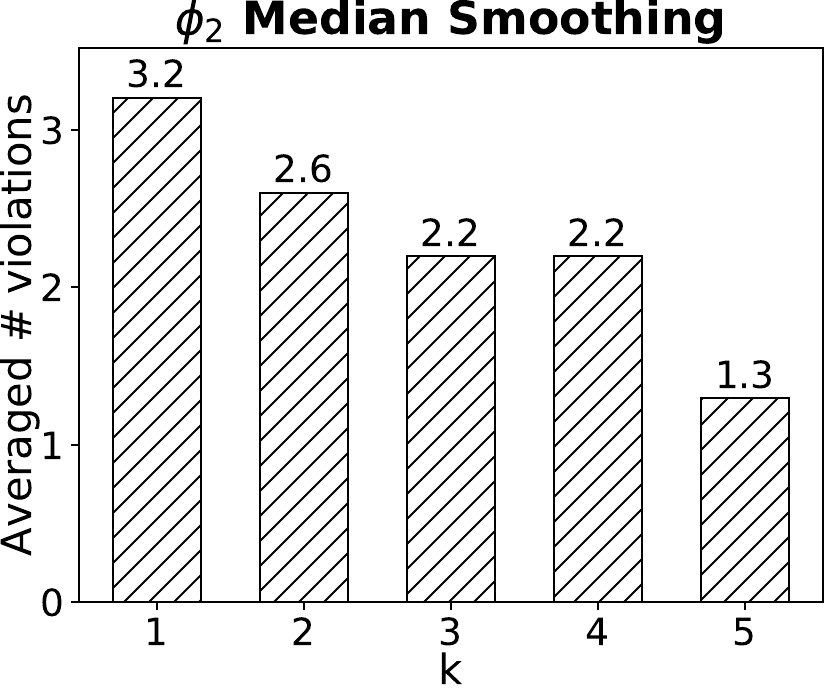}
\label{subfig:k_violation2}}
\hfill
\subfloat[]{
\includegraphics[width=0.15\textwidth, height=0.14\textwidth]{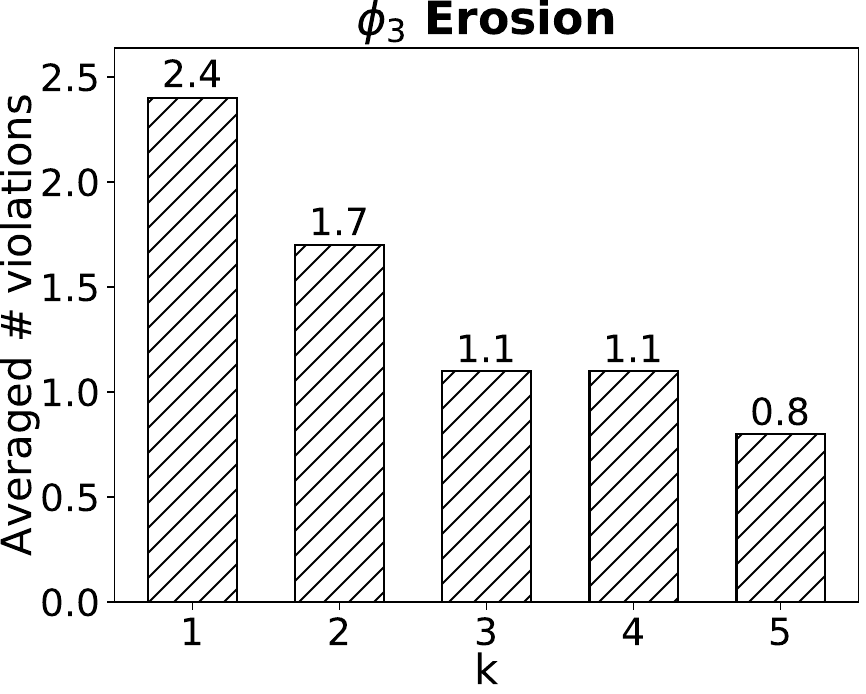}
\label{subfig:k_violation3}}
\hfill
\subfloat[]{
\includegraphics[width=0.15\textwidth, height=0.14\textwidth]{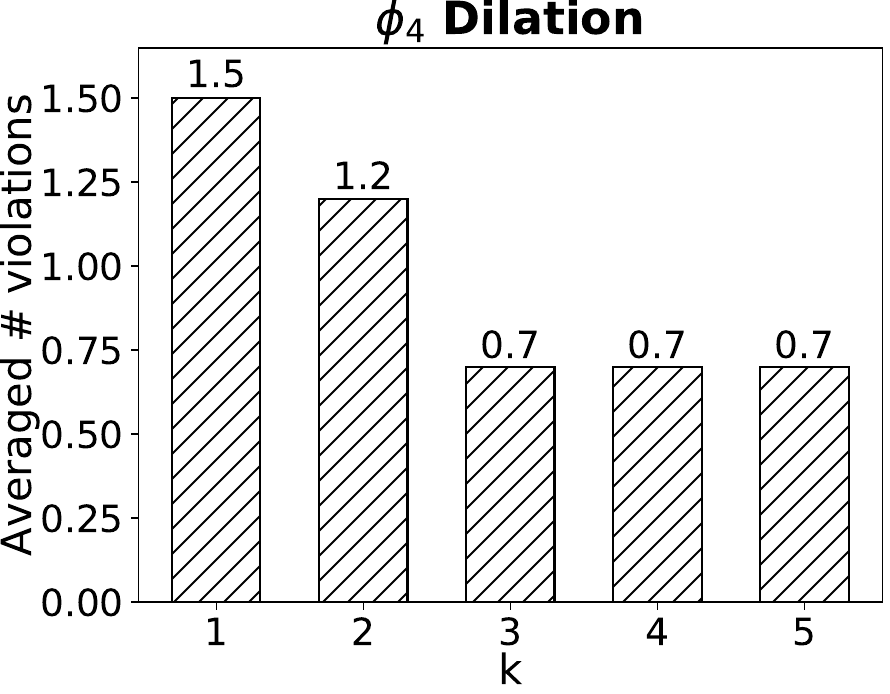}
\label{subfig:k_violation4}}
\hfill
\subfloat[]{
\includegraphics[width=0.15\textwidth, height=0.14\textwidth]{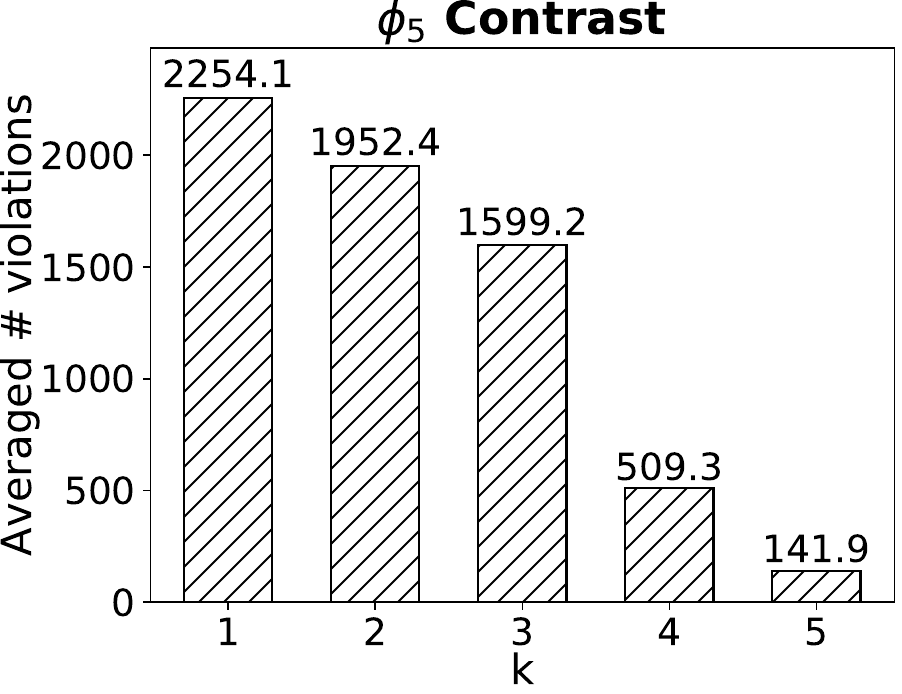}
\label{subfig:k_violation5}}
\hfill
\subfloat[]{
\includegraphics[width=0.15\textwidth, height=0.14\textwidth]{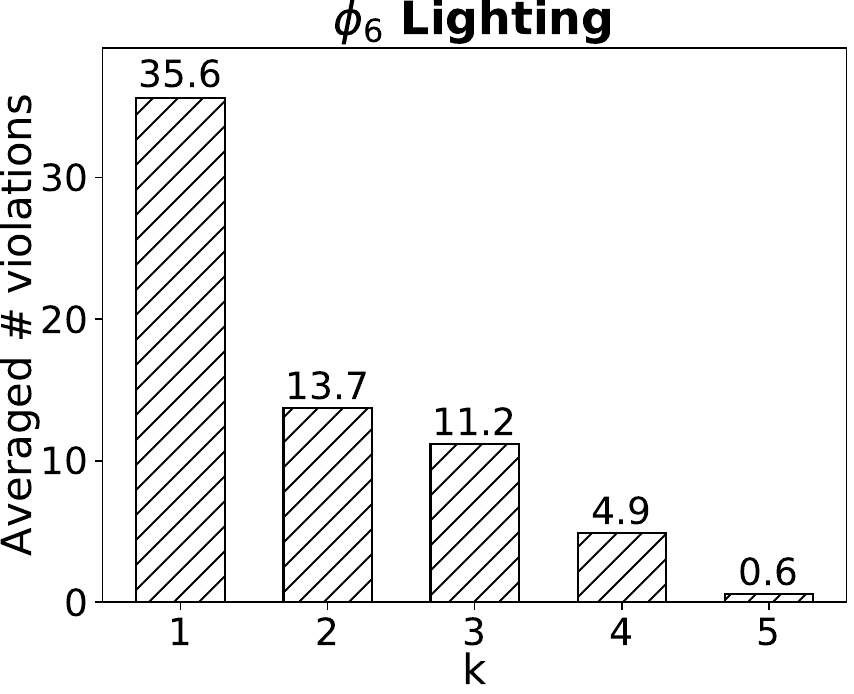}
\label{subfig:k_violation6}}
\vspace{-.5cm}
\subfloat[]{
\includegraphics[width=0.15\textwidth, height=0.14\textwidth]{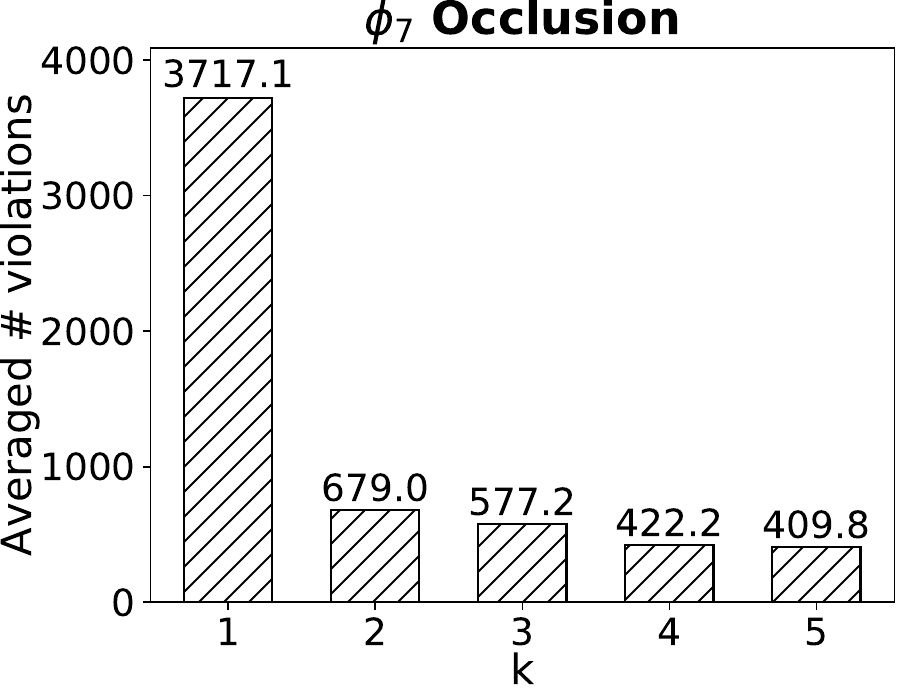}
\label{subfig:k_violation7}}
\hfill
\subfloat[]{
\includegraphics[width=0.15\textwidth, height=0.14\textwidth]{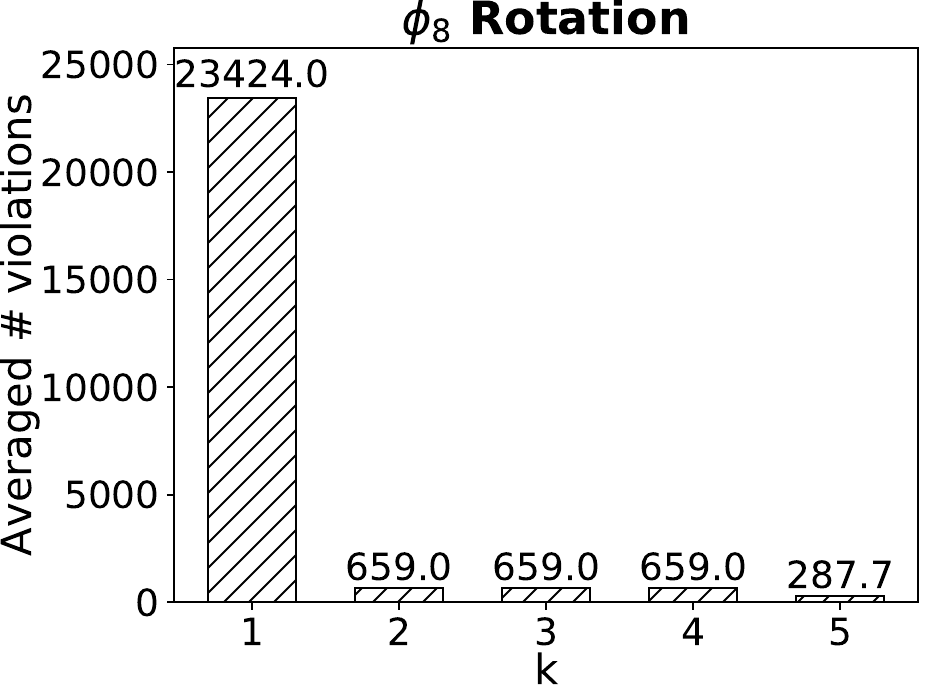}
\label{subfig:k_violation8}}
\hfill
\subfloat[]{
\includegraphics[width=0.15\textwidth, height=0.14\textwidth]{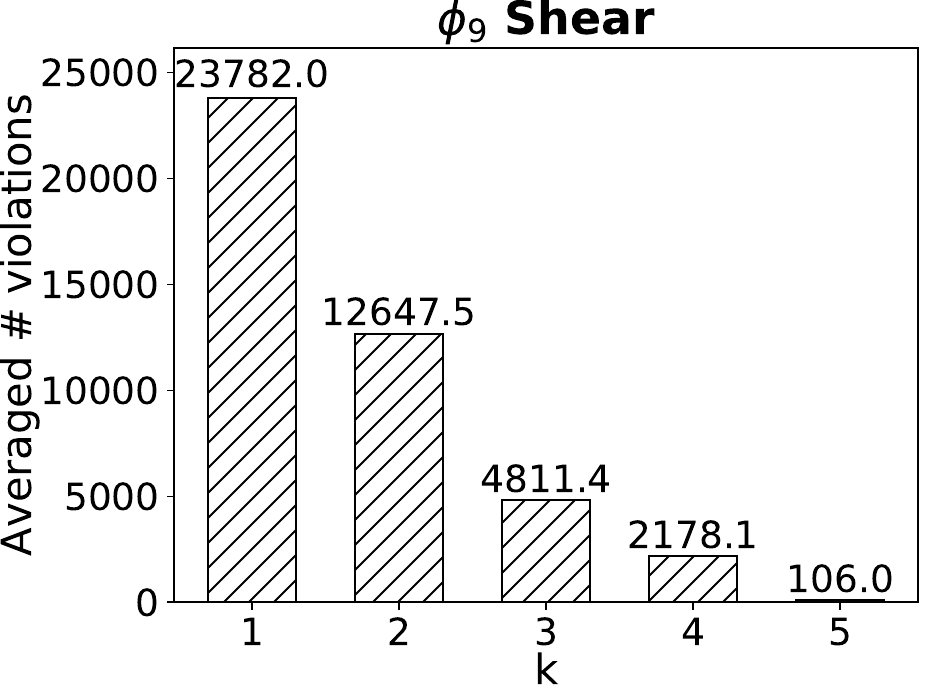}
\label{subfig:k_violation9}}
\hfill
\subfloat[]{
\includegraphics[width=0.15\textwidth, height=0.14\textwidth]{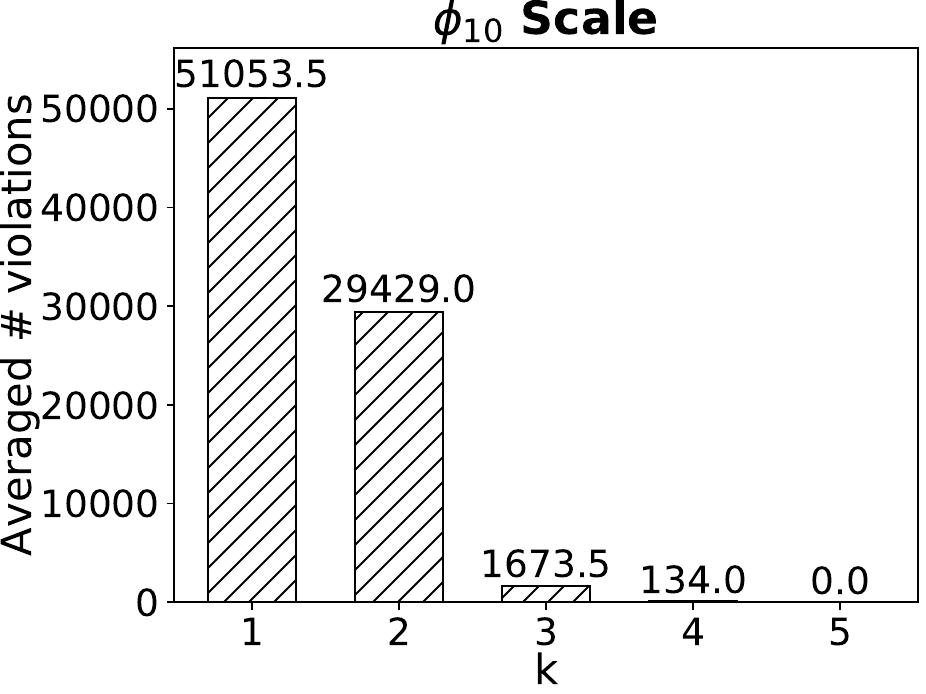}
\label{subfig:k_violation10}}
\hfill
\subfloat[]{
\includegraphics[width=0.15\textwidth, height=0.14\textwidth]{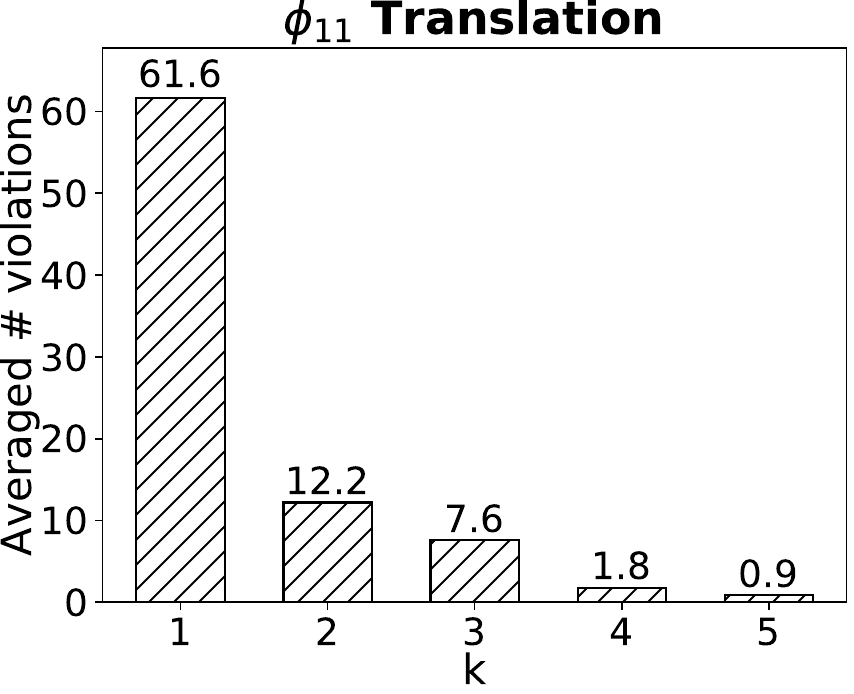}
\label{subfig:k_violation11}}
\hfill
\subfloat[]{
\includegraphics[width=0.15\textwidth, height=0.14\textwidth]{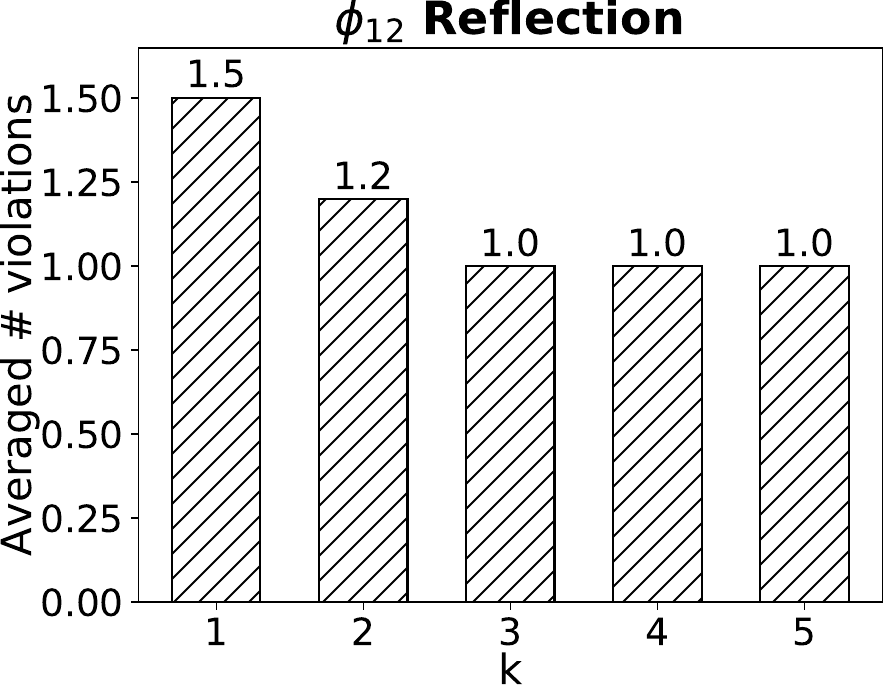}
\label{subfig:k_violation12}}

\vspace{-.18cm}{\bf \scriptsize \hspace{0.025in} Altering $t$:\vspace{-.3cm}}

\subfloat[]{
\includegraphics[width=0.15\textwidth, height=0.14\textwidth]{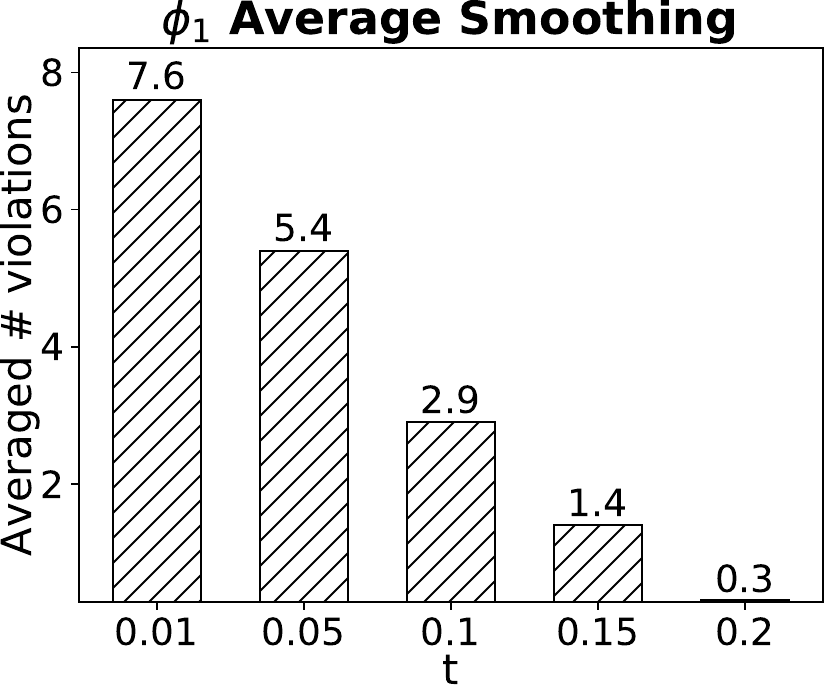}
\label{subfig:t_violation1}}
\hfill
\subfloat[]{
\includegraphics[width=0.15\textwidth, height=0.14\textwidth]{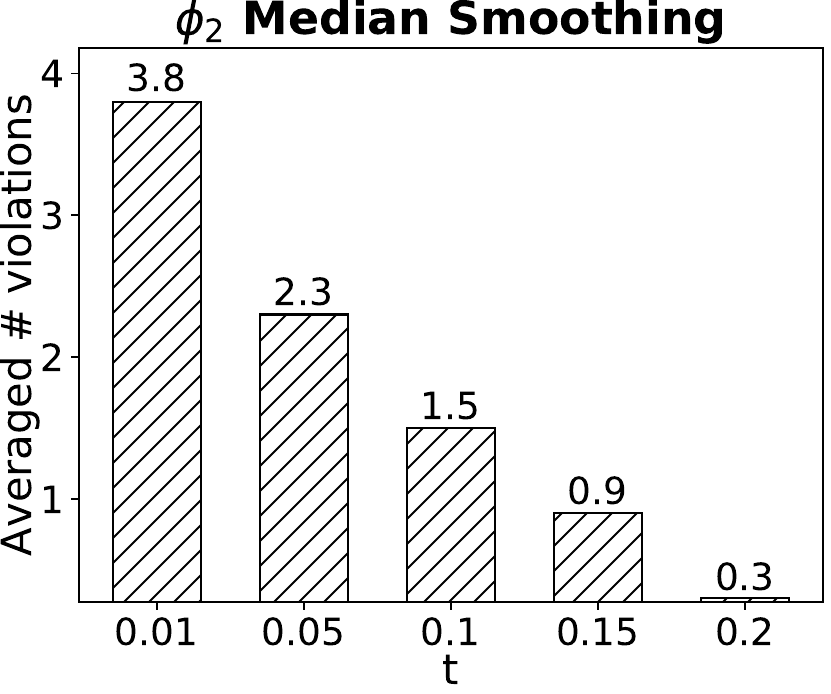}
\label{subfig:t_violation2}}
\hfill
\subfloat[]{
\includegraphics[width=0.15\textwidth, height=0.14\textwidth]{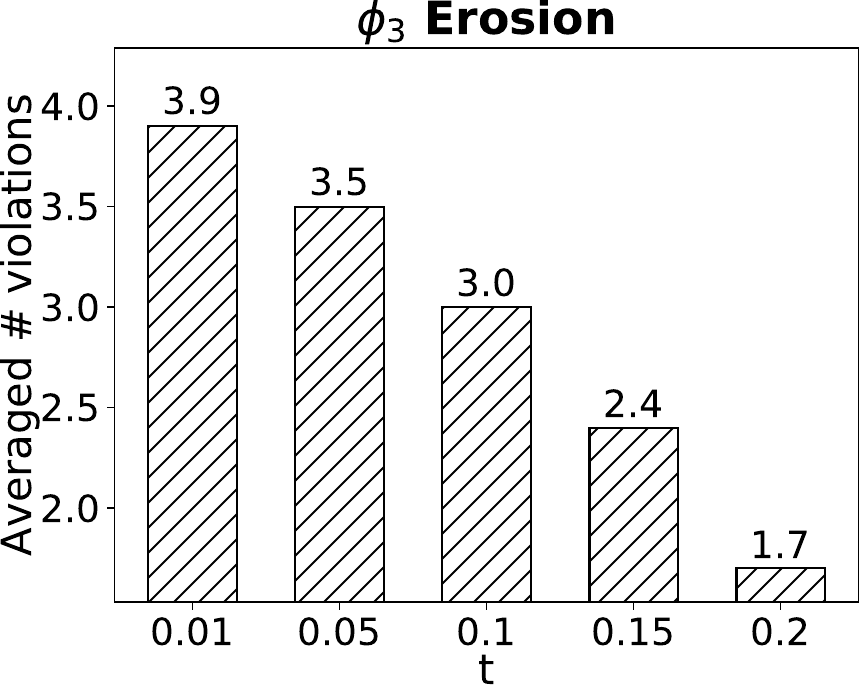}
\label{subfig:t_violation3}}
\hfill
\subfloat[]{
\includegraphics[width=0.15\textwidth, height=0.14\textwidth]{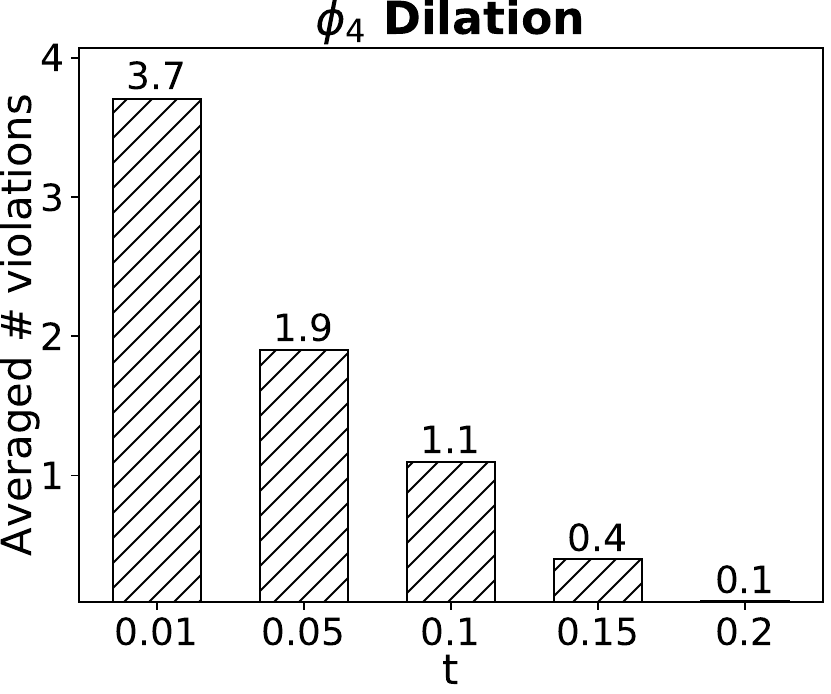}
\label{subfig:t_violation4}}
\hfill
\subfloat[]{
\includegraphics[width=0.15\textwidth, height=0.14\textwidth]{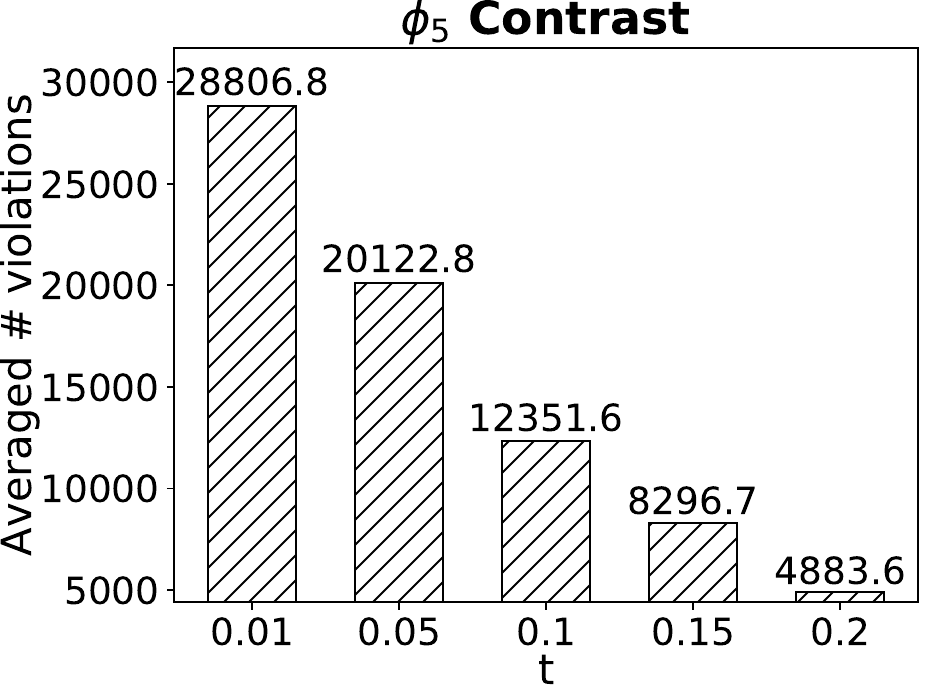}
\label{subfig:t_violation5}}
\hfill
\subfloat[]{
\includegraphics[width=0.15\textwidth, height=0.14\textwidth]{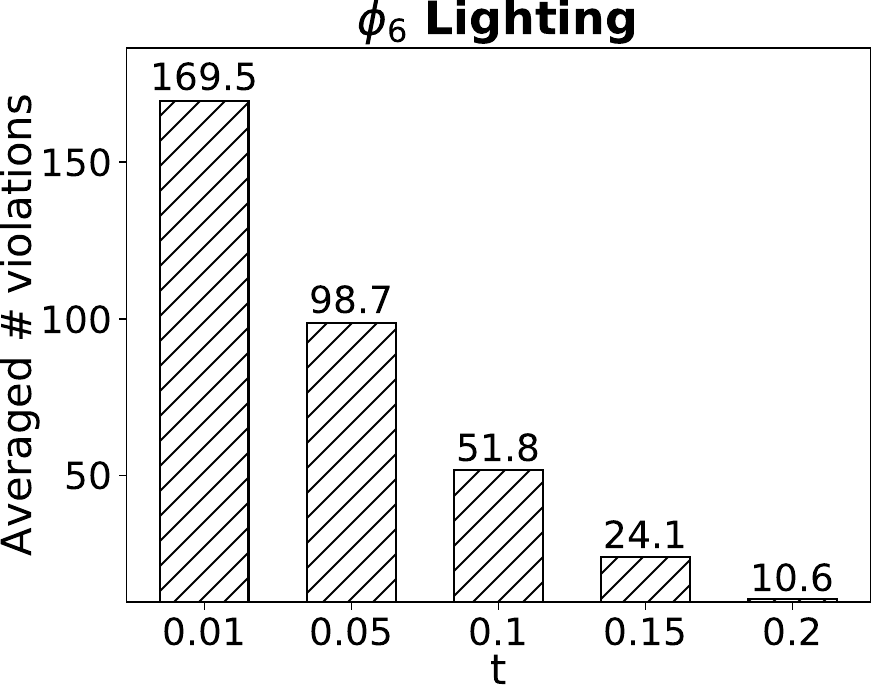}
\label{subfig:t_violation6}}
\vspace{-.5cm}
\subfloat[]{
\includegraphics[width=0.15\textwidth, height=0.14\textwidth]{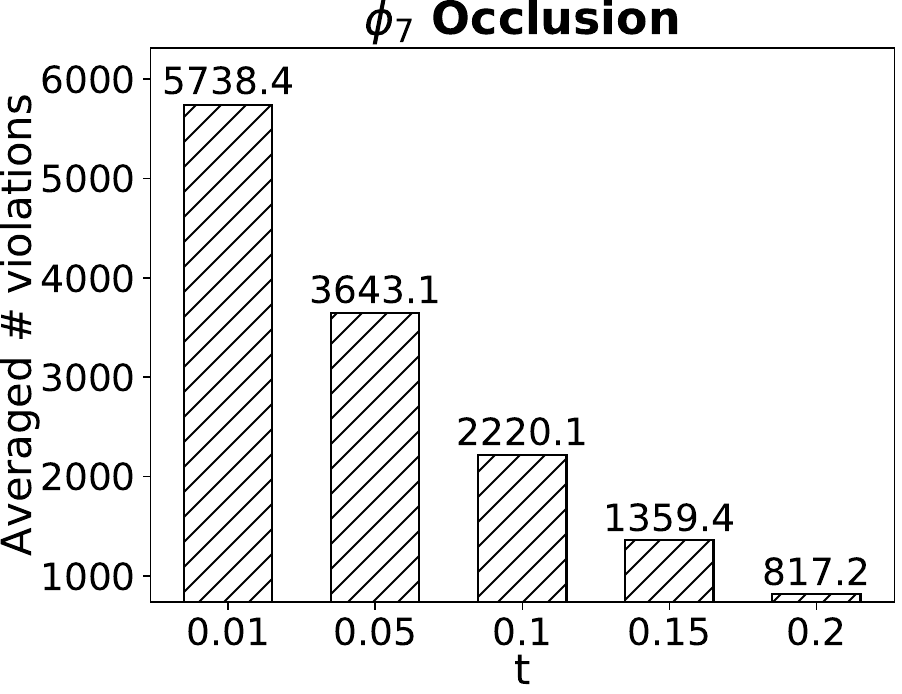}
\label{subfig:t_violation7}}
\hfill
\subfloat[]{
\includegraphics[width=0.15\textwidth, height=0.14\textwidth]{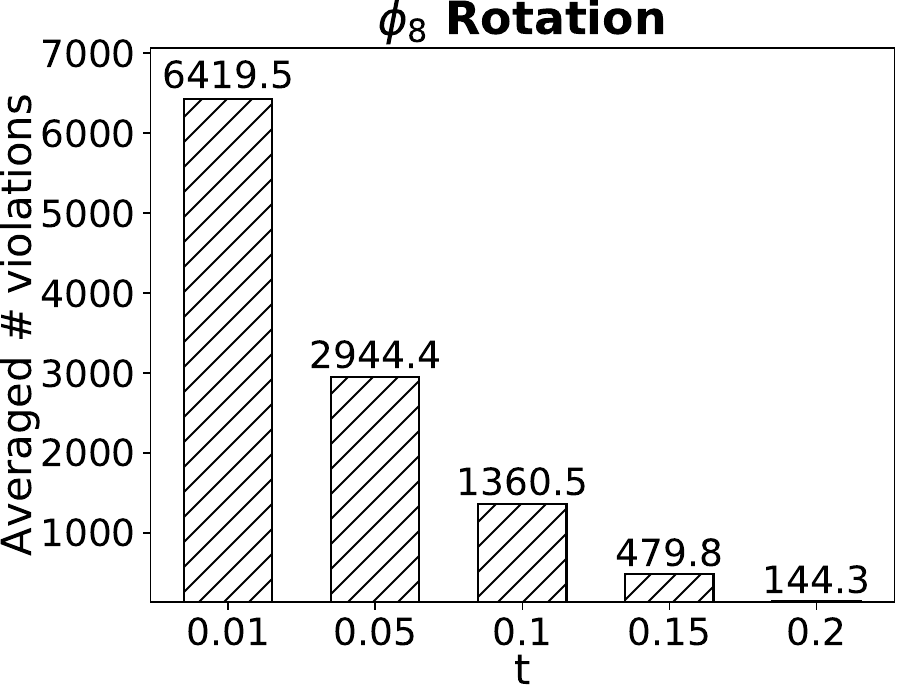}
\label{subfig:t_violation8}}
\hfill
\subfloat[]{
\includegraphics[width=0.15\textwidth, height=0.14\textwidth]{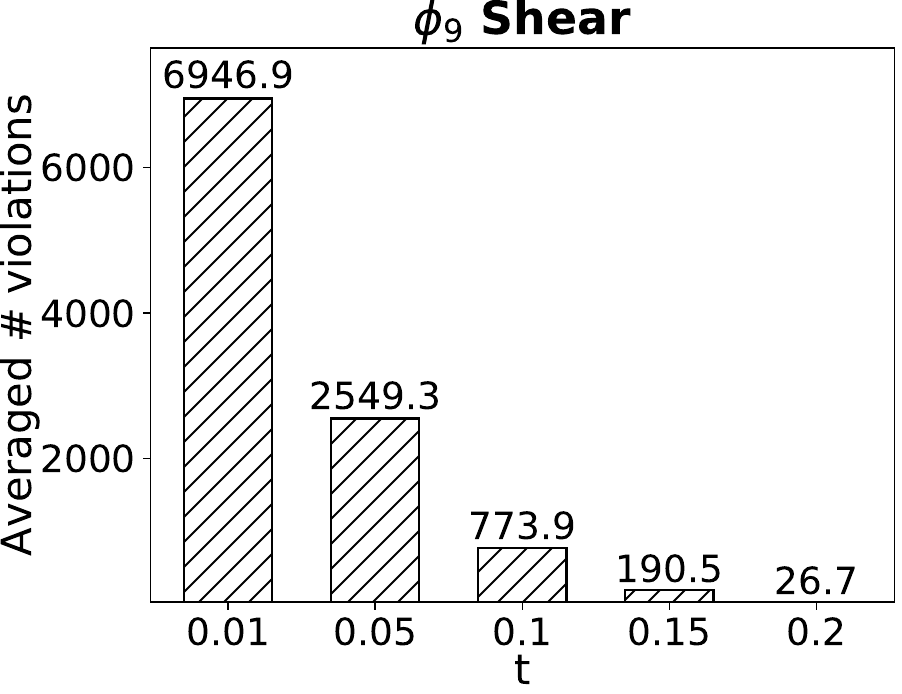}
\label{subfig:t_violation9}}
\hfill
\subfloat[]{
\includegraphics[width=0.15\textwidth, height=0.14\textwidth]{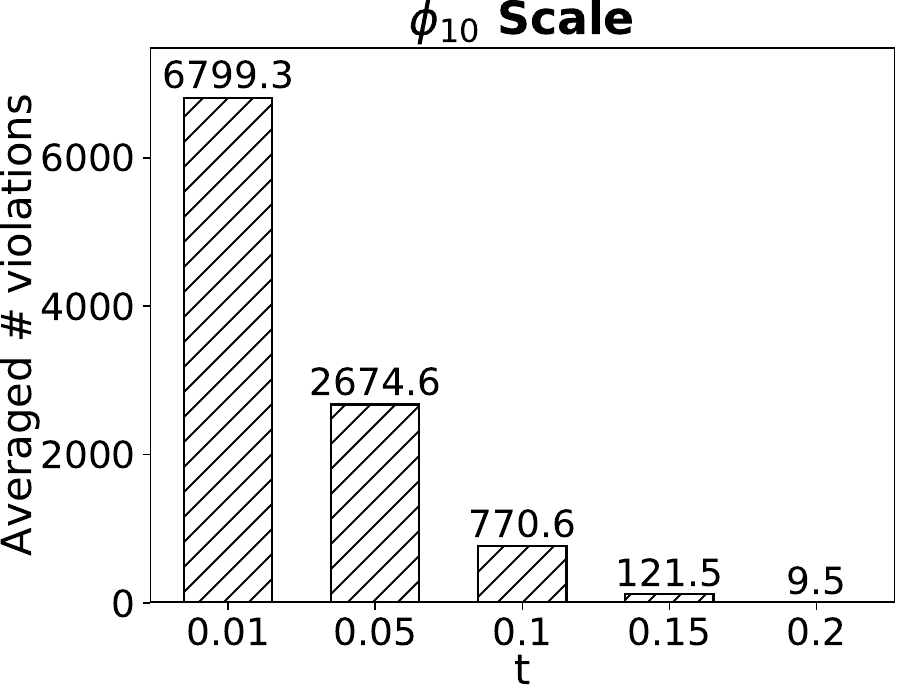}
\label{subfig:t_violation10}}
\hfill
\subfloat[]{
\includegraphics[width=0.15\textwidth, height=0.14\textwidth]{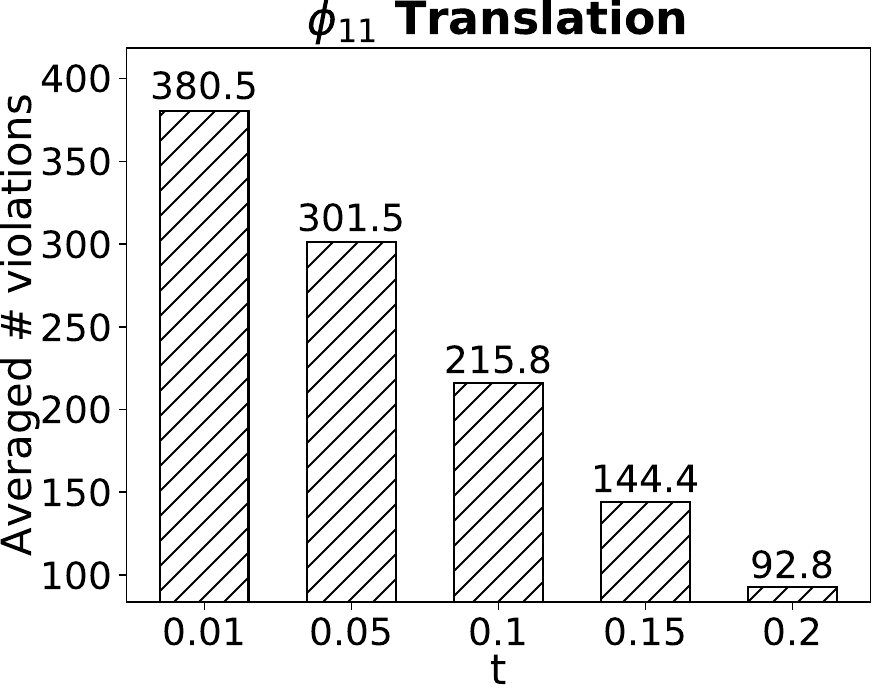}
\label{subfig:t_violation11}}
\hfill
\subfloat[]{
\includegraphics[width=0.15\textwidth, height=0.14\textwidth]{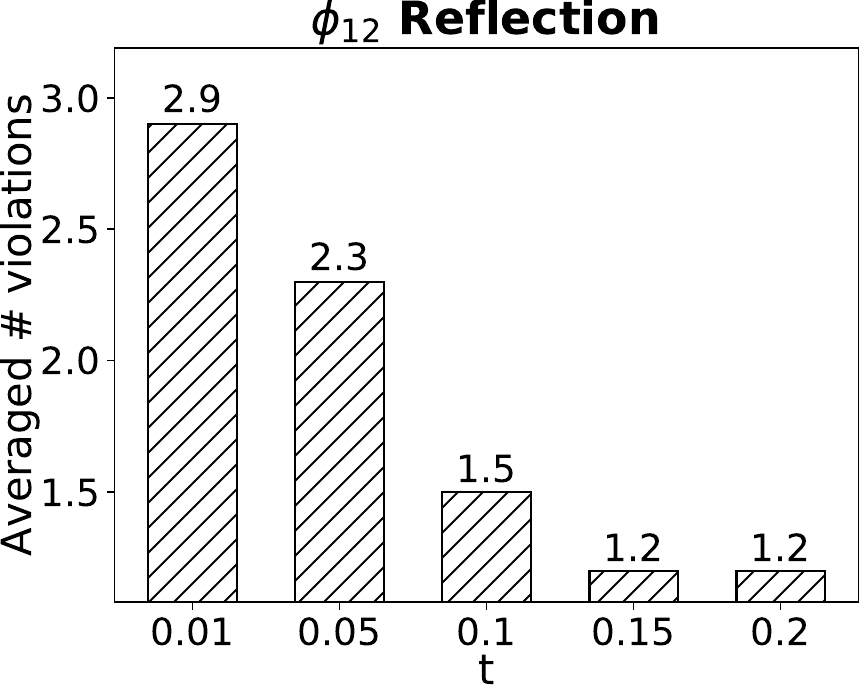}
\label{subfig:t_violation12}}

\caption{The change in the number of errors of different invariance properties as we increase $k$ (upper two rows) and $t$ (lower two rows). The number of errors tend to decrease with increasing $k$ or $t$. The number above each bar shows the actual number of errors (averaged over 10 inputs) found for each $k$ or $t$ in IMG\_C3 (MobileNet) and DRV\_C2 (Dave-orig) respectively}
\label{fig:kt_violation}
\end{figure*}

\subsection{Errors for complex transformations}

Several real-world phenomena (\eg fog, rain, etc.) that may affect input images are hard to emulate using the simple transformations described in the paper. However, custom transformations can be designed to mimic such effects and use \sys to check invariances with these transformations. As an example of this approach, we demonstrate how a simple parameterized fog-simulating transformation can be designed and verified with \sys. For this transformation, we start with a fog mask, apply average smoothing on the mask, and apply the mask to the input image. By controlling the smoothing kernel size, we simulate different amounts of fog.

We use \sys to enumerate and check all critical parameter values for the fog transform described above,  \ie different convolution sizes for average smoothing. \sys is able to find hundreds of errors in IMG\_C3 (MobileNet), API\_C1, and DRV\_C2 (dave-orig). Figure~\ref{fig:fog} shows three sample errors found by \sys. 

\begin{figure*}
\centering
\captionsetup[subfloat]{labelformat=empty}

\subfloat[cougar]{
\includegraphics[width=0.25\textwidth]{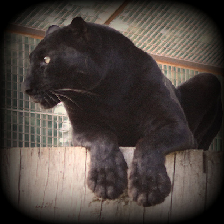}
\label{subfig:fog1}}
\subfloat[honeycomb]{
\includegraphics[width=0.25\textwidth]{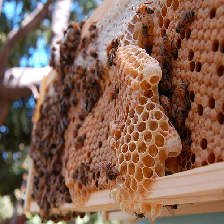}
\label{subfig:fog2}}
\subfloat[turn right]{
\includegraphics[width=0.25\textwidth]{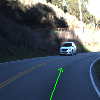}
\label{subfig:fog3}}

\subfloat[elephant]{
\includegraphics[width=0.25\textwidth]{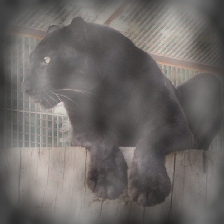}
\label{subfig:fog4}}
\subfloat[invertebrate]{
\includegraphics[width=0.25\textwidth]{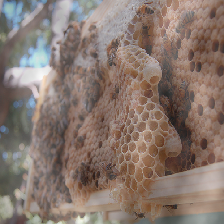}
\label{subfig:fog5}}
\subfloat[go straight]{
\includegraphics[width=0.25\textwidth]{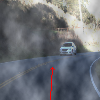}
\label{subfig:fog6}}

\caption{Errors found by \sys for a fog-simulating transformation in IMG\_C3 (MobileNet), API\_C1, and DRV\_C2 (dave-orig). The first and second rows show the original images and the foggy images that result in errors, respectively}
\label{fig:fog}
\end{figure*}

\subsection{Transferability}
\begin{figure*}
\captionsetup[subfloat]{labelformat=empty}

\subfloat[Average smoothing]{
\includegraphics[width=0.32\textwidth]{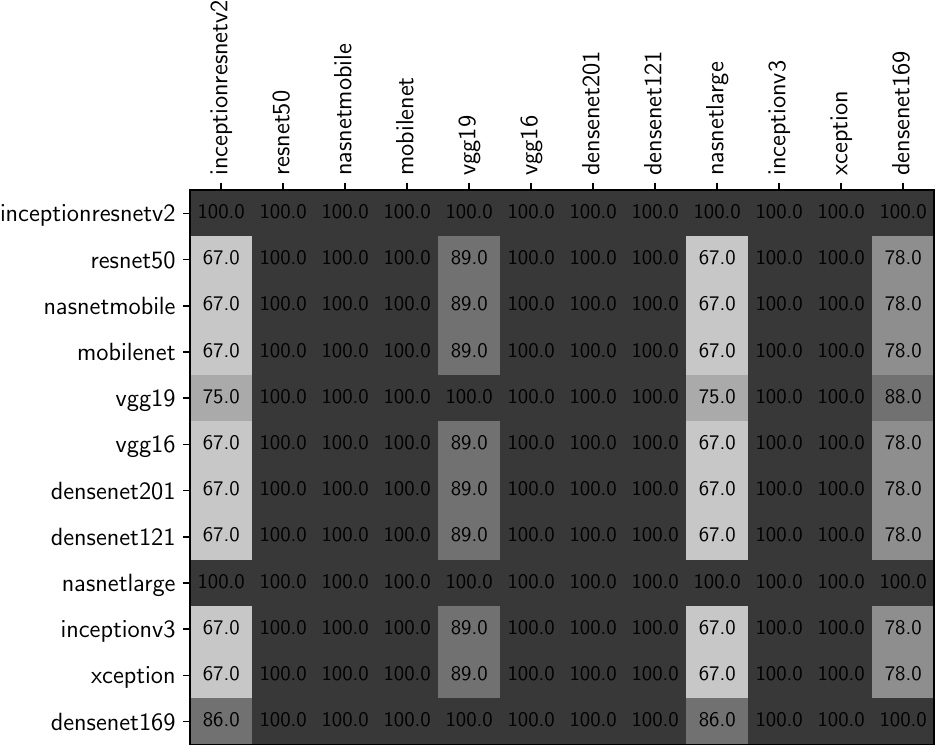}}%
\hfill
\subfloat[Median smoothing]{
\includegraphics[width=0.32\textwidth]{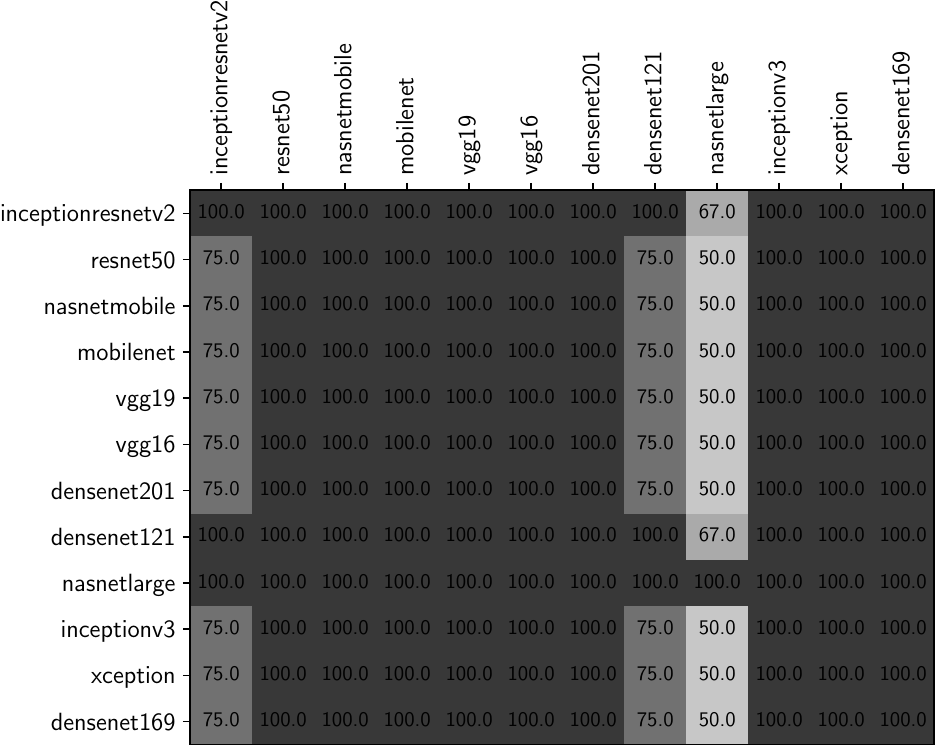}}%
\hfill
\subfloat[Erosion]{
\includegraphics[width=0.32\textwidth]{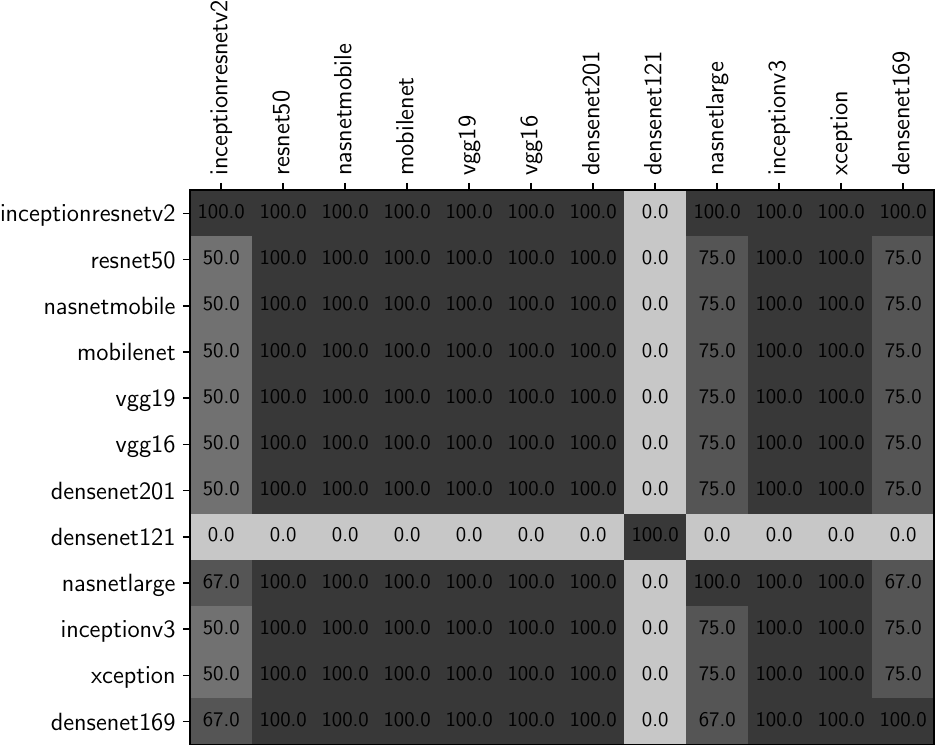}}

\subfloat[Dilation]{
\includegraphics[width=0.32\textwidth]{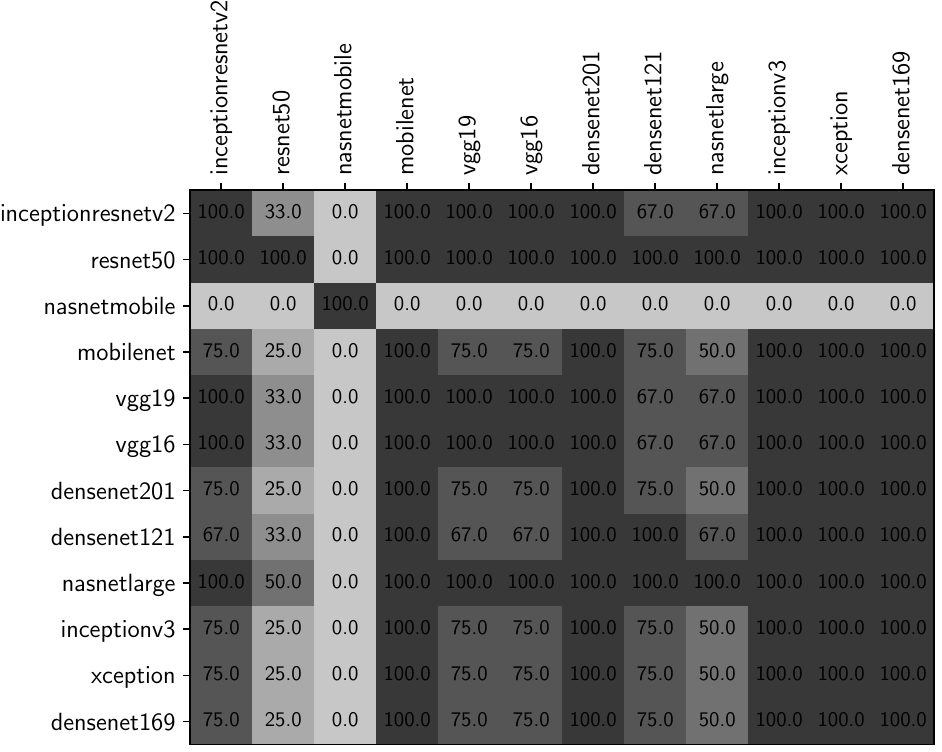}}%
\hfill
\subfloat[Contrast]{
\includegraphics[width=0.32\textwidth]{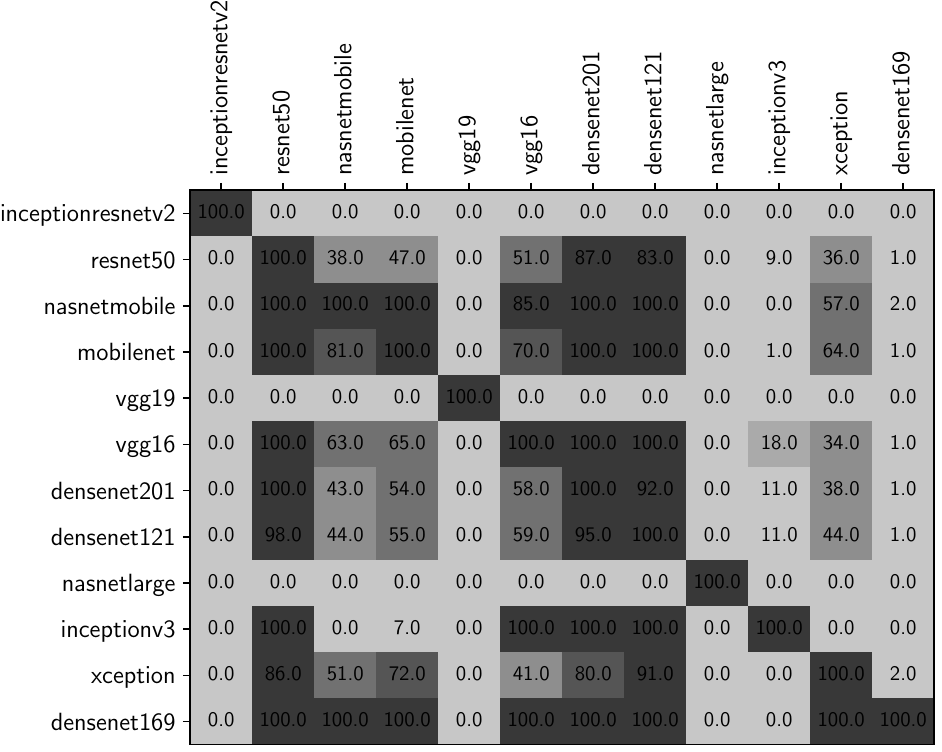}}%
\hfill
\subfloat[Brightening]{
\includegraphics[width=0.32\textwidth]{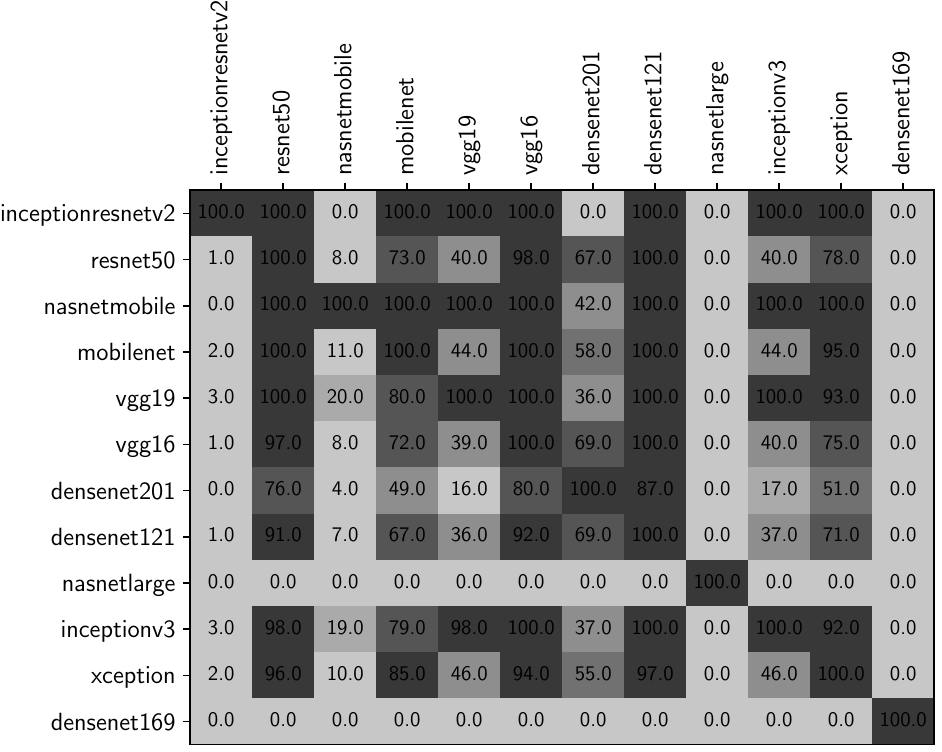}}

\subfloat[Occlusion]{
\includegraphics[width=0.32\textwidth]{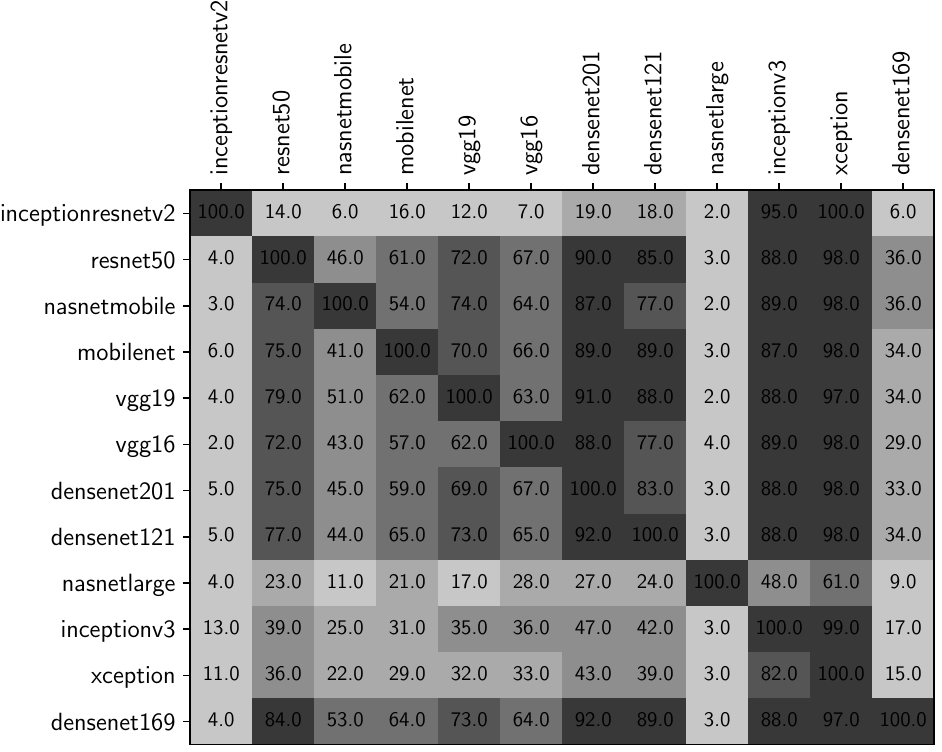}}%
\hfill
\subfloat[Rotation]{
\includegraphics[width=0.32\textwidth]{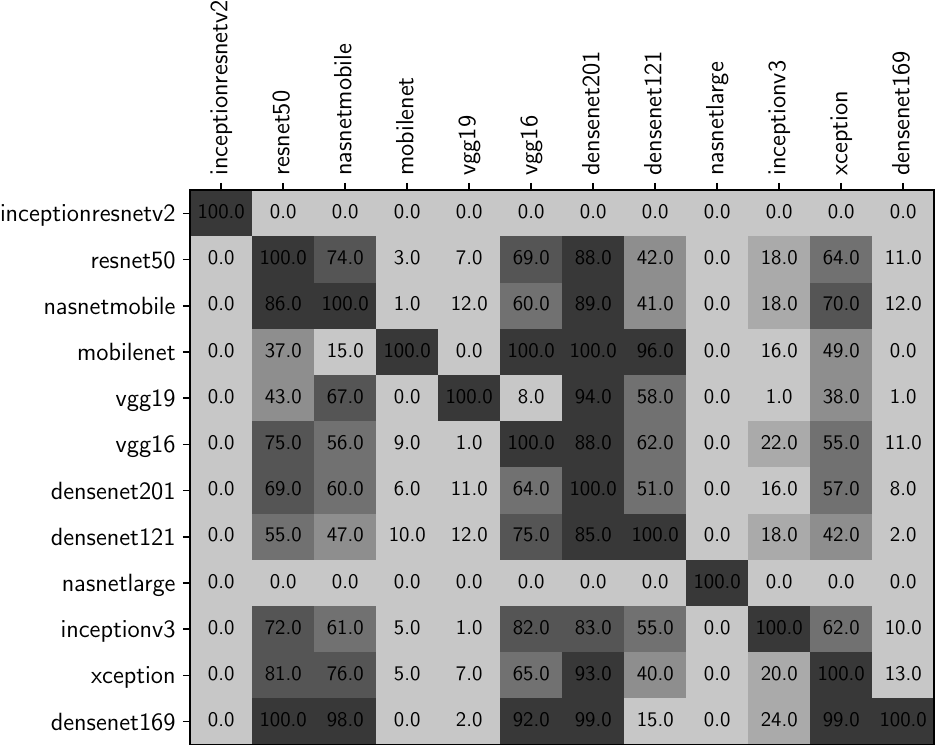}}%
\hfill
\subfloat[Shear]{
\includegraphics[width=0.32\textwidth]{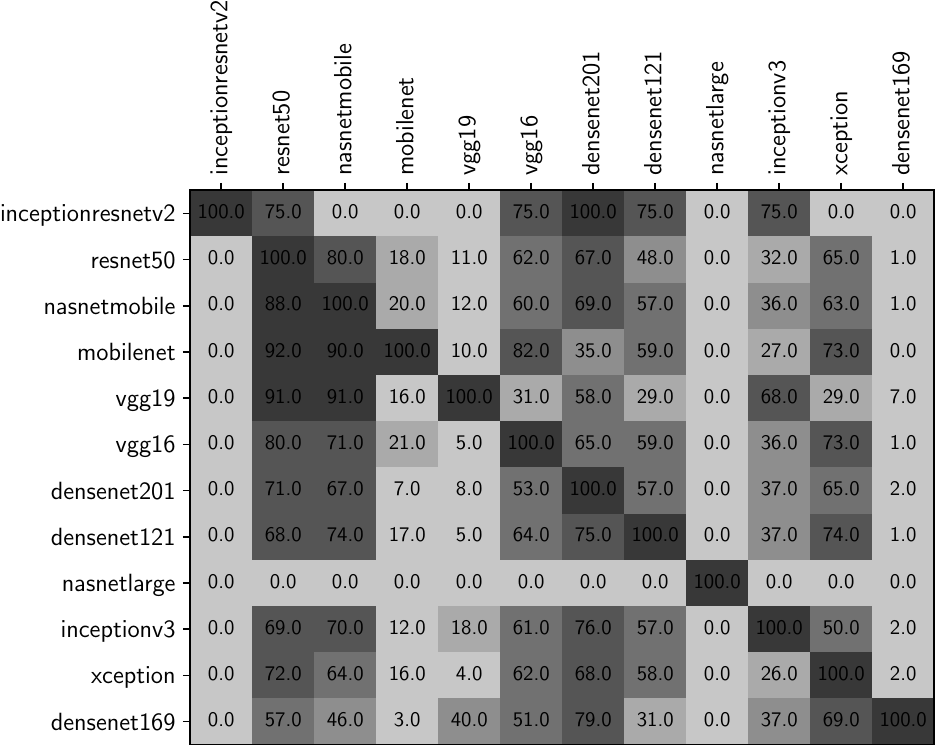}}

\subfloat[Scale]{
\includegraphics[width=0.32\textwidth]{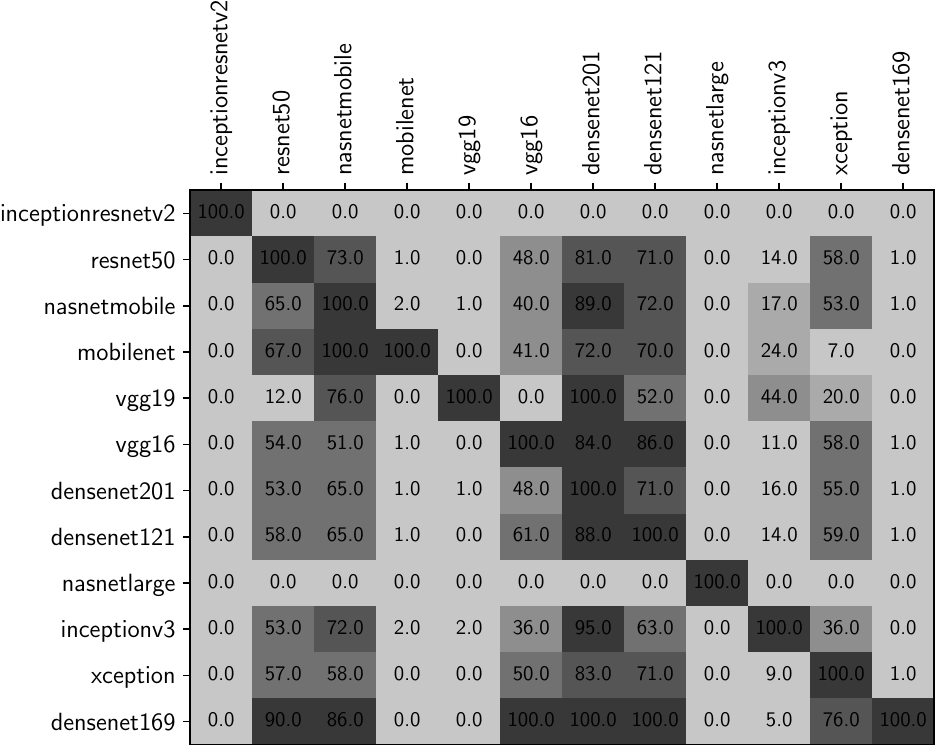}}%
\hfill
\subfloat[Translation]{
\includegraphics[width=0.32\textwidth]{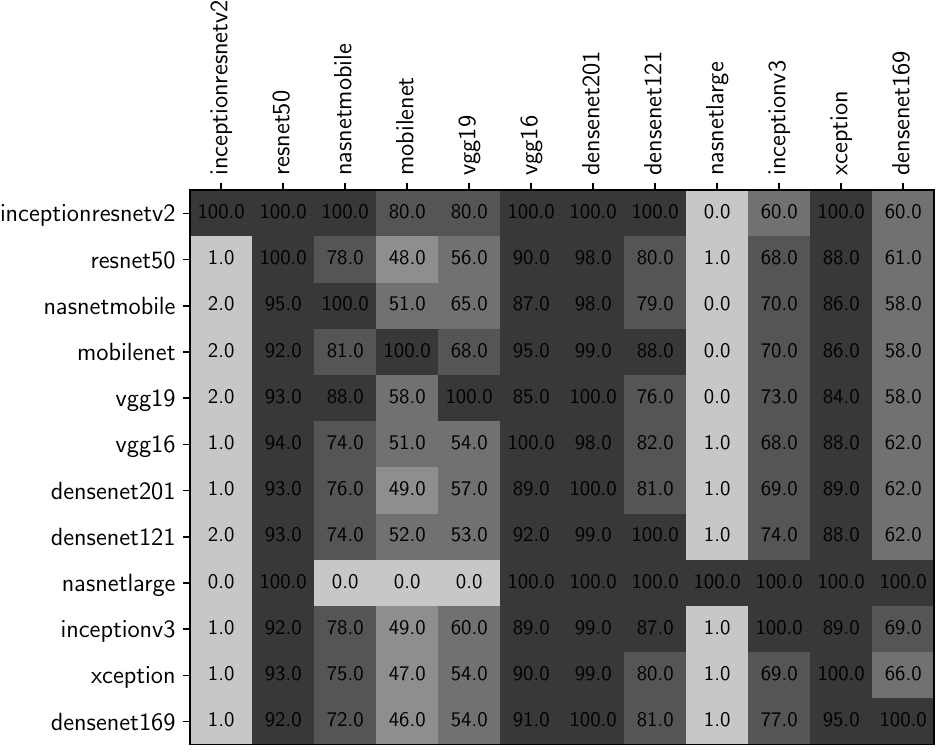}}%
\hfill
\subfloat[Reflection]{
\includegraphics[width=0.32\textwidth]{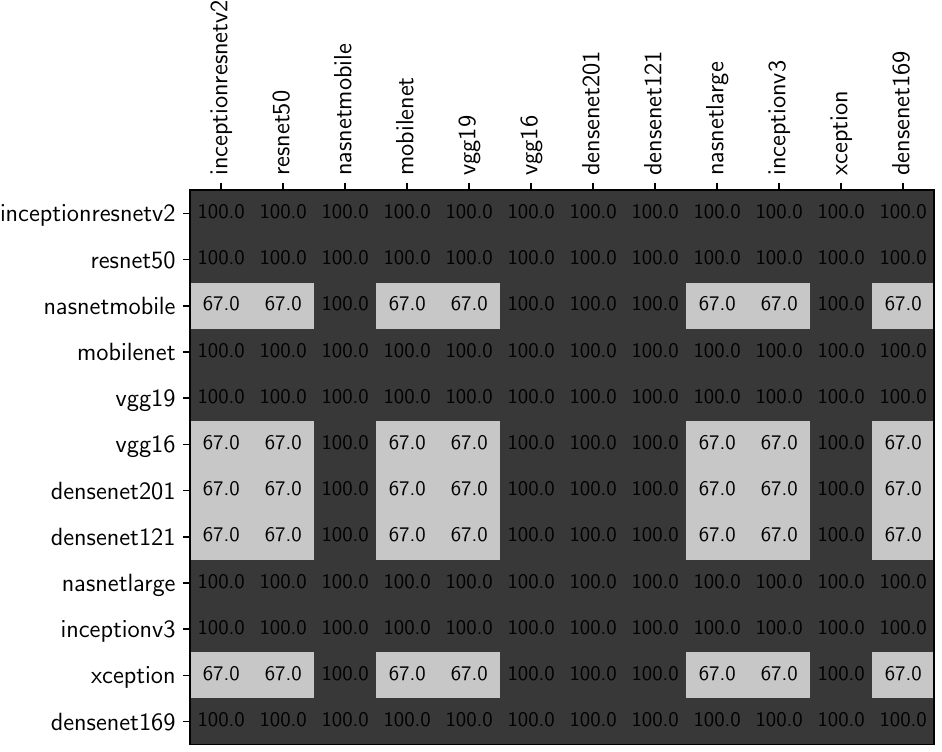}}

\caption{Transferability of errors across models for each transformation in the invariance property. Darker color indicates higher transfer rate}
\label{fig:transferability}
\end{figure*}

We also study if the errors for one model also transfer to other models. 
Specifically, we compute all transformed inputs that induce errors for a specific model and check if they also induce errors for another model. 
Figure~\ref{fig:transferability} shows the results of transferring rate between every pair of models (ImageNet twelve models) under all twelve transformations, demonstrating that a large fraction of errors \textit{do transfer between different models}. 
Therefore, using ensemble models is not likely help in fixing such incorrect predictions on slightly transformed inputs.
In fact, we have also tested the online commercial APIs that are expected to adopt ensemble strategies and found that they still suffer from high error rate under transformed inputs (Table~\ref{tab:violations_imagenet}).

\end{appendix}

\end{document}